\documentclass[rmp,aps,twocolumn]{revtex4}

\usepackage{graphics}
\usepackage{graphicx}
\usepackage{bm}
\usepackage{latexsym}
\usepackage{color}
\usepackage{mathrsfs}
\usepackage{braket}
\usepackage{enumerate}
\usepackage[usenames,dvipsnames]{xcolor}
\usepackage[colorlinks=true,citecolor=blue,linkcolor=blue,urlcolor=blue,backref=true]{hyperref}
\usepackage{float}
\usepackage[raggedright,tight]{subfigure}  %side-by-side figures
\usepackage{amsmath}
\usepackage{amsfonts}
\usepackage{braket}
\usepackage[normalem]{ulem}
\usepackage{cancel}
\usepackage{mathtools}
\usepackage{appendix}
\makeatletter
\newsavebox{\@brx}
\newcommand{\llangle}[1][]{\savebox{\@brx}{\(\m@th{#1\langle}\)}%
  \mathopen{\copy\@brx\kern-0.5\wd\@brx\usebox{\@brx}}}
\newcommand{\rrangle}[1][]{\savebox{\@brx}{\(\m@th{#1\rangle}\)}%
  \mathclose{\copy\@brx\kern-0.5\wd\@brx\usebox{\@brx}}}
\makeatother

\begin{document}
%\UseRawInputEncoding

\title{Implications of gauge freedom for nonrelativistic quantum electrodynamics}
\author{Adam Stokes}\email{adamstokes8@gmail.com}
\affiliation{Department of Physics and Astronomy, University of Manchester, Oxford Road, Manchester M13 9PL, United Kingdom}
\author{Ahsan Nazir}\email{ahsan.nazir@manchester.ac.uk}
\affiliation{Department of Physics and Astronomy, University of Manchester, Oxford Road, Manchester M13 9PL, United Kingdom}

%\ead{a.stokes@leeds.ac.uk}

\begin{abstract}
Gauge freedom in quantum electrodynamics (QED) outside of textbook regimes is reviewed. It is emphasized that QED subsystems are defined relative to a choice of gauge. Each definition uses different gauge-invariant observables. This relativity is eliminated only if a sufficient number of Markovian and weak-coupling approximations are employed. All physical predictions are gauge invariant, including subsystem properties such as photon number and entanglement. However, subsystem properties naturally differ for different physical subsystems. Gauge ambiguities arise not because it is unclear how to obtain gauge-invariant predictions, but because it is not always clear which physical observables are the most operationally relevant. The gauge invariance of a prediction is necessary but not sufficient to ensure its operational relevance. It is shown that, in controlling which gauge-invariant observables are used to define a material system, the choice of gauge affects the balance between the material system’s localization and its electromagnetic dressing. Various implications of subsystem gauge relativity for deriving effective models, for describing time-dependent interactions, for photodetection theory, and for describing matter within a cavity are reviewed.
\end{abstract}

\maketitle

\tableofcontents
\makeatletter
\let\toc@pre\relax
\let\toc@post\relax
\makeatother

\section{Introduction}

Traditional regimes of light-matter physics involve relatively small values of a ratio $r$ that compares, in a qualitative sense, the interaction
strength to the energies characterising the bare light and matter subsystems \cite{devoret_circuit-qed_2007}. Over the past two decades however, much more extreme light-matter interaction regimes have become an important topic in both applied and fundamental physics. In the simplest case of a two-level emitter coupled to a single photonic mode, the so-called ultrastrong coupling regime, $r\gtrsim 0.1$, is typically taken as the point at which the rotating-wave approximation certainly breaks down, incurring a departure from Jaynes-Cummings \cite{jaynes_comparison_1963} physics. This regime has now been realised in a relatively large range of experimental platforms [recent reviews include \cite{forn-diaz_ultrastrong_2019,kockum_ultrastrong_2019}]. Even values $r\gtrsim 1$, which define the so-called deepstrong coupling regime have now been realised in both superconducting circuits \cite{yoshihara_superconducting_2017} and via plasmonic nanoparticle crystals \cite{mueller_deep_2020}.

Beyond those systems in which only a few photonic modes dominate, there now exist diverse multi-mode photonic systems in which non-Markovian effects may become significant. These platforms include materials within dielectric and metallic environments, which may be uniform or nanostructured  \cite{ma_engineering_2021}, superconducting circuits coupled to transmission lines \cite{forn-diaz_ultrastrong_2017}, solid-state systems \cite{de_vega_dynamics_2017,nazir_modelling_2016}, and cavity-molecule systems that offer a promising means by which to control chemical processes \cite{hertzog_strong_2019}. Experimental progress in ultrafast light-matter interactions is also continuing steadily. Femto-second laser pulses offer the potential to control bare charges on ultrafast timescales \cite{ciappina_attosecond_2017}, while sub-cycle ultrastrong light-matter interaction switching was achieved some time ago \cite{gunter_sub-cycle_2009}.

Recent reviews \cite{forn-diaz_ultrastrong_2019,kockum_ultrastrong_2019,boite_theoretical_2020} of light-matter physics outside of weak-coupling regimes have focussed on effective models and new theoretical methods, which are required because standard weak-coupling quantum optics cannot be applied. Despite new methods, our understanding continues to be based on processes involving real and virtual bare quanta, which can vary significantly with the form of the model considered. Non-standard regimes where weak-coupling theory breaks down,  are precisely where effective models that are only superficially motivated are liable to fall short. This necessitates an appraisal of the fundamental physics from first principles, as will be the focus of the current article. We focus specifically on the implications of QED's gauge-theoretic aspects.

Gauge freedom in ultrastrong and deepstrong coupling QED has recently been investigated in a number of contexts, including the truncation of a material subsystem to a finite number of energy levels \cite{stokes_gauge_2019,stokes_ultrastrong_2021,stokes_uniqueness_2020,stefano_resolution_2019,roth_optimal_2019,de_bernardis_breakdown_2018,settineri_gauge_2021,de_bernardis_cavity_2018,stokes_gauge_2020,taylor_resolution_2020,garziano_gauge_2020,ashida_cavity_2021}, time-dependent interactions \cite{stokes_ultrastrong_2021,stefano_resolution_2019,settineri_gauge_2021}, Dicke model superradiance \cite{de_bernardis_cavity_2018,stokes_uniqueness_2020,garziano_gauge_2020}, and photodection theory \cite{settineri_gauge_2021}.

Gauge-freedom in QED implies a relativity in the assignment of physical meaning to the vectors and operators that represent states and observables. This is akin to the relativity encountered in theories of space and time. For example, the time interval $\Delta t_X$ between two events $x$ and $y$, as measured by a clock at rest in frame $X$ does not predict the outcome $\Delta t_Y$ of measuring the time between $x$ and $y$ in a co-moving frame $Y$. We have $\Delta t_X\approx \Delta t_Y$ only if the relativistic mixing incurred by the Lorentz transformation from $X$ to $Y$ can be ignored. Otherwise, we must recognise that we have two different predictions, $\Delta t_X$ and $\Delta t_Y$, for two different experiments; one in frame $X$ and one in frame $Y$. We do know however, which prediction corresponds to which experiment, that is, we always know which prediction is {\em relevant}. This is determined by the rest frame of the clock, i.e., it is determined by the apparatus.

In the same way that intervals in space and time can only be defined relative to an inertial frame in Minkowski spacetime, light and matter quantum subsystems can only be defined relative to a gauge-frame in Hilbert space. Unlike in special relativity, where it is straightforward to identify which predictions of space and time intervals are relevant in which situations, in QED there are a number of conceptual subtleties regarding the identification of the most relevant theoretical subsystems. The problem is closely related to the interpretation of virtual processes and particles, an aspect of light-matter physics that already possesses a long history of theoretical studies predominantly confined so far to the weak-coupling regime. Such studies possess significant overlap with the quantum theory of measurement \cite{stokes_extending_2012,drummond_unifying_1987,dalibard_vacuum_1982,passante_cloud_1985,compagno_detection_1988,compagno_dressed_1988,compagno_dressed_1990,compagno_bare_1991,compagno_atom-field_1995} as well as with the identification of local fields and causal signal propagation \cite{fermi_quantum_1932,cohen-tannoudji_photons_1989,biswas_virtual_1990,milonni_photodetection_1995,power_analysis_1997,power_time_1999,power_time_1999-1,sabin_fermi_2011,stokes_noncovariant_2012,buchholz_there_1994}.

The primary purpose of the present article is to identify what gauge ambiguities occur beyond the regimes traditionally considered in quantum optics and to clarify how they arise. In Sec.~\ref{sec2} we begin with a pedagogical introduction to gauge freedom. We then provide a rigorous derivation of arbitrary gauge (nonrelativistic) QED using the principles of modern gauge-field theory, showing that the implications of gauge freedom discussed in Secs.~\ref{minc} onward are a fundamental feature. They are not in any way an artefact of approximations or simplifications. In particular, we emphasize that gauge ambiguities arise not because it is unclear how to obtain gauge invariant predictions, but because it is not always clear which gauge invariant subsystems are {\em operationally relevant}. In Sec.~\ref{minc} we address a number of common pitfalls related to gauge freedom in QED.

In Sec.~\ref{s0} we introduce the notion of {\em subsystem gauge relativity}. We explain its relation to gauge invariance, identify the regimes within which it is important, and discuss its implications. In Sec. \ref{s1} we review theoretical background for the implementation of material level truncations \cite{stokes_gauge_2019,stokes_ultrastrong_2021,stokes_uniqueness_2020,stefano_resolution_2019,roth_optimal_2019,de_bernardis_breakdown_2018,stokes_gauge_2020,taylor_resolution_2020,ashida_cavity_2021}, noting that the resulting gauge noninvariance is prosaic, because it can always be avoided by avoiding the truncation. We review various proposals for obtaining two-level models, along with their varying degree of accuracy in different regimes, as well as their significance for understanding gauge ambiguities.

In Sec. \ref{s6} we discuss time-dependent interactions. We first review the QED $S$-matrix formalism. Here subsystem gauge relativity does not occur due to the condition of adiabatic interaction-switching which implies strict conservation of the bare-energy $h$, where $H=h+V$ is the full Hamiltonian and $V$ is the interaction Hamiltonian. We show directly that conventional weak-coupling and Markovian approximations mimic the $S$-matrix, enforcing the conservation of $h$ and thereby eliminating subsystem gauge relativity. In this sense, these traditional regimes are {\em gauge nonrelativistic}. In contrast, it is shown that when describing non-Markovian and strong-coupling effects subsystem gauge relativity cannot be ignored. 

In Sec. \ref{pd} we consider photodetection theory. We emphasize that gauge ambiguities arise because it is not always clear that any one definition of ``photon" is always the most operationally relevant. For example, the Coulomb gauge definition has recently been preferred in ultrastrong-coupling light-matter physics literature \cite{stefano_resolution_2019,settineri_gauge_2021}. However, as has been known for some time, certain predictions, such as the natural lineshape of spontaneous emission, have  been found to be closer to experiment if photons are defined relative to the multipolar gauge \cite{power_coulomb_1959,fried_vector_1973,davidovich_theory_1980,milonni_natural_1989,woolley_gauge_2000,stokes_gauge_2013}. 

We identify how the definitions of the subsystems, as controlled by the choice of gauge, are related to photodetection divergences \cite{drummond_unifying_1987,stokes_extending_2012}. We determine the relation between subsystem gauge relativity and electromagnetic dressing. We extend these considerations to cavity QED beyond standard regimes, and discuss how subsystem gauge relativity relates to weak measurements of intra-cavity subsystems and to ground state superradiance. We briefly mention outlook for predictions regarding extra-cavity fields. Finally, we summarise in Sec. \ref{conc}.
  
Throughout this paper we use natural units, such that $\hbar=c=\epsilon_0=\mu_0=1$. The elementary electric charge is $e=\sqrt{4\pi\alpha_{\rm fs}}$ where $\alpha_{\rm fs}$ is the fine structure constant. Unless otherwise stated, latin characters $i,j,k,\dots$ denote cartesian components of vectors whenever they appear as subscripts, and we adopt the summation convention for repeated cartesian indices. The imaginary unit is also denoted $i$ (not a subscript). We use the notation ${\dot f}(t)$ as shorthand for the total derivative $df(t)/dt$.

\section{Gauge freedom and gauge fixing}\label{sec2}

Quantum electrodynamics is the underpinning theory that describes all physical interactions occurring from the atomic scale upwards, until gravitation becomes significant. Modern light-matter physics encompasses an extremely broad and diverse range of natural and artificial systems with numerous interactions that span a large parameter space. Dividing composite systems into quantum subsystems that emit, absorb and exchange photons, remains the basic conceptual framework used to understand light-matter physics, but beyond traditional regimes new challenges arise, both conceptual and technical. QED's gauge freedom becomes important because the choice of gauge controls the physical nature of the adopted theoretical quantum subsystem decomposition.

\begin{figure}[t]
\begin{minipage}{\columnwidth}
\begin{center}
\hspace*{-1mm}\includegraphics[scale=0.19]{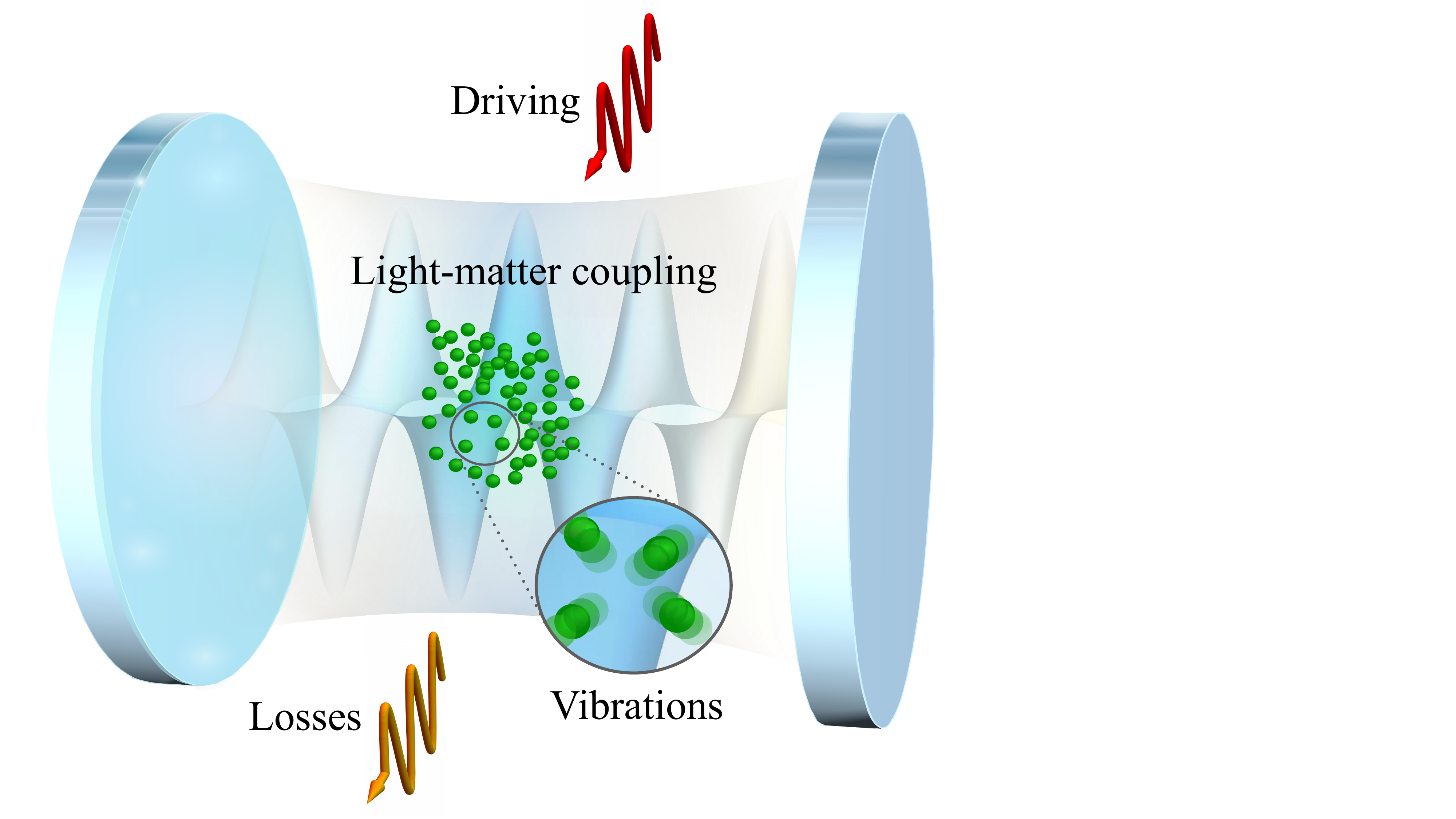}
\vspace*{-4mm}\caption{Material systems, such as atoms or molecules, confined within an electromagnetic cavity, which enhances the light-matter coupling. Internal vibrational interactions may also be strong and non-Markovian. Driving via laser light may take many forms including the use of ultrafast and strong pulses. Losses within such systems may be complex including direct emission to external modes, as well as leakage through the cavity mirrors.}\label{intropic}
\vspace*{-4mm}
\end{center}
\end{minipage}
\end{figure}

Weak-coupling theory will typically breakdown when dealing with complex or artificial systems of the type depicted in Fig.~\ref{intropic}. However, in order to identify and understand the challenges faced in as simple a setting as possible, we begin by revisiting the case of elementary charged particles in free space. Although sound treatments can be found in various textbooks (e.g. Refs.~\cite{cohen-tannoudji_photons_1989,craig_molecular_1998}), the role and significance of gauge freedom is less widely understood and has even been debated recently \cite{rousseau_quantum-optics_2017,vukics_gauge-invariant_2021,andrews_perspective_2018,rousseau_reply_2018}. This motivates a collation of present understanding and the provision of a coherent overview. Sec.~\ref{sum} summarise the results of a rigorous derivation of arbitrary gauge nonrelativistic QED that uses the principles of modern gauge-field theory, with further details given in Supplementary Note~II. We define the gauge principle, gauge freedom, gauge symmetry transformations, gauge fixing transformations, and gauge-invariance. We address conceptual issues and common pitfalls.

\subsection{A single stationary atom in standard gauges}\label{sp}

Consider a single charge $q$ with position ${\bf r}$ bound to a fixed charge $-q$ at the origin ${\bf 0}$ of our chosen inertial frame. The charge and current densities are 
\begin{align}
&\rho({\bf x})=-q\delta({\bf x})+q\delta({\bf x}-{\bf r}),\label{rho11}\\ 
&{\bf J}({\bf x})={q\over 2}[{\dot {\bf r}}\delta({\bf x}-{\bf r})+\delta({\bf x}-{\bf r}){\dot {\bf r}}],\label{J}
\end{align}
such that ${\partial_t  \rho}=-\nabla\cdot {\bf J}$. Note that in quantum theory $[r_i,{\dot r}_j]\neq0$, so the expression for the current must be symmetrised. The above fields together with electric and magnetic fields ${\bf E}$ and ${\bf B}$, exhaustively assign material and electromagnetic properties to each event $x=(t,{\bf x})$ in spacetime. Gauge freedom can be understood as a many-to-one correspondence between auxiliary mathematical objects used to express the theory and the physical observables $\rho,\,{\bf J}, \,{\bf E},\,{\bf B}$. It is hailed by the occurrence of non-dynamical constraints, $\nabla \cdot {\bf B}=0$ and $\nabla \cdot {\bf E}=\rho$, which imply redundancy within the formalism. Scalar and vector potentials $A_0$ and ${\bf A}$ are defined by
\begin{align}
{\bf E}&=-\nabla A_0 - {\partial_t  {\bf A}}\label{elecf},\\
{\bf B}&=\nabla\times {\bf A}\label{magf},
\end{align}
which imply that the homogeneous Maxwell equations, $\nabla \cdot {\bf B}=0$ and ${\partial_t {\bf B}} = -\nabla\times {\bf E}$, are automatically satisfied. The inhomogeneous constraint $\nabla \cdot {\bf E}=\rho$ (Gauss' law) must be imposed within the theory while the remaining inhomogeneous equation is dynamical ${\partial_t  {\bf E}}=\nabla \times {\bf B}-{\bf J}$ (Maxwell-Ampere law). This is an equation of motion that must be produced by any satisfactory Lagrangian or Hamiltonian description. 
\begin{figure}[t]
\begin{minipage}{\columnwidth}
\begin{center}
\hspace*{-0.5mm}\includegraphics[scale=0.13]{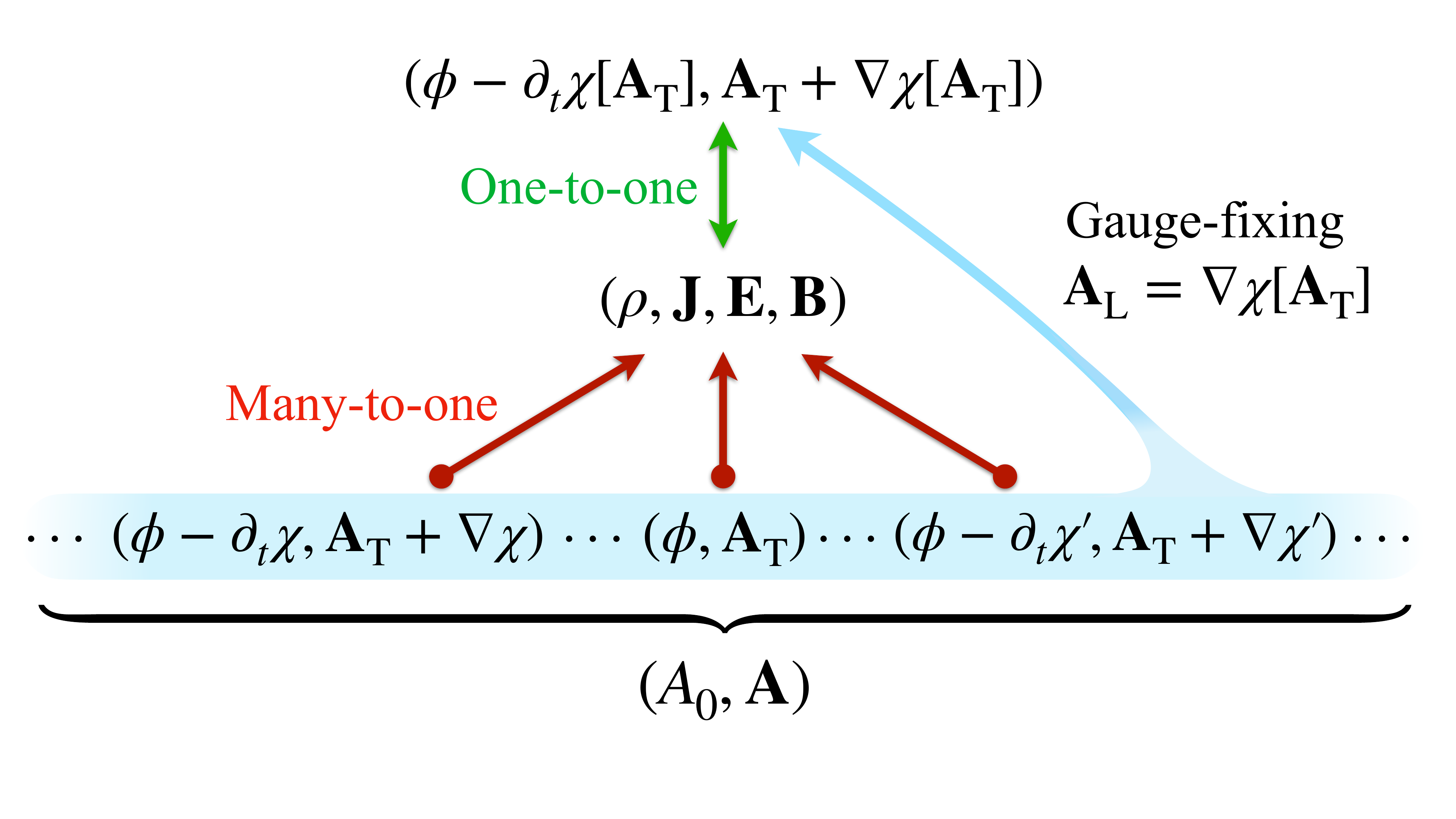}
\vspace*{-9mm}\caption{A schematic representation of gauge-redundancy in electrodynamics. The central potential pair is $(\phi,{\bf A}_{\rm T})$ (Coulomb gauge). The blue (shaded) band represents an uncountable infinity of potential pairs all of which produce the same physical fields, and all of which are related to each other by gauge transformation. Gauge fixing can be achieved by setting the redundancy that causes this many-to-one correspondence, ${\bf A}_{\rm L}$, equal to a known functional, $\chi$, of the fixed and gauge invariant object ${\bf A}_{\rm T}=(\nabla \times)^{-1}{\bf B}$. Afterwards, the map between the chosen fixed potential pair and the physical fields is invertible (one-to-one). The fixed potentials can be written as known functions of $(\rho,\, {\bf J},\, {\bf E},\,{\bf B})$, while ${\bf E}$ and ${\bf B}$ are also known functions of the fixed potentials [Eqs.~(\ref{elecf}) and (\ref{magf})].}\label{gauge_fix}
\end{center}
\end{minipage}
\end{figure}
The electric and magnetic fields are invariant under the {\em gauge transformation}
\begin{align}
{\bf A}'&={\bf A}+\nabla\chi,\label{gt11}\\
A_0'&=A_0-\partial_t\chi\label{gt12}
\end{align}
where $\chi$ is arbitrary.

An unconstrained Hamiltonian description in terms of potentials $A_0$ and ${\bf A}$ requires elimination of gauge-redundancy. Recall that the Helmholtz decomposition of a (square-integrable) vector-field ${\bf V}$ into transverse and longitudinal fields, ${\bf V}={\bf V}_{\rm T}+{\bf V}_{\rm L}$, is unique. The transverse and longitudinal components satisfy $\nabla\cdot {\bf V}_{\rm T}=0$ and $\nabla\times {\bf V}_{\rm L}={\bf 0}$. Transverse and longitudinal $\delta$-functions (dyadics) are defined by the non-local conditions
\begin{align}\label{tldelt}
{\bf V}_{\rm L,T}({\bf x}) = \int d^3 x' \, \delta^{\rm L,T}({\bf x}-{\bf x}')\cdot {\bf V}({\bf x}').
\end{align}
The process of {\em gauge fixing} eliminates the mathematical redundancy within the formalism by specifying all freely choosable objects as known functions of objects that cannot be freely chosen (Fig.~\ref{gauge_fix}). Since the curl of the gradient is identically zero, the transverse vector potential ${\bf A}_{\rm T}$ is gauge invariant, that is, if ${\bf A}'={\bf A}+\nabla\chi$ then ${\bf A}'_{\rm T}\equiv {\bf A}_{\rm T}$, which cannot be freely chosen. Gauge freedom is therefore the freedom to choose the longitudinal vector potential ${\bf A}_{\rm L}=\nabla \chi$ where ${\bf A}={\bf A}_{\rm T}+\nabla \chi$. In Supplementary Note~I, this gauge-freedom is related to the $U(1)$-phase of material wave-functions and electromagnetic wave-functionals.

One of the most commonly chosen gauges is the Coulomb gauge defined by the choice ${\bf A}_{\rm L}={\bf 0}$, such that ${\bf A}={\bf A}_{\rm T}$. From Gauss' law $\nabla \cdot {\bf E}=\rho$ and Eq.~(\ref{elecf}) it follows that in the Coulomb gauge the scalar potential $A_0$ coincides with the Coulomb potential defined by
\begin{align}\label{sccoul}
\phi({\bf x}) = -\nabla^{-2}\rho({\bf x}) = \int d^3 x' {\rho({\bf x'})\over 4\pi|{\bf x}-{\bf x}'|}
\end{align}
where the kernel $1/(4\pi|{\bf x}|)$ is the Green's function for the Laplacian; $\nabla^2[1/(4\pi|{\bf x}|)]=-\delta({\bf x})$. Specifying ${\bf A}={\bf A}_{\rm T}$ and $A_0 =\phi$ is an example of gauge fixing.

The other commonly chosen gauge in nonrelativistic electrodynamics is the Poincar\'e (multipolar) gauge defined by ${\bf x}\cdot {\bf A}({\bf x})=0$. This is the Coulomb gauge condition applied in reciprocal space. More generally, we may define the arbitrary-gauge potential
\begin{align}
{\bf A}_\alpha({\bf x}) = {\bf A}_{\rm T}({\bf x}) -\alpha\nabla \int_0^1 d\lambda\, {\bf x} \cdot {\bf A}_{\rm T}(\lambda {\bf x})\label{aa}
\end{align}
where the value of $\alpha$ selects the gauge by specifying ${\bf A}_{\rm L}$. The Coulomb and multipolar gauges are now simply special cases given by $\alpha=0$ and $\alpha=1$ respectively \cite{stokes_gauge_2019,stokes_ultrastrong_2021,stokes_uniqueness_2020}. Eq.~(\ref{aa}) can be written
\begin{align}
{\bf A}_\alpha = {\bf A}_{\rm T} + \nabla \chi_\alpha \label{aa2}
\end{align}
where
\begin{align}
&\chi_\alpha({\bf x})= \int d^3 x'\, {\bf g}_{\rm T\alpha}({\bf x'},{\bf x}) \cdot {\bf A}_{\rm T}({\bf x}')\label{chialph} \\
&{\bf g}_{\rm T\alpha}({\bf x}',{\bf x}) =-\alpha \int_0^1 d\lambda \,{\bf x}\cdot \delta^{\rm T}({\bf x}'-\lambda{\bf x}).\label{gal}
\end{align}
For each value of $\alpha$, all freely choosable objects are known functions of objects that cannot be freely chosen. More precisely, the theory has been expressed entirely in terms of ${\bf A}_{\rm T}$ (Fig.~\ref{gauge_fix}), which serves as an elementary dynamical coordinate. Different values of $\alpha$ provide different choices of ${\bf A}_{\rm L}$ as different fixed functionals of the coordinate ${\bf A}_{\rm T}$. Parametrisation via $\alpha$ in this way obviously does not exhaust all possible gauge choices. It does, however, allow us to provide clear definitions of {\em gauge invariance} and {\em gauge relativity}, as will be done in Sec.~\ref{gamb}. In Supplementary Note~II we provide a more general encoding of gauge freedom and the results are summarised in Sec.~\ref{sum}.

It is useful to define the polarisation field ${\bf P}$ by the equation $-\nabla\cdot {\bf P}= \rho$, which specifies ${\bf P}_{\rm L}$ uniquely, but leaves ${\bf P}_{\rm T}$ an essentially arbitrary transverse field. We are free to define the field ${\bf P}_\alpha:={\bf P}_{\rm L}+{\bf P}_{\rm T\alpha}$ where ${\bf P}_{\rm T\alpha}$ is called the $\alpha$-gauge transverse polarisation defined by the condition
\begin{align}\label{ptchi}
\int d^3 x\, \rho({\bf x})\chi_\alpha({\bf x}) =  -\int d^3 x\,  {\bf P}_{\rm T\alpha}({\bf x})\cdot {\bf A}_{\rm T}({\bf x}).
\end{align}
It follows from Eqs.~(\ref{chialph}) and (\ref{gal}) that we may set
\begin{align}
{\bf P}_{\rm T\alpha}({\bf x})&=-\int d^3 x' \,{\bf g}_{\rm T\alpha}({\bf x},{\bf x}')\rho({\bf x}') \nonumber \\ & =\alpha q \int_0^1 d\lambda\, {\bf r}\cdot \delta^{\rm T}({\bf x}-\lambda{\bf r})=\alpha {\bf P}_{\rm T}({\bf x}),
\label{poltalph}
\end{align}
where ${\bf P}_{\rm T}:={\bf P}_{\rm T1}$ is the multipolar transverse polarisation. According to these definitions, in the Coulomb gauge we have ${\bf P}_{\rm T0}={\bf 0}$ and therefore ${\bf P}_0={\bf P}_{\rm L}$. In the multipolar gauge we have
\begin{align}\label{mulp}
{\bf P}_1({\bf x}):= {\bf P}_{\rm T1}({\bf x})+{\bf P}_{\rm L}({\bf x})=q\int_0^1 d\lambda\,{\bf r} \delta({\bf x}-\lambda{\bf r}).
\end{align}
This field specifies a straight line of singular dipole moment density, that stretches from the charge $-q$ at ${\bf 0}$ to the dynamical charge $q$ at ${\bf r}$. 

We now provide a canonical (Hamiltonian) quantum description. Typically this would be derived from a suitable Lagrangian and the gauge would be fixed from the outset. However, our only requirement is that the theory produces the correct Maxwell-Lorentz system of equations and it can therefore be obtained through a series of ansatzes. A rigorous and more general derivation of arbitrary-gauge QED is given using modern gauge-field theory in Supplementary Notes~II-IV.

We proceed by writing down the total energy of the system as the sum of kinetic and electromagnetic energies;
\begin{align}\label{enH}
E =&{1\over 2}m{\dot {\bf r}}^2 +{1\over 2}\int d^3 x \left({\bf E}^2+{\bf B}^2\right)\nonumber \\
=&{1\over 2}m{\dot {\bf r}}^2 + U({\bf r}) + V_{\rm self}({\bf r}) +{1\over 2}\int d^3 x \left({\bf E}_{\rm T}^2+{\bf B}^2\right),
\end{align}
where ${\bf E}_{\rm T}=-{\partial_t  {\bf A}}_{\rm T}$ and
\begin{align}\label{coulpots}
U({\bf r}) +V_{\rm self} = {1\over 2}\int d^3 x\, {\bf E}_{\rm L}^2 \equiv {1\over 2}\int d^3 x\, {\bf P}_{\rm L}^2.
\end{align}
Here $U({\bf r})=-q^2/(4\pi|{\bf r}|)$ is the Coulomb energy binding the charges $q$ and $-q$, while $V_{\rm self}$ is the sum of the infinite Coulomb self-energies of each individual charge. Eq.~(\ref{coulpots}) is obtained by solving Gauss' law $\nabla\cdot {\bf E}=\rho$, which yields ${\bf E}_{\rm L}=-{\bf P}_{\rm L}=-\nabla\phi$ with $\phi$ defined in Eq.~(\ref{sccoul}).

The canonical operators ${\bf y}=\{{\bf r},\,{\bf A}_{\rm T},\,{\bf p},\,{\bf \Pi}\}$ in terms of which we will express the theory satisfy
\begin{align}
[r_i,p_j]&=i\delta_{ij}\label{co1},\\
[A_{{\rm T},i}({\bf x}),\Pi_{{\rm T},j}({\bf x}')]&=i\delta_{ij}^{\rm T}({\bf x}-{\bf x}')\label{co2}
\end{align}
while all other commutators between canonical operators vanish. Here we will assume these commutators and show that they yield the correct result. A systematic derivation is given in   Supplementary Note~IX. Since energy generates translations in time, the Hamiltonian that we seek must equal the total energy expressed in terms of the canonical operators; $H({\bf y})=E$. Given this constraint we must now make suitable ansatzes for the velocities ${\dot {\bf r}}\equiv {\dot {\bf r}}({\bf y})$ and ${\partial_t  {\bf A}}_{\rm T}\equiv {\partial_t  {\bf A}}_{\rm T}({\bf y})$. We require that upon substitution into the right-hand-side of Eq.~(\ref{enH}) our ansatzes define a Hamiltonian $H({\bf y})$, for which the Heisenberg equation, $\partial_tO=-i[O,H]$, together with Eqs.~(\ref{co1}) and (\ref{co2}) yields the correct Maxwell-Lorentz equations.

Since we wish to provide a Hamiltonian description in an arbitrary gauge we make the arbitrary-gauge minimal coupling ansatzes
\begin{align}
m{\dot {\bf r}} &= {\bf p}-q{\bf A}_\alpha({\bf r}),\label{min0}\\
{\partial_t  {\bf A}}_{\rm T} &= {\bf \Pi}+{\bf P}_{\rm T\alpha}.\label{min0b}
\end{align}
Note that minimal coupling is not synonymous with any one gauge, and in particular it is not synonymous with the Coulomb gauge despite that the Coulomb gauge Hamiltonian is often called the minimal coupling Hamiltonian. This point is discussed in more detail in Sec.~\ref{minc2}. From Eqs.~(\ref{enH}), (\ref{min0}), and (\ref{min0b}) we obtain
\begin{align}\label{HA}
E=&{1\over 2m}\left[{\bf p}-q{\bf A}_\alpha({\bf r})\right]^2 + U({\bf r}) + V_{\rm self} \nonumber \\ &+{1\over 2}\int d^3 x \left[({\bf \Pi}+{\bf P}_{\rm T\alpha})^2+(\nabla\times {\bf A}_{\rm T})^2\right]=:H_\alpha({\bf y}).
\end{align}
This defines the arbitrary gauge Hamiltonian $H_\alpha$, which coincides with the one derived in Refs. \cite{stokes_gauge_2019,stokes_ultrastrong_2021,stokes_uniqueness_2020}. The canonical commutation relation (CCR) algebra, Eqs.~(\ref{co1}) and (\ref{co2}), yield
\begin{align}
{\bf p}-q{\bf A}_\alpha({\bf r})& = -im\left[{\bf r},H_\alpha\right]\label{sc1a},\\
{\bf \Pi}({\bf x})+{\bf P}_{\rm T\alpha}({\bf x}) &= -i[{\bf A}_{\rm T}({\bf x}),H_\alpha]\label{sc2}.
\end{align}
This shows that the ansatzes in Eqs.~(\ref{min0}) and (\ref{min0b}) are self-consistent, because they are re-obtained using the Heisenberg equation. It is a straightforward exercise to verify that $H_\alpha$ does indeed yield the correct Maxwell-Lorentz system of equations for {\em any} choice of gauge $\alpha$.

It is readily verified that Hamiltonians of different fixed gauges $\alpha$ and $\alpha'$ are unitarily equivalent;
\begin{align}\label{halp}
H_{\alpha'} = R_{\alpha\alpha'}H_{\alpha}R_{\alpha\alpha'}^\dagger
\end{align}
where $R_{\alpha\alpha'}$ is called a {\em gauge fixing transformation} and is defined by \cite{stokes_gauge_2019,lenz_quantum_1994,stokes_noncovariant_2012,chernyak_gauge_1995,stokes_uniqueness_2020}
\begin{align}\label{pzw}
R_{\alpha\alpha'} &:= \exp \left(i\int d^3 x \, \left[{\bf P}_{\rm T\alpha}({\bf x})-{\bf P}_{\rm T\alpha'}({\bf x})\right] \cdot {\bf A}_{\rm T}({\bf x})\right) \nonumber \\ &=\exp\left(-iq[\chi_\alpha({\bf r})-\chi_{\alpha'}({\bf r})]\right)
\end{align}
in which the second equality follows from Eq.~(\ref{ptchi}). We emphasize that the {\em definition} of gauge freedom continues to be the freedom to choose $\alpha$, which specifies ${\bf A}_{\rm L}$. It therefore constitutes the freedom to transform between distinct minimal coupling prescriptions as
\begin{align}
&R_{\alpha\alpha'}\left[{\bf p}-q{\bf A}_\alpha({\bf r})\right]R_{\alpha\alpha'}^\dagger = {\bf p}-q{\bf A}_{\alpha'}({\bf r})\label{mina}\\
&R_{\alpha\alpha'}\left({\bf \Pi}+{\bf P}_{\rm T\alpha}\right)R_{\alpha\alpha'}^\dagger ={\bf \Pi}+{\bf P}_{\rm T\alpha'}\label{minb}
\end{align}
from which Eq.~(\ref{halp}) follows. The effect of the transformation has been the replacement $({\bf A}_\alpha,{\bf P}_{\rm T\alpha}) \to ({\bf A}_{\alpha'},{\bf P}_{\rm T\alpha'})$, which clearly constitutes a gauge transformation from the fixed gauge $\alpha$ to the fixed gauge $\alpha'$. The reason Eq.~(\ref{minb}) occurs is that in Eq.~(\ref{aa}) we chose to fix the gauge ${\bf A}_{\rm L}$ as a functional of ${\bf A}_{\rm T}$, which generates translations in ${\bf \Pi}$. The gauge freedom inherent in the polarisation is discussed further in Supplementary Note~I. 2. Note that since $Uf(O)U^\dagger=f(UOU^\dagger)$ for any unitary transformation $U$, suitably well-defined function $f$, and operator $O$, Eqs.~(\ref{mina}) and (\ref{minb}) are necessary and sufficient to define how arbitrary functions of the canonical operators transform under a gauge transformation. 

We remark that in order to implement the gauge transformation ${\bf p}-q{\bf A}({\bf r}) \to {\bf p}-q[{\bf A}({\bf r})+\nabla \chi({\bf r})]$ the canonical momentum must transform as $e^{iq\chi({\bf r})}{\bf p}e^{-iq\chi({\bf r})}={\bf p} - q\nabla \chi({\bf r})$, which states that ${\bf r}$ generates translations in ${\bf p}$. This property relies upon the canonical commutation relation in Eq.~(\ref{co1}). Eq.~(\ref{mina}) in particular, features the gauge fixing transformation $R_{\alpha\alpha'}=e^{-iq[\chi_\alpha({\bf r})-\chi_{\alpha'}({\bf r})]}$.   As recognised early by Weyl \cite{weyl_quantenmechanik_1927},  the CCR algebra cannot be supported by a finite-dimensional Hilbert space. Thus, retaining only a finite number of material energy levels will ruin gauge invariance. Material truncation is discussed in detail in Sec.~\ref{s1}.

\subsection{Electric dipole approximation}\label{eda}

The electric-dipole approximation (EDA) of the theory presented in Sec.~\ref{sp} can be performed preserving all kinematic and algebraic relations of the theory such that gauge invariance is also preserved. We define the Fourier transform of a field $f$ by ${\tilde f}({\bf k}) := \int d^3x f({\bf x})e^{-i{\bf k}\cdot {\bf x}}/\sqrt{(2\pi)^3}$. Considering the charge and current densities in Eqs.~(\ref{rho11}) and (\ref{J}), the EDA (also known as the long-wavelength approximation) is defined by retaining only the leading contributions after performing the expansion $e^{-i{\bf k}\cdot {\bf r}} = 1-i{\bf k}\cdot {\bf r}+...$. This gives ${\tilde \rho}({\bf k}) \approx -iq{\bf k}\cdot {\bf r}/\sqrt{(2\pi)^3}$, $\rho({\bf x}) \approx -q{\bf r}\cdot \nabla \delta({\bf x})$, ${\bf J}({\bf x})\approx q{\dot {\bf r}}\delta({\bf x})$, and ${\bf A}_{\rm T}({\bf r}) \approx {\bf A}_{\rm T}({\bf 0})$, and in turn
\begin{align}
&P_{{\rm T}\alpha,i}({\bf x})\approx\alpha qr_j\delta^{\rm T}_{ij}({\bf x}), \label{PTeda}\\
&\chi_\alpha({\bf r}) \approx -\alpha{\bf r}\cdot {\bf A}_{\rm T}({\bf 0}),\label{chieda}\\
&{\bf A}_\alpha({\bf r}) \approx (1-\alpha){\bf A}_{\rm T}({\bf 0}).\label{Aeda}
\end{align}
When these approximate equalities are substituted into Eq.~(\ref{HA}) the $\alpha$-gauge Hamiltonian in the EDA is obtained. Similarly, the unitary gauge fixing transformation $R_{\alpha\alpha'}$ in Eq.~(\ref{pzw}) becomes
\begin{align}\label{Rgteda}
R_{\alpha\alpha'}=\exp\left[i(\alpha-\alpha')q{\bf r}\cdot {\bf A}_{\rm T}({\bf 0})\right].
\end{align}
Since unitarity is preserved, so too is gauge invariance [see Sec.~\ref{gamb}]. Hamiltonians belonging to different gauges continue to be unitarily equivalent as in Eq.~(\ref{halp}).

Certain (non-fundamental) properties hold within (and only within) the EDA \cite{stokes_gauge_2020}. In particular, the gauge function $\chi_\alpha$ in Eq.~(\ref{chialph}) becomes that in Eq.~(\ref{chieda}), which gives $\nabla \chi_1({\bf r}) = -{\bf A}_{\rm T}({\bf 0})$, such that ${\bf p}-q{\bf A}_1({\bf r}) \approx {\bf p}$. Thus, letting $\alpha=1$ on the left-hand-side of Eq.~(\ref{mina}) we obtain $R_{1\alpha}{\bf p}R_{1\alpha}^\dagger 
={\bf p}-q{\bf A}_\alpha$ where ${\bf A}_\alpha := (1-\alpha){\bf A}_{\rm T}({\bf 0})$ is the EDA of ${\bf A}_\alpha({\bf r})$. Within the full 3-dimensional setting and without the EDA this is impossible, because for any differentiable function $f$ we have $e^{-if({\bf r})}{\bf p}e^{if({\bf r})} = {\bf p}+\nabla f({\bf r})$. The gradient $\nabla f$ is a longitudinal field, such that we cannot have $\nabla f({\bf r}) =-q{\bf A}({\bf r})$ for all ${\bf r}$, because ${\bf A}_{\rm T}({\bf r})$ is non-vanishing. The gauge transformation $e^{iqf({\bf r})}[{\bf p}-q{\bf A}({\bf r})]e^{-iqf({\bf r})} = {\bf p}-q[{\bf A}+\nabla f({\bf r})]$ is fundamental and yields the result $R_{1\alpha}{\bf p}R_{1\alpha}^\dagger 
={\bf p}-q{\bf A}_\alpha$ as an approximate special case in which we let $f=\chi_\alpha-\chi_1$, and perform the EDA.

\subsection{Generalisations and the gauge principle}\label{sum}

Modern gauge-field theories are understood to result from the {\em gauge principle} applied to a material field $\psi$. The principle states that
\begin{itemize}
\item{The form of electromagnetic and other interactions should be invariant under the local action of  a group ${\cal G}$ on the matter field $\psi$, written $\psi'(x)=g(x)\cdot \psi(x)$. In QED ${\cal G}=U(1)$ and $\psi'(x) = e^{iq\chi(x)}\psi(x)$ where $\chi$ is arbitrary.}
\end{itemize}
In   Supplementary Note~I we review how gauge invariance can be understood as $U(1)$-phase invariance. In Supplementary Notes~II-V we provide a  general derivation of arbitrary gauge nonrelativistic QED using the principles of modern gauge-field theory. The main results are summarised below. A sufficiently general expression of the theory that is suitable for our purposes results from encoding gauge freedom into the arbitrary transverse component, ${\bf g}_{\rm T}$, of the Green's function ${\bf g}$ for the divergence operator;
\begin{align}\label{Gg}
&\nabla \cdot {\bf g}({\bf x},{\bf x}') \equiv \nabla \cdot {\bf g}_{\rm L}({\bf x},{\bf x}') = \delta({\bf x}-{\bf x}'),\\
&{\bf g}_{\rm L}({\bf x},{\bf x}')=-\nabla{1\over 4\pi|{\bf x}-{\bf x}'|}
\end{align}
such that ${\bf g}_{\rm T}={\bf g}-{\bf g}_{\rm L}$ is arbitrary. 

We refer to the gauge specified by ${\bf g}_{\rm T}$ as the gauge $g$. The associated vector potential and polarisation are 
\begin{align}
{\bf A}_g({\bf x}) &= {\bf A}_{\rm T}({\bf x})+\nabla\int d^3x' {\bf g}({\bf x}',{\bf x})\cdot{\bf A}_{\rm T}({\bf x}')\nonumber \\ 
&= {\bf A}_{\rm T}({\bf x})+\nabla\chi_g({\bf x},[{\bf A}_{\rm T}]),\\
{\bf P}_{g}({\bf x}) &= -\int d^3 x' {\bf g}({\bf x},{\bf x}')\rho({\bf x}')\label{PTg0}
\end{align}
where 
\begin{align}\label{chi0}
\chi_g({\bf x},[{\bf A}_{\rm T}]) = \int d^3x' \,{\bf g}({\bf x}',{\bf x})\cdot{\bf A}_{\rm T}({\bf x}').
\end{align}
The Hamiltonian in gauge $g$ is
\begin{align}\label{hggg0}
H_g=H({\bf g}_{\rm T})=&{1\over 2m}\left[{\bf p}-q{\bf A}_g({\bf r})\right]^2 + U({\bf r}) + V_{\rm self} \nonumber \\ &+{1\over 2}\int d^3 x \left[({\bf \Pi}+{\bf P}_{\rm Tg})^2+(\nabla\times {\bf A}_{\rm T})^2\right]
\end{align}
Hamiltonians $H_g$ and $H_g'$ are unitarily related by
\begin{align}\label{hgg0}
H_{g'}=U_{gg'}H_g U_{gg'}^\dagger
\end{align}
where
\begin{align}\label{ugg'0}
U_{gg'} &:= \exp\left(-i\int d^3x \, \big[\chi_g({\bf x},{\hat {\bf A}}_{\rm T})-\chi_{g'}({\bf x},{\hat {\bf A}}_{\rm T})\big]\rho({\bf x}) \right) \nonumber \\ 
&=\exp\left(i\int d^3x \, \big[{\bf P}_g({\bf x})-{\bf P}_{g'}({\bf x})\big]\cdot {\bf A}_{\rm T}({\bf x})\right)
\end{align}
is a unitary {\em gauge fixing transformation} from gauge $g$ to gauge $g'$.

The theory is simplified by restricting ${\bf g}_{\rm T}$ as in Eq.~(\ref{gal}) in terms of the gauge-parameter $\alpha$. The theory remains exact but reduces to the form presented in Sec.~\ref{sp} (see Supplementary Note~V for details). Gauge freedom becomes the freedom to choose the parameter $\alpha$ which specifies ${\bf P}_{\rm T\alpha}$ and ${\bf A}_\alpha$ as in Eqs.~(\ref{poltalph}) and (\ref{aa}) respectively \cite{stokes_gauge_2019,stokes_ultrastrong_2021,stokes_uniqueness_2020}. The Hamiltonian $H_g$ in Eq. (\ref{hggg0}) becomes $H_\alpha$ given in Eq.~(\ref{HA}) and the gauge fixing transformation $U_{gg'}$ in Eq.~(\ref{ugg'0}) becomes $R_{\alpha\alpha'}$ in Eq.~(\ref{pzw}). Hamiltonians belonging to different gauges are unitarily related as in Eq.~(\ref{halp}).

Finally we remark that the primary use of nonrelativistic QED lies in describing collections of charges partitioned into certain groups that we call atoms and molecules. The formalism above describes a single hydrogen atom in which the positive charge $-q$ is assumed fixed (non-dynamical). This is equivalent to describing the system using relative and centre-of-mass coordinates instead of the charge coordinates themselves, and assuming that the centre-of-mass is fixed, all centre-of-mass couplings being ignored. In   Supplementary Note~VI, we provide the extension of this formalism to arbitrary charge distributions in the vicinity of fixed molecular centres \cite{craig_molecular_1998}. In Supplementary Note~VII we review the extension to linear dispersing and absorbing (macroscopic) dielectric media, which is a valuable tool in describing cavity QED systems \cite{knoll_resonators_1991,gruner_green-function_1996,dung_three-dimensional_1998,knoll_QED_2003,khanbekyan_qed_2005,viviescas_field_2003} (see also~Supplementary Note~XVII). Concerning further extensions, we note that Ref.~\cite{wei_quantization_2009} considers an anisotropic medium, and Ref.~\cite{ judge_canonical_2013} considers a linear magnetoelectric medium. Finally we note that the use of the above formalism in providing a microscopic description of electrons in crystal lattices is given in Supplementary Note~VIII.

\subsection{Physical nature of the gauge function ${\bf g}_{\rm T}$}\label{natgt}

We now seek to understand the ways in which different fixed-gauge formulations of QED differ. The gauge is selected by choosing $\chi_g$. If $\chi_g$ is restricted in form as in Eq.~(\ref{chi0}) then the gauge is selected by choosing a concrete transverse function ${\bf g}_{\rm T}$. The gauge choice directly specifies two basic quantities, ${\bf A}_g$ and ${\bf P}_{{\rm T}g}$. This in turn specifies the physical nature of the canonical momenta ${\bf p}$ and ${\bf \Pi}$, which together with ${\bf r}$ and ${\bf A}_{\rm T}$ define the quantum subsystems conventionally termed ``matter" and ``light". The importance of this fact will be described in detail throughout sections~\ref{share}-\ref{rel}. 

\subsubsection{Path-dependent solution}

The Green's function ${\bf g}$ is defined by Eq.~(\ref{Gg}). We have seen that the two most commonly chosen gauges of nonrelativistic QED can be linearly interpolated between via a parameter $\alpha$, with $\alpha=1$ specifying the multipolar gauge that possesses a straight-line of singular polarisation stretching between the charges. This is a special case of the following more general  path- and origin-dependent solution discussed by Woolley \cite{woolley_gauge_1998,woolley_power-zienau-woolley_2020}
\begin{align}\label{dipo}
{\bf g}({\bf x},{\bf x}') = {\bf g}_{\rm L}({\bf x},{\bf o}) - \int_{C({\bf o},{\bf x}')} d{\bf z}\, \delta({\bf z}-{\bf x})
\end{align}
where $C({\bf o},{\bf x}')$ is any curve starting at the arbitrary origin ${\bf o}$ and ending at ${\bf x}'$. Verification of the solution is most easily achieved in Fourier space whereby Eq.~(\ref{Gg}) becomes $i{\bf k}\cdot {\tilde {\bf g}}({\bf k},{\bf x}) = e^{-i{\bf k}\cdot {\bf x}}/\sqrt{(2\pi)^3}$. Using Eq.~(\ref{dipo}) we obtain independent of $C$ and ${\bf o}$
\begin{align}\label{dipo2}
\sqrt{(2\pi)^3}i{\bf k}\cdot {\tilde {\bf g}}({\bf k},{\bf x}) = e^{-i{\bf k}\cdot{\bf o}} -i \int_{\bf k\cdot o}^{\bf k \cdot  x} du\, e^{-iu} = e^{-i{\bf k}\cdot {\bf x}}
\end{align}
as required. 

Substituting Eq.~(\ref{Gg}) into Eq.~(\ref{PTg0}) we obtain the $g$-gauge polarisation field
\begin{align}
{\bf P}_g({\bf x}) =- Q{\bf g}_{\rm L}({\bf x},{\bf o}) + \int d^3 x' \int_{C({\bf o},{\bf x}')} d{\bf z}\,\delta({\bf z}-{\bf x})\rho({\bf x}')
\end{align}
where the first term vanishes for a globally neutral system defined by $Q=\int d^3 x \rho({\bf x}) = 0$. An important class of solutions is given by the straight line $C({\bf o},{\bf x}') = \{{\bf z}(\sigma)={\bf x}' + \sigma {\hat {\bf n}}:  \, {\bf z}(\sigma_0)={\bf o}\}$, which starts at the origin ${\bf o}$ specified by value $\sigma_0$, is directed along ${\hat {\bf n}}=({\bf z}(\sigma)-{\bf x}')/\sigma$, and ends at ${\bf x}'$. For example, if we choose the origin ${\bf o}$ as the coordinate origin ${\bf 0}$, which in Eq.~(\ref{rho11}) is the position of the charge $-q$, then the associated polarisation field is
\begin{align}\label{polep}
{\bf P}({\bf x}) &= -q\int^0_{\sigma_0} d\sigma\,{{\bf r}\over \sigma_0} \delta\left({\bf x}-{\bf r}\left[1-{\sigma \over \sigma_0}\right]\right)\nonumber \\
&=q\int_0^1 d\lambda\, {\bf r} \delta\left({\bf x}-\lambda{\bf r}\right),
\end{align}
which we recognise as the multipolar gauge polarisation. Using Eq.~(\ref{ugg'0}) we can also express the Power-Zienau-Woolley transformation as
\begin{align}
U_{01} = \exp\left[-i q\Lambda_C \right]
\end{align}
where
\begin{align}
 \Lambda_C := \int_{C({\bf 0},{\bf r})} d{\bf z} \cdot {\bf A}_{\rm T}({\bf z})
\end{align}
is a Wilson line operator \cite{wilson_confinement_1974}. This expression provides an analogy with quark confinement \cite{woolley_power-zienau-woolley_2020}. Specifically, in Ref.~\cite{woolley_power-zienau-woolley_2020}, Woolley finds that for the multipolar gauge choice of path, i.e., for the straight-line path between two charges at ${\bf r}_1$ and ${\bf r}_2$, the polarisation energy 
\begin{align}
E_{\rm P} = {1\over 2}\int d^3x \,{\bf P}_g({\bf x})^2 = {q\over 2}\int_{{\bf r}_1}^{{\bf r}_2} d{\bf z}\cdot {\bf P}_g({\bf z})
\end{align}
possesses a contribution that increases with increasing separation. Analogously, in a state involving $e^{-iq \Lambda_C}$ as a phase factor, the electric-field energy of two oppositely charged quarks increases linearly with separation, which is interpreted as the cause of confinement. The energy $E_P$ also includes a $\delta$-function contribution and a term that diverges as $1/a$, where $a \to 0$ specifies the point charge limit \cite{woolley_power-zienau-woolley_2020}. We note that the polarisation in Eq.~(\ref{polep}) does not require specifying an arbitrary fixed centre of the charge distribution. As we describe briefly in Supplementary Note~VI. 2, it is possible to extend this treatment to arbitrary numbers of charges and this provides a description of atoms and molecules that unlike conventional molecular QED (see, for example, \cite{craig_molecular_1998}), does not depend on (arbitrary) fixed molecular centres.

Following arguments due to Belinfante \cite{belinfante_consequences_1962}, the solution in Eq.~(\ref{Gg}) can also be used to provide a novel derivation of the Coulomb gauge polarisation ${\bf P}_{\rm L} = -{\bf E}_{\rm L}$. For concreteness, we again consider the atomic charge density in Eq.~(\ref{rho11}). We consider the straight line $C({
\bf o},{\bf x})$ and choose the origin ${\bf o}$ as a point at spatial infinity, which yields the polarisation
\begin{align}\label{Pinf}
{\bf P}({\bf x}) =-q\int_{-\infty}^0 d\sigma \,{\hat {\bf n}}\left[\delta(\sigma{\hat {\bf n}}-{\bf x})-\delta({\bf r}+\sigma{\hat {\bf n}}-{\bf x})\right].
\end{align}
Letting ${\bf y}=-\sigma {\hat {\bf n}}$ with $|{\bf y}|=-\sigma$, and expressing the associated volume element as $d^3 y = dy d\Omega |{\bf y}|^2$, the average of Eq.~(\ref{Pinf}) over all directions ${\hat {\bf n}}$ is
\begin{align}\label{Pinf2}
\int {d\Omega\over 4\pi}\, {\bf P}({\bf x}) &= -{q\over 4\pi}\int d^3 y {{\bf y}\over |{\bf y}|^3}\left[\delta({\bf y}+{\bf x})-\delta({\bf y}-{\bf r}+{\bf x})\right] \nonumber \\ &= {q\over 4\pi}\left[{{\bf x}\over |{\bf x}|^3}-{{\bf x}-{\bf r}\over |{\bf x}-{\bf r}|^3}\right] = -{\bf E}_{\rm L}({\bf x}).
\end{align}
We see therefore, that the Coulomb gauge specifies a delocalised polarisation, in which polarisations localised along the straight line with direction ${\hat {\bf n}}$ stretching between the charges and spatial infinity, are then averaged over {\em all} directions ${\hat {\bf n}}$. 

Quite generally, the solution in Eq.~(\ref{dipo}) suggests an interpretation of the paths on which the polarisation field is localised as ``lines of force" in the sense of Faraday \cite{faraday_experimental_1846}. For a single charge at position ${\bf r}$, Dirac has interpreted the path $C({\bf o},{\bf r})$ as a single line of force between the charge and the origin ${\bf o}$ \cite{dirac_gauge-invariant_1955}. It has been suggested that a novel QED might be constructed in which the paths on which the polarisation field is localised are themselves taken as the dynamical variables of the theory. A suitable averaging procedure over all paths would be required to eliminate the dependence on any particular choice of path \cite{dirac_gauge-invariant_1955,woolley_power-zienau-woolley_2020}.

 \subsubsection{Fourier transform}\label{FT}

To further understand the significance of the freedom to choose ${\bf g}_{\rm T}$ it is convenient to introduce the unconstrained function ${\bf G}$, which is essentially completely arbitrary, as 
\begin{align}
{\tilde {\bf g}}_{\rm T}({\bf k},{\bf x}) = \sum_\sigma {\bf e}_\sigma({\bf k})[{\bf e}_\sigma({\bf k})\cdot {\tilde {\bf G}}({\bf k},{\bf x})]
\end{align}
where ${\bf e}_\sigma({\bf k}),~\sigma=1,2$ are orthonormal vectors spanning the plane orthogonal to ${\bf k}$. Restricting our attention to the $\alpha$-gauges of Sec.~\ref{sp} amounts to restricting ${\bf G}$ as 
\begin{align}\label{unregG}
{\tilde {\bf G}}_\alpha ({\bf k},{\bf x}) = \alpha{\tilde {\bf G}}_1({\bf k},{\bf x}) = -{\alpha {\bf x}\over \sqrt{(2\pi)^3}}\int_0^1 d\lambda \,e^{-i{\bf k}\cdot \lambda{\bf x}}
\end{align}
where now only $\alpha$ is freely choosable. The multipolar gauge $\alpha=1$ specifies polarisation ${\bf P}_{\rm T1}$ that is singular at the origin and which is therefore often regularised at small distances \cite{cohen-tannoudji_photons_1989,vukics_fundamental_2015,grieser_depolarization_2016}. This is achieved through the introduction of a form factor such as a Lorentzian with frequency cut-off $k_M$, to give
\begin{align}\label{Galph}
{\tilde {\bf G}}_{\alpha M} ({\bf k},{\bf x}) = -{\alpha {\bf x}\over \sqrt{(2\pi)^3}}{k_M^2\over k^2+k_M^2} \int_0^1 d\lambda \,e^{-i{\bf k}\cdot \lambda{\bf x}}.
\end{align}
For $k_M$ finite the field ${\bf P}_{\rm T\alpha}$ is no longer singular at ${\bf 0}$. This regularisation of ${\bf P}_{\rm T\alpha}$ actually constitutes a choice of gauge, that is, we now have a two-parameter gauge function uniquely specified by a gauge vector $(\alpha, k_M)$. Only for  $\alpha=0$ do we have ${\bf P}_{\rm T\alpha}={\bf 0}$ and $\chi_0=0$, such that regularisation of ${\bf P}_{\rm T}$ has no effect on the Hamiltonian. Note that if ${\bf P}_{\rm L}$ is similarly regularised then for $\alpha=1$ the ensuing total polarisation ${\bf P}_1$ is no longer point-localised, but exponentially localised instead. Regularisation of ${\bf P}_{\rm L} = -\nabla^{-2}\rho$ is not, however, a choice of gauge. The procedure instead amounts to a relaxation of the strict point-particle limit of $\rho({\bf x})$, given by $k_M \to \infty$.

More generally than Eq.~(\ref{Galph}), we may let
\begin{align}\label{Galph2}
{\tilde {\bf G}}_{\{\alpha\}} ({\bf k},{\bf x}) = -{\alpha({\bf k})^* {\bf x}\over \sqrt{(2\pi)^3}} \int_0^1 d\lambda \,e^{-i{\bf k}\cdot \lambda{\bf x}},
\end{align}
which from Eqs.~(\ref{chi0}) and (\ref{PTg0}) yields
\begin{align}
&\chi_g({\bf x}) = \int d^3k \sum_\sigma \alpha({\bf k}) {\bf e}_\sigma ({\bf k})\cdot  {\tilde {\bf A}}_{\rm T}({\bf k}){\bf e}_\sigma({\bf k})\cdot{\tilde {\bf G}}_1({\bf k},{\bf x})^*,\\
&{\tilde {\bf P}}_{{\rm T}g}({\bf k})^* =- \int d^3 x \sum_\sigma  \alpha({\bf k}) {\bf e}_\sigma({\bf k}) {\bf e}_\sigma({\bf k})\cdot {\tilde {\bf G}}_1({\bf k},{\bf x})^*\rho({\bf x})
\end{align}
where ${\tilde {\bf G}}_1$ is given in Eq.~(\ref{unregG}). The field $\chi_g$ depends on photonic degrees of freedom through ${\tilde {\bf A}}_{\rm T}$ and couples to the material momentum ${\bf p}$ within the Hamiltonian, while the field ${\tilde {\bf P}}_{{\rm T}g}$ depends on the material degrees of freedom through $\rho$ and couples to the photonic momentum ${\tilde {\bf \Pi}}$ within the Hamiltonian. Thus, Eq.~(\ref{Galph2}) enables broad control over the physical nature of the light-matter coupling, because while it is restricted in its ${\bf x}$-dependence, $\alpha({\bf k})=\alpha(-{\bf k})^*$ is essentially arbitrary. As an example, we will see in Sec.~\ref{pd} that the gauge $\alpha({\bf k}) = \omega_m/(\omega+\omega_m)$ where $\omega_m$ is a material frequency, is noteworthy. It can be interpreted as defining a canonical harmonic dipole that automatically subsumes the virtual photons dressing the system ground state \cite{stokes_gauge_2019,stokes_extending_2012,drummond_unifying_1987}. It is clear that ${\bf g}_{\rm T}({\bf x},{\bf x'})$ may be yet more general than the forms listed above. In particular, the specification of the above fixed dependence on the second argument ${\bf x}'$ stems from the line-integral solution in Eq.~(\ref{dipo}), which is not the most general form of ${\bf g}_{\rm T}$, as shown by Healy \cite{healy_representation_1977}. Furthermore, the gauge function $\chi_g$ need not even be restricted as in Eq.~(\ref{chi0}). This broad generality warrants further study, but will not be considered here.%This is discussed in more detail in Sec.~\ref{osc}.

\subsection{Sharing out the constrained degrees of freedom: Regularisation and localisation}\label{share}

The choice of ${\bf G}$ determines the physical meaning of the canonical degrees of freedom. To see how, we will focus on the simple choices given by Eqs.~(\ref{unregG}) and (\ref{Galph}). Let us begin by considering the ``unregularised" one-parameter gauges with ${\bf G}_\alpha$ defined by Eq.~(\ref{unregG}). First we consider the potential ${\bf A}_\alpha$ and the momentum ${\bf p}$ determined physically by ${\bf A}_\alpha$. According to Eq.~(\ref{aa}) ${\bf A}_\alpha$ is a function of ${\bf A}_0 = {\bf A}_{\rm T}=(\nabla\times )^{-1}{\bf B}$, so it can be expressed as a convex sum of the extremal potentials ${\bf A}_0$ and ${\bf A}_1$;
\begin{align}\label{amag}
{\bf A}_\alpha({\bf x}&) = (1-\alpha){\bf A}_0({\bf x}) +\alpha {\bf A}_1({\bf x})\nonumber \\ =&\int d^3x' {(1-\alpha)\nabla'\times {\bf B}({\bf x}')\over 4\pi|{\bf x}-{\bf x}'|} - \alpha \int_0^1 d\lambda \, \lambda{\bf x} \times {\bf B}(\lambda{\bf x}).
\end{align}
Eq.~(\ref{amag}) shows that the potential ${\bf A}_\alpha({\bf r})$, as appears in the Hamiltonian, is non-local in any gauge, but it is most localised in the multipolar gauge, $\alpha=1$, because all points ${\bf x}$ for which ${\bf A}_1({\bf r})$ depends on the local field ${\bf B}({\bf x})$ are {\em inside} the atom; $|{\bf x}|\leq|{\bf r}|$. More precisely, ${\bf A}_1({\bf r})$ depends on ${\bf B}$ only at points on the straight line connecting ${\bf 0}$ to ${\bf r}$. The value of $\alpha$ within the vector potential ${\bf A}_\alpha$, dictates the balance between this local contribution and the non-local contribution $(1-\alpha){\bf A}_0$ given by the ${\bf x}'$ integral in Eq.~(\ref{amag}). The quantity $q{\bf A}_0({\bf r})=q{\bf A}_{\rm T}({\bf r})$ is the momentum associated with the longitudinal electric field of the charge $q$ at ${\bf r}$, viz. \cite{cohen-tannoudji_photons_1989}
\begin{align}\label{elr}
{\bf K}_{\rm long}:=  \int d^3 x\, {\bf E}_{\rm L{\bf r}}\times {\bf B}
=q{\bf A}_{\rm T}({\bf r})
\end{align}
where ${\bf E}_{\rm L{\bf r}}({\bf x}) := -q\nabla(4\pi |{\bf x}-{\bf r}|)^{-1}$, consistent with Eq.~(\ref{amag}). 

To see most clearly how ${\bf A}_\alpha$ determines the physical nature of ${\bf p}$, which defines the canonical atom, we consider the EDA implemented as
\begin{align}\label{GEDA}
{\tilde {\bf G}}_\alpha({\bf k},{\bf r}) \approx -{\alpha{\bf r} \over \sqrt{(2\pi)^3}},
\end{align}
which implies
\begin{align}\label{avan}
{\bf A}_\alpha({\bf r})&:={\bf A}_{\rm T}({\bf r}) -\alpha \nabla_{\bf r} \int_0^1 d\lambda\, {\bf r} \cdot {\bf A}_{\rm T}(\lambda {\bf r}) \nonumber  \\& \approx {\bf A}_{\rm T}({\bf 0}) -\alpha \nabla_{\bf r} [{\bf r} \cdot {\bf A}_{\rm T}({\bf 0})] = (1-\alpha){\bf A}_{\rm T}({\bf 0}).
\end{align}
According to Eq.~(\ref{avan}), the multipolar vector potential at the position of the dipole, ${\bf A}_1({\bf 0})$, vanishes at dipole order. The dipole canonical momentum is defined by ${\bf p}=m{\dot {\bf r}}+q{\bf A}_\alpha({\bf 0})$ where ${\bf A}_\alpha({\bf 0})=(1-\alpha){\bf A}_{\rm T}({\bf 0})$ [Eq.~(\ref{avan})]. For $\alpha=1$ we have ${\bf p}=m{\dot {\bf r}}$, such that ${\bf E}_{\rm L}$ makes no contribution to the canonical pair $\{{\bf r},{\bf p}\}$, which is therefore ``bare". For $\alpha=0$, the momentum ${\bf p}=m{\dot {\bf r}}+{\bf K}_{\rm long}$ is fully dressed by ${\bf E}_{\rm L \bf r}$. Thus, the gauge $\alpha$ controls the extent to which the canonical dipole is dressed by the electrostatic field of the dynamical charge $q$ at ${\bf r}$.

Let us now repeat the above analysis in the case of the other quantity that is determined by the gauge $\alpha$, namely ${\bf P}_{\rm T\alpha}$. We will then see how this quantity determines the second canonical momentum ${\bf \Pi}$. The {\em total} $\alpha$-gauge polarisation is ${\bf P}_\alpha = {\bf P}_{\rm L}+\alpha{\bf P}_{\rm T1}$, where ${\bf P}_{\rm L}=-{\bf E}_{\rm L}={\bf P}_0$ defines the non-local Coulomb gauge polarisation and where ${\bf P}_{\rm T1}$ is the transverse part of the multipolar polarisation. The total multipolar polarisation ${\bf P}_1$ is given in Eq.~(\ref{mulp}), showing that it is a line integral that vanishes at all points ${\bf x}$ not on the straight line from ${\bf 0}$ to ${\bf r}$. Therefore, outside the atom $(|{\bf x}|>|{\bf r}|)$ we have ${\bf P}_{\rm T1} =-{\bf P}_{\rm L}={\bf E}_{\rm L}$. The $\alpha$-gauge polarisation can be written analogously to Eq.~(\ref{amag}) as a convex sum of local and non-local extremal polarisations ${\bf P}_0$ and ${\bf P}_1$;
\begin{align}
{\bf P}_\alpha = (1-\alpha){\bf P}_0+\alpha {\bf P}_1.
\end{align}
The polarisation ${\bf P}_\alpha$ is non-local in any gauge, but it is most localised in the multipolar gauge, $\alpha=1$, because all points ${\bf x}$ for which ${\bf P}_1({\bf x})\neq {\bf 0}$ are {\em inside} the atom; $|{\bf x}|\leq|{\bf r}|$. Within ${\bf P}_\alpha$, the value of $\alpha$ dictates the balance between this local contribution and the non-local contribution $(1-\alpha){\bf P}_0=-(1-\alpha){\bf E}_{\rm L}$.

As before, we can approximate the stationary atom as a dipole at the origin ${\bf 0}$ using Eq.~(\ref{GEDA}) to obtain ${\bf P}_{\rm T1}({\bf x})=q{\bf r}\cdot \delta^{\rm T}({\bf x})$ where $q{\bf r}$ is the dipole moment. Within the fixed gauge $\alpha$, the field canonical momentum operator is defined by ${\bf \Pi}=-{\bf E}_{\rm T}-\alpha {\bf P}_{\rm T1}=-{\bf E}_{\rm T}-\alpha {\bf E}_{\rm L}$ where the second equality holds for ${\bf x}\neq {\bf 0}$. Thus, the value of $\alpha$ controls the extent to which the canonical pair $\{{\bf A}_{\rm T},{\bf \Pi}\}$, includes the electrostatic field ${\bf E}_{\rm L}={\bf E}-{\bf E}_{\rm T}$. For $\alpha=0$ we have ${\bf \Pi}=-{\bf E}_{\rm T}$, so ${\bf E}_{\rm L}$ is completely absent from the field canonical degrees of freedom. For $\alpha=1$ we have ${\bf \Pi}=-{\bf E}$ for ${\bf x}\neq {\bf 0}$, so the situation is reversed; ${\bf E}_{\rm L}$ is fully included in the field canonical degrees of freedom for all ${\bf x}\neq {\bf 0}$. This holds beyond the EDA, but the condition ${\bf x}\neq {\bf 0}$ must be replaced by $|{\bf x}|>|{\bf r}|$ specifying all points outside the atom. Gauss' law implies gauge-redundancy by {\em constraining} ${\bf E}$ and this lies at the heart of gauge ambiguities in ultrastrong coupling QED. The gauge $\alpha$ controls the weight with which ${\bf E}_{\rm L}$ is shared between the two canonical pairs $\{{\bf A}_{\rm T},{\bf \Pi}\}$ and $\{{\bf r},{\bf p}\}$. 

We can also consider the regularisation of the above theory at short distances around the distribution centre ${\bf 0}$ using ${\bf G}_{\alpha M}$ in Eq.~(\ref{Galph}), which within the EDA is
\begin{align}\label{GEDA2}
{\tilde {\bf G}}_{\alpha M}({\bf k},{\bf r}) \approx -{\alpha{\bf r} \over \sqrt{(2\pi)^3}} {k_M^2\over k^2+k_M^2}.
\end{align}
The transverse $(\alpha,k_M)$-gauge polarisation within the EDA is therefore
\begin{align}
{\bf P}_{{\rm T}\alpha M}({\bf x}) = \alpha q {\bf r} \cdot \delta_M^{\rm T}({\bf x})
\end{align} 
where $\delta_M^{\rm T}({\bf x})$ denotes the regularised transverse $\delta$-function \cite{cohen-tannoudji_photons_1989}
\begin{align}
&\delta_{M,ij}^{\rm T}({\bf x}) = {2\over 3}\delta_{ij}\delta({\bf x}) - {\beta(x) \over 4\pi x^3}(\delta_{ij}-3{\hat x}_i{\hat x}_j),\\
&\beta(x) = 1 - \left(1+k_M x+{1\over 2}k_M^2 x^2\right)e^{-k_M x}.
\end{align}
The function $\beta(x)$ controls the singularity at ${\bf 0}$, but is unity for $x\gg 1/k_M$. The transverse $\delta$-function $\delta^{\rm T}({\bf x})$ is strictly recovered in the limit $k_M\to \infty$. In the $(\alpha,k_M)$-gauge the parameter $\alpha$ functions as before while the additional gauge-parameter $k_M$ controls the rate of exponential localisation of what was previously the singular point-like multipolar dipole. It is now the case that only for $x\gg 1/k_M$ do we have ${\bf P}_1({\bf x})={\bf 0}$. Thus, there are now many ``multipolar gauges" specified by the gauge vectors $(1,k_M)$, each of which possesses a different degree of exponential dipolar localisation.

The $(\alpha,k_M)$-gauge vector potential is within the EDA
\begin{align}
{\bf A}_{\alpha M}({\bf r}) \approx {\bf A}_{\rm T}({\bf 0}) -\alpha \int {d^3 k \over \sqrt{(2\pi)^3}} {k_M^2\over k^2+ k_M^2} {\tilde {\bf A}}_{\rm T}({\bf k}),
\end{align}
such that ${\bf A}_1({\bf 0})={\bf 0}$ is recovered in the limit $k_M\to \infty$. More generally, vanishing of ${\bf A}_{1 M}({\bf r})$ to dipole order requires that ${\tilde {\bf A}}_{\rm T}({\bf k})\approx {\bf 0}$ for $k\geq k_M$. In order for this to be the case the modes $k \geq k_M$ must not be populated. This is the case if the bare atom (as occurs in the free theory) is small compared to the characteristic wavelengths of the populated modes. In other words, the EDA places a lower bound on the cut-off $k_M$ in order that gauges $(1,k_M)$ possess the property ${\bf A}_{1M}({\bf 0})={\bf 0}$ that at dipole order characterises the usual multipolar gauge $(\alpha,k_M)=(1,\infty)$.

\subsection{Discussion: gauge fixing, forms of rotation, forms of coupling, and common pitfalls}\label{minc}

\subsubsection{Gauge freedom and gauge fixing}\label{gfandgf}

We have defined the gauge principle according to modern gauge-field theory and we have given a formulation of canonical (Hamiltonian) nonrelativistic QED in an arbitrary gauge. One of the main objectives of the present article is to clarify what gauge freedom, gauge fixing, and gauge ambiguities are, within this theory:
\begin{itemize}
\item{{\em Gauge freedom} in electrodynamics is a freedom to choose ${\bf A}_{\rm L}$. Once ${\bf A}_{\rm L}$ is fixed then the scalar potential $\phi_{{\bf A}_{\rm L}}$ is also fixed up to a constant by $-\nabla\phi_{{\bf A}_{\rm L}} = {\bf E}_{\rm L}+{\partial_t {\bf A}_{\rm L}}$. {\em Gauge fixing} means specifying ${\bf A}_{\rm L}$ in terms of gauge invariant quantities.}
\end{itemize}
We have provided a formulation of QED in which ${\bf A}_{\rm L}$ is fixed by Eq.~(36) (Supplementary Note~II) as ${\bf A}_{\rm L}({\bf x})=\nabla\chi_g({\bf x},{\bf A}_{\rm T})$ meaning that it is fixed up to a choice of the non-operator-valued function ${\bf g}_{\rm T}$. The corresponding vector and scalar potentials are given in accordance with their fundamental definitions by ${\bf A}_g ={\bf A}_{\rm T}+\nabla \chi_g$ and $\phi_g=\phi -\partial_t\chi_g$ where ${\bf E}_{\rm L}=-\nabla\phi$.
  
\subsubsection{Equality of multipolar and Poincar\'e gauges} 

QED in multipolar form and its relation to the Poincar\'e gauge has been a recent topic of debate \cite{rousseau_quantum-optics_2017,andrews_perspective_2018,vukics_gauge-invariant_2021,rousseau_reply_2018}.  Ref.~\cite{rousseau_quantum-optics_2017} by Rousseau and Felbacq employs Dirac's constrained quantisation procedure to derive the nonrelativistic QED Hamiltonian in the Poincar\'e gauge. The authors claim that the multipolar Hamiltonian will not produce the same results as the Coulomb gauge Hamiltonian and that it does not coincide with the correct Poincar\'e gauge Hamiltonian. Refs.~\cite{vukics_gauge-invariant_2021} and \cite{andrews_perspective_2018} dispute this, concluding that criticisms of the multipolar framework in Ref.~\cite{rousseau_quantum-optics_2017} are not valid. A reply to Ref.~\cite{vukics_gauge-invariant_2021} is offered by Rousseau and Felbacq in Ref.~\cite{rousseau_reply_2018}, which argues that the conclusions of Ref.~\cite{vukics_gauge-invariant_2021} are not correct, maintaining their conclusion that the Poincar\'e gauge Hamiltonian does not coincide with the multipolar Hamiltonian.

We have clarified the relation between the multipolar theory and the Poincar\'e gauge in Supplementary Note~VI. We show in Supplementary Notes~II and IX that Dirac's quantisation procedure does yield the well-known multipolar theory. We have also showed that the latter can be obtained via a gauge fixing transformation from the Coulomb gauge, and that even in the case of multiple charge distributions, the multipolar theory is obtained by choosing $\zeta$-Poincar\'e gauge fixing conditions. Complete reconciliation of our results with those of Refs.~\cite{rousseau_quantum-optics_2017,vukics_gauge-invariant_2021,andrews_perspective_2018,rousseau_reply_2018} is provided   in Ref.~\cite{stokes_identification_2021} and in Supplementary Note~IX,  through the construction of Dirac brackets within the theory of a single electron atom. This reveals precisely where misunderstanding has occurred, while also fully clarifying the status of the multipolar (Poincar\'e gauge) theory.\\

\subsubsection{Dipolar coupling}

An aspect of light-matter interactions, which is especially poorly understood, concerns the field that a dipole couples to within the multipolar gauge. Common misidentifications are exacerbated by the development of the theory via semi-classical treatments as features heavily in textbook quantum optics (e.g. textbooks \cite{schleich_quantum_2001,gerry_introductory_2004,scully_quantum_1997}). In such treatments the gauge principle implies that the EDA of the semi-classical PZW transformation applied to ${\bf p}-q{\bf A}_{\rm T}({\bf 0})$ within the Scr\"odinger equation yields the bilinear coupling $-q{\bf r}\cdot {\bf E}_{\rm T}({\bf 0})$ where ${\bf E}_{\rm T}=-{\partial_t {\bf A}}_{\rm T}$ \cite{schleich_quantum_2001,gerry_introductory_2004,scully_quantum_1997}. However, according to the fully quantum description the correct bilinear component of the coupling is $-q{\bf r}\cdot {\bf D}_{\rm T}({\bf 0})$ where ${\bf D}_{\rm T}= {\bf E}_{\rm T}+{\bf P}_{\rm T}$. The field ${\bf P}_{\rm T}$ is singular at ${\bf 0}$ so the fully quantum description provides a coupling that is {\em infinitely} different from the result of a semi-classical approach. Nevertheless, the notation $-q{\bf r}\cdot {\bf E}_{\rm T}({\bf 0})$ remains prevalent even in textbooks that employ fully quantum treatments (e.g. \cite{loudon_quantum_2000}). Further confusion stems from the fact that ${\bf E}_{\rm T}$ is often simply written as ${\bf E}$ even when ${\bf E}_{\rm L}\neq {\bf 0}$, such that the notation $-q{\bf r}\cdot {\bf E}({\bf 0})$ is also encountered in textbooks (e.g. \cite{agarwal_quantum_2012}) and more recently, in ultrastrong coupling light-matter physics literature (e.g. Ref. \cite{settineri_gauge_2021}).

Further still, it is not commonly recognised that within the EDA ${\bf D}_{\rm T}={\bf E}$, but only for ${\bf x}\neq {\bf 0}$ (see Sec. \ref{share}). The unfortunate interchanging of fields ${\bf D}_{\rm T}$, ${\bf E}_{\rm T}$, and ${\bf E}$, which are related but not equal, may lead to the misidentification of fields both at, and away from the dipole's position. We emphasize that neither $-q{\bf r}\cdot {\bf E}_{\rm T}({\bf 0})$ nor $-q{\bf r}\cdot {\bf E}({\bf 0})$ are correct interactions, and neither is it true that ${\bf E}_{\rm T}={\bf E}$ whereas it is true that ${\bf D}_{\rm T}={\bf E}$ at points ${\bf x}$ outside of the charge distribution, which within the EDA means for ${\bf x}\neq {\bf 0}$. In the weak-coupling regime one can often afford to misidentify the physical fields involved in light-matter interactions, but this may lead to erroneous results in sufficiently strong-coupling regimes.

Similarly, confusion can arise in nonrelativistic QED, due to claims that a dipolar-coupling such as $-q{\bf r}\cdot {\bf E}_{\rm T}({\bf 0})$ may be preferable to a Coulomb gauge coupling, because it is expressible solely in terms of a gauge invariant electric field. For example, Ref.~\cite{scully_quantum_1997} provides a typical semi-classical derivation of $-q{\bf r}\cdot {\bf E}_{\rm T}({\bf 0})$, and refers to the semi-classical multipolar gauge as the $E$-gauge. It is stated that the $E$-gauge interaction is gauge invariant in contrast to a linear ${\bf p}\cdot {\bf A}$ interaction as found in the Coulomb gauge. It is argued that only in the $E$-gauge is the unperturbed dipolar Hamiltonian 
\begin{align}\label{Hmdef}
H_m={{\bf p}^2\over2m}+V({\bf r}),
\end{align}
a physical quantity. However, in the Coulomb gauge, for example, both ${\bf p}=m{\dot {\bf r}}+{\bf K}_{\rm long}$ and ${\bf A}_{\rm T}$ are also gauge invariant. Indeed, there is no means by which the requirement of gauge-invariance can be leveraged as an argument to prefer one gauge over another. The theory in any gauge can be expressed entirely in terms of gauge invariant quantities, by definition of gauge fixing, as explained in Sec.~\ref{gfandgf} . 

We have already seen that the Hamiltonian always represents the total energy [Eq.~(54) in Supplementary Note~III and Eq.~(\ref{enH})] \cite{stokes_gauge_2019}. If one prefers to eliminate only ${\bf A}_\alpha$ from the expression for the Hamiltonian, but retain its explicit dependence on the canonical momenta, then this is easily achieved in {\em any} gauge $\alpha$, using Eq.~(\ref{amag}). In particular, the Coulomb gauge theory for which ${\bf \Pi} = -{\bf E}_{\rm T}$ can be expressed solely in terms of electric and magnetic fields. The latter property is not unique to the multipolar theory.

\subsubsection{Active and passive perspectives of unitary rotations}\label{active}

A generic feature of linear spaces is that rotations therein may be implemented in an active or passive way. A vector ${\bf v}=\sum_iv_i{\bf e}_i$ within Hermitian inner-product space $V$ may be actively rotated by a unitary transformation $R$ into a new vector ${\bf v}'=R{\bf v} = \sum_i v'_i{\bf e}_i$. Expressed in the same basis $\{{\bf e}_i\}$ the new vector has components $v'_i = \sum_j R_{ij}v_j$ where $R_{ij}=\langle {\bf e}_i,R{\bf e}_j\rangle$. Alternatively, the original vector ${\bf v}$ may be expressed in a rotated basis $\{{\bf e}'_i=R^\dagger{\bf e}_i\}$ to give ${\bf v}=\sum_iv'_i{\bf e}'_i$. In both cases the same numerical components, $\{v'_i\}$, are obtained from the rotation. Note that the passive rotation $R^\dagger$ of basis vectors ${\bf e}_i$ is opposite to the active rotation $R$ of ${\bf v}$.

The same considerations apply when unitarily rotating a Hamiltonian expressed in a canonical operator ``basis". In Sec.~\ref{sp} and   Supplementary Note~III an active perspective of unitary rotations has been adopted, whereby the canonical operators ${\bf y}=(\psi,\psi^\dagger,{\bf A}_{\rm T},{\bf \Pi})$ are viewed as fixed, while the Hamiltonian can be rotated to a new but equivalent form using a gauge fixing transformation as $H_{g'}({\bf y})= U_{gg'}H_g({\bf y})U_{gg'}^\dagger$ (in the particle-based $\alpha$-gauge formalism of Sec.~\ref{sp} we instead have ${\bf y}=({\bf r},{\bf p},{\bf A}_{\rm T},{\bf \Pi})$ and $H_{\alpha'}({\bf y})=R_{\alpha\alpha'}H_\alpha({\bf y})R_{\alpha\alpha'}^\dagger$). The transformation of the Hamiltonian can of course be implemented via transformation of the canonical operators in the sense that $U_{gg'}H_g({\bf y})U_{gg'}^\dagger=H_g(U_{gg'} {\bf y}U_{gg'}^\dagger)$ [see Eqs.~(51) and (52) in Supplementary Note~III and also Eqs.~(\ref{mina}) and (\ref{minb})]. 

The active perspective is commonly found, and is adopted for example in the textbook by Cohen-Tannoudji et al. \cite{cohen-tannoudji_photons_1989}. 
From this point of view, any operator that does not commute with gauge fixing transformations, such as ${\bf \Pi}$, will represent a different physical observable before and after such a transformation \cite{cohen-tannoudji_photons_1989}. Conversely, a given physical observable will be represented by a different operator before and after transformation. For example, the energy $E$ is represented by $H_g({\bf y})$ in gauge $g$ and by $H_{g'}({\bf y})$ in gauge $g'$. The eigenvalue equation $H_g({\bf y})\ket{E_g^n}=E^n\ket{E_g^n}$ implies that the vector $\ket{E^n_g}$ represents, within the gauge $g$, the physical state ${\cal S}^n$ in which the system definitely possesses energy $E^n$. Meanwhile, in the gauge $g'$ the same state ${\cal S}^n$ is represented by the different vector $\ket{E_{g'}^n}=U_{gg'}\ket{E_g^n}$, because the energy is represented by the different operator $H_{g'}({\bf y})$.

Alternatively, a passive perspective of rotations may be adopted whereby different canonical operators are associated with different gauges as ${\bf y}_g=U_{gg'}{\bf y}_{g'}U_{gg'}^\dagger$. Notice that the rotation between canonical operators associated with different gauges $g$ and $g'$ is opposite to the rotation between the Hamiltonians associated with $g$ and $g'$ obtained via tha active perspective. Nevertheless, the same relationship between Hamiltonian functions is obtained within the passive viewpoint by noting that $H_g({\bf y}_g)= H_g(U_{gg'}{\bf y}_{g'}U_{gg'}^\dagger)= U_{gg'}H_g({\bf y}_{g'})U_{gg'}^\dagger = H_{g'}({\bf y}_{g'})$. The passive perspective is also commonly found within the literature, for example, in the works of Power and Thirunamachandran \cite{craig_molecular_1998,power_quantum_1983,power_quantum_1983-1,power_quantum_1983-2,power_quantum_1992,power_quantum_1993,power_time_1999,power_time_1999-1}. Therein, the Hamiltonian $H_g({\bf y}_g)=H_{g'}({\bf y}_{g'})$ is unique and it uniquely represents the energy $E$. Similarly, the eigenvector $\ket{E^n}$ uniquely represents the physical state ${\cal S}^n$ of definite energy $E^n$. Conversely, each different set of canonical operators ${\bf y}_g$ explicitly represents a different set of physical observables. This again contrasts the active perspective wherein the physical difference between the same canonical operators ${\bf y}$ in different gauges was implicit.

Obviously, either an active or a passive perspective can be chosen, but the associations between operators and observables and between vectors and states will generally depend on the perspective adopted. The importance of such associations and their relation to gauge freedom is discussed in Secs.~\ref{qsr} and \ref{gamb}. Here, unless otherwise stated, we adopt an active perspective of unitary rotations. 

\subsubsection{Gauge symmetry transformations versus gauge fixing transformations}

Confusion can stem from the fact that the PZW transformation $R_{01}$ commutes with  $A_0=\phi$ and ${\bf A}_0={\bf A}_{\rm T}$, so it cannot {\em directly} implement a gauge transformation [see Eqs.~(61) and (62) of Supplementary Note~IV] as noted, for example, in Ref.~\cite{andrews_perspective_2018}. The situation becomes clear upon recognising that the PZW transformation is not a gauge-{\em symmetry} transformation $S_\chi$, but an example of a gauge-{\em fixing} transformation $U_{gg'}$. The distinction between these types of gauge transformation was recognised some tine ago in relativistic physics \cite{lenz_quantum_1994}, but it is perhaps less well-known in quantum optics and atomic physics. Within the final unconstrained theory all gauge-{\em symmetry} transformations have been reduced to the identity, expressing the fact that once the gauge has been fixed there is no longer any redundancy within the state space or operator algebra. The redundant degrees of freedom ${\bf A}_{\rm L}$ have been fixed as known functions of the gauge invariant degrees of freedom. The gauge fixing transformation $U_{gg'}$ transforms between alternative isomorphic realisations of the physical state space that result from {\em different} choices of gauge ${\bf A}_{\rm L} = \nabla\chi_g$ and ${\bf A}_{\rm L}=\nabla\chi_{g'}$. 

Although $U_{gg'}$ cannot transform $(\phi_g,{\bf A}_g)$ directly, it does so indirectly. To see this note that $H_g({\bf y})$ is shorthand for $H({\bf g}_{\rm T},{\bf y})$ where the function $H$ is unique. The concrete choice of function ${\bf g}_{\rm T}$ used to evaluate $H$ is left {\em open}. In other words, $H_{g'}({\bf y})$ defined by $H_{g'}({\bf y}):= U_{gg'}H_g({\bf y})U_{gg'}^\dagger$ is given by $H_{g'}({\bf y})\equiv H({\bf g}_{\rm T}',{\bf y})$. By construction the functional form of the Hamiltonian in terms of ${\bf g}_{\rm T}$, as well as all resulting dynamical equations written in terms of $(\phi_g,{\bf A}_g)$, are the same for every possible concrete choice of ${\bf g}_{\rm T}$ (gauge). Thus, in the final unconstrained theory:
\begin{itemize}
\item{Gauge freedom is the freedom to transform between different Hamiltonians $H_g$ and $H_{g'}$ resulting from different fixed choices of gauge ${\bf g}_{\rm T}$ and ${\bf g}_{\rm T}'$.}
\end{itemize}
Gauge invariance means that formulations corresponding to different choices of ${\bf g}_{\rm T}$ must be physically equivalent. The {\em unitarity} of gauge fixing transformations $U_{gg'}$ ensures that this is the case, because the quantum-theoretic definition of physical equivalence is unitary equivalence (see Sec.~\ref{qsr}).

\subsubsection{Minimal coupling}\label{minc2}

A final common pitfall that we wish to address concerns the nature of the minimal coupling prescription and its relation to the Coulomb gauge. In Sec.~\ref{sp} we saw that $R_{\alpha\alpha'}$ implements a gauge change within the Hamiltonian by transforming between distinct minimal coupling prescriptions [Eqs.~(\ref{mina}) and (\ref{minb})]. This shows that the minimal coupling replacement is {\em not} synonymous with the Coulomb gauge.

It is unfortunate that the term ``minimal coupling" has so often been reserved exclusively for the Coulomb gauge Hamiltonian $H_0$, because this nomenclature is in direct opposition to the fundamental meaning of minimal coupling. The gauge principle implies the existence of a potential whose gauge ${\bf A}_{\rm L}$  can be chosen {\em freely}. Different fixed gauges correspond to different fixed minimal coupling replacements, as is clearly shown by Eqs. (\ref{HA}) and (\ref{mina}). This fact is obscured by the almost universal practice of expressing the multipolar potential ${\bf A}_1$ in terms of ${\bf B}$ within the Hamiltonian via Eq. (\ref{amag}). It is then not obvious that the multipolar Hamiltonian does result from the minimal coupling replacement ${\bf p}\to {\bf p}-q{\bf A}_1({\bf r})$. Meanwhile, despite it being possible to express the Coulomb gauge potential ${\bf A}_0$ in terms of ${\bf B}$, the Coulomb gauge Hamiltonian is nearly always left as a function of ${\bf A}_0$. The minimal coupling prescription ${\bf p}\to {\bf p}-q{\bf A}_0({\bf r})$ is therefore immediately apparent therein. The combined effect of these conventions may be the false impression that {\em only} the Coulomb gauge Hamiltonian results from minimal coupling replacement. In fact, in {\em any} gauge $\alpha$, the Hamiltonian includes a minimal coupling replacement ${\bf p}\to {\bf p}-q{\bf A}_\alpha({\bf r})$ and the potential ${\bf A}_\alpha$ is expressible as a function of the magnetic field ${\bf B}$.

Yet further obfuscation occurs within the EDA which states that ${\bf A}_{\rm T}({\bf x})\approx {\bf A}_{\rm T}({\bf 0})$ whenever $|{\bf x}|\leq |{\bf r}|$. This implies that $\chi_\alpha$ in Eq.~(\ref{chialph}) is approximated as in Eq.~(\ref{chieda}). Thus, choosing the multipolar gauge, ${\bf A}_1$, means choosing ${\bf A}_{\rm L} = \nabla\chi_1$ such that ${\bf A}_{\rm L}({\bf r})= -{\bf A}_{\rm T}({\bf r})$ within the EDA, giving ${\bf A}_1({\bf r})\approx{\bf 0}$ [see Eq.~(\ref{avan})].  The position ${\bf r}$ is of course where the potential ${\bf A}_1$ is evaluated within the Hamiltonian [see Eq.~(\ref{HA})]. Thus, the dipole approximation of the kinetic energy part of the multipolar gauge Hamiltonian is independent of the potential and the canonical momentum ${\bf p}$ becomes purely mechanical; ${\bf p}=m{\dot {\bf r}}$. This again, may lead to the false impression that the multipolar Hamiltonian is not a minimal coupling Hamiltonian.

Crucially, according to the gauge principle {\em all} Hamiltonians $H_\alpha$ are equally valid, and {\em any one of them} can be taken as the starting point for a canonical description of QED. It is certainly not the case that only one particular gauge's Hamiltonian, such as $H_0$, is compatible with the gauge principle. Indeed, such a conclusion would contradict the gauge principle. In particular, it is not the case that $H_0(t)$ is a fundamentally preferable starting point when considering time-dependent interactions and that any other Hamiltonian must be obtained from it via a time-dependent gauge transformation. This fact appears to contradict recent articles \cite{stefano_resolution_2019,settineri_gauge_2021}. Time-dependent interactions are discussed in detail in Sec.~\ref{s6}.
`
\section{Subsystem gauge relativity}\label{s0}

Quantum theory provides postulates for the association of physical states and observables with their mathematical representations, and for the calculation of predictions of observable properties. The notion of a quantum system is an inherently relative one \cite{zanardi_virtual_2001,zanardi_quantum_2004,barnum_subsystem-independent_2004,viola_entanglement_2007,harshman_tensor_2007,harshman_observables_2011}. Understanding quantum subsystem properties in light of this remains a topic of current interest (e.g. \cite{ali_ahmad_quantum_2022,cai_entanglement_2021}). The partition of a quantum system into subsystems is dictated by the set of operationally accessible interactions and measurements \cite{zanardi_quantum_2004}. The importance of this fact in QED beyond traditional regimes is addressed in this section.

\subsection{Quantum subsystem relativity}\label{qsr}

We begin by examining fundamental concepts relating to composite quantum systems and subsequently relate them to gauge freedom. In quantum theory all predictions are obtained from the inner-product, therefore the following associations
\begin{align}
{\rm physical~state}~{\cal S} &\leftrightarrow {\rm vector}\,\ket{\psi} \nonumber \\
{\rm physical~observable}~{\cal O} &\leftrightarrow {\rm operator}~O \nonumber
\end{align}
are equivalent to the associations
\begin{align}
{\rm physical~state}~{\cal S} &\leftrightarrow  {\rm vector}\,\ket{\psi'} =U\ket{\psi} \nonumber \\
{\rm physical~observable}~{\cal O} &\leftrightarrow  {\rm operator}~O'=UOU^\dagger \nonumber
\end{align}
where $U$ is any unitary operator \cite{isham_lectures_1995}. In other words, the associations $\{${\em state} $\leftrightarrow$ {\em vector}$\}$ and $\{${\em observable} $\leftrightarrow$ {\em operator}$\}$, can only be made {\em relative to a Hilbert space frame}. The unitary group is the symmetry group of the inner-product, $\braket{\cdot | \cdot}$, defined over ${\cal H}$, meaning that $U$ transforms between two distinct Hilbert space frames (bases). This is analogous to moving between frames within, for example, Minkowski spacetime, $E^{1,3}$, using a Lorentz transformation, $\Lambda$, belonging to the Lorentz group, which is the symmetry group of the (indefinite) Minkowski inner-product. The definition of a composite quantum system uses the tensor-product $\otimes$, which extends the inner-product in the way required in order that probabilities associated with independent subsystems are statistically independent. Specifically, $(\bra{\psi_{\rm A}}\otimes \bra{\psi_{\rm B}})(\ket{\varphi_{\rm A}}\otimes \ket{\varphi_{\rm B}})\equiv \braket{\psi_{\rm A}|\varphi_{\rm A}}\braket{\psi_{\rm B}|\varphi_{\rm B}}$.

To understand how the relativity of associations between operators and observables affects the meaning of quantum subsystems, let us consider a composite system of two spins ${\rm A}$ and ${\rm B}$ with Hilbert space ${\cal H}={\cal H}_{\rm A}\otimes {\cal H}_{\rm B}$. We denote spin observables in some specified directions for A and B by ${\cal O}_{\rm A}$ and ${\cal O}_{\rm B}$ respectively and we let these observables be represented in frame $X$ by operators $\sigma_{\rm A}\otimes I_{\rm B}$ and $I_{\rm A}\otimes \sigma_{\rm B}$ where $I_{\rm A}$ and $I_{\rm B}$ are identity operators over ${\cal H}_{\rm A}$ and ${\cal H}_{\rm B}$ respectively. For reasons of notational economy one often writes $O_{\rm A}\otimes I_{\rm B}$ ($I_{\rm A}\otimes O_{\rm B}$) simply as $O_{\rm A}$ ($O_{\rm B}$). Spin can take two values denoted $s=+,-$. The (eigen)state ${\cal S}_{s}$ is the {\em physical state} in which the {\em physical observables} ${\cal O}_{\rm A}$ and ${\cal O}_{\rm B}$ simultaneously possess value $s$. It is represented in frame $X$ by the vector $\ket{s_{\rm A}}\otimes \ket{s_{\rm B}}=:\ket{s_{\rm A},s_{\rm B}}$ where $\sigma_{\rm Z}\ket{s_{\rm Z}}=s\ket{s_{\rm Z}}$ with ${\rm Z}={\rm A,\,B}$. Now consider the unitary transformation $U$ for which
\begin{align}\label{vec}
U\ket{+_{\rm A},+_{\rm B}} = {1\over \sqrt{2}}(\ket{+_{\rm A},+_{\rm B}}+\ket{-_{\rm A},-_{\rm B}}),
\end{align}
which connects frame $X$ to a new frame $Y$, but does not have the form $U_{\rm A}\otimes U_{\rm B}$. In frame $Y$ the observables ${\cal O}_{\rm A}$ and ${\cal O}_{\rm B}$ are represented by the operators $\sigma'_{\rm A}:=U\sigma_{\rm A}U^\dagger$ and $\sigma'_{\rm B}:=U\sigma_{\rm B}U^\dagger$ respectively, such that the state ${\cal S}_s$ is represented by the vector $U\ket{s_{\rm A},s_{\rm B}}$. This ensures that the physical prediction $\langle {\cal O}_{\rm Z} \rangle_{{\cal S}_s}$ is frame-independent. The frame can therefore be chosen freely. However, in frame $Y$ the operator $\sigma_{\rm Z}\neq \sigma'_{\rm Z}$ evidently does {\em not} represent the observable ${\cal O}_{\rm Z}$. It must therefore represent some other physical observable, which we will denote by ${\mathsf O}_{\rm Z}$.

Let us now define the subalgebras ${\cal A}_{\rm Z} :=\{O_{\rm Z}:O_{\rm Z}~{\rm Hermitian}\}$ with $\rm Z=A,B$. The {\em mathematical quantum subsystem} Z may be defined as the following pair ${\rm Z} = ({\cal H}_{\rm Z},{\cal A}_{\rm Z})$. {\em Operationally}, meanwhile, any physical system must be specified through a collection of observable properties. And yet, whether or not a given observable property belongs to the set of observables that defines the quantum subsystem $Z$ depends on the Hilbert space frame. For example, observable ${\cal O}_{\rm Z}$ is represented by $\sigma_z\in {\cal A}_{\rm Z}$ in frame $X$ and by $U\sigma_Z U^\dagger \not\in{\cal A}_Z$ in frame $Y$ whereas observable ${\mathsf O}_{\rm Z}$ is represented by $\sigma_Z\in {\cal A}_{\rm Z}$ in frame $Y$ and by $U^\dagger \sigma_Z U \not\in{\cal A}_Z$ in frame $X$. It follows that ${\rm Z} = ({\cal H}_{\rm Z},{\cal A}_{\rm Z})$ represents a distinct collection of states and observables in the two different frames $X$ and $Y$

Any question about the physics of the system must be posed in terms of {\em states} and {\em observables}. For example, we can ask; is the physical state ${\cal S}_+$ an entangled state, to which there are two answers: Yes ${\cal S}_+$ is entangled with respect to the observables ${\mathsf O}_{\rm A}$ and ${\mathsf O}_{\rm B}$, and no ${\cal S}_+$ is not entangled with respect to the observables ${\mathcal O}_{\rm A}$ and ${\mathcal O}_{\rm B}$. The first answer is deduced using frame $Y$ wherein we have the representations ${\mathsf O}_{\rm Z}\leftrightarrow \sigma_{\rm Z}$ with ${\rm Z=A,B}$ and in terms of the eigenvectors of the $\sigma_{\rm Z}$ the state ${\cal S}_+$ is represented by the {\em entangled} vector $(\ket{+_{\rm A},+_{\rm B}}+\ket{-_{\rm A},-_{\rm B}})/\sqrt{2}$. The second answer is deduced using frame $X$ wherein we have ${\mathcal O}_{\rm Z}\leftrightarrow \sigma_{\rm Z}$ and in terms of the eigenvectors of the $\sigma_{\rm Z}$ the state ${\cal S}_+$ is represented by the vector $\ket{+_{\rm A},+_{\rm B}}$.

Importantly, both answers to the question are physically meaningful and they are certainly compatible statements regarding {\em states} and {\em observables}. The same physical state ${\cal S}_+$ is simultaneously entangled and not entangled because the term ``entanglement" is referring to different physical observable properties within the two different answers to the question. We can further ask; is the entanglement in the state ${\cal S}_+$ physically relevant? The answer is yes if we are able to access observables ${\mathsf O}_{\rm A}$ and ${\mathsf O}_{\rm B}$, and the answer is no if we are only able to access the observables ${\mathcal O}_{\rm A}$ and ${\mathcal O}_{\rm B}$. This again, is a statement about physical {\em states} and {\em observables}, but it also concerns which observables are actually measurable in a given experiment. 
   
In QED gauge fixing transformations are unitary, so a gauge can be understood as a {\em frame} within the Hilbert space. We have seen that the choice of frame can be labelled by a parameter $\alpha$, such that the Hilbert space has the form ${\cal H}[\alpha]={\cal H}_{\rm matter}[\alpha]\otimes {\cal H}_{\rm light}[\alpha]$. Gauge transformations mix the matter and light canonical operators of the theory which possess the forms $O_{\rm matter }\otimes I_{\rm light} \in {\cal A}_{\rm matter}[\alpha]$ and $I_{\rm matter}\otimes O_{\rm light}\in {\cal A}_{\rm light}[\alpha]$ respectively. Thus, the ``matter" and ``light" mathematical subsystems defined as the pairs $({\cal H}_{\rm matter}[\alpha],{\cal A}_{\rm matter}[\alpha])$ and $({\cal H}_{\rm light}[\alpha],{\cal A}_{\rm light}[\alpha])$ respectively, are defined by {\em physically different collections of observables} for each different gauge $\alpha$. The ``matter" subsystem constitutes a different operational subsystem in each different gauge, as does the ``light" subsystem. 

\subsection{Gauge ambiguities and gauge invariance}\label{gamb}
     
Quantum theory provides predictions for {\em observables} and the unitarity of gauge fixing transformations $U_{gg'}$ ($R_{\alpha\alpha'}$) guarantees the gauge invariance of these predictions. We define {\em gauge invariance} as follows:
\begin{itemize}
\item{A prediction is gauge invariant if it is independent of the gauge in which it is calculated. If all predictions pertaining to an observable are gauge invariant then the observable is gauge invariant.}
\end{itemize}
In general, an observable ${\cal O}$ is represented in the fixed-gauge $\alpha$ by a generally $\alpha$-dependent function $o_\alpha$ of the canonical operators ${\bf y} = \{{\bf r},{\bf A}_{\rm T},{\bf p},{\bf \Pi}\}$. A physical state ${\cal S}$ is represented by an $\alpha$-dependent vector $\ket{\psi_\alpha}$. In the gauge $\alpha'$, the same observable ${\cal O}$ is represented by the operator $o_\alpha(R_{\alpha\alpha'}{\bf y}R_{\alpha\alpha'}^\dagger)\equiv R_{\alpha\alpha'}o_\alpha({\bf y})R_{\alpha\alpha'}^\dagger =: o_{\alpha'}({\bf y})$ and the same state ${\cal S}$ is represented by the vector $\ket{\psi_{\alpha'}}=R_{\alpha\alpha'}\ket{\psi_\alpha}$. Clearly the average $\langle {\cal O}\rangle_{\cal S}$ can be calculated in any gauge
\begin{align}\label{gi}
\bra{\psi_\alpha}o_\alpha({\bf y})\ket{\psi_\alpha} = \langle {\cal O}\rangle_{\cal S} = \bra{\psi_{\alpha'}}o_{\alpha'}({\bf y})\ket{\psi_{\alpha'}}.
\end{align} 
This gauge invariance holds as a consequence of the unitarity of gauge fixing transformations and so it should be clear that it will hold independently of any restriction on the form of the gauge. %Thus, if gauge fixing transformations are unitary, then
%\begin{itemize}
%\item{In canonical nonrelativistic QED, all physical observables and predictions are gauge invariant.}
%\end{itemize}
An example of a gauge invariant observable is the total energy ${\cal O}=E$, which in the gauge $\alpha$ is represented by the Hamiltonian $H_\alpha({\bf y})$.

Although QED is fundamentally gauge invariant, the task remains of deciding which observables are relevant to us. %The transformation $U_{gg'}$ does not have the form $U_{gg'}=U_{\rm A}\otimes U_{\rm B}$ with respect to any tensor-product structure that can be imposed on the theory's Hilbert space. Thus, the mathematical ``light" and ``matter" subsystems are physically distinct in each different gauge. To understand the implications of this 
By way of example, let us consider the {\em observables} ${\bf E}_{\rm T}$ and ${\bf P}_{\rm T}$ where hereafter we use use ${\bf P}:={\bf P}_1$ to denote the multipolar polarisation. The transformation $R_{\alpha\alpha'}$ commutes with ${\bf P}_{\rm T}$, so this observable possesses the same operator representation in every gauge \cite{cohen-tannoudji_photons_1989}. The same is not true for ${\bf E}_{\rm T}$. Consider the physical observable ${\cal O}:=-{\bf E}_{\rm T}-\alpha {\bf P}_{\rm T}$ where $\alpha$ denotes a {\em fixed} real number. As a fixed linear combination of gauge invariant observables, ${\cal O}$ is gauge invariant. If we now choose our gauge-parameter to have the same fixed value $\alpha$, then the {\em observable} ${\cal O}$ is represented by the {\em operator} ${\bf \Pi}$. 

We emphasize that gauge freedom is not a freedom to define ${\cal O}$. It is a freedom to decide whether the parameter that fixes the redundancy ${\bf A}_{\rm L}$ within our description, equals the number $\alpha$ that defines ${\cal O}$. If the gauge parameter is instead chosen to have value $\alpha'\neq \alpha$, then the observable ${\cal O}$ is represented by the operator ${\bf \Pi}' = R_{\alpha\alpha'}{\bf \Pi} R_{\alpha\alpha'}^\dagger = {\bf \Pi} -(\alpha-\alpha'){\bf P}_{\rm T}$. The operator ${\bf \Pi}$ represents the different gauge invariant physical observable ${\cal O}':=-{\bf E}_{\rm T}-\alpha' {\bf P}_{\rm T}$. %Similarly, in the gauge $\alpha$ the physical observable $O'$ is represented by the operator ${\bf \Pi}'' = R_{\alpha'\alpha}{\bf \Pi}R_{\alpha'\alpha}^\dagger={\bf \Pi} -(\alpha'-\alpha){\bf P}_{\rm T}$.
A physical state ${\cal S}$ is represented by the vectors $\ket{\psi}$ and $\ket{\psi'}=R_{\alpha\alpha'}\ket{\psi}$ in the gauges $\alpha$ and $\alpha'$ respectively. Thus, the averages of ${\cal O}$ and ${\cal O}'$ in the state ${\cal S}$ are $\langle {\cal O}\rangle_S = \bra{\psi}{\bf \Pi} \ket{\psi}$ and $\langle {\cal O}'\rangle_S=\bra{\psi'}{\bf \Pi}\ket{\psi'}$. The same operator ${\bf \Pi}$ represents different observables ${\cal O}$ and ${\cal O}'$ in the two averages, whereas different vectors represent the same physical state ${\cal S}$. Of crucial importance is to recognise that both of the above predictions satisfy gauge invariance as defined by Eq.~(\ref{gi}). %However, they are clearly {\em different} physical predictions even though both are averages of an operator ${\bf \Pi}$, which defines the ``light" quantum subsystem.

For fixed $\alpha$ the combination ${\bf \Pi}=-{\bf E}_{\rm T}-\alpha {\bf P}_{\rm T}$ is a gauge invariant observable, but by definition of ${\bf \Pi}$, here $\alpha$ {\em is} the gauge parameter. Thus, while it is true that in each gauge ${\bf \Pi}$ represents a physical observable and while it is also true that every observable possesses unique physical predictions that can be calculated in any gauge, it is {\em not} true that the operator ${\bf \Pi}$ represents the {\em same} physical observable in any two different gauges, and predictions pertaining to different observables are different; for example, two different observables will not generally possess the same average value.
%The physical predictions in Eqs.~(\ref{gi1}) and (\ref{gi2}) are both found using ${\bf \Pi}$ and both are gauge invariant, but they are clearly {\em different}.
%  For example, in a given physical situation, which of the two distinct predictions in Eqs.~(\ref{gi1}) and (\ref{gi2}), both of which are averages in the state ${\cal S}$ of an observable represented by the mathematical ``light" operator $\Pi \equiv I_{\rm matter}\otimes \Pi$, might be the more relevant one?   
As will be discussed throughout the present article, the task of determining which gauge invariant predictions are {\em relevant} in which situations is not necessarily straightforward, because it depends on the interpretation of virtual processes, dressing, and localisation. Thus: \cite{stokes_ultrastrong_2021}:
\begin{itemize}
\item{\em Gauge ambiguities arise not because it is unclear how to obtain gauge invariant predictions, but because it is not always clear which gauge invariant observables are operationally relevant. The gauge invariance of a prediction is necessary but not sufficient to ensure its operational relevance.}
\end{itemize}

On a practical level, simply verifying the fundamental gauge invariance of predictions does not imply that gauge freedom can be ignored. For example, Ref.~\cite{settineri_gauge_2021} (Sec.~V) notes that ``of course detectable subsystem excitations and correlations have to be gauge invariant, since the results of experiments cannot depend on the gauge. On this basis we can define gauge invariant excitations and qubit-field entanglement". We note however, that providing gauge invariant definitions is straightforward and this has never been a problem. Indeed, given the unitarity of gauge fixing transformations, gauge invariance is automatic. ``Ambiguities" occur not because gauge invariance breaks down, but because there are {\em many different} gauge invariant definitions of ``excitations and qubit-field entanglement". The latter can be defined {\em relative} to any gauge (see Sec.~\ref{rel}). Gauge invariance is necessary, but it is not a {\em sufficient} ``basis" for providing physically {\em relevant} theoretical definitions. Any conceptual ambiguities that result from the availability of many different physical definitions can be called ``gauge ambiguities", but they are not due to a breakdown of gauge invariance, which is a fundamental requirement.

\subsection{Definition of subsystem gauge relativity}\label{rel}

We adopt the viewpoint that the relevant definition of any system is determined by experimental capability. Operationally, a ``system" comprises a set of observable properties that can be measured. On the other hand, theoretically there exists a continuous infinity of different gauge invariant transverse fields, all of which are represented by the operator ${\bf \Pi}$. Any of these fields can be used to define a boson called a photon. Mathematically, ``photons" are defined directly in terms of ${\bf \Pi}$ via
\begin{align}\label{phot}
a_\lambda({\bf k}):= {1\over \sqrt{2\omega}} {\bf e}_\lambda({\bf k})\cdot [\omega {\tilde {\bf A}}_{\rm T}({\bf k})+i{\tilde {\bf \Pi}}({\bf k})]
\end{align}
where $\omega:=|{\bf k}|$ and ${\bf e}_\lambda({\bf k})$ is a unit polarisation vector orthogonal to ${\bf k}$ (Fourier transforms are denoted with a tilde). From ${\tilde {\bf \Pi}}= - {\tilde {\bf E}}_{\rm T}-\alpha {\tilde {\bf P}}_{\rm T}$, it is clear that for each different fixed value of $\alpha$ the photon number operator $n=\sum_{{\bf k}\lambda}a_\lambda^\dagger({\bf k})a_\lambda({\bf k}) $ represents a different gauge invariant observable:
\begin{itemize}
\item{Photons defined using the gauge invariant observable ${\cal O}=-{\bf E}_{\rm T}-\alpha {\bf P}_{\rm T}$, which in the gauge $\alpha$ is represented by the operator ${\bf \Pi}$, are said to be defined {\em relative} to the gauge $\alpha$.}
\end{itemize}
The eigenstates of the corresponding number operator $n$ are a basis for the ``light" Hilbert space, which is therefore defined relative to a choice of gauge. We can express this relativity symbolically by writing the subsystem label ``light" or ``photons" as a function of the observable that defines it, for instance, in the gauge $\alpha$ ``light"=${\rm light}({\bf E}_{\rm T}+\alpha{\bf P}_{\rm T})=:{\rm light}_\alpha$ and ``photons"=photons$_\alpha$. As an example, suppose that in a given experiment the observable ${\bf E}_{\rm T}$ is measurable, then in this situation ${\rm light}_0$ is a relevant mathematical subsystem.    It is clear that the relativity described above in the case of photons, applies to any subsystem property defined in terms of the canonical momenta.  To summarise, according to the postulates of quantum theory, QED subsystems are defined relative to a choice of gauge \cite{stokes_gauge_2019}.
%However, doing so does not mean that  gauge ambiguities do not arise.%, contrary to the claim of Ref. \cite{settineri_gauge_2021}.

%Ambiguities arise because it is not always clear that any one {\em definition} of photon, for example the definition using ${\bf E}_{\rm T}$, is the most physically relevant definition. These conceptual difficulties can be avoided in traditional weak-coupling and Markovian regimes, as will be clearly demonstrated in Sec.~\ref{s6}, but they cannot generally be ignored in, for example, the ultrastrong-coupling regime. %Ref.~\cite{stokes_ultrastrong_2021} shows that when dealing with fast time-dependent interaction switching, the relevant definition depends strongly on the experimental context.
%Time-dependent interactions and photodetection theory are discussed in detail in Sec.~\ref{s6}.

\subsection{Implications of subsystem gauge relativity}\label{imps}

Predictions are necessarily gauge-invariant when they pertain entirely to gauge-invariant objects. An example is the mechanical momentum $m{\dot {\bf r}}={\bf p}-q{\bf A}({\bf r})$, which is represented by the gauge-covariant derivative $-i\nabla-q{\bf A}({\bf r})$ when acting on position space wave-functions. \cite{scully_quantum_1997} therefore argue that only this momentum is physical unlike the canonical momentum ${\bf p}$. Similarly, \cite{schwinger_gauge_1951}, favoured the use of only gauge-covariant quantities in the calculation of relativistic vacuum effects. Yet once the gauge has been fixed, every operator within the theory represents an observable that is a known function of manifestly gauge invariant observables (Fig.~\ref{gauge_fix}). Physical predictions will therefore be gauge invariant [see~Eq.~(\ref{gi})] provided approximations that ruin gauge invariance are avoided and that they are calculated properly. For example, when dealing with time-dependent interactions one must of course take into account the time-dependence of gauge transformations, as noted in Refs. \cite{stokes_ultrastrong_2021,settineri_gauge_2021}.  %This is another way to understand the fundamental gauge-invariance of predictions established in Eq.~(\ref{gi}).

\begin{figure}[t]
\begin{minipage}{\columnwidth}
\begin{center}
\includegraphics[scale=0.13]{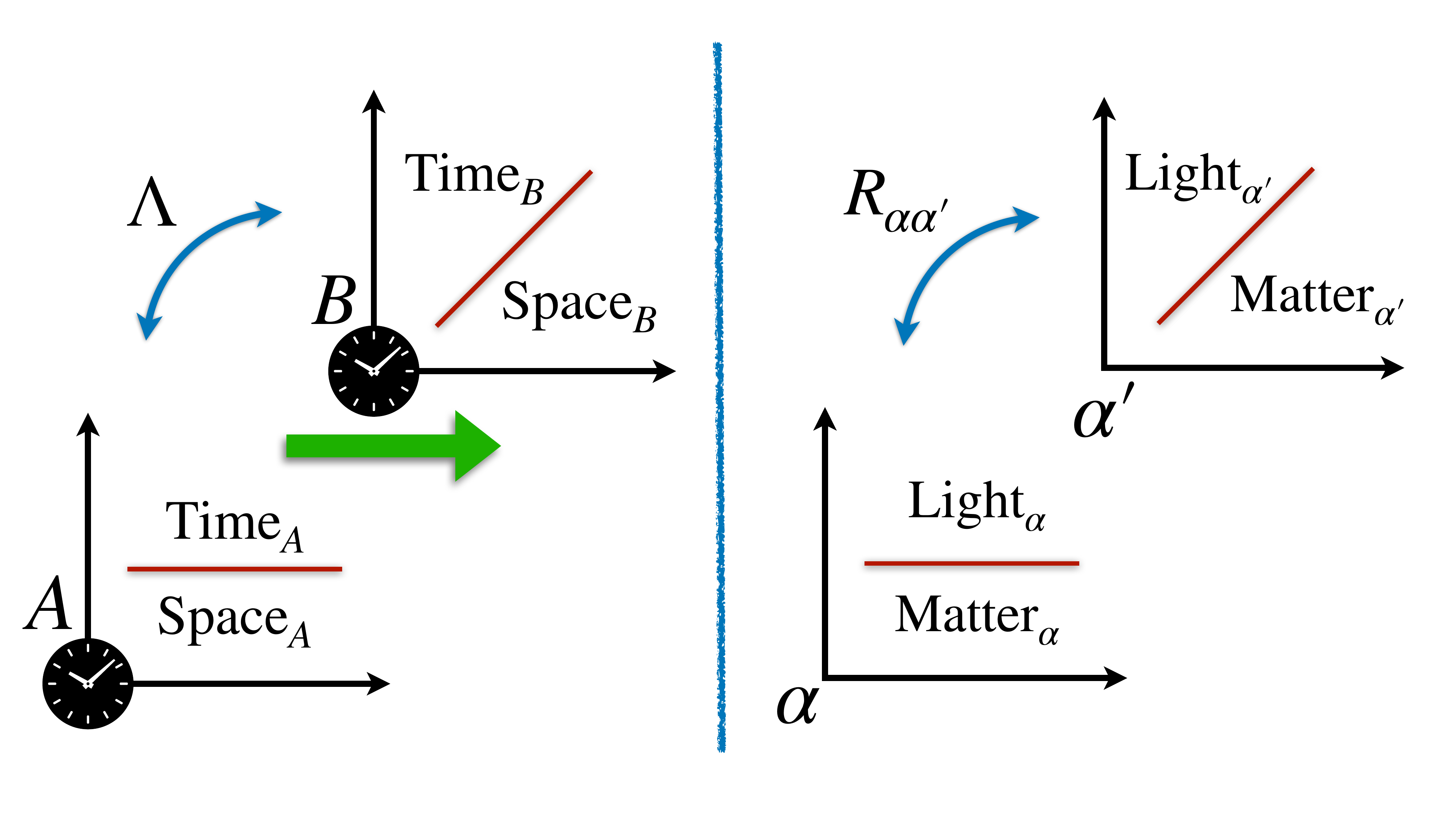}
\vspace*{-9mm}\caption{The analogy between the relativity of space and time when partitioning spacetime, and the relativity of QED subsystems when partitioning the QED Hilbert space. \textbf{Left}: The Lorentz transformation $\Lambda$ mixes space$_X$ and time$_X$ in transforming to the co-moving frame $Y$. The relevant definition of time for the prediction of time intervals measured by a clock at rest in frame $X$, is time$_X$. \textbf{Right}: The unitary gauge fixing transformation $R_{\alpha\alpha'}$ mixes matter$_\alpha$ and light$_\alpha$ in moving to frame $\alpha'$. %As in special relativity, the most relevant definitions will generally depend on the experiment being considered.
}\label{relpic}
\vspace*{-2mm}
\end{center}
\end{minipage}
\end{figure}

Subsystem gauge relativity means that the ``light" and ``matter" quantum subsystems are defined by {\em different} gauge invariant observables in each different gauge. A subsystem property such as the degree of light-matter entanglement constitutes two different gauge-invariant physical predictions when calculated in two different gauges. This is a form of linear-space relativity analogous to that encountered in theories of space and time (Fig.~\ref{relpic}).

%For example, the definition of time provided by inertial frame $X$ is useful in predicting the time interval $\Delta t_X$ between two events as measured by a clock at rest in $X$, but it is not relevant in predicting $\Delta t_Y$ defined as the time interval between the same two events as measured by a clock at rest in inertial frame $Y$. The two predictions coincide, $\Delta t_X\approx \Delta t_Y$, only in the non-relativistic regime wherein time and space intervals can be considered frame-independent, i.e., when their mixing due to Lorentz transformation can be ignored.  Similarly, the definition ${\rm light}_\alpha$ is not directly relevant for predicting the outcome of a measurement of the photons that define ${\rm light}_{\alpha'}$ except in a regime where the mixing of ``light" and ``matter" due to the gauge transformation $R_{\alpha\alpha'}$ can be ignored. This latter situation is often encountered in the regime of weakly-coupled, nearly-resonant, and Markovian systems \cite{stokes_ultrastrong_2021}, as will be demonstrated directly in Sec.~\ref{s6}.

%As will be shown in Sec.~\ref{s6}, all definitions of the subsystems are approximately the same in traditional weak-coupling and Markovian regimes.
Within sufficiently strong-coupling or non-Markovian regimes, the relativity of light and matter quantum subsystems cannot be ignored. %There will only generally exists a subset of theoretical definitions of, for example, ``photon" that yield predictions consistent with values measured in a given experiment. 
Unlike in special relativity, determining which theoretical definition of, for example, a ``photon" is the most relevant one for predicting experimental outcomes is not necessarily straightforward, because the task is intimately related to the interpretation of virtual processes and spacetime localisation properties. It is also far from clear that the most relevant definition of photon is independent of the given experiment.  Ref.~\cite{settineri_gauge_2021}, for example, assumes that a photodetector registers photons defined by the gauge invariant transverse electric field ${\bf E}_{\rm T}$. Given this {\em assumption} about which physical observable is relevant, one can of course calculate the rate of photodetection as a unique physical prediction in any gauge for both time-dependent and time-independent interactions. In Glauber's original theory however, the {\em total} electric field ${\bf E}={\bf E}_{\rm T}+{\bf E}_{\rm L}$ was used \cite{glauber_quantum_1963,milonni_photodetection_1995} and this field is only transverse when there are no charges present. Indeed, %As we have already noted, recognising the distinction between ${\bf E}_{\rm T}$ and ${\bf E}$ is essential and lies at the heart of gauge ambiguities.
it has been argued in the past that the transverse displacement field ${\bf D}_{\rm T}={\bf E}_{\rm T}+{\bf P}_{\rm T}$ provides the a more relevant definition, because its source-component equals the source-component of ${\bf E}$ away from the source, and it is therefore local, unlike ${\bf E}_{\rm T}$ \cite{cohen-tannoudji_photons_1989,biswas_virtual_1990,milonni_photodetection_1995,power_analysis_1997,power_time_1999,power_time_1999-1,sabin_fermi_2011,stokes_noncovariant_2012}. In particular, it has been known for six decades that photons defined relative to the multipolar gauge, i.e., in terms of ${\bf D}_{\rm T}$, are able to provide a natural lineshape prediction that is in sufficient agreement with early experiments to rule out the corresponding prediction for the same experiments when photons are defined using ${\bf E}_{\rm T}$ \cite{power_coulomb_1959,fried_vector_1973,davidovich_theory_1980,milonni_natural_1989,woolley_gauge_2000,stokes_gauge_2013}. For these specific experiments the multipolar gauge subsystems are evidently more operationally relevant than the Coulomb gauge subsystems. %   Clearly the most relevant definitions can always be determined through comparison of predictions with experiment. 
Predictions of radiation spectra are discussed further in Secs.~\ref{line} and \ref{specs}.

%In classical electrodynamics, the most relevant definitions of the subsystems are uncontentious. ``Matter" is such that a material particle is defined in terms of its position and momentum variables ${\bf r}$ and $m{\dot {\bf r}}$, whereas ``light" is defined in terms of the electric field ${\bf E}$ and magnetic field ${\bf B}$. In nonrelativistic QED almost the same ``classical" definitions can be obtained, but only within the EDA and only by using the multipolar gauge whereby ${\bf p}=m{\dot {\bf r}}$ and ${\bf \Pi}=-{\bf D}_{\rm T}=-{\bf E}$ for all ${\bf x}\neq {\bf 0}$. The clarifier ``almost the same"  refers to the fact that at the dipole's own position ${\bf 0}$ we have ${\bf D}_{\rm T}\neq {\bf E}$. We should expect this fact to possess non-trivial implications for radiation-reaction phenomena, which are closely related to virtual processes and vacuum effects \cite{dalibard_vacuum_1982,milonni_quantum_1994}. In particular, the multipolar gauge suffers relatively severely  from ultraviolet photodetection divergences \cite{drummond_unifying_1987,stokes_extending_2012}.

The multipolar gauge, $\alpha=1$ defines a dipole$_1$ that is purely mechanical, i.e., completely ``bare" (see Sec.~\ref{share}). However, one often views physical atoms as being dressed by virtual photons and this is more consistent with definitions provided by $\alpha\neq 1$ whereby the dipole is instead a delocalised dressed object. Only the localised dipole$_1$ does not respond instantaneously to a test charge placed away from its centre at ${\bf 0}$ \cite{cohen-tannoudji_photons_1989,biswas_virtual_1990,milonni_photodetection_1995,power_analysis_1997,power_time_1999,power_time_1999-1,sabin_fermi_2011,stokes_noncovariant_2012}. In gauges $\alpha\neq 1$, the extent of the apparently instantaneous, but typically small response of a test charge distribution to the field of the $\alpha$-gauge dipole could simply be interpreted as a measure of the dressed dipole's delocalisation due to its own virtual cloud of photons \cite{hegerfeldt_causality_1994}. These points are discussed in the context of photodetection theory in Sec.~\ref{pd}.

For given values of the remaining model parameters, it is often possible to choose an intermediate value of $\alpha$ denoted $\alpha_{\rm JC}$, which lies between $0$ and $1$, and for which ground state virtual photons are highly suppressed \cite{stokes_gauge_2019}.   This representation is defined in Secs.~\ref{opt}, \ref{osc}, and \ref{dressingmulti}, where the choice of notation $\alpha_{\rm JC}$ is explained . The representation can be interpreted as one in which virtual photons have been absorbed into the definitions of the quantum subsystems. The physical meanings of the different mathematical definitions of ``light" and ``matter" are evidently closely related to virtual photons and processes.

Finally we note that a prosaic implication of subsystem gauge relativity is that approximations performed on the subsystems can ruin the gauge invariance of the theory. A well-known example is the truncation of the material system to a finite number of levels \cite{stokes_gauge_2019,stefano_resolution_2019,roth_optimal_2019,de_bernardis_breakdown_2018}. Because ``matter" is defined differently in different gauges, the truncation generally constitutes a significantly different physical procedure in different gauges. This is discussed in detail in Sec.~\ref{s1}.

\subsection{Canonical transformations in quantum field theory and unitary inequivalence}

In the preceding development of non-relativistic QED, the gauge-fixing transformations $U_{gg'}$ defined in Eq.~(\ref{ugg'0}) possesses the form $e^{iS}$ with $S$ Hermitian, and it is easy to verify that $U_{gg'}$ preserves the canonical commutation relations. However, the gauge-invariance identified in Sec.~\ref{gamb} comes with a certain caveat, this being that the transformation $U_{gg'}$ is only unitary {\em in form}. By this we mean that although formally $U_{gg'}U_{gg'}^\dagger =I$, establishing rigorously the unitarity of canonical transformations in quantum field theory is non-trivial, because one often encounters generators $S$ that are too poorly behaved to avoid the occurrence of infinite terms during the course of formal manipulations. This point is directly relevant when considering $U_{gg'}$ and so it is discussed briefly below. We follow the intuitive (heuristic) discussion found in \cite{umezawa_advanced_1995}.

Consider the formally unitary transformation 
\begin{align}\label{Uthe}
U[\theta] =\exp\left[\int d^3 k  \sum_\lambda \left(\theta_\lambda ({\bf k})^*a_\lambda ({\bf k})  -\theta_\lambda ({\bf k})a_\lambda^\dagger({\bf k}) \right)\right]
\end{align}
where $\theta$ is an arbitrary function ${\bf k}\to \theta({\bf k})$. Assuming that $[\theta_\lambda({\bf k}),\theta_{\lambda'}({\bf k}')]=0$, then since $a'_\lambda ({\bf k}) = U[\theta] a_\lambda ({\bf k})U[\theta]^\dagger = a_\lambda ({\bf k}) + \theta_\lambda ({\bf k})$, the transformation is canonical, that is, $[a_\lambda ({\bf k}),a^\dagger_\lambda ({\bf k}')]=\delta_{\lambda\lambda'} \delta({\bf k}-{\bf k}') \Leftrightarrow [a_\lambda'({\bf k}),a'^\dagger_\lambda ({\bf k}')]=\delta_{\lambda\lambda'} \delta({\bf k}-{\bf k}')$. Denoting the vacuum annihilated by $a_\lambda({\bf k})$ by $\ket{0}$ and assuming that $\braket{0|0}=1$, we see that according to Eq.~(\ref{Uthe}) the vacuum $\ket{0'}=U[\theta]\ket{0}$ annihilated by $a'({\bf k})$ is also formally normalised, $\braket{0'|0'} = 1$. One finds in addition, however, that
\begin{align}\label{0'}
&\ket{0'} = \tau \exp\left[-\int d^3 k \sum_\lambda \theta_\lambda({\bf k})a_\lambda^\dagger({\bf k})\right]\ket{0}
\end{align}
where
\begin{align}
\tau = \exp\left[-{1\over 2}\int d^3k\sum_\lambda |\theta_\lambda({\bf k})|^2\right],
\end{align}
such that if $\int d^3k |\theta_\lambda({\bf k})|^2 = \infty$, then the prefactor $\tau =\braket{0|0'}$ is vanishingly small. It would then follow that $\braket{\psi|0'}=0$ where $\ket{\psi}$ is any Fock state generated by applying the operators $a^\dagger_\lambda({\bf k})$ to the vacuum $\ket{0}$. From this it would follow that the vacuum $\ket{0'}$ and the Fock states generated from it using the operators $a'^\dagger_\lambda({\bf k})$, cannot be expressed as linear combinations of the Fock states generated using the $a^\dagger_\lambda({\bf k})$ and $\ket{0}$. The two bases are then said to be inequivalent \cite{umezawa_advanced_1995}. 

Let us now turn our attention to the PZW gauge-fixing transformation $R_{01}$, which can be written in the form in Eq.~(\ref{Uthe}) with $\theta_\lambda({\bf k}) = i{\bf e}_\lambda({\bf k}) \cdot {\tilde {\bf P}}({\bf k})/\sqrt{2\omega}$,
where ${\tilde {\bf P}}$ denotes the Fourier transform of the multipolar polarisation. In this case $-\ln \tau = {1\over 2} \bra{0'}\int d^3k \sum_\lambda a_\lambda^\dagger({\bf k})a_\lambda({\bf k}) \ket{0'} = \int d^3k {|{\tilde {\bf P}}_{\rm T}({\bf k})|^2/(4\omega)}$ is half the average number of photons$_0$ in the vacuum $\ket{0'}$. Via the same analysis as is presented above, Woolley finds that for a two-charge system $\tau \to 0$ in the point-charge limit, that is, the vacua of the Coulomb and multipolar-gauges do indeed become orthogonal \cite{woolley_power-zienau-woolley_2020}. Physically, the vacua of the Coulomb-gauge ($\alpha=0$) and multipolar-gauge ($\alpha=1$) must contain an infinite number of photons$_1$ (multipolar photons) and photons$_0$ (Coulomb-gauge photons) respectively, so the two vacua cannot be simultaneously meaningful. It is worth noting that according to a simple second order perturbation theory calculation the ground state of the Hamiltonian in Eq.~(\ref{HA}) contains both photons$_0$ and photons$_1$. Identifying the ground state of the Hamiltonian as the physical vacuum is the underlying idea of the JC-gauge mentioned in Sec.~\ref{imps}. This is discussed further in Secs.~\ref{opt}, \ref{osc}, and \ref{dressingmulti}.

Unitary inequivalence results from the singular nature of ${\bf P}({\bf x})$ within the PZW transformation. This localisation of ${\bf P}({\bf x})=\int d^3k {\tilde {\bf P}}({\bf k})e^{i{\bf k}\cdot {\bf x}}/\sqrt{(2\pi)^3}$ requires that all wavevectors ${\bf k}$ are retained within the Fourier transformation. However, in non-relativistic QED, one can argue {\em a priori} that relativistic modes are not properly described \cite{cohen-tannoudji_photons_1989}. The multiplication of ${\tilde {\bf P}}_{\rm T}({\bf k})$ by a form factor, such as $\ell_M(k)=k_M^2/(k^2+k_M^2)$, as described in Secs.~\ref{FT} and \ref{share}, removes the contributions of relativistic wavevectors, such that $\int d^3 k |{\tilde {\bf P}}_{\rm T}({\bf k})|^2<\infty$. Similarly, $\braket{0|0'}$ only vanishes in the point-charge limit \cite{woolley_power-zienau-woolley_2020}, yet the elimination of relativistic wavevector contributions to the point-charge density $\rho$ is equivalent to considering extended charge distributions, which yield a more rigorously well-defined quantum theory \cite{spohn_dynamics_2004}. %As noted in Sec.~\ref{share}, using form factors to temper ${\bf P}_{\rm T}$ and $\rho$ are not in general two equivalent procedures, but either procedure can be used to regularise the generator of a gauge-fixing transformation, and in particular to render $\braket{0|0'}$ non-zero.
In this article we consider formally unitary gauge-fixing transformations and assume that gauge-invariance as defined in Sec.~\ref{gamb} holds, but with the understanding that when dealing with quantum fields strict unitarity may require invoking suitable regularisation procedures.

We note finally that although after suitable regularisation that gives $\braket{0|0'} \neq 0$, the vacua $\ket{0}$ and $\ket{0'}$ contain only a {\em finite} number of the photons$_1$ and photons$_0$ respectively, the two vacua clearly remain physically distinct and it remains to determine which, if either, is relevant in a given situation. We remark also that as is discussed in more detail in Sec.~\ref{s6}, the gauge non-relativistic property of the QED $S$-matrix under only formally unitary gauge-fixing transformations $U=e^{iS}$ can be proved quite generally \cite{craig_molecular_1998,woolley_gauge_2000}. These points demonstrate that while the occurrence of unitarily inequivalent representations of the CCR algebra is of importance with regards to the technical challenge of establishing the strict {\em gauge-invariance} of predictions, it is of far less importance with regards to the occurrence or otherwise of {\em gauge-relativity}.

\subsection{Modal restrictions and transversality}

Restrictions on the number of photonic modes are extremely common in light-matter physics. However, retaining all modes is necessary to maintain spacetime localisation and causal wave propagation. In particular, the Green's function for the wave operator receives contributions from all ${\bf k}$-space-modes. A modal restriction should be understood as a statement about which particular frequencies are dominant within a given light-matter interaction Hamiltonian.%We briefly discuss the implementation of such restrictions.

\subsubsection{Significance of transversality}

We begin by noting that the transversality of canonical fields is closely related to gauge freedom. Only transverse fields can be used to define unconstrained physical photons as in Eq.~(\ref{phot}). This feature is fundamental and it persists in the presence of background media (see Supplementary Note~VII), as are relevant in numerous artificial photonic systems that realise large coupling strengths \cite{todorov_ultrastrong_2010,ciuti_input-output_2006,ciuti_quantum_2005,bamba_dissipation_2012,bamba_recipe_2014,bamba_system-environment_2013}.   Relativistic particles can be specified via the unitary representations of the Poincar\'e group \cite{bargmann_group_1948}, which are labelled by two numbers, ``mass" $m\geq 0$ and integer or half odd integer ``spin" $s$.  Massless fields possess only two independent helicities $-s,+s$ obtained from the projection of the spin $s$ onto the axis of particle motion \cite{hassani_mathematical_2013}. In particular, the massless spin-1 Maxwell field supports the two independent polarisations of a photon. Scalar and longitudinal photons can also be defined, as in the Lorenz gauge \cite{lorenz1867}, but such photons are not unconstrained. They satisfy a non-dynamical constraint (Lorenz subsidiary condition), whose derivative in time is Gauss' law \cite{cohen-tannoudji_photons_1989}.

Gauss' law generates gauge symmetry transformations and its derivative in time is the continuity equation for electric charge, which is the conserved quantity associated with gauge symmetry. It specifies ${\bf E}_{\rm L}$ as a function of $\rho$, telling us that longitudinal photons are not independent. Specifically, an analog of Eq.~(\ref{phot}) may be written
\begin{align}\label{photL}
a_{\rm L}({\bf k})&:= -{i\over \sqrt{2\omega}} {\hat {\bf k}}\cdot {\tilde {\bf E}}({\bf k})  = -{{\tilde \rho}({\bf k})\over \sqrt{2\omega^3}}.
\end{align}

Although ${\bf E}_{\rm T}$ is the part of the electric field not constrained by Gauss' law, it is by fundamental assumption that the {\em total} electric field ${\bf E}$ is local. %The fundamental assumption of Maxwell electrodynamics is that the fields ${\bf E}$ and ${\bf B}$ assign electromagnetic properties to each (local) event $(t,{\bf x})$ in spacetime.
It follows that the fields ${\bf E}_{\rm L}$ and notably ${\bf E}_{\rm T}={\bf E}-{\bf E}_{\rm L}$, are both non-local [see Eq.~(\ref{tldelt})] and away from a localised source they respond instantaneously to changes in the source \cite{cohen-tannoudji_photons_1989,craig_molecular_1998}. The multipolar gauge momentum ${\bf \Pi}=-{\bf D}_{\rm T}$ offers the best possible representation of the non-transverse local field ${\bf E}$ by an unconstrained transverse field that can then be used to define unconstrained photons \cite{cohen-tannoudji_photons_1989}. Specifically, ${\bf P}_{\rm L}=-{\bf E}_{\rm L}$ implies that ${\bf D}:={\bf E}+{\bf P}= ({\bf E}_{\rm T}+{\bf E}_{\rm L})+({\bf P}_{\rm T}-{\bf E}_{\rm L})\equiv{\bf D}_{\rm T}$ and since ${\bf P}$ vanishes outside of a charge distribution we have ${\bf D}\equiv {\bf D}_{\rm T}={\bf E}$ at all such points. It is certainly not the case however that ${\bf E}={\bf E}_{\rm T}$ nor that ${\bf P}={\bf P}_{\rm T}$.

In the case of a dipole at ${\bf 0}$ the multipolar polarisation is ${\bf P}=q{\bf r}\delta({\bf x})$ whereas ${\bf P}_{\rm T}({\bf x})=q{\bf r}\cdot \delta^{\rm T}({\bf x})$. The transverse dyadic $\delta^{\rm T}({\bf x})$ is not purely singular, rather it decays as $1/x^3$ away from ${\bf 0}$. %The transverse and longitudinal components of a field are non-local projections as defined in Eq.~(\ref{tldelt}).
From elementary electrostatics we know that ${\bf E}_{\rm L}$ decays as $1/x^3$ away from a dipole at ${\bf 0}$ and for a dipole we do indeed have ${\bf P}_{\rm T}={\bf E}_{\rm L}$ for ${\bf x}\neq {\bf 0}$ (i.e., ${\bf P}({\bf x})=q{\bf r}\delta({\bf x})=0$ for ${\bf x}\neq {\bf 0}$). %The Coulomb gauge field canonical momentum is ${\bf \Pi}=-{\bf E}_{\rm T}=-{\bf E}+{\bf E}_{\rm L}\neq {\bf E}$.
For any $\alpha$ the field ${\bf \Pi}$ can be expanded in terms of photons using Eq. (\ref{phot}). Crucially however, for different $\alpha$ these fields are related by the {\em non-local} field ${\bf P}_{\rm T}$. 

For a transverse field, the mode functions ${\bf f}_\lambda({\bf k},{\bf x}) = {\bf e}_\lambda({\bf k})e^{i{\bf k}\cdot {\bf x}}/\sqrt{(2\pi)^3}$ of a canonical mode-expansion are not complete with respect to the usual inner-product in $L^2({\mathbb R}^3)$, because $\{{\bf e}_\lambda({\bf k})\}$ is an orthonormal basis in the two-dimensional plane orthogonal to ${\bf k}$. They instead furnish a representation of the {\em transverse} delta function;
\begin{align}
\int d^3 k \sum_{\lambda=1,2} {\bf f}_\lambda({\bf k},{\bf x})^*{\bf f}_\lambda({\bf k},{\bf x}') = \delta^{\rm T}({\bf x}-{\bf x}').
\end{align}
To obtain a representation of $\delta({\bf x}-{\bf x}')$ one must include the vector ${\hat {\bf k}}$ in Fourier space, to obtain the 3-dimensional basis $\{{\hat {\bf k}},{\bf e}_\lambda({\bf k})\}$. %such that subsequently summing over modes provides a representation of $\delta({\bf x}-{\bf x}')$. and then define the longitudinal mode function ${\bf f}_{\rm L}({\bf k},{\bf x}) = {\hat {\bf k}}e^{i{\bf k}\cdot {\bf x}}/\sqrt{(2\pi)^3}$ such that
%\begin{align}
%\int d^3 k \sum_{\lambda=1,2,{\rm L}} {\bf f}_\lambda({\bf k},{\bf x})^*{\bf f}_\lambda({\bf k},{\bf x}') = {\bf I}\delta({\bf x}-{\bf x}') 
%\end{align}
%where ${\bf I}$ is the identity in ${\mathbb R}^3$.
If the longitudinal eigenfrequency is set to vanish $\omega_{\rm L}\equiv 0$ then one can of course expand ${\bf \Pi}$ using the complete set of mode functions. 
% as
%\begin{align}
%{\bf \Pi}({\bf x}) = i\int d^3 k& \sum_{\lambda=1,2,{\rm L}} \sqrt{\omega_\lambda\over 2} \nonumber \\ \times &\left[{\bf f}_\lambda({\bf k},{\bf x})^*a^\dagger_\lambda ({\bf k})+{\bf f}_\lambda({\bf k},{\bf x})a_\lambda({\bf k})\right]
%\end{align}
%where $\omega_{1,2}\equiv \omega$. 
However, the operators $a_{\rm L}({\bf k})$ have completely arbitrary definition and cannot contribute to physical predictions.

\subsubsection{Modal restriction}\label{modalres}

Ultrastrong-coupling between light and matter arises in artificial systems in which the set of photonic modes is altered and often restricted. Theoretically, care must be taken when carrying out such restrictions. To demonstrate this we choose the multipolar gauge, such that ${\bf \Pi}=-{\bf D}_{\rm T}$, implying that the Coulomb gauge momentum $-{\bf E}_{\rm T}$ is represented by the operator ${\bf \Pi}'=-{\bf E}_{\rm T}=R_{10}{\bf \Pi}R_{10}^\dagger={\bf \Pi}+{\bf P}_{\rm T}$. Coulomb and multipolar gauge {\em transverse} photonic operators $a'_\lambda({\bf k})$ and $a_\lambda({\bf k})$ are defined as in Eq.~(\ref{phot}) using ${\bf \Pi}'=-{\bf E}_{\rm T}$ and ${\bf \Pi}=-{\bf D}_{\rm T}$ respectively. They are therefore related by
\begin{align}\label{phrel}
a'_\lambda({\bf k})=R_{10}a_\lambda ({\bf k})R_{10}^\dagger= a_\lambda ({\bf k}) +i{q{\bf r}\cdot {\bf f}_\lambda({\bf k},{\bf 0})\over  \sqrt{2\omega_\lambda}}.
\end{align}
For the unphysical longitudinal mode operators any relation can be specified. We note however, that the right-hand-side of Eq.~(\ref{phrel}) would be undefined for $\lambda={\rm L}$, because $\omega_{\rm L}\equiv 0$. %, therefore Eq. (A13) of Ref.~\cite{settineri_gauge_2021} is ill-defined for longitudinal modes.
The total electric field is given by ${\bf E}={\bf D}_{\rm T}-{\bf P}=-{\bf \Pi}-{\bf P} = -{\bf \Pi}'-{\bf P}_{\rm L}$ and ${\bf P}({\bf x})=q{\bf r}\delta({\bf x})$ is fully localised. The electric field ${\bf E}$ is completely independent of the $a_{\rm L}({\bf k})$, as any physical field must be.

When the modes are confined to a volume $v$ with periodic boundary conditions the mode functions become discrete ${\bf f}_\lambda({\bf k},{\bf x})\to{\bf f}_{{\bf k}\lambda}({\bf x})$ such that factors of $(2\pi)^3$ are replaced by $v$. For a field ${\bf F}$ the component associated with the wavevector ${\bf k}$ or mode ${\bf k}\lambda$ can be read-off by expressing ${\bf F}$ as ${\bf F}({\bf x}) = \sum_{\bf k} {\bf F}_{\bf k} = \sum_{{\bf k}\lambda} {\bf F}_{{\bf k}\lambda}$. For the transverse and longitudinal polarisation fields we have
%\begin{align}\label{ET}
%{\bf \Pi}_{\bf k}({\bf x}) =&-{\bf D}_{\rm T{\bf k}}({\bf x})=  i\sum_{\lambda=1,2}\sqrt{\omega \over 2}\left({\bf f}_{{\bf k}\lambda} ({\bf x})^*a^\dagger_{{\bf k}\lambda} -{\rm H.c.}\right),\\
%{\bf \Pi}'_{\bf k}({\bf x}) =&-{\bf E}_{\rm T{\bf k}}({\bf x})=  i\sum_{\lambda=1,2}\sqrt{\omega \over 2}\left({\bf f}_{{\bf k}\lambda} ({\bf x})^*a'^\dagger_{{\bf k}\lambda} -{\rm H.c.}\right)\\
${\bf P}_{\rm T{\bf k}}({\bf x}) = q{\bf e}_{{\bf k}\lambda}{{\bf e}_{{\bf k}\lambda}\cdot {\bf r} \over v}\cos{{\bf k}\cdot {\bf x}}$ and
 ${\bf P}_{\rm L{\bf k}}=-{\bf E}_{\rm L{\bf k}} = q{\hat {\bf k}}{({\hat {\bf k}}\cdot {\bf r})\over v}\cos({\bf k}\cdot {\bf x})$ respectively, such that the restricted total polarisation is ${\bf P}_{{\bf k}}({\bf x}) = {q{\bf r}\over v}\cos({\bf k}\cdot {\bf x})$. For the total electric field we have ${\bf E}_{\bf k}({\bf x}) =-{\bf \Pi}_{\bf k}({\bf x})-{\bf P}_{\bf k}({\bf x})$. These single-mode restriction can be implemented at the position ${\bf 0}$ of a single dipole via the $\alpha$-gauge theory presented in Secs.~\ref{s0} and \ref{eda}. Since all algebraic and kinematic relations are preserved so too is gauge invariance. The dipole approximated fields in Eqs.~(\ref{Aeda}) and (\ref{PTeda}) are assumed to point in the direction ${\bm \varepsilon}$ of the mode polarisation and in this direction have components \cite{stokes_gauge_2019,stokes_uniqueness_2020}
\begin{align}
&A_\alpha = (1-\alpha)A = {1-\alpha\over \sqrt{2\omega v}}(a^\dagger +a)\label{Aa}\\
&P_{\rm T\alpha} ={\alpha qx\over v}.\label{1mpt}
\end{align}
Here $x={\bm \varepsilon}\cdot {\bf r}$ and $A= {\bm \varepsilon}\cdot {\bf A}_{\rm T}({\bf 0})$ where ${\bm \varepsilon}$ is the unit polarisation vector of the single transverse mode retained. The Hamiltonian reduces to a simple form that has now been used in a number of works \cite{stokes_gauge_2019,stokes_ultrastrong_2021,stokes_uniqueness_2020,stefano_resolution_2019,roth_optimal_2019} (see Sec.~\ref{smapp}). The gauge fixing transformations in Eq.~(\ref{Rgteda}) remain unitary, becoming $R_{\alpha\alpha'} = \exp (i [\alpha-\alpha']q xA)$ \cite{stokes_gauge_2019,stokes_ultrastrong_2021}.
%The {\em definition} of gauge freedom continues to be given by Eqs. (\ref{mina}) and (\ref{minb}), which now read
%\begin{align}
%&R_{\alpha\alpha'}pR_{\alpha\alpha'}^\dagger = p-(\alpha-\alpha')qA,\label{min1}\\
%&R_{\alpha\alpha'}\Pi R_{\alpha\alpha'}^\dagger =\Pi-(\alpha-\alpha'){qr\over v}\label{min2}
%\end{align}
%where $\Pi = {\bm \varepsilon}\cdot {\bf \Pi}_{\rm T}$. Note also that $Uf(O)U^\dagger=f(UOU^\dagger)$ for any unitary transformation $U$, suitably well-defined function $f$, and operator $O$. Therefore, Eqs. (\ref{min1}) and (\ref{min2}) are necessary and sufficient to define how arbitrary (suitably well-behaved) functions of $p$ and $\Pi$ must transform under a gauge transformation.

The restriction to a finite-number of modes within the Hamiltonian of a light-matter system must evidently be understood as an assumption about which modes are dominant within the dipole-field interaction. This may be valid at the position of the dipole centre, ${\bf 0}$, in the form ${\bf V}({\bf 0})=\sum_{\bf k} {\bf V}_{\bf k}({\bf 0})\approx {\bf V}_{\bf k}({\bf 0})$. However, the dipole's centre ${\bf 0}$ is also where the field cannot be measured by an external detector. For any ${\bf x}$ the field ${\bf E}_{\bf k}$ equals neither $-{\bf E}_{\rm T{\bf k}}$ nor $-{\bf D}_{\rm T{\bf k}}$. Due to Gauss' law the electric-field, whether restricted or not, cannot be expressed solely in terms of physical (transverse) photons. In particular, since ${\bf \Pi}_{\bf k}$ is orthogonal to ${\bf k}$, one cannot obtain ${\bf E}_{\bf k}$ by means of a unitary operator acting on ${\bf \Pi}_{\bf k}$.

Obviously the fully localised physical polarisation ${\bf P}({\bf x})= \sum_{\bf k} {\bf P}_{\bf k}({\bf x})$ cannot be elicited in a restricted space of wavevectors. A modal restriction at an arbitrary point ${\bf x}\neq {\bf 0}$ will therefore violate the property ${\bf P}_{\rm T}=-{\bf P}_{\rm L}$ of the full theory. Naively restricting the polarisation and electric fields to only one transverse mode ${\bf k}\lambda$ means ${\bf P}_{\rm L{\bf k}}({\bf x})\equiv {\bf 0}$ and we obtain ${\bf E}_{{\bf k}}({\bf x})\equiv -{\bf \Pi}'_{{\bf k}\lambda}({\bf x})=-{\bf \Pi}_{{\bf k}}({\bf x})-{\bf P}_{\rm T{\bf k}}({\bf x})$. This yields a theory without ${\bf E}_{\rm L}$, that can therefore only be valid in the far-field. Of course, in the far-field  where ${\bf E}_{\rm L}={\bf P}_{\rm T}$ vanishes, we have $-{\bf \Pi}'={\bf E}_{\rm T}\approx {\bf E}=-{\bf \Pi}$ whether or not the modes are restricted. If we instead use the fact that ${\bf \Pi}({\bf x})=-{\bf E}({\bf x})$ for ${\bf x}\neq {\bf 0}$ and then restrict our attention to one transverse mode, we obtain the different result ${\bf E}_{{\bf k}}({\bf x})\equiv -{\bf \Pi}_{{\bf k}}({\bf x}) = -{\bf \Pi}'_{{\bf k}}({\bf x})+{\bf P}_{\rm T{\bf k}}({\bf x})$. This single-mode limit respects the equalities ${\bf E}=-{\bf \Pi}=-{\bf \Pi}'+{\bf P}_{\rm T}$ holding for ${\bf x}\neq {\bf 0}$ in the unrestricted theory. Within the light-matter interaction Hamiltonian fields are evaluated at ${\bf x}={\bf 0}$ so these considerations do not apply.

Evidently, different implementations of a modal restriction can result in altogether different identifications of the same physical field, such that care must be taken. In the above case of the electric field ${\bf E}$ we have fundamentally that at all points ${\bf x}$ outside of a charge distribution, which is where the field can be measured by an external detector, the multipolar polarisation vanishes, implying that at such points ${\bf \Pi}({\bf x})=-{\bf E}({\bf x})$ in and only in the multipolar gauge. We should not expect a modal restriction in which this is no longer the case to offer a generally robust approximation of the unrestricted theory for describing measurements involving ${\bf E}({\bf x})$. In particular, the Glauber intensity at $(t,{\bf x})$ is given within the single-mode limit that respects the fundamental equalities of the multi-mode theory by
\begin{align}
\langle {\bf E}^{(-)}_{{\bf k}\lambda}(t,{\bf x})\cdot {\bf E}^{(+)}_{{\bf k}\lambda}(t,{\bf x})\rangle = {\omega\over 2v}\langle a^\dagger_{{\bf k}\lambda}(t)a_{{\bf k}\lambda}(t)\rangle
\end{align}
where $a_{{\bf k}\lambda}$ is the multipolar gauge photonic operator. Irrespective of modal restrictions, the Glauber intensity is not proportional to the photon number operator defined relative to the Coulomb gauge except in the far-field where ${\bf E}\approx {\bf E}_{\rm T}$. Photodetection is discussed in more detail in Sec.~\ref{s6}.

\subsection{Simple extension to superconducting circuits}

The arbitrary-gauge formalism is readily adapted to describe circuit QED systems, which we now briefly review.   Ref.~\cite{vool_introduction_2017} provides an introductory review of circuit QED, while a more recent review is Ref.~\cite{blais_circuit_2021}.  Conventional descriptions of superconducting circuits employ the lumped-element model, which results from Kirchoff's assumptions applied to Maxwell theory. Consider a node defined as the meeting point of $N$ conducting wire branches outside of which there is no current. Bounding the node is a closed surface ${\mathscr S}$ containing a region $v$ with outward normal ${\hat {\bf n}}$. The continuity equation $\partial_\nu j^\nu=0$ and divergence theorem yield
\begin{align}
&\sum_{\mu=1}^N I_\mu(t) \equiv \sum_{\mu=1}^N \int_{{\mathscr S}_\mu} dS\,  {\hat {\bf n}} \cdot {\bf J}(t,{\bf x}) = - {dQ(t)\over dt},\label{kirch1}\\ &Q(t) = \int_v d^3x \, \rho(t,{\bf x}),
\end{align}
where ${\mathscr S}_\mu$ is the subsurface of ${\mathscr S}$ intersecting the $\mu$'th wire, $I_\mu$ is the current entering $v$ through the $\mu$'th wire, and $Q(t)$ is the total charge within the region $v$ containing the node. Eq.~(\ref{kirch1}) assumes that ${\bf J}(t,{\bf x})=0$ for all ${\bf x} \in {\mathscr S}/ \bigcup_\mu {\mathscr S}_\mu$ (there is no current outside the conducting wires). Kirchoff assumed further a local steady-state current condition within $v$, namely, $d Q(t)/dt = 0$, yielding the current law
\begin{align}\label{kirch2}
\sum_{\mu=1}^N I_\mu(t) = 0.
\end{align}

\begin{figure}[t]
\begin{minipage}{\columnwidth}
\begin{center}
\hspace*{-1mm}\includegraphics[scale=0.4]{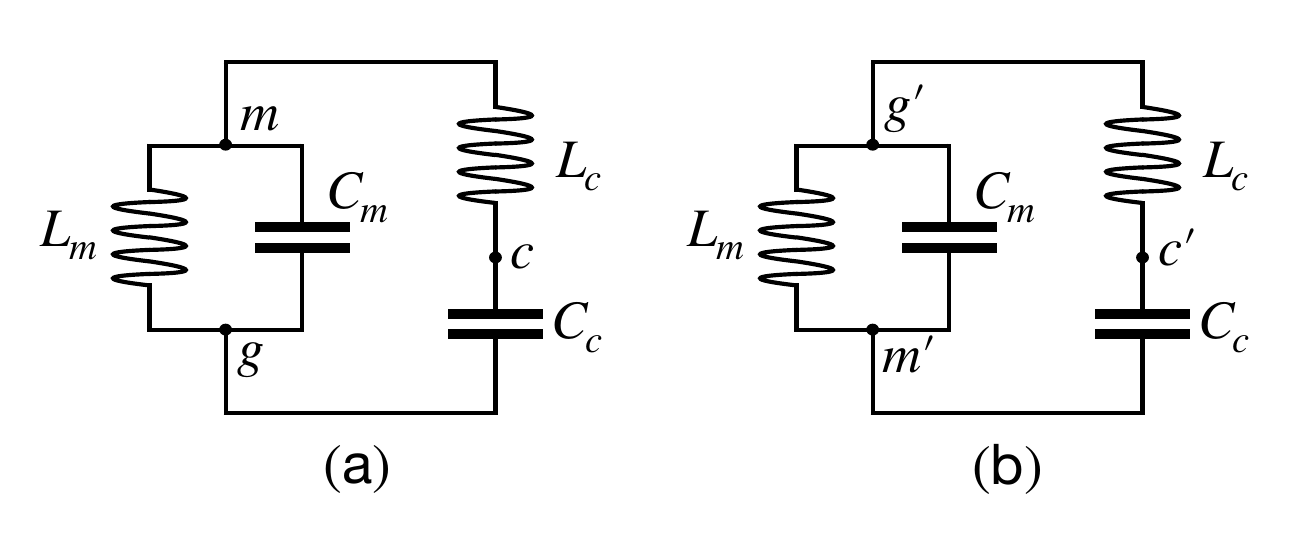}
\caption{Circuit diagram for a parallel $LC$-oscillator coupled to a series $LC$-oscillator. There are three nodes within the circuit. The subfigures each provide a different labelling of the nodes corresponding to different specifications of the ground flux. As a result, they depict two different divisions of the circuit into subsystems. Specifically, these are the two extreme cases of (\textbf{a}) fully inductive coupling whereby the ground flux is specified as the flux associated with the node that is labelled $g$, and (\textbf{b}) fully capacitive coupling whereby the ground flux is specified as the flux associated with the node that is labelled by $g'$.}\label{circ}
\vspace*{-4mm}
\end{center}
\end{minipage}
\end{figure}

Arbitrary lumped-element circuits can be considered as collections of nodes joined by (super)conducting branches, with Kirchoff's law, Eq,~(\ref{kirch2}), satisfied at each node. As a non-trivial example we consider the coupled $LC$-oscillator circuit depicted in Fig.~\ref{circ}. As basic dynamical variables we take the node fluxes denoted $\phi_k$. The current into node $k$ through a branch $j\to k$ with an inductor connecting node $k$ to node $j$ is $I_{j\to k} = (\phi_k-\phi_j)/L$ where $L$ is the inductance of the inductor.  The current into node $k$ through a branch $j\to k$ with a capacitor connecting node $k$ to node $j$ is $I_{j\to k} = C({\ddot \phi}_k-{\ddot \phi}_j)$ where $C$ is the capacitance of the capacitor. Since only flux differences are of importance we can specify the flux zero-point arbitrarily. This is the so-called ground flux such that $\phi_g=0$. As particular special cases, we can choose this flux zero-point to be the flux of one of the circuit nodes depicted in Fig.~\ref{circ} wherein subfigures (\textbf{a}) and (\textbf{b}) give two different specifications of which node possesses the ground flux.

%%%%%%%%%%%%%%%%%%%%%%%%%%%%%%%%%%%%%%%%%%%%%%%%%%%%%%%%%
%%%%%%%%%%%%%%%%%%%%%%%%%%%%%%%%%%%%%%%%%%%%%%%%%%%%%%%%%

In the circuit of Fig.~\ref{circ}\,(\textbf{a}) there are two non-ground nodes labelled $m$ and $c$. Kirchoff's law, Eq,~(\ref{kirch2}), yields the equations of motion
\begin{align}
0&= I_{g\to m}+I_{c \to m} = C_m {\ddot \phi}_m + {\phi_m \over L_m} + {\phi_m-\phi_c \over L_c}, \label{k1}\\
0&= I_{g\to c}+I_{m \to c} = C_c {\ddot \phi}_c + {\phi_c-\phi_m \over L_c}.\label{k2}
\end{align} 
These equations of motion are obtained from the Lagrangian
\begin{align}
L = {1\over 2}\left[C_m {\dot \phi}_m^2 -{\phi_m^2\over L_m} + C_c{\dot \phi}_c^2 -{(\phi_c-\phi_m)^2\over L_c}\right]\label{Lc1}
\end{align}
or corresponding Hamiltonian
\begin{align}
H = {1\over 2}\left[{q_m^2\over C_m} +{\phi_m^2\over L_m} + {q_c^2\over C_c} + {(\phi_c-\phi_m)^2\over L_c}\right]\label{Hc1}
\end{align}
where $q_x = \partial L/\partial {\dot \phi}_x$ are the node charges conjugate to the $\phi_x$ with $x=m,\,c$. A node flux and its conjugate charge satisfy a canonical Lie bracket relation, which generates the dynamics in conjunction with the Hamiltonian. In particular, in the quantum theory, $[\phi_x,q_{x'}]=i\delta_{xx'}$.

Let us now consider a relabelling of the nodes as depicted in Fig.~\ref{circ}\,(\textbf{b}). The ground node has flux $\phi_{g'}=0$ and the non-ground nodes $m'$ and $c'$ are now connected by the capacitance $C_c$ rather than by the inductance $L_c$. Since the physical currents through the branches must stay the same we obtain the coordinate relations
\begin{align}
\phi_{m'}&=-\phi_m,\\
\phi_{c'}&= \phi_c -\phi_m.
\end{align}
Either $\phi_m$ or $\phi_{m'}$ can be used as a coordinate with $\phi_{c'}$. We choose $\phi_m$. The sum of Eqs.~(\ref{k1}) and (\ref{k2}) can be expressed as
\begin{align}
0=C_m {\ddot \phi}_m +C_c ({\ddot \phi}_m+{\ddot \phi}_{c'}) + {\phi_m\over L_m},
\end{align}
and this equation together with Eq.~(\ref{k2}) is obtained from the Lagrangian (\ref{Lc1}) or Hamiltonian (\ref{Hc1}) with $\phi_m$ and $\phi_{c'}$ taken as dynamical coordinates. At the Hamiltonian level the primed and unprimed canonical operators are related by a gauge fixing transformation as
\begin{align}
q_{m'} &= R_{10}q_m R_{10}^\dagger,\\
\phi_{c'}&= R_{10}\phi_c R_{10}^\dagger
\end{align}
where $R_{10} := e^{-i q_c \phi_m}$. This is analogous to the PZW transformation between the charge (Coulomb)-gauge and the flux (multipolar)-gauge.

Note that within the above derivation we have adopted a passive view of rotations within the operator algebra (see Sec.~\ref{active}) , by which we mean that the same Hamiltonian has been expressed in terms of alternative canonical operators that belong to different gauges. Equivalently, we may adopt an active perspective as in previous sections, whereby the Hamiltonian $H$ is actively rotated using gauge fixing transformations yielding new Hamiltonians which are all expressed in terms of the same canonical operators. The extension to arbitrary gauges is straightforward via the the gauge fixing transformation $R_{\alpha\alpha'} := e^{-i(\alpha-\alpha') q_c \phi_m}$. We note that gauges specified by $\alpha\neq 0,\,1$ do not correspond to a definite specification of one of the nodes within Fig.~\ref{circ} as possessing the ground flux $\phi_g=0$. Instead the ground flux is specified as some combination of the fluxes associated with the three nodes.

The basic non-linear element in superconducting circuits is the Josephson junction   \cite{josephson_possible_1962}. These junctions are typically realised using two conducting materials separated by a thin gap of insulator. Quantum mechanically, electron tunnelling across the junction is possible, with the tunnelling charge flowing in units of Cooper pairs as $Q=2qN$ where $N$ denotes the number of Cooper pairs on one side of the junction. The junction Hamiltonian is
\begin{align}
H_J = -{E_J\over 2}\sum_N (\ket{N}\bra{N+1}+\ket{N+1}\bra{N})
\end{align}
where the energy $E_J$ determines the coupling strength across the junction. Introducing the phase variable $\phi_m$ conjugate to $Q$ through Fourier transformation as $\ket{\phi_m}=\sum_{N=-\infty}^\infty e^{2iq\phi_m N}\ket{N}$, one can express the junction Hamiltonian as $H_J=-E_J\cos[2q\phi_m]$.

The formalism above is easily extended to arbitrary circuits constructed from capacitors, inductors and Josephson junctions. For example, by adding a Josephson junction connecting the ground node $g$ to the node $m$ in Fig.~\ref{circ}, one obtains the light-matter Hamiltonian
\begin{align}
H' = H - E_J \cos [2q \phi_m].
\end{align}
The Hamiltonian $H'$ possesses the same structure as the cavity QED Hamiltonian considered in Sec.~\ref{sp} in which the material potential is arbitrary.

We have seen that the choice of gauge is determined by the choice of ground flux, and that arbitrary choices of gauge selected by a parameter $\alpha$ may be considered. Gauge fixing transformations are directly analogous to those encountered in conventional QED, and as such, they are non-local with respect to Hilbert space tensor-product structure. A circuit may be dividing into physically distinct canonical sub-circuits arbitrarily and this division is directly controlled by the choice of gauge.

\section{Material truncation and gauge noninvariance}\label{s1}
 
Material energy level truncation is a commonly adopted procedure, that nevertheless breaks the gauge invariance of QED by fundamentally modifying the algebra of material operators. This has been discussed in the context of strong and ultrastrong-coupling in Refs. \cite{stokes_gauge_2019,stokes_ultrastrong_2021,stokes_uniqueness_2020,stefano_resolution_2019,roth_optimal_2019,de_bernardis_breakdown_2018,settineri_gauge_2021,de_bernardis_cavity_2018,stokes_gauge_2020,taylor_resolution_2020,ashida_cavity_2021}. Here we review the implications of the resulting gauge noninvariances, which as was explained in Sec.~\ref{gamb} are not synonymous with gauge ambiguities. We review various proposed theoretical approaches for obtaining truncated models.
 
\subsection{Single dipole interacting with a single cavity mode}%: the Hamiltonian via unitary transformations}
\label{smapp}

%Ref.~\cite{stefano_resolution_2019} noted that in the case of a single atom and a single radiation mode in one spatial dimension, within the EDA the Hamiltonian can be obtained by combining unitary transformations of the free material and photonic Hamiltonians. This property was then used to derive a certain class of two-level models. We review various classes of two-level model in Sec.~\ref{mt}. 

The EDA and single-mode approximation can be performed preserving all algebraic properties of the theory, thereby preserving gauge invariance \cite{stokes_gauge_2019,stokes_ultrastrong_2021,stokes_uniqueness_2020}. The dipole is assumed to be located at the origin ${\bf 0}$ and for simplicity the canonical operators are assumed to point in the direction ${\bm \varepsilon}$ of polarisation of the the single mode. %They are specified entirely by scalar operator components in this direction. 
We define $x={\bm \varepsilon}\cdot {\bf r}$ %for ease of comparison with Ref.~\cite{stefano_resolution_2019}
and $A={\bm \varepsilon}\cdot {\bf A}_{\rm T}$ and denote by $p$ and $\Pi$ the corresponding dipole and cavity canonical momenta, such that $[x,p]=i$ and $[A,\Pi]=i/v$ with $v$ the cavity volume. Details of the EDA and single-mode restriction are given in Secs. \ref{eda} and \ref{modalres} respectively.

The $\alpha$-gauge continues to be specified by its vector potential ${\bf A}_\alpha={\bm \varepsilon}A_\alpha$ and material polarisation ${\bf P}_{\rm T\alpha}={\bm \varepsilon}P_{\rm T\alpha}$ which are given by Eqs.~(\ref{Aa}) and (\ref{1mpt}) respectively. %The unitary gauge fixing transformation $R_{\alpha\alpha'}$ between gauges $\alpha$ and $\alpha'$ is given by Eq.~(\ref{trans}).
The definition of gauge freedom given by Eqs.~(\ref{mina}) and (\ref{minb}) now reads
\begin{align}
&R_{\alpha\alpha'}pR_{\alpha\alpha'}^\dagger = p-(\alpha-\alpha')qA,\label{min1c}\\
&R_{\alpha\alpha'}\Pi R_{\alpha\alpha'}^\dagger =\Pi-(\alpha-\alpha'){qx\over v}\label{min2c}.
\end{align}
Since gauge fixing transformations remain unitary the gauge invariance of the theory is preserved. %Were Eqs. (\ref{min1c}) and (\ref{min2c}) not satisfied, then $R_{\alpha\alpha'}$ would not effect the replacement $(A_\alpha,P_{\rm T\alpha})\to (A_{\alpha'},P_{\rm T\alpha'})$, meaning it would not be a gauge transformation from $\alpha$ to $\alpha'$.
The Hamiltonian is as ever the total energy %which is the sum of material mechanical and transverse electromagnetic energies 
\cite{stokes_gauge_2019,stokes_ultrastrong_2021};
\begin{align}
&H_\alpha = {\cal H}_{m}(A_\alpha)+{\cal H}_{\rm ph,\alpha}\label{edaham} \\
&{\cal H}_{m}(A_\alpha):={1\over 2}m{\dot x}^2+V(x) = {1\over 2m}\left(p-qA_\alpha\right)^2+V(x),\label{mechen1} \\ 
&{\cal H}_{\rm ph,\alpha}:= {v\over 2}(E_{\rm T}^2+\omega A^2) = {v\over 2}\left[(\Pi+P_{\rm T\alpha})^2+\omega^2A^2\right],\label{photen1}
\end{align}
where ${\dot x}=-i[x,H_\alpha]$ and $E_{\rm T}=-{\dot A}_{\rm T}= i[A_{\rm T},H_\alpha]$. All three energies are gauge invariant;
\begin{align}
&{\cal X}_{\alpha'}=R_{\alpha\alpha'}{\cal X}_{\alpha}R_{\alpha\alpha'}^\dagger,\label{mechg}
%&{\cal H}_{\rm ph,\alpha'}=R_{\alpha\alpha'}{\cal H}_{\rm ph,\alpha}R_{\alpha\alpha'}^\dagger.\label{phg}
\end{align}
where ${\cal X}_\alpha = H_\alpha,\,{\cal H}_{m}(A_\alpha),\,{\cal H}_{\rm ph,\alpha}$. 
%Hamiltonians of different gauges continue to be unitarily related as in Eq.~(\ref{halp}). 
Note also that as discussed in Sec.~\ref{eda}, within (and only within) the EDA the $\alpha$-gauge mechanical momentum may be obtained from the canonical momentum ${\bf p}$ using $R_{1\alpha}$. For $\alpha=1$, Eq.~(\ref{mechg}) with ${\cal X}={\cal H}_{m}(A_\alpha)$ then has the appearance of a unitary transformation applied to the free material Hamiltonian, Eq.~(\ref{Hmdef}), as \cite{stefano_resolution_2019}
\begin{align}\label{halpm2}
{\cal H}_{m}(A_\alpha) = R_{1\alpha}H_m R_{1\alpha}^\dagger.
\end{align}
This holds in and only in the EDA.

The transverse electromagnetic energy can be written similarly as ${\cal H}_{\rm ph,\alpha} = R_{0\alpha}H_{\rm ph}R_{0\alpha}^\dagger$ where $H_{\rm ph}={\cal H}_{\rm ph,0}={v\over 2}(\Pi^2+\omega^2A^2)$. We see therefore that within the present simplified setting the Hamiltonian can be written
\begin{align}\label{halp2}
H_\alpha = R_{1\alpha}H_m R_{1\alpha}^\dagger + R_{0\alpha}H_{\rm ph} R_{0\alpha}^\dagger.
\end{align}
This is an approximate special case of the more general and fundamental expression
\begin{align}\label{halp3}
H_\alpha = R_{\alpha'\alpha}{\cal H}_{m}(A_{\alpha'}) R_{\alpha'\alpha}^\dagger + R_{\alpha''\alpha}{\cal H}_{\rm ph,\alpha''} R_{\alpha''\alpha}^\dagger,
\end{align}
which follows immediately from Eqs.~(\ref{mina}) and (\ref{minb}). Eq.~(\ref{halp3}) reduces to Eq.~(\ref{halp2}) when we choose $\alpha'=1$ and $\alpha''=0$, and we make use of ${\cal H}_{\rm ph,0} =H_{\rm ph}$ and ${\cal H}_{m}(A_1)=H_m$, which holds only because of the approximations and simplifying assumptions made. It should be noted that without the latter, the derivation of ${\cal H}_{m}(A_\alpha)$ via unitary transformation of $H_m$ is impossible.

%The $\alpha$-gauge Hamiltonian in Eq.~(\ref{HA}) retains the same basic structure and is now given by \cite{stokes_gauge_2019,stokes_ultrastrong_2021}
%\begin{align}
%H_\alpha = {1\over 2m}\left(p-qA_\alpha\right)^2+U(x) +{v\over 2}\left[(\Pi+P_{\rm T\alpha})^2+\omega^2A^2\right],
%\end{align}
%such that Hamiltonians of different gauges continue to be unitarily related as in Eq.~(\ref{halp}).
%The EDA accommodates a further property noted in Ref.~\cite{stefano_resolution_2019}, which is that the Hamiltonian of any gauge $\alpha$ can be derived by applying gauge fixing transformations to the free Hamiltonians $H_m$ and $H_{\rm ph}$;
%\begin{align}\label{halp2}
%H_\alpha = R_{1\alpha}H_m R_{1\alpha}^\dagger + R_{0\alpha}H_{\rm ph} R_{0\alpha}^\dagger
%\end{align}
%where
%\begin{align}\label{bareen}
%H_m&={p^2\over 2m}+U(x),\\
%H_{\rm ph}&={v\over 2}(\Pi^2+\omega^2A^2).
%\end{align}
%We note that this procedure will only work within the EDA, so Eq.~(\ref{halp2}) is not a fundamental feature of QED.

\subsection{Material truncation}\label{mt}

Let us now consider truncating the material Hilbert space \cite{stokes_gauge_2019,stokes_ultrastrong_2021,stokes_uniqueness_2020,stefano_resolution_2019,roth_optimal_2019,de_bernardis_breakdown_2018,settineri_gauge_2021,de_bernardis_cavity_2018,stokes_gauge_2020,taylor_resolution_2020}. Since the canonical momentum $p$ represents a different physical observable for each different value of $\alpha$, the same is true of $H_m$. Therefore, projecting onto a finite number of eigenstates of $H_m$ is a gauge-dependent procedure. Eigenvalues of $H_m$ are denoted $\epsilon_n$. The projection $P$ onto the first two-levels $\ket{\epsilon^0},~\ket{\epsilon^1}$ of $H_m$ gives $P H_m P= \omega_m \sigma^+ \sigma^- +\epsilon_0$ and $P qxP =d\sigma^x$ where $\sigma^+ = \ket{\epsilon^1}\bra{\epsilon^0}$, $\sigma^- = \ket{\epsilon^0}\bra{\epsilon^1}$ and $\sigma^x=\sigma^+ +\sigma^-$. The first transition energy is denoted $\omega_m=\epsilon_1-\epsilon_0$, and the transition dipole moment $d=\bra{\epsilon^0}qx\ket{\epsilon^1}$ is assumed to be real. More generally, $P$ may project onto any finite number of levels.

There are many ways to define two-level models. In general, truncation of $H_\alpha$ is a $P$-dependent map $M_P: H_\alpha\to M_P(H_\alpha)$, such that $M_P(H_\alpha) : P{\cal H} \to P{\cal H}$ is an Hermitian operator on $P{\cal H}$ \cite{stokes_gauge_2020}. If, unlike the $H_\alpha$, the $M_P(H_\alpha)$ are not equivalent for different $\alpha$, then truncation has broken the gauge invariance of the theory. To obtain what we will refer to as the standard $\alpha$-gauge two-level model one replaces $x$ and $p$ with their projected counterparts $P x P$ and $P p P$ to obtain 
\begin{align}\label{2lmst}
 M_P(H_\alpha)=H_\alpha^2=  PH_mP+PH_{\rm ph}P +V^\alpha(PxP,PpP)
\end{align}
where $V^\alpha(x,p)=H_\alpha-H_m-H_{\rm ph}$ is the interaction Hamiltonian. The terminology ``standard" is used because this definition of $M_P$ is capable of yielding the standard quantum Rabi model (QRM) that is ubiquitous in light-matter physics. Specifically, a standard QRM is obtained by choosing $\alpha=1$ in Eq.~(\ref{2lmst}). More generally, for distinct values of $\alpha$ the Hamiltonians $H^2_\alpha $ are not equivalent to each other \cite{stokes_gauge_2019,de_bernardis_breakdown_2018,stokes_ultrastrong_2021}, because $P$ represents a different physical projection in each different gauge.

Of crucial importance when defining two-level models is recognition that for a Hermitian operator $O$, projection $P\neq I$, and non-linear function $f$ we have
\begin{align}\label{lem2}
Pf(O)P \neq f(POP).
\end{align}
Thus, for a general material operator $O(x,p)$ we have $PO(x,p)P \neq O(PxP,PpP)$. This becomes an equality if and only if $O$ is linear in $x$ and $p$ \cite{stokes_gauge_2019}. As a result, various alternative truncating maps have been identified within the literature \cite{stokes_gauge_2019,stefano_resolution_2019,de_bernardis_breakdown_2018,settineri_gauge_2021,de_bernardis_cavity_2018,stokes_gauge_2020,taylor_resolution_2020}.

Two further methods have been proposed in Ref.~\cite{stefano_resolution_2019} [see also Ref.~\cite{taylor_resolution_2020}]. Both methods require the EDA and involve replacing the unitary transformation $R_{\alpha\alpha'}$ in Eq. (\ref{halp2}) with a two-level model counterpart. %However, the argumentation of Ref.~\cite{stefano_resolution_2019} is incorrect, because it erroneously equates what are actually two quite different two-level model versions of $R_{\alpha\alpha'}$. The error is identified by observing that for a Hermitian operator $O$, projection $P\neq I$, and non-linear function $f$ we have
%\begin{align}\label{lem2}
%Pf(O)P \neq f(POP).
%\end{align} 
There are two {\em different} two-level model versions of $R_{\alpha\alpha'}:{\cal H}\to {\cal H}$, which are defined as
\begin{align}
&{\cal G}_{\alpha\alpha'} = PR_{\alpha\alpha'}P = P\exp[iq(\alpha-\alpha')xA]P \\
&{\cal T}_{\alpha\alpha'} =  \exp[iq(\alpha-\alpha')PxPA]\neq {\cal G}_{\alpha\alpha'} \label{tcal}
\end{align}
where the final inequality holds because $e^{PxP}\neq Pe^xP$ [see Eq.~(\ref{lem2})]. Moreover, we cannot expect this inequality to become an approximate equality even for highly anharmonic material systems. An arbitrary operator O that is not necessarily diagonal in momentum space is defined by
\begin{align}
[O\psi](p,A)= \int dp' dA' \,O(p,p',A,A')\psi(p',A')
\end{align}
where $\psi$ is the wave function of the composite system represented in momentum space for the matter subsystem and in position space ($A$ space) for the photonic mode. It is straightforward to show that $R_{\alpha\alpha'}$ enacts a gauge transformation of the momentum arguments of $O$ as
\begin{align}
[R_{\alpha\alpha'}OR_{\alpha\alpha'}^\dagger \psi](p,A)= &\int dp' dA' O(p-q[\alpha-\alpha']A, p'\nonumber \\ &-q[\alpha-\alpha']A',A,A')\psi(p',A')
\end{align}
We may write this more succinctly using the shorthand notation $R_{\alpha\alpha'}OR_{\alpha\alpha'}^\dagger = p-q(\alpha-\alpha')A$ in which it is to be understood that the gauge transformation applies to both momentum arguments of a generally nondiagonal operator. Since here both $O$ and $\psi$ are arbitrary, these results apply in particular to a projected operator $F=POP$ and a projected vector $P\psi$. Furthermore, since ${\cal G}_{\alpha\alpha'} = PR_{\alpha\alpha'} P$ and $P = P^2$, it follows again using shorthand notation that
\begin{align}
{\cal G}_{\alpha\alpha'}F(p){\cal G}_{\alpha\alpha'}^\dagger = PF( p-q(\alpha-\alpha')A)P
\end{align}
Therefore, ${\cal G}_{\alpha\alpha'}$ implements a gauge transformation [as defined by Eqs.~(\ref{min1c}) and (\ref{min2c})] within a projected operator and then reprojects the result. By replacing $R_{\alpha\alpha'}$ in Eq.~(\ref{halp2}) [or (\ref{halp3})] with ${\cal G}_{\alpha\alpha'}$ one obtains a new kind of two-level model
\begin{align}\label{halpP}
{\tilde H}_\alpha^2 =& {\cal G}_{1\alpha}PH_mP{\cal G}_{1\alpha}^\dagger+  {\cal G}_{0\alpha}PH_{\rm ph}P {\cal G}_{0\alpha}^\dagger.
\end{align}
These models are not equivalent for different $\alpha$. 
%The $\alpha=0$ case
%\begin{align}\label{cg0}
%{\tilde H}_0^2 = {\cal G}_{10}PH_mP{\cal G}_{10}^\dagger + PH_{\rm ph}P
%\end{align}
%in particular, is claimed to be the ``correct" Coulomb gauge two-level model in Ref. \cite{stefano_resolution_2019}. This derivation of ${\tilde H}_0^2$ is described as constituting the ``correct application of the gauge principle", the idea being that non-localities introduced by the projection of the atomic potential in $H_m$ have now been properly accounted for. 

The other two-level model transformation ${\cal T}_{\alpha\alpha'}$ which is given in Eq. (\ref{tcal}) is clearly unitary (unlike ${\cal G}_{\alpha\alpha'}$), but it does not implement a gauge change   [in the sense of Eqs.~(\ref{min1c}) and (\ref{min2c})]  even when considering a projected operator $F(p)=POP$;
\begin{align}\label{nprojmin}
{\cal T}_{\alpha\alpha'}F(p) {\cal T}_{\alpha\alpha'} \neq PF(p-(\alpha-\alpha')qA)P.
\end{align}
A two-level model unitary transformation cannot implement the minimal coupling replacement $p\to p-qA$, because the required operator algebra cannot be supported by the truncated space    \cite{weyl_quantenmechanik_1927}. In general, the unitary transformations $R_{\alpha\alpha'}$, ${\cal G}_{\alpha\alpha'}$, and ${\cal T}_{\alpha\alpha'}$ (trivially) coincide in (and only in) the limit $P\to I$, which is the limit of no truncation.  

%The operator ${\cal T}_{\alpha\alpha'}$ has the form of a polaron transformation.
By replacing $R_{\alpha\alpha'}$ in Eq.~(\ref{halp2}) [or (\ref{halp3})] with ${\cal T}_{\alpha\alpha'}$ one obtains the two-level models
\begin{align}\label{halpP2}
h_1^2(\alpha) = {\cal T}_{1\alpha}PH_mP{\cal T}_{1\alpha}^\dagger+  {\cal T}_{0\alpha}PH_{\rm ph}P {\cal T}_{0\alpha}^\dagger ={\cal T}_{1\alpha}H_1^2{\cal T}_{1\alpha}^\dagger
\end{align}
where the second equality shows that these models are equivalent to the standard multipolar gauge QRM $H_1^2$. In particular, $h_1^2(1)=H_1^2$.   We note that the entire class $\{h_1^2(\alpha)\}$ results from truncation within the multipolar gauge \cite{stokes_gauge_2020} (see also Sec.~\ref{ss3}) and so we refer to this class as a multipolar gauge equivalence class. As will be discussed in Sec.~\ref{ss3}, the transformations ${\cal T}_{\alpha\alpha'}$ refer to a phase-invariance principle defined entirely {\em within} a truncated space in terms of $x_P=PxP\neq x$.  

Although it is clear that ${\cal T}_{\alpha\alpha'} \neq {\cal G}_{\alpha\alpha'}$ it is instructive to consider how the associated two-level models in Eqs.~(\ref{halpP}) and (\ref{halpP2}) differ. Defining dimensionless coupling parameter $\eta=d/\sqrt{2\omega v}$, and ${\bar x}=\bra{\epsilon^0}x\ket{\epsilon^1}=d/q$, if we assume that $PxQ \ll PxP$ where $Q=I-P$, and we neglect terms $PxQ$ and $QxP$ in the exponent of $R_{10}$ then we obtain
\begin{align}\label{ap2}
{\cal G}_{10}&\approx P \exp\left[i\eta(\sigma^x+QxQ/{\bar x})(a^\dagger +a)\right]P \nonumber \\ &= P \exp\left[i\eta\sigma^x(a^\dagger +a)\right]P ={\cal T}_{10}
\end{align}
However, as already noted, such a naive approximation cannot be justified, even for a sufficiently anharmonic material system. To see this note that by employing this approximation and then following exactly the same steps as above one obtains ${\cal T}_{\alpha\alpha'} \approx PR_{\alpha\alpha'}$. From this one obtains $H_0^2\approx h_1^2(0)$ where the left-hand-side is the {\em standard} Coulomb gauge Rabi model and the right-hand-side is equivalent to the standard multipolar gauge Rabi model $H_1^2$. Since it is known that the spectra of $H_0^2$ and $H_1^2$ are markedly different \cite{stefano_resolution_2019,stokes_gauge_2019,de_bernardis_breakdown_2018}, it follows that in general, one cannot neglect terms $PxQ$ and $QxP$ in the exponent of $R_{\alpha\alpha'}$ even for highly anharmonic material systems. The multipolar gauge models $h_1^2(\alpha)$ are indeed very different from ${\tilde H}_0^2$, exemplifying the importance of inequality (\ref{lem2}) \cite{stokes_gauge_2020}.

The approximate equality ${\cal T}_{\alpha\alpha'} \approx {\cal G}_{\alpha\alpha'}$ does result if the exponentials on both sides are expanded to linear order in $q$. In this case the two-level models ${\tilde H}_\alpha^2$ are then the same as the models $h_1^2(\alpha)$ and they must be equivalent to each other for different $\alpha$. However, a first order expansion of the model $h_1^2(\alpha)$ simply gives back the standard two-level model $H_\alpha^2$ with quadratic terms neglected. It follows that in the weak-coupling regime {\em all} two-level models are the same ${\tilde H}_\alpha^2=h_1^2(\alpha)=H_\alpha^2$. This is the only regime in which such an equivalence can generally be obtained. 

As the coupling strength increases the first order expansion in $q$ becomes progressively worse, so ${\cal T}_{\alpha\alpha'}$ and ${\cal G}_{\alpha\alpha'}$ become progressively different. Thus, if a particular gauge's truncation were found to be accurate for some particular observable in some particular situation, then as the coupling strength increases, truncation in any other gauge could be expected to become progressively less accurate by comparison. The relative optimality of different two-level models is discussed in Sec ~\ref{opt}.

\subsection{Phase invariance with respect to truncated position}\label{ss3}

Supplementary Note 1 of Ref.~\cite{stefano_resolution_2019} provides an alternative derivation of the multipolar equivalence class $\{h_1^2(\alpha)\}$ via the imposition of a phase invariance principle defined using the truncated operator $x_P:=PxP$. More generally, as shown in Ref.~\cite{stokes_gauge_2020}, this principle can be applied in any gauge $\alpha$ and it yields an equivalence class $\{h_\alpha^2(\alpha')\}$.

In the first quantised-setting the gauge principle asserts that the mechanical energy ${\cal H}_{m}(A_\alpha)$ in Eq.~(\ref{mechen1}) satisfies local phase invariance (gauge invariance)
\begin{align}\label{pic2b}
\bra{\psi}{\cal H}_{m}(A_\alpha)\ket{\psi} = \bra{\psi'}{\cal H}_{m}(A'_\alpha)\ket{\psi'}
\end{align}
where $\ket{\psi'}=e^{iq\chi}\ket{\psi}$ and $A'_\alpha=A_\alpha+\nabla \chi$. In particular, the equality $\bra{\psi_\alpha}{\cal H}_{m}(A_\alpha)\ket{\psi_\alpha} = \bra{\psi_{\alpha'}}{\cal H}_{m}(A_{\alpha'})\ket{\psi_{\alpha'}}$ in which $\ket{\psi_{\alpha'}}=R_{\alpha\alpha'}\ket{\psi_\alpha}$, expresses gauge invariance within the $\alpha$-gauge framework and is a special case of Eq.~(\ref{pic2b}) obtained by letting $\chi = \chi_{\alpha'}-\chi_\alpha$.

To define the class $\{h_1^2(\alpha)\}$, the gauge fixing transformation $R_{1\alpha}$ was replaced with ${\cal T}_{1\alpha}$ in Eq.~(\ref{halpm2}) and the multipolar gauge mechanical energy ${\cal H}_m(A_1)=H_m$ was replaced with its projection $P{\cal H}_m(A_1)P$. More generally however, Eqs.~(\ref{halpm2}) and ${\cal H}_m(A_1)=H_m$ are special cases of Eqs.~(\ref{mechg}) and (\ref{mechen1}) respectively. If we replace $R_{\alpha\alpha'}$ with ${\cal T}_{\alpha\alpha'}$ and ${\cal H}_{m,\alpha}(A)$ with ${\cal H}^2_{m}(A_\alpha):=P{\cal H}_{m}(A_\alpha)P$ on the right-hand-side of Eq.~(\ref{mechg}), then we obtain a truncated $\alpha'$-``gauge" mechanical energy;
\begin{align}\label{tpi1}
{\mathscr H}^2_{m,\alpha}(A_{\alpha'}):={\cal T}_{\alpha\alpha'}{\cal H}^2_{m}(A_\alpha){\cal T}_{\alpha\alpha'}^\dagger.
\end{align}
This truncated energy satisfies a form of phase invariance analogous to Eq.~(\ref{pic2b}) but defined with respect to the truncated position operator $x_P:=PxP$. The phase transformation is defined by
\begin{align}\label{phas}
U_{x_P}=e^{iq\chi(x_P)} =e^{i\beta} e^{id\Lambda \sigma^x}
\end{align}
where $\beta$ and $\Lambda$ are constants depending on the choice of function $\chi$. The global phase $e^{i\beta}$ can be ignored. Letting $\ket{\psi_2}=P\ket{\psi}$ denote an arbitrary truncated state we have
\begin{align}\label{tpi2}
&\bra{\psi_2}{\mathscr H}^2_{m,\alpha}(A_{\alpha'})\ket{\psi_2} = \bra{{\psi_2}'}{\mathscr H}^2_{m,\alpha}(A'_{\alpha'})\ket{{\psi_2}'}
\end{align}
where $A'_{\alpha'}=A_{\alpha'}+\partial_{x_P}\chi(x_P)=A_{\alpha'}+\Lambda$ and $\ket{{\psi_2}'}=U_{x_P}\ket{\psi_2}$. Thus, we see that ${\mathscr H}^2_{m,\alpha}(A_{\alpha'})$ is the mechanical energy of the $\alpha'$-``gauge" where here the term ``gauge" does not possess the same meaning as in the non-truncated theory but instead refers to $x_P$-phase invariance within the $\alpha$-gauge truncated mechanical energy. Subsequently, a ``gauge"-transformation of $A_{\alpha'}$ under this principle is $A_{\alpha'}'=A_{\alpha'}+\Lambda$.

To obtain the complete $\alpha'$-dependent Hamiltonian one adds the transverse electromagnetic energy, ${\cal H}_{\rm ph,\alpha'}$, defined in Eq.~(\ref{photen1}), to the mechanical energy. This gives the total energy. Noting that $E_{\rm T}=-\Pi -\alpha' d \sigma^x/v =- \Pi-P_{\rm T\alpha'}$ is the transverse electric field after truncation, the truncated transverse electromagnetic energy ${\mathscr H}_{\rm ph,\alpha'}^2$ may be defined as
\begin{align}\label{phtf}
{\mathscr H}^2_{{\rm ph},\alpha'} &:= {v\over 2}\left[\left(\Pi+{\alpha' d \sigma^x\over v}\right)^2+\omega^2 A^2\right] \nonumber \\ &= {\cal T}_{\alpha\alpha'}{\mathscr H}^2_{{\rm ph},\alpha}{\cal T}_{\alpha\alpha'}^\dagger = {\cal T}_{0\alpha'}H_{\rm ph}{\cal T}_{0\alpha'}^\dagger.
\end{align} 
The second equality in Eq.~(\ref{phtf}) follows from the fact that unlike when acting on $p$, the transformation ${\cal T}_{\alpha\alpha'}$ has the same effect as a gauge transformation when acting on $\Pi$, because truncation does not alter the algebra of photonic operators. Combining Eqs.~(\ref{tpi1}) and (\ref{phtf}) we may now define the full $\alpha'$-dependent two-level model as the total energy 
\begin{align}
h_\alpha^2(\alpha') = {\mathscr H}^2_{m,\alpha}(A_{\alpha'}) + {\mathscr H}^2_{{\rm ph},\alpha'} = {\cal T}_{\alpha\alpha'}H_\alpha^2 {\cal T}_{\alpha\alpha'}^\dagger.
\end{align}
Thus, the equivalence class $\{h_\alpha^2(\alpha')\}$ can be obtained as the class of Hamiltonians satisfying $x_P$-phase invariance after truncation within the $\alpha$-gauge. The particular class $\{h_1^2(\alpha)\}$ derived in Refs.~\cite{stefano_resolution_2019,settineri_gauge_2021,taylor_resolution_2020} is the special case resulting when the $x_P$-phase invariance principle is applied to the multipolar gauge truncated theory. This has the appearance of an application to the free theory only due to approximations that have implied that $A_1 \equiv 0$ so that $p-qA_1=p$, and therefore that ${\cal H}_m(A_\alpha) \equiv R_{1\alpha}{\cal H}_m(A_1)R_{1\alpha}^\dagger =R_{1\alpha} H_m R_{1\alpha}^\dagger$. 
\subsection{Relating models belonging to different equivalence classes}

Further insight into the nature of the models $h_\alpha^2(\alpha')$ may be obtained by asking how any given standard two-level model must be modified in order that it coincides with the standard two-level model found using a different gauge. For example, let us consider the term $q^2{\bf A}^2/(2m)$ of the Coulomb gauge Hamiltonian. The coefficient $q^2/(2m)$ satisfies the Thomas-Reiche-Kuhn (TRK) sum rule 
\begin{align}\label{id}
\sum_n \omega_{nl}d_{nl}^i d_{ln}^j = i{q^2\over 2m}\bra{\epsilon^l}[p_i,r_j]\ket{\epsilon^l} = \delta_{ij}{q^2\over 2m}.
\end{align}
This result rests directly on the CCR algebra which as already noted can only be supported in an infinite-dimensional Hilbert space. Eq.~(\ref{id}) is independent of the level $l$ appearing on the left-hand-side. However, if on the left-hand-side we restrict ourselves to two levels $n,\,l =0,\,1$ with energy difference $\omega_m$, then for $l=1$ Eq.~(\ref{id}) reads
\begin{align}\label{a}
\sum_n \omega_{n1 } d_{n1}^id_{1n}^j = -\omega_m d_{10}^id_{01}^j
\end{align}
whereas for $l=0$ Eq.~(\ref{id}) reads
\begin{align}\label{b}
\sum_n \omega_{n0} d_{n0}^id_{0n}^j = +\omega_m d_{10}^id_{01}^j.
\end{align}
The result obtained now clearly depends on whether $l$ is the ground or excited state. As first noted in Refs.~\cite{stokes_gauge_2019,stokes_master_2018}, if one takes the two-level projection of the Coulomb gauge self-energy term, namely, $q^2{\bf A}^2(\ket{\epsilon^0}\bra{\epsilon^0}+\ket{\epsilon^1}\bra{\epsilon^1})/2m$, and one applies Eqs.~(\ref{a}) and (\ref{b}) to the excited state projection $q^2\ket{\epsilon^1}\bra{\epsilon^1}/(2m)$ and the ground state projection $q^2\ket{\epsilon^0}\bra{\epsilon^0}/(2m)$ respectively, then one arrives at the following modified term, which now constitutes a non-trivial light-matter interaction;
\begin{align}\label{modAsq}
{q^2\over 2m}{\bf A}^2   \leftrightarrow  -\omega_m ({\bf d}\cdot {\bf A})^2 \sigma^z
\end{align}
where ${\bf d}:={\bf d}_{10}$ and $\sigma^z = \ket{\epsilon^1}\bra{\epsilon^1}-\ket{\epsilon^0}\bra{\epsilon^0}$. As noted in Ref.~\cite{stefano_resolution_2019} the modification (\ref{modAsq}) is ad hoc. It results in a model that no longer has the interaction of the Coulomb gauge. However, to order $q^2$ the model obtained does coincide with the multipolar gauge model $h_1^2(0)$ \cite{stefano_resolution_2019}. In this sense the truncated ``gauge"-principle can reveal what non-unitary modifications are required in order to relate non-equivalent truncated theories.

As already noted, at order $q$ all two-level models are equivalent without any modification. At order $q^2$, forcing equivalence requires a non-unitary modification of at least one of the models involved.    The modification (\ref{modAsq}) suffices to give the Coulomb gauge model $H_0^2$ from the model $h_1^2$, if and only if all higher order terms in the expansion of $h_1^2$ in powers of the coupling strength are neglected. This shows  that as the coupling strength increases, increasingly drastic non-unitary modifications will be needed to transform a given model into one that belongs to a different equivalence class. This perspective is another way to understand the increasing difference with increasing coupling strength, between the transformations $PR_{\alpha\alpha'}$ and ${\cal G}_{\alpha\alpha'}$, and the rotation ${\cal T}_{\alpha\alpha'}$.

\subsection{Representing observables after truncation}\label{notgt}
 
It has been argued within the literature that the transformation ${\cal T}_{10}$ constitutes a two-level model gauge transformation and that since ${\cal T}_{10}$ is unitary, this resolves any gauge noninvariance due to truncation \cite{stefano_resolution_2019,settineri_gauge_2021,taylor_resolution_2020}. However, the inequality (\ref{nprojmin}) states that ${\cal T}_{1\alpha}$ does not implement a gauge change,   as defined by Eqs.~(\ref{min1c}) and (\ref{min2c}),  when acting on (projected) functions of $p$. The action of ${\cal T}_{\alpha\alpha'}$ coincides with that of the gauge transformation $R_{\alpha\alpha'}$ followed by projection $P$, only when acting on operators that commute with $R_{\alpha\alpha'}$ (functions of $x$ and $A$) and linear functions of $\Pi$, for which it is clear that $PR_{\alpha\alpha'}\Pi R_{\alpha\alpha'}^\dagger P = {\cal T}_{\alpha\alpha'}\Pi {\cal T}_{\alpha\alpha'}^\dagger$. As first shown in Ref.  \cite{stokes_gauge_2020} (reviewed in Sec.~\ref{ss3}), the invariance of the models related by ${\cal T}_{\alpha\alpha'}$ is $x_P$-phase invariance   [as defined by Eq.~(\ref{pic2b})]  rather than gauge invariance   [as defined by Eq.~(\ref{gi})]. This is not merely a matter of semantics but an important mathematical distinction, as discussed below. 

\begin{figure}[t]
\begin{minipage}{\columnwidth}
\begin{center}
\includegraphics[scale=0.13]{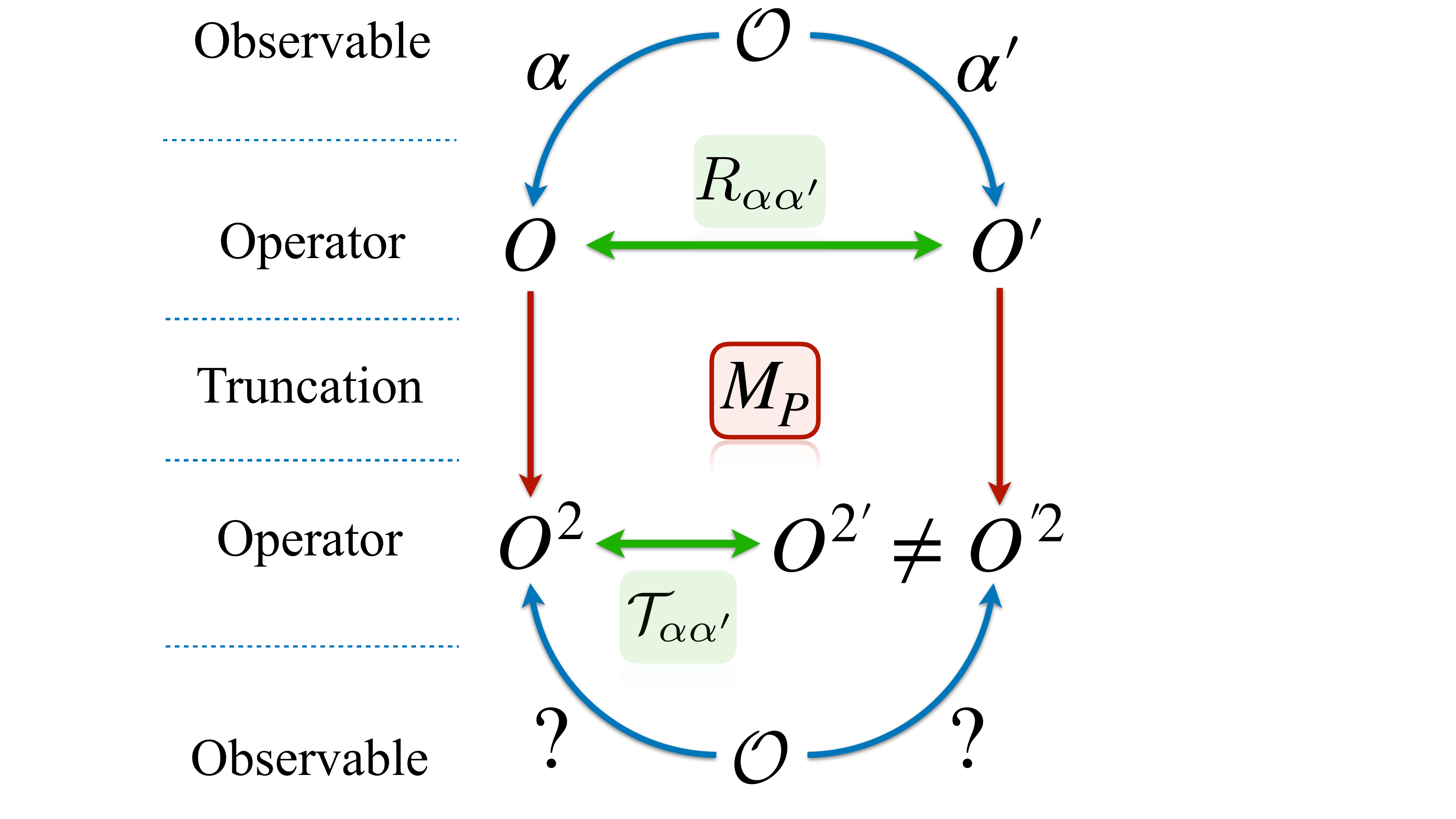}
\vspace*{-4mm}\caption{The breakdown of gauge invariance under a truncating map $M_P$. Equivalent representations $O$ and $O'$ of the same observable ${\cal O}$ are not equivalent after application of $M_P$. A two-level unitary such as ${\cal T}_{\alpha\alpha'}$ cannot produce from the truncated $\alpha$-gauge theory, the same observable $\leftrightarrow$ operator association as is defined by a distinct gauge $\alpha'$. The ``correct" association ${\cal O}\leftrightarrow O^2$ after truncation, can only be defined by identifying a gauge in which the truncation $O^2$ is accurate. Subsequently, {\em any} two-level unitary operator can be used to define an equivalent truncated representation.}\label{2lt}
\vspace*{-2mm}
\end{center}
\end{minipage}
\end{figure}

According to the general quantum postulates given in Sec.~\ref{qsr} for the identification of states and observables with vectors and operators, different gauges constitute different such associations within the starting theory (pre-truncation). If we assume that in gauges $\alpha$ and $\alpha'$ the observable ${\cal O}$ is represented by operators $o_\alpha$ and $o_{\alpha'}$, and if we assume that after truncation ${\cal  O}$ is represented by $M_P(o_\alpha)$ and $M_P(o_{\alpha'})$, then these truncated representations of ${\cal O}$ are not connected by a unitary operator in general (Fig.~\ref{2lt}).  

A truncating map $M_P$ does not preserve the algebra of material operators and so it cannot preserve the unitary relation between distinct associations of operators with observables (gauges) made within the starting theory. The particular word or words used to label the freedom to choose among unitarily equivalent representations of an observable within quantum theory is, of course, immaterial. In particular, the label ``gauge freedom" has been used for this purpose within truncated theories \cite{stefano_resolution_2019,settineri_gauge_2021}. 
Specifically, within a starting theory, the different representations of observable ${\cal O}$ that comprise the equivalence class
\begin{align}
C({\cal O})=\{R OR^\dagger: R~{\rm unitary}\}
\end{align}
can be referred to as different ``gauges", and similarly, in a truncated theory obtained subsequently using a map $M_P$, the truncated representations of ${\cal O}$ belonging to
\begin{align}
&C^2(M_P,{\cal O}) \nonumber \\ &=\{U^2O^2(U^2)^\dagger: O^2=M_P(O)~{\rm and}~U^2~{\rm unitary}\}
\end{align}
could also be referred to as different ``gauges". Crucially however, given a rotation $U^2$ within the truncated space, in general we have that
\begin{align}\label{mpg}
U^2 M_P(O)(U^2)^\dagger \neq M_P(ROR^\dagger),
\end{align}
for any rotation $R$. In other words,
\begin{align}\label{ineqg}
M_P(C({\cal O}))\neq C^2(M_P,{\cal O}).
\end{align}
This proves that identifying the equivalences that occur within the truncated and non-truncated theories would be erroneous. Specifically, one must not surreptitiously and incorrectly equate the left and right-hand-sides of inequality (\ref{ineqg}) simply because one has chosen to refer to both the elements of $C({\cal O})$ and to the elements of $C^2(M_P,{\cal O})$ using the same label, ``gauges".

The definition of the class $C^2(M_P,{\cal O})$ relies upon an accurate truncation $O^2=M_P(O)$ having first been found, that is, $C^2(M_P,{\cal O})$ cannot be defined until a map $M_P$ has first been applied to one of the elements of $C({\cal O})$ to give $O^2$. Yet, applying $M_P$ to different elements of $C({\cal O})$ will give different (non-equivalent) operators $O^2$, that is, the left-hand-side of inequality (\ref{ineqg}) is not a unitary equivalence class. Thus, given a map $M_P$, every different (but equivalent) element of $C({\cal O})$ defines a {\em different} equivalence class $C^2(M_P,{\cal O})$. These different equivalence classes are not equivalent, which constitutes gauge noninvariance. The fact that each $C^2(M_P,{\cal O})$ is an equivalence class constitutes $x_P$-phase invariance. {\em Within} $C^2(M_P,{\cal O})$ any two elements are connected by an $x_P$-phase transformation. Thus, gauge noninvariance and $x_P$-phase invariance are necessarily simultaneously satisfied by truncated models. It follows that the two invariances cannot coincide and exhibiting one of these invariances cannot resolve a breakdown of the other. We note that although we have focussed on two-level truncations the general analysis above holds for any $P\neq I$.

In summary, the possibility of applying unitary rotations after truncation, does not eliminate the problem of first determining a gauge and a map $M_P$ that combined provide an accurate representation, $O^2$, of the observable of interest ${\cal O}$. This problem arises because a truncating map $M_P$ breaks gauge invariance.

\subsection{Optimality of truncations}\label{opt}

We briefly discuss which two-level models are known to be accurate in which situations. Subsequently we discuss the importance of two-level model predictions for gauge ambiguities. Material tuncation should be expected to offer a robust approximation when the material system is sufficiently anharmonic that the orthogonal subspace $Q{\cal H}$ is sufficiently well separated from $P{\cal H}$, where $P{\cal H}\oplus Q{\cal H}={\cal H}$ is the full Hilbert space. Such regimes may or may not be of experimental importance when considering specific implementations of light-matter physics models.

\begin{figure}[t]
\begin{minipage}{\columnwidth}
\begin{center}
\hspace*{-0.5mm}\includegraphics[scale=0.38]{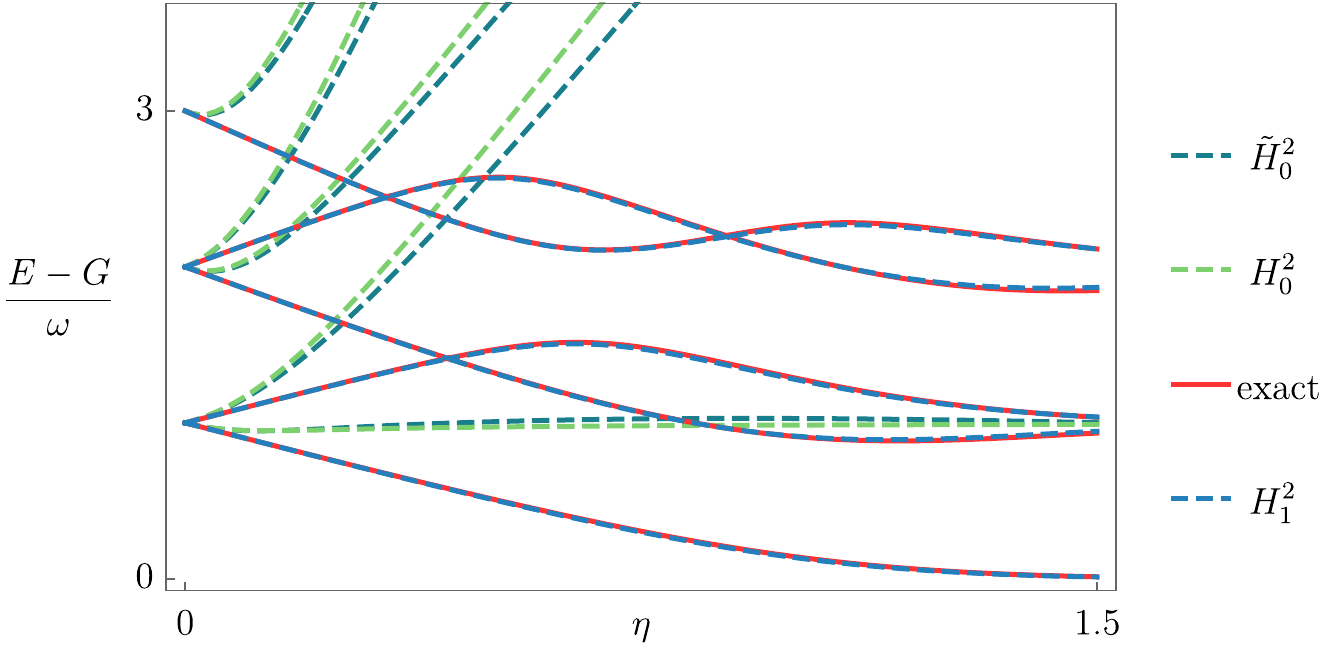}
\caption{The transition spectra (relative to the ground energy $G$) of two-level models are compared with the exact transition spectrum (points), assuming a material anharmonicity of $\mu:=(\omega_{21}-\omega_m)/\omega_m=70$ and resonance $\delta:=\omega/\omega_m=1$. The multipolar gauge QRM (black curves) is {\em generally} accurate in this regime, in the sense that one must go to very high energy levels before discrepancies with the exact spectrum are found. The two Coulomb gauge two-level models $H_0^2$ (lighter curves) and ${\tilde H}_0^2$ (dashed curves) are generally inaccurate, and are qualitatively very similar.}\label{fcomp}
\vspace*{1mm}
\end{center}
\end{minipage}
\end{figure}

Let us first suppose we have a highly anharmonic system at arbitrary coupling strength and only a single radiation mode. The Coulomb gauge coupling involves the canonical momentum ${\bf p}$, which possesses matrix elements in the material basis $\{\ket{\epsilon^n}\}$ that scale with material transition frequencies as
\begin{align}
q{\bf p}_{nl} = i m \omega_{nl}{\bf d}_{nl}.
\end{align}
As first explained in Ref.~ \cite{de_bernardis_breakdown_2018} transitions to higher states are not suppressed within the Coulomb gauge, because the increasing energy gap is compensated by an increasing coupling matrix element. In contrast, the multipolar coupling involves only the dipole moment. Therefore, for sufficiently strong coupling where two-level models are not equivalent, the Coulomb gauge truncation will generally perform poorly in comparison to the multipolar gauge truncation as a {\em general} approximation of the non-truncated theory. These points were also elaborated in Ref.~\cite{stokes_gauge_2019} via a Schrieffer-Wolff-type analysis. As an illustrative example we take a double-well dipole with potential
$V(\theta,\phi) = -\theta r^2/2 + \phi r^4/4$ where $\theta$ and $\phi$ control the shape of the double-well  \cite{de_bernardis_breakdown_2018,stefano_resolution_2019,stokes_uniqueness_2020}. The material Hamiltonian is therefore \cite{de_bernardis_breakdown_2018}
\begin{align}
H_m^\alpha = {{\cal E}\over 2}\left(-\partial_\zeta^2 -\beta\zeta^2+{\zeta^4 \over 2} \right)
\end{align} 
where we have defined the dimensionless variable $\zeta = r/r_0$ with $r_0=(1/[m\phi])^{1/6}$, along with ${\cal E}=1/(mr_0^2)$ and $\beta=\theta m r_0^4$. We first consider the case of resonance $\delta=\omega/\omega_m=1$ together with a high anharmonicity $\mu=(\omega_{21}-\omega_m)/\omega_m$ of $\mu=70$. We compare the unique spectrum of the non-truncated Hamiltonian $H_\alpha$, with the different approximations given by the QRMs $H_1^2$ and $H_0^2$, as well as with the non-standard Coulomb gauge model ${\tilde H}_0^2$ defined by Eq.~(\ref{halpP}). We note that for each $\alpha$ the standard two-level model $H_\alpha^2$ can be selected as the representative of its unitary equivalence class $\{h_\alpha^2(\alpha')\}$ without loss of generality. As shown in Fig.~\ref{fcomp}, the multipolar gauge QRM $H_1^2$ is very accurate for predicting transition spectra in this regime while the Coulomb gauge models $H_0^2$ and ${\tilde H}_0^2$ are qualitatively similar and very inaccurate for strong enough couplings. 
%%%%%%%%%%%%%%%%%%%%%%%%%%%%%%%%%%%%%%%%%%%%%%%%%%%%%%%%%
%%%%%%%%%%%%%%%%%%%%%%%%%%%%%%%%%%%%%%%%%%%%%%%%%%%%%%%%%

There are a number of factors determining the optimality of a truncation. For example, when the detuning $\delta=\omega/\omega$ is large (small) the Coulomb gauge two-level model coupling $\eta'=(\omega_m/\omega)d/\sqrt{2\omega v}$ is weaker (stronger) than the corresponding multipolar gauge coupling $\eta=d/\sqrt{2\omega v} = \delta \eta'$. The two-level model Hamiltonian $PH_\alpha P$ constitutes the first order (in $V^\alpha$) contribution to a more general effective Hamiltonian defined over the two-level subspace $P{\cal H}$ \cite{wilson_brillouin-wigner_2010}. If the model $PH_\alpha P$ is found to be inaccurate, then higher order corrections can be calculated perturbatively using various forms of perturbation theory \cite{wilson_brillouin-wigner_2010}. In particular, the second order contribution is straightforwardly obtainable for a two-level system and single-mode and should yield a two-level model with improved accuracy. In a single-mode theory, such higher order contributions will tend to be larger towards the Coulomb gauge value $\alpha=0$, because as noted the energy gap to the orthogonal subspace $Q{\cal H}$ is compensated by the form of the Coulomb gauge coupling. 

When more radiation modes are considered the optimal gauge may often be shifted away from the multipolar gauge towards the Coulomb gauge \cite{roth_optimal_2019}. The multipolar and Coulomb gauge linear interactions scale as $\sqrt{\omega}$ and $1/\sqrt{\omega}$ respectively. The introduction of more radiation modes causes the multipolar gauge truncation to become sub-optimal because the effects of non-resonant modes are more pronounced in this gauge, as will be discussed further in Sec.~\ref{glauber}. Results illustrating this effect within the strong-coupling regime have been given in Ref.~\cite{roth_optimal_2019}. When more dipoles are considered, but only a single radiation mode is retained the multipolar gauge truncation is again typically optimal at sufficiently large anharmonicity, and accuracy increases with the number of dipoles considered \cite{stokes_uniqueness_2020}. 

Ref.~\cite{rouse_avoiding_2021} addresses the issue of gauge noninvariance due to truncation using a novel description in terms of dual coordinates. This is reviewed briefly in   Supplementary Note~X. It is found that approximations within the multipolar gauge, $\alpha=1$, will typically most accurately represent the physics of small, bound dipoles interacting with a single mode. A wide range of system types is considered along with the effects of both material truncation and the EDA.
  
In Ref.~\cite{ashida_cavity_2021} the authors identify a Pauli-Fierz-type representation obtained from the Coulomb gauge by unitary transformation. For a one-dimensional material system coupled to a single cavity mode with frequency $\omega$, the transformation is defined by $U=e^{-i\xi_g p\pi}$. Here $\pi = i(c^\dagger-c)$, with $c$ a renormalised cavity annihilation operator for a photon with frequency ${\tilde \omega}$ where ${\tilde \omega} = \sqrt{\omega^2+g^2}$ and $x_{\tilde \omega} g$ is a bare coupling strength defined using the Coulomb gauge Hamiltonian with $x_{\tilde \omega}=1/\sqrt{m{\tilde \omega}}$. The renormalised coupling $\xi_g: = gx_{\tilde \omega}/ {\tilde \omega}$ is a non-constant function of the bare coupling parameter $g$ with maximum value close to $g = 1$. The idea of the Pauli-Fierz representation is to eliminate the component of the transverse field tied to material charges \cite{cohen-tannoudji_atom-photon_2010}. The Hamiltonian within the transformed frame is $H^U =H_m(p,r + \xi_g \pi) + {\tilde \omega} c^\dagger c$ where $H_m(p,r) := p^2/ (2m_{\rm eff})+ V (r)$ and the effective mass is defined by $m_{\rm eff} = m[1 +2(g/\omega)^2]$. For increasing $g$ the renormalised frequency ${\tilde \omega}$ is increasingly dominant while the coupling $\xi_g$ eventually begins to decrease. For sufficiently large $g$ the eigenvectors of $H^U$ become approximately separable despite remaining highly entangled in the Coulomb gauge.

If $V$ has local minima near to which it can be expanded as $\delta V \propto r^2$, then since $m_{\rm eff}$ increases quadratically with $g$, the eigenfunctions of $H_m(p,r)$ are increasingly localized around the potential minima and the low-lying spectrum of $H^U$ is that of a harmonic oscillator with narrowing level spacing $\delta E \propto 1/g$. It is argued further in Ref.~\cite{ashida_cavity_2021} that truncation is increasingly well-justified within $H^U$ at larger g, due to increased localization of the eigenstates of $H_m(p,r)$ that results from the dependence on $m_{\rm eff}$. Ref.~\cite{ashida_cavity_2021} studies a double-well dipole as an example application. It is found that for a shallow double-well, even the multipolar gauge truncation fails quite severely at extreme coupling strengths, and even in the case of only a single mode, whereas truncation in the Pauli-Fierz frame remains accurate. \cite{ashida_cavity_2021} also provide a multimode generalization of their Pauli Fierz-type transformation.
 
\begin{figure}[t]
\begin{minipage}{\columnwidth}
\begin{center}
\hspace*{-4mm}\includegraphics[scale=0.39]{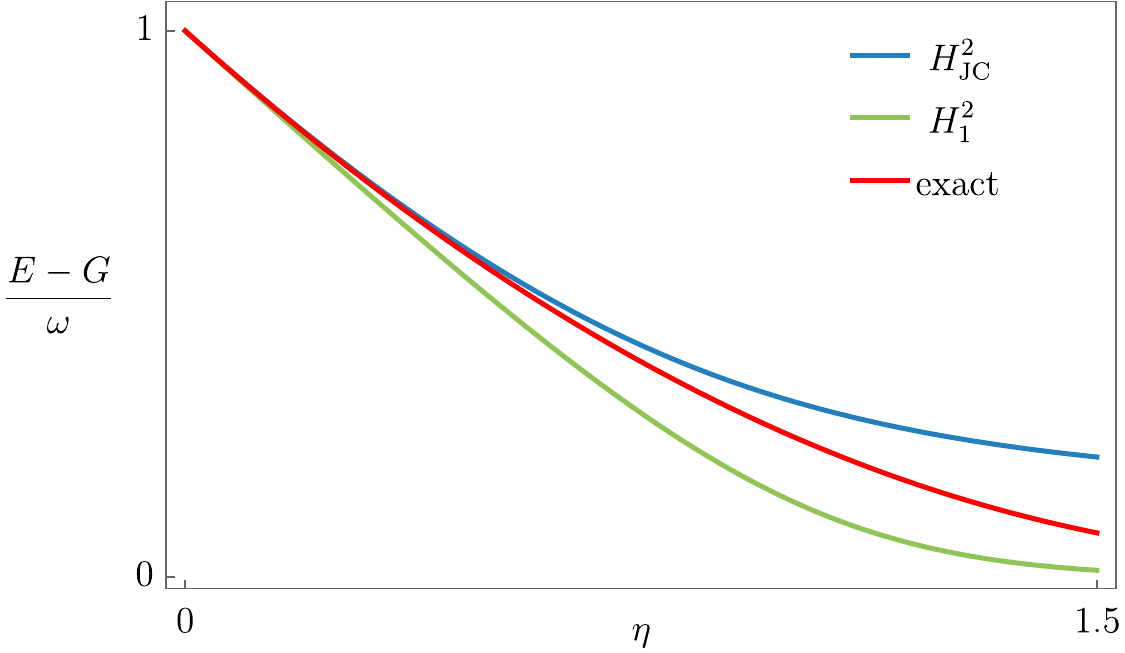}
\caption{The first transition energies of the two-level models $H^2_1$ (lower curve) and $H^2_{\rm JC}$ (upper curve), are compared with the exact transition energy (middle curve), assuming a material anharmonicity of $\mu \approx 3$ and resonance $\delta=1$. The $\alpha_{\rm JC}$-gauge two-level model can be more accurate than the multipolar gauge QRM in the ultrastrong-coupling regime.}\label{transJC1}
\vspace*{-4mm}
\end{center}
\end{minipage}
\end{figure}
%%%%%%%%%%%%%%%%%%%%%%%%%%%%%%%%%%%%%%%%%%%%%%%%%%%%%%%%%

Via the literature reviewed above the relative accuracy of material truncations performed in different regimes and gauges is now well understood, at least for simple light-matter systems. In particular, truncation will obviously break down as a general approximation for sufficiently harmonic material systems. However, as will be emphasized below, in simple models the accuracy of a given truncation is of limited importance, because the truncation is straightforwardly avoidable. 

Truncation is most significant in its capacity to reveal important qualitative physical implications. In particular, the onset of USC has often been identified through a departure from Jaynes-Cummings physics, due to the breakdown of the rotating-wave approximation (RWA). In the USC regime the qualitative low energy physics of the Jaynes-Cummings model (JCM) is markedly different from that of the quantum Rabi model (QRM). For example, the JCM predicts no ground state entanglement and no ground state photon population for all coupling strengths. The contrary predictions of the QRM have previously been regarded as definitive of ultrastrong-coupling phenomenology. However, Ref.~\cite{stokes_gauge_2019} shows that there exists a gauge choice that yields a Jaynes-Cummings model without performing the RWA. The corresponding gauge-parameter $\alpha_{\rm JC}$ varies with the coupling and detuning parameters of the theory, but this is certainly permissible, it simply amounts to choosing a non-constant gauge function (see Sec~\ref{natgt}).

For a material harmonic oscillator two-level truncation is essentially as poor a {\em general} approximation as it can ever be, yet for this system the ground state of the truncated model is {\em exact} in the JC-gauge; $P\ket{G_{\rm JC}}=\ket{G_{\rm JC}}$ (see Sec.~\ref{osc}). As a result, there exist gauges $\alpha\neq 1$ in which two-level truncation of material systems with low anharmonicity remains accurate for low energy states, despite truncation in any gauge generally breaking down for higher levels. Ref. \cite{stokes_gauge_2019} exemplifies an experimentally realistic regime of a fluxonium $LC$-oscillator system with anharmonicity $\mu \approx 3.15$, such that two-level models remain accurate for predictions up to the first excited state, and for which the JC-gauge two-level model is usually more accurate. It follows that low energy weak-coupling phenomenology can persist even within the USC regime, such that the phenomenology previously viewed as definitive of the USC regime need not hold even within gauge invariant non-truncated models. Essentially the same findings are obtained for a double-well dipole, as illustrated in Fig.~\ref{transJC1}. 
 
%%%%%%%%%%%%%%%%%%%%%%%%%%%%%%%%%%%%%%%
\begin{figure}[t]
  \centering
\hspace*{-2mm}\includegraphics[scale=0.43]{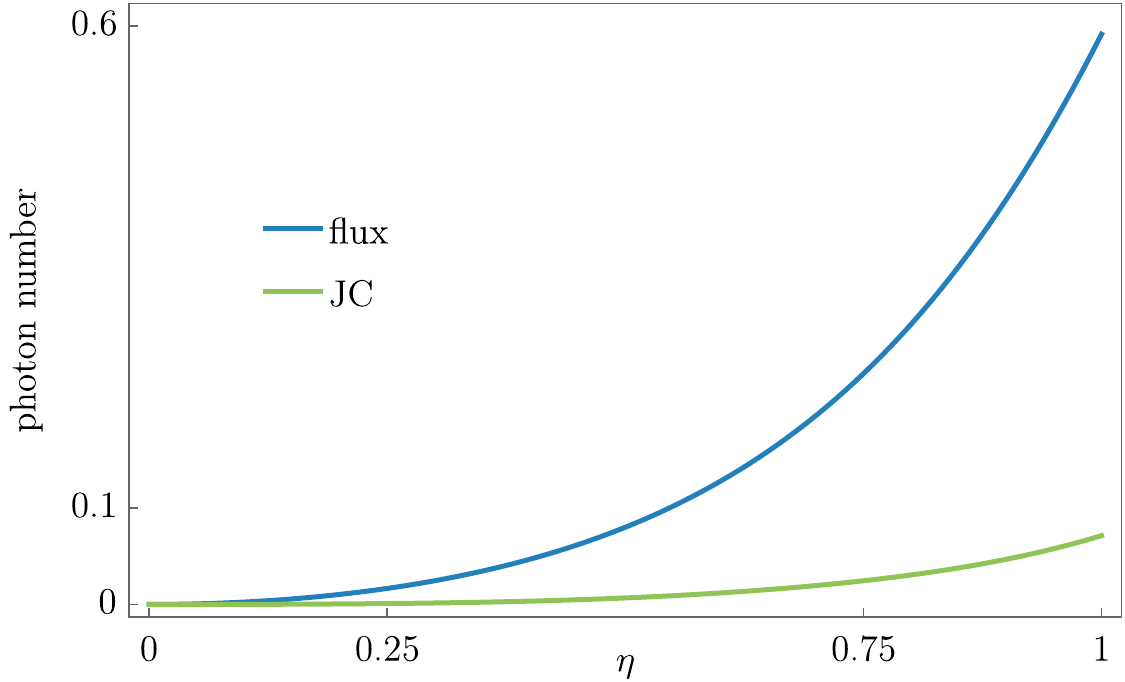}
  \vspace*{0.3cm}
 \caption{The exact ground state average numbers of flux-gauge (upper curve) and JC-gauge (lower curve) photons with coupling strength $\eta$ for a fluxonium system assuming an anharmonicity of $\mu\approx 3$ and resonance $\delta=1$. The number of JC-gauge photons is much lower than the number of flux-gauge photons. Appreciable JC-gauge photon population only occurs for very large couplings approaching the deepstrong limit $\eta=1$.}
  \label{fluxonium_ph}
\end{figure}
%%%%%%%%%%%%%%%%%%%%%%%%%%%%%%%%%%%%%%%%
\begin{figure}[t]
  \centering
\hspace*{-2mm}\includegraphics[scale=0.43]{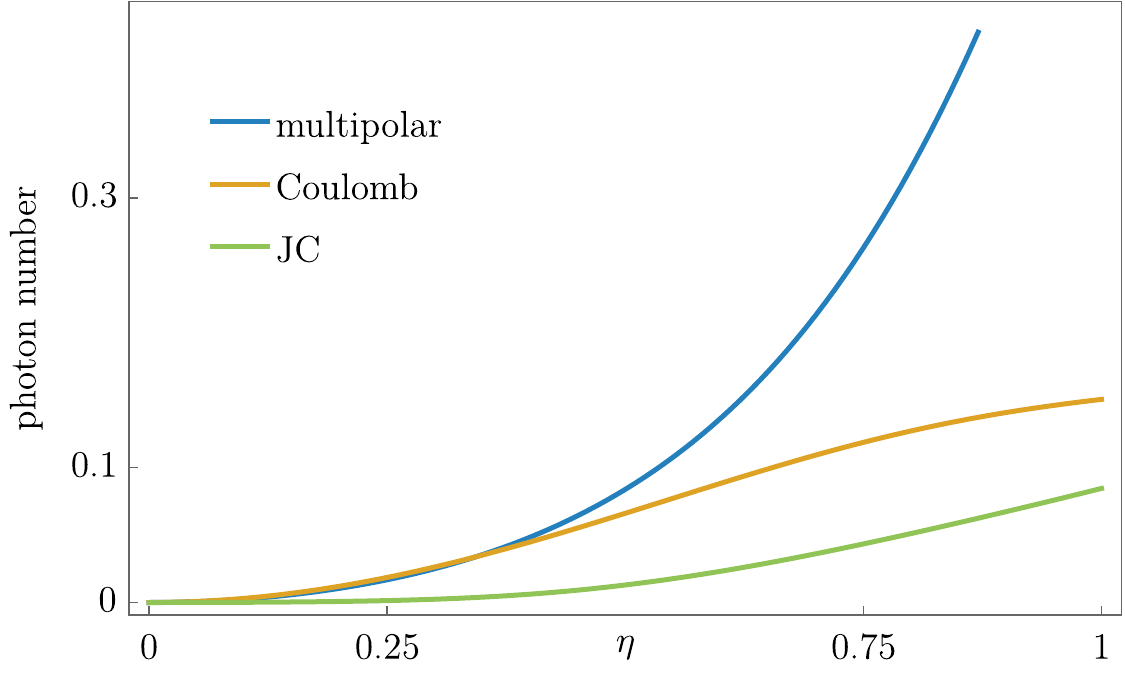}
  \vspace*{0.3cm}
 \caption{The exact ground state average numbers of multipolar gauge (upper curve), Coulomb gauge (middle curve), and JC-gauge (lower curve) photons for a double-well dipolar system assuming an anharmonicity of $\mu\approx 70$ and resonance $\delta=1$. The number of JC-gauge photons is only appreciable well into the USC regime $\eta > 1/2$.}
  \label{dw_ph}
\end{figure} 

We now turn our attention to photon number observables. The dipole-cavity Hamiltonian in Eq.~(\ref{edaham}) possesses a cavity self-energy term $q^2A_\alpha^2/2m$, which only vanishes for $\alpha=1$ ($A_1=0$). Since it acts non-trivially only within the photonic Hilbert space, this term is unaffected by material truncation. It can be absorbed into the cavity Hamiltonian using a local Bogoliubov transformation. Thus, each gauge $\alpha\neq 1$ possesses two possible definitions of photon number which do and do not include this renormalisation respectively. In the JC-gauge the renormalised photon number predicted by the JC-gauge two-level model is identically zero in the ground state, because in terms of the corresponding photonic operators the JC-gauge two-level model has Jaynes-Cummings form. On the other hand, the ground state average of the ``bare" JC-gauge photon number (which does not include the $A_{\rm JC}^2$-term) can possess non-zero values for sufficiently large coupling strengths, even when the average is found using the JC-gauge truncated model. Moreover, when the two-level truncation is avoided, both the renormalised and non-renormalised (``bare") JC-gauge photon numbers can be nonzero in the ground state, due to counter-rotating terms into dipole levels above the first.

Figs.~\ref{fluxonium_ph} and \ref{dw_ph} show the exact, i.e., non-truncated, ground state ``bare" photon numbers defined relative to the multipolar (flux), Coulomb (charge) and JC gauges for fluxonium and double-well dipole systems respectively. In the cases of the Coulomb and JC-gauges these photon numbers do not include in their definitions the $A_\alpha^2$-type terms. In particular, for sufficiently large coupling strengths, the predicted JC-gauge photon number average is non-zero both due to the $A_{\rm JC}^2$-term, as well as due to counter-rotating terms to higher dipolar-levels.    To illustrate different regimes of anharmonicity, we have assumed $\mu\approx 3$ for the fluxonium system and $\mu\approx 70$ for the double-well dipole. In both cases the ground state photon$_{\rm JC}$ population is highly suppressed when compared with the ground state photon$_0$ and photon$_1$ populations. All of these predictions are gauge invariant having been obtained from the non-truncated theory.

\section{Time-dependent interactions and adiabatic switching}\label{s6}

Time-dependent interactions arise in a number of contexts in light-matter physics. Herein, the notion of a {\em process} in which material charges exchange photons, is elementary. The concept arises from scattering theory wherein the interaction $V=H-h$, where $h$ is called the unperturbed Hamiltonian, is adiabatically switched on and off over an infinite duration. Such an idealisation may not however, be applicable in extreme light-matter interaction regimes. Gauge freedom in scattering theory has been discussed extensively in the context of atomic lineshape and level-shift phenomena \cite{lamb_fine_1952,low_natural_1952,power_coulomb_1959,fried_vector_1973,bassani_choice_1977,kobe_question_1978,cohen-tannoudji_photons_1989,baxter_gauge_1990,woolley_gauge_1998,woolley_gauge_2000,stokes_gauge_2013}. We explain why subsystem gauge relativity can be ignored in calculating the $S$-matrix \cite{cohen-tannoudji_photons_1989}. We then directly demonstrate that conventional quantum optical approximations mimic the $S$-matrix, and thereby eliminate subsystem gauge relativity. Only within such approximations do ``atoms" and ``photons" defined as quantum subsystems, become at least ostensibly unique concepts for a given definition of $h$. It should also be noted however, that different definitions of $h$ are available and might be considered, as we briefly discuss. We also discuss non-adiabatic switching of ultrastrong couplings whereby subsystem gauge relativity becomes important quite generally.

\subsection{Adiabatic switching and a unique invariance property of the $S$-matrix}\label{adiabatic}

As explained in Secs.~\ref{gamb}-\ref{imps}, the task we are faced with is the determination of which gauge invariant subsystem definitions are {\em relevant} in which situations. However, if the $S$-matrix is applicable in providing all physical predictions then we are able to completely ignore this question. The subsystems become ostensibly unique within scattering theory because of the adiabatic interaction switching condition therein. Feynman diagrams can be used as a mnemonic when calculating the terms in a perturbative expansion of the Hamiltonian resolvent used to define the $S$-matrix, which is the primary source of predictions in particle physics. This gives rise to the notions of ``real" and ``virtual" processes. 

The $\alpha$-gauge Hamiltonian can be partitioned as $H_\alpha=h+V^\alpha$ where $h=H_m+H_{\rm ph}$ is the unperturbed Hamiltonian and $V^\alpha$ is the interaction. The unperturbed energy eigenvalues and eigenvectors are defined by $h\ket{\epsilon^n}=\epsilon^n\ket{\epsilon^n}$. The vectors $\{\ket{\epsilon^n}\}$ are each a tensor product of an eigenvector of $H_m$ and an eigenvector of $H_{\rm ph}$ (photon number state).   
Suppose that physical states ${\cal S}_n$ and ${\cal S}_m$ are represented in gauge $\alpha$ by vectors $\ket{\epsilon^n}$ and $\ket{\epsilon^m}$. The same states are represented in gauge $\alpha'$ by vectors $\ket{\epsilon'^n}=R_{\alpha\alpha'}\ket{\epsilon^n}$ and $\ket{\epsilon'^n}=R_{\alpha\alpha'}\ket{\epsilon^m}$ respectively, therefore the bare eigenvectors of $h$ represent different physical states in each gauge (subsystems are gauge-relative). The evolution operator generated by $H_\alpha$ between times $t_i$ and $t_f$ is denoted $U_\alpha(t_i,t_f)$. Evolutions in different gauges are related by $U_{\alpha'}(t_i,t_f)=R_{\alpha\alpha'}U_{\alpha}(t_i,t_f) R_{\alpha\alpha'}^\dagger$. The probability amplitude, $A({\cal S}_n,t_f;{\cal S}_m,t_i)$, to find the system at time $t_f$ in state ${\cal S}_n$ given that at time $t_i$ its state was ${\cal S}_m$, is given by the corresponding evolution operator matrix element, and is a {\em gauge invariant} preditction;  
\begin{align}\label{eqt}
&A({\cal S}_n,t_f;{\cal S}_m,t_i) \nonumber \\ &= \bra{\epsilon^n}U_\alpha(t_i,t_f)\ket{\epsilon^m} = \bra{\epsilon'^n} U_{\alpha'}(t_i,t_f)\ket{\epsilon'^m}.\\& ~~~~~~~~~~~~~~~~   ({\rm gauge~invariance}) \nonumber 
\end{align}
It is equally clear that for $\alpha\neq \alpha'$ we have
\begin{align}\label{ineqt}
\bra{\epsilon^n}U_\alpha(&t_i,t_f)\ket{\epsilon^m} \neq \bra{\epsilon^n} U_{\alpha'}(t_i,t_f) \ket{\epsilon^m}. \\& ~~~   ({\rm gauge~relativity}) \nonumber 
\end{align}
Inequality~(\ref{ineqt}) simply exemplifies the expected result that an eigenvector of $h$ represents a different physical state in each different gauge.   The left-hand-side is $\alpha$-dependent while the right-hand-side is $\alpha'$-dependent, but both predictions are gauge invariant amplitudes of the form specified by Eq.~(\ref{eqt}). We refer to this $\alpha$-dependence despite the gauge invariance of both predictions as {\em gauge relativity}. 

In scattering theory it is assumed that $V^\alpha=0$ in the remote past and distant future $t=\pm\infty$, such that at these times $H=h$ and so the unperturbed energy eigenvectors {\em uniquely} represent the total energy eigenstates. It is then assumed that the interaction is switched-on and -off adiabatically between $t=\pm\infty$. Subsequently, the $S$-matrix is formally defined by \cite{cohen-tannoudji_photons_1989,weinberg_quantum_2005}
\begin{align}\label{smat2}
S_{nm} = \lim_{t\to\infty}\bra{\epsilon^n} U_{\alpha I}(-t,t) \ket{\epsilon^m}
\end{align}
where $U_{\alpha I}$ denotes the corresponding evolution operator in the interaction picture defined by $h$. In contrast to inequality (\ref{ineqt}), the $S$-matrix possesses the remarkable property that it is independent of $\alpha$ despite being defined in terms of the {\em same} unperturbed vectors for every $\alpha$.   In other words, a special property of the $S$-matrix is that it is {\em gauge non-relative} . In calculating $S_{nm}$ we do not have to transform the eigenvectors of $h$ in order to ensure that we are using the same physical states in each gauge, as in Eq.~(\ref{eqt}) \cite{cohen-tannoudji_photons_1989}. ``Photonic" and ``material" excitations represented by the eigenvectors of $h$ become ostensibly unique in scattering theory, so we do not have to confront the question of which subsystem definitions are the most relevant.

A general proof of this unique invariance property of the $S$-matrix has been given for nonrelativistic QED by Woolley \cite{woolley_gauge_1998,woolley_gauge_2000}. Essential for the proof is that the unperturbed operator $h$ is kept the same in each gauge. The $S$-matrix can also be expressed in the form \cite{cohen-tannoudji_atom-photon_2010,woolley_gauge_1998,woolley_gauge_2000}
\begin{align}\label{Sma}
S_{nm} =\delta_{nm}-2\pi i T_{nm}\delta(\epsilon_n-\epsilon_m)
\end{align}
where $T$ is called the transition matrix whose elements in the unperturbed basis naturally depend on $\alpha$. However, when it is evaluated on-energy-shell as expressed by the $\delta$-function in the $S$-matrix element $S_{nm}$, all $\alpha$-dependence drops out \cite{woolley_gauge_1998,woolley_gauge_2000}. This bare-energy conservation property is thereby seen to be crucial in ensuring that the gauge relativity of the subsystems can be ignored when calculating the $S$-matrix.

We can define any process that conserves $h$ as real. A virtual process is then one that is not real. In the $S$-matrix, the latter can only occur as intermediate processes constituting part of a real process. More generally however, the $S$-matrix can be understood as an infinite-time limit of the more general matrix given by \cite{cohen-tannoudji_atom-photon_2010}
\begin{align}\label{stau}
S^{(\tau)}_{nm} =\delta_{nm}-2\pi i T_{nm}\delta^{(\tau)}(\epsilon_n-\epsilon_m)
\end{align}
The function $\delta^{(\tau)}(\epsilon_n-\epsilon_m)$ has a peak at $\epsilon_n=\epsilon_m$ with width on the order $1/\tau$, which is often taken as expressing the conservation of bare energy to within $1/\tau$ \cite{cohen-tannoudji_atom-photon_2010}. This is the heuristic energy-time uncertainty relation, which it should be noted is quite different from the rigorous Heisenberg uncertainty relation for conjugate operators.

It is clear that the processes described by the matrix $S^{(\tau)}$ are not purely real (zero energy-uncertainty) unless $\tau\to \infty$. It is widely regarded that physical processes are ``real". However, although the total energy $E$ represented by the operator $H_\alpha$ is automatically conserved, there is nothing in quantum or classical theory that requires a physical process to conserve only part of this energy, such as the part represented by $h$. This is required and does occur in the $S$-matrix only because $H_\alpha=h$ at the beginning and the end of a scattering process. And yet, the limit of infinite times with adiabatic switching is clearly an idealisation, such that purely ``real" processes cannot truly occur. In this sense the term ``real" is a misnomer. Further still, it is clear that only when a process {\em is} ``real", i.e., is a scattering process, can the gauge relativity of the subsystems necessarily be ignored. In other words, scattering theory is gauge nonrelativistic.

All predictions are fundamentally gauge invariant in the sense of Eq.~(\ref{eqt}). Thus, both sides of inequality (\ref{ineqt}) are gauge invariant predictions, but beyond scattering theory, i.e., over finite-times, we must recognise that they are {\em different} gauge invariant predictions and we are confronted with the task of determining which (if either) is more {\em relevant}.

\subsection{Partitioning the Hamiltonian}

Although the $S$-matrix is gauge non-relative in the sense defined above, it can only be defined relative to a partition of the Hamiltonian into unperturbed and interacting parts as $H =h+V$, which is of course non-unique. Naively, one might  attempt to define $V$ and $h$ as the components that respectively do and do not depend on a ``coupling" parameter, of which the only obvious choice is the electric charge $q$. According to this definition, $h$ would consist of the free photonic Hamiltonian $H_{\rm ph} = \int d^3 k \sum_\lambda \omega[ a_\lambda^\dagger({\bf k})a_\lambda({\bf k})+1/2]$ together with particle energies $H_m =\sum_{\rm charges} {\bf p}^2 /(2m)$. The unperturbed vectors would therefore be incapable of representing bound material states. This definition would be of little use in applications of QED at low energies whereby a separation of near-field interactions is advantageous in allowing bound charge systems to emerge as the constituents of ``unperturbed" stable matter. In particular, the most commonly used definition of $h$ in non-relativistic QED, namely, the definition suggested by the Coulomb-gauge which reads
\begin{align}\label{ch0}
h=\sum_{\rm charges}{ {\bf p}^2\over 2m} +V_{\rm Coul} + H_{\rm ph}
\end{align}
where $V_{\rm Coul} = \int d^3 x\, {\bf E}_{\rm L}^2/2$ is the Coulomb energy, would be ruled-out, because $V_{\rm Coul}$ depends on $q$. 

A different criterion to define $V$ would be that it must not include any terms that act exclusively within the ``matter" Hilbert space or exclusively within the ``photonic" Hilbert space. In particular, $V$ must not include any ``self-interaction" terms, which although dependent on $q$, are of the form $O_m\otimes I_{\rm ph}$ or $I_m\otimes O_{\rm ph}$. In the Coulomb-gauge this criterion does indeed lead to the commonly used definition of $h$ given in Eq.~(\ref{ch0}) and concurrently to the familiar Coulomb-gauge interaction Hamiltonian of the form $-q{\bf p}\cdot {\bf A}_{\rm T}({\bf r})/m + q^2{\bf A}_{\rm T}({\bf r})^2/(2m)$ for each charge $q$. 

However, this method does not in general yield the same definition of $h$ when it is applied in other gauges. In the gauge $g$, for example, the material Hamiltonian $H_m$ would include the total polarisation energy $\int d^3x \,{\bf P}_g^2/2$, which in addition to $V_{\rm Coul}$ includes the transverse polarisation ``self-term" $\int d^3x\,{\bf P}_{g\rm T}^2/2$. In the multipolar-gauge this additional term is divergent. It cannot contribute to processes in which the number of photons change and otherwise it is often ignored until such a point that its contributions can be ``renormalised-out" of final predictions. This, for example, is how on-energy-shell $T$-matrix elements for bound-state level-shifts are typically calculated using the multipolar-gauge \cite{craig_molecular_1998}. Predictions obtained in this way are identical to those found using the Coulomb-gauge because they result from having employed the same definition of $h$.

If the multipolar transverse polarisation is instead regularised, as described in Secs.~\ref{FT} and \ref{share}, then $\int d^3x\,{\bf P}_{g\rm T}^2/2$ is finite but its relative magnitude depends on the cut-off $k_M$. It can be considered a weak perturbation of $V_{\rm Coul}$ provided $k_M$ is chosen appropriately \cite{vukics_fundamental_2015}. Importantly, in this case, and more generally whenever $H_m$ includes terms additional to $V_{\rm Coul}$, a different $S$-matrix is obtained to that obtained when using $h$ in Eq.~(\ref{ch0}).

One might also consider relative magnitudes to be a guide in determining appropriate definitions of $h$ and $V$. In order that weak-coupling methods are applicable the interaction $V$ must obviously be a weak-perturbation of $h$. For example, when considering multiple systems of interest within a common reservoir, if direct inter-system interactions are sufficiently strong then they should be included within $h$ rather than within the system-reservoir interaction $V$ \cite{stokes_master_2018,santos_master_2014}. Subsequently applying weak-coupling theory yields a reduced description in the form of a Lindblad master equation whose coefficients are $S$-matrix elements and an example of this is given in Sec.~\ref{master}, but it should be noted that the particular $S$-matrix obtained is specific to whatever definition of $h$ is adopted. Similarly, when dealing with strong system-reservoir couplings analytic methods such as polaron transformations \cite{pollock_multi-site_2013,nazir_modelling_2016} and Hamiltonian mapping techniques \cite{iles-smith_environmental_2014,strasberg_nonequilibrium_2016} work by redrawing the system-reservoir boundary so as to obtain a weak perturbation $V$. 

Physically, when subsystem interactions are strong it is unclear to what extent the subsystems should be considered operationally accessible. A given experiment may (or may not) only be capable of granting access to a dressed composite rather than to the individual subsystems that comprise it. The balance between localisation and dressing within the context of nonrelativistic QED is discussed throughout Sec.~\ref{pd}. In the context of open quantum systems theory, this topic is closely related to the distinction between local and global approaches to deriving reduced descriptions, which is discussed briefly in Sec.~\ref{excavfs}. 

In conclusion, we note that while the $S$-matrix is gauge non-relative in the sense defined in Sec.~\ref{adiabatic}, this property does not necessarily circumvent the challenge of determining a gauge relative to which one is to obtain physical predictions, even within scattering theory itself. Indeed, the myriad existing scattering-theoretic predictions of nonrelativistic QED found using low-order perturbation theory, rely on the specific definition of $h$ given in Eq.~(\ref{ch0}). The prospect of deriving alternative QED $S$-matrices that result from different definitions of $h$, for example, that include a  ``weak" self-term $\int d^3x\,{\bf P}_{g\rm T}^2/2$, warrants further study.

\subsection{Quantum optical approximations: Mimicking the $S$-matrix}\label{mimic}

We now show directly that subsystem gauge relativity can be eliminated after a sufficient number of weak-coupling approximations are performed.

\subsubsection{Toy model: material oscillator and a single mode}\label{osc}

We begin by again considering a simple toy model consisting of a material harmonic oscillator and a single radiation mode, such that Eqs.~(\ref{poltalph}) and (\ref{aa}) become
\begin{align}
&P_{\rm T\alpha} ={\alpha qx\over v},\\
&A_\alpha = (1-\alpha)A
\end{align}
where $v$ is the cavity volume. The cavity canonical operators are $A = (a^\dagger +a)/\sqrt{2\omega v}$ and $\Pi = i\sqrt{{\omega/2v}}(a^\dagger -a)$, such that $[A,\Pi]=i/v$. We assume that the material oscillator points in the same direction as the mode. The theory is gauge invariant because gauge fixing transformations remain unitary; $R_{\alpha\alpha'}=e^{iq(\alpha-\alpha')rA}$.

The $\alpha$-gauge Hamiltonian in Eq. (\ref{HA}) can be written $H^\alpha = h+V^\alpha$ where $h=\omega(a^\dagger a+1/2)+\omega_m(b^\dagger b +1/2)$ and
\begin{align}\label{hamtim0}
V^\alpha=\,&{\eta^2\omega \over 4}\left[(1-\alpha)^2(a^\dagger+a)^2 + \delta \alpha^2(b^\dagger+b)^2\right] \nonumber \\ &+ iu_\alpha^-(ab^\dagger - a^\dagger b) + iu_\alpha^+(a^\dagger b^\dagger - ab)
\end{align}
with $\eta = -q/(\omega\sqrt{mv})$ a dimensionless coupling parameter, $\delta =\omega/\omega_m$, and 
\begin{align}
u_\alpha^\pm={\eta\omega_m\over 2}\sqrt{\delta}[(1-\alpha)\mp \delta\alpha].
\end{align}
Clearly the value of $\alpha$, which determines the physical definitions of the two oscillator subsystems, can have a profound affect on the form of $V^\alpha$. This is completely eliminated however, if we assume weakly-coupled nearly-resonant oscillators. We can then let $\omega_m=\omega$, and we can neglect terms quadratic in $\eta$. We can also perform the rotating-wave approximation by setting $u^+_\alpha = 0$. The final result is the $\alpha$-independent Hamiltonian $H=h+V$ where $h=\omega(a^\dagger a + b^\dagger b +1)$ and
\begin{align}\label{cons}
V^\alpha=V = {i\over 2}\omega \eta(ab^\dagger-a^\dagger b).
\end{align}
This Hamiltonian satisfies bare-energy conservation
\begin{align}
[h,H]=0,
\end{align}
which we saw in the context of the $S$-matrix was crucial in eliminating subsystem gauge relativity. We have obtained the same result here in a very direct manner. We can now pretend that the two oscillators represent unique physical subsystems.

Outside of the regime of validity of weak-coupling approximations, it is typically thought that one cannot let $u^+_\alpha\approx0$. In general, this is true, by which we mean that one can only use this approximation {\em independent of the value of $\alpha$} in the weak-coupling regime. However, whether $V^\alpha$ includes counter-rotating terms depends on the value of $\alpha$, so there exists a range of values for which the rotating-wave approximation will remain valid well into the ultrastrong coupling regime. For a specific choice of $\alpha$ the rotating-wave approximation is exact \cite{drummond_unifying_1987,stokes_ultrastrong_2021,stokes_extending_2012,stokes_master_2018,stokes_gauge_2019}. Specifically, by choosing
\begin{align}\label{alJC}
\alpha(\omega)=\alpha_{\rm JC}(\omega) :={\omega_m\over \omega_m+\omega}
\end{align}
we obtain $u_\alpha^+\equiv 0$, so the counter-rotating terms in the bilinear component of $V^\alpha$ in Eq.~(\ref{hamtim0}) are automatically eliminated. As before, by performing {\em non-mixing} Bogoliubov transformations within the separate ${\rm light}_{\rm JC}$ and ${\rm matter}_{\rm JC}$ Hilbert spaces, we can eliminate terms quadratic in $\eta$ via modes $c$ and $d$ such that
\begin{align}
&{p^2\over 2m} +{m\omega_m^2 \over 2}x^2 +{q^2 \over 2v}\alpha_{\rm JC}^2r^2= {\tilde \omega}_{m}\left(d^\dagger d +{1\over 2}\right),\label{selfpren}\\
&{v\over 2}(\Pi^2 +\omega^2 A^2) +{q^2\over 2m}(1-\alpha_{\rm JC})^2A^2 = {\tilde \omega}\left(c^\dagger c +{1\over 2}\right)
\end{align}
where ${\tilde \omega}_{m}^2 = \omega_m^2\mu$ and ${\tilde \omega}^2= \omega^2\mu$ in which $\mu = 1+\left({\eta \omega /(\omega_m+\omega})\right)^2$. In the single-mode case this elimination of self-energy terms is exact. It follows that $\alpha_{\rm JC}$ can be written $\alpha_{\rm JC}={{\tilde \omega}_m/(\tilde \omega}+{\tilde \omega}_m)$. The corresponding Hamiltonian is
\begin{align}\label{nc}
H_{\rm JC}=&{\tilde \omega}_m \left(d^\dagger d +{1\over 2}\right)+{\tilde \omega}\left( c^\dagger c +{1\over 2}\right)\nonumber \\ &- iq \sqrt{\omega \omega_m \over mv} {1\over \omega_m+\omega}(d^\dagger c -d c^\dagger).
\end{align}
The ground state is represented by the the vacuum of the $c$ and $d$ modes; $\ket{G_{\rm JC}}=\ket{0_d,0_c}$. We emphasize that at no point have we made use of any approximations or assumptions that ruin the gauge invariance of the theory. Neither however, have we performed a diagonalising transformation of the Hamiltonian. We have simply considered a particular {\em definition} of the subsystems specified by a value $\alpha_{\rm JC}$ {\em in between} the commonly chosen values $\alpha=0$ and $\alpha=1$, and within this gauge we have only performed {\em non-mixing} Bogoliubov transformations of the form $U_m\otimes U_{\rm ph}$. Whether or not the latter transformations are employed counter-rotating terms are absent, because $u_{\rm JC}^+\equiv 0$. Thus,
\begin{itemize}
\item{It is premature to define the paradigm of extreme-coupling light-matter physics through properties such as high-levels of ground-state light-matter entanglement and photon population, which result from terms appearing in commonly chosen gauges, but which are not necessarily present.}
\end{itemize}

There are no ground state virtual excitations in the modes $c$ and $d$ when they are defined relative to the gauge $\alpha_{\rm JC}$. We will see in Secs.~\ref{loc_caus} (see also   Supplementary Note~XIII) that as a result, ``matter" cannot be fully localised in this gauge. Finally we remark that although in this example  a projection $P=\ket{0_d}\bra{0_d}+\ket{1_d}\bra{1_d}$ onto the first two levels of the material oscillator is as ill-justified as it can ever be as an approximation (because the matter system is harmonic), such a projection nevertheless yields the {\em exact} ground state; $P\ket{G_{\rm JC}}=\ket{G_{\rm JC}}$. This fact is relevant to our discussion of material truncation in Sec.~\ref{s1}.

\subsubsection{Quantum optical master equation}\label{master}

We now turn our attention to a more realistic setting by deriving the quantum optical master equation for the ${\rm dipole}_\alpha$, which can be viewed as a detector for the corresponding $\alpha$-gauge radiation field. We will show that the weak-coupling approximations comprising the traditional quantum optics paradigm, have the effect of mimicking the $S$-matrix and they thereby cause all $\alpha$-dependence to drop out of the final result. More precisely, they ensure that all master equation coefficients are well-known second-order QED matrix elements. A similar demonstration has been given for a pair of two-level dipoles in Ref.~\cite{stokes_master_2018}. Here we consider only one dipole (the detector), but we do not restrict our attention to only two dipolar energy levels. The Hamiltonian reads
\begin{align}\label{HAd}
H&=h+V^\alpha_1+V^\alpha_2\\
h&= \sum_n \epsilon^n\ket{\epsilon^n}\bra{\epsilon^n} + \int d^3 k \sum_\lambda \omega \left(a^\dagger_\lambda({\bf k})a_\lambda({\bf k})+{1\over 2}\right),\\
V^\alpha_1 &= -(1-\alpha){q\over m}{\bf p}\cdot {\bf A}_{\rm T}({\bf 0}) + \alpha q{\bf r}\cdot {\bf \Pi}({\bf 0}),\\
V^\alpha_2 &=(1-\alpha)^2{q^2\over 2m}{\bf A}_{\rm T}({\bf 0})^2 + {\alpha^2q^2\over 2}{\bf r}\cdot \delta^{\rm T}({\bf 0})\cdot {\bf r}
\end{align}
where $h$, $V^\alpha_1$ and $V^\alpha_2$ are zeroth, first and second order in $q$ respectively.

We make the following weak-coupling approximations concerning the state of the ${\rm detector}_\alpha$ represented by the density operator $\rho(t)$ in a suitable interaction picture:
\begin{enumerate}
\itemsep0em
\item{Born approximation: The dipole and reservoir are uncorrelated over the relevant timescale.}
\item{Second order perturbation theory: The coupling is much smaller than the unperturbed energies.}
\item{Markov approximation A: The system dynamics are memoryless; $\rho(s)\approx \rho(t)$ for all $s\in [0,t]$.}
\item{Markov approximation B: The temporal limit of the integrated Von-Neumann equation is $t\approx \infty$.}
\item{Secular (rotating-wave) approximation: Rapidly oscillating contributions are negligible.}
\end{enumerate}
The Markov approximations mimic the adiabatic switching condition of the $S$-matrix and together with the secular approximation they enforce bare-energy conservation.

The derivation of the quantum optical master equation is well-known \cite{breuer_theory_2007}, but we repeat it in Supplementary Note~XI using an arbitrary gauge $\alpha$ in order to show how approximations 1-5 cause all $\alpha$-dependence to drop out. Specifically, approximation 1 ensures that the master equation coefficients can be calculated using the photonic vacuum at any time $t$. Approximation 2 ensures that they are second order in $q$. Approximation 3 ensures they can be calculated independent of $\rho$. Approximation 4 ensures that the expected energy denominators are obtained as in the $T$-matrix, and approximation 5 ensures that they are evaluated on-energy-shell. By reducing all master equation coefficients to well-known QED matrix elements the approximations 1-5 ensure $\alpha$-independence.

In the Schr\"odinger picture the final result reads
\begin{align}\label{me}
{\dot \rho} = i[\rho,{\bar H}_m]+\sum_{\substack{n,m \\ n>m}} \Gamma_{nm}\left(L_{nm}\rho L_{nm}^\dagger -{1\over 2}\left\{L_{nm}^\dagger L_{nm},\rho\right\} \right)
\end{align}
where ${\bar H}_m=H_m+\Delta$ and where $\Delta$ and $\Gamma_{nm}$ are $\alpha$-independent QED matrix elements, namely, the Lamb-shift and the Fermi-golden-rule spontaneous emission rate respectively (see Supplementary Note~XI). The Lindblad operators are $L_{nm}=\ket{\epsilon^m}\bra{\epsilon^n}$. The master equation (\ref{me}) is readily extended to a finite temperature reservoir \cite{breuer_theory_2007}. Clearly:
\begin{itemize}
\item{The reduced description of the ${\rm detector}_\alpha$ is $\alpha$-independent within the approximations 1-5, such that ``detector" becomes an ostensibly unique theoretical concept.}
\end{itemize}
The stationary state $\rho_0$ of this detector is the bare ground state $\rho_0 = \ket{\epsilon^0}\bra{\epsilon^0}$, according to which the probability of excitation of the detector initially in the ground state is ${\cal P}_{{\rm d},0}(t)=0$ for all $t$. Within the approximations made photon emission requires a downward dipolar transition and absorption an upward one. Furthermore, the energies of any photons involved must be exactly equal to the energies of the corresponding dipolar transitions involved. The processes captured by the master equation (\ref{me}) are precisely those captured by the $S$-matrix wherein $h$ is strictly conserved.

Outside of the approximations 1-5 emission and absorption can occur without preserving the number of bare quanta, but evidently such (``virtual") processes are not perfectly bare-energy conserving and they are non-secular and/or non-Markovian inasmuch that they are only eliminated when both Markov and secular approximations are performed. These processes are allowed (not only as intermediates) by the more general matrix $S^{(\tau)}$ defined in Eq.~(\ref{stau}) and although they are viewed as unphysical in scattering theory (except as intermediates), in open quantum systems theory the opposite is true; they are allowed unless they have been suppressed by {\em approximations} whose avoidance must provide a {\em more} accurate description. Moreover, these approximations have a relatively narrow regime of validity \cite{breuer_theory_2007}. There is presently considerable interest within open quantum systems theory in understanding strong-coupling and non-Markovian effects using both numerical and analytical methods \cite{ishizaki_quantum_2010,breuer_colloquium_2016,de_vega_dynamics_2017,nazir_modelling_2016,nazir_reaction_2018}. From this perspective, when the approximations 1-5 break down the idealisations used to define the $S$-matrix must be interpreted as no longer realistic.

\subsection{Time-dependent interactions and ground state photons}\label{nonadiabatic}

We now turn our attention to non-adiabatic interaction switching whereby the gauge relativity of subsystems cannot be ignored. It is sometimes argued that the Coulomb gauge must be used to describe residual photon population left after a sufficiently fast interaction switch-off (e.g. Ref. \cite{stefano_resolution_2019,settineri_gauge_2021}). In fact, the correct description depends on the experimental context \cite{stokes_ultrastrong_2021} as will be discussed in detail below.

The ground state of a light-matter system is gauge invariant, but its representation using a vector differs between gauges (see Sec.~\ref{gamb}). This gives rise to different photon number predictions all of which are physical. The different predictions within one and the same physical state correspond to different gauge invariant definitions of a photon. The task remains of determining which prediction is most relevant in which situations. For our purpose it is sufficient to consider the simple $\alpha$-gauge framework, but it should be borne in mind that the gauge function is completely arbitrary and the following considerations apply generally.

For each fixed $\alpha$ the Hamiltonian operator $H_\alpha$ represents the same total energy observable $E$. The total energy eigenvectors are defined by $H_\alpha\ket{E_\alpha^n}=E^n\ket{E_\alpha^n}$ where the eigenvalues $E^n$ are manifestly $\alpha$-independent (unitary transformations are isospectral). According to the postulates of quantum theory, the vector $\ket{E^n_\alpha}$ represents, within the gauge $\alpha$, the unique physical state in which the system definitely possesses energy $E^n$. Consider now the average
\begin{align}\label{Nalph}
N_\alpha =\bra{G_\alpha}\sum_{{\bf k}\lambda} a^\dagger_\lambda({\bf k})a_\lambda({\bf k}) \ket{G_\alpha} = \bra{G_\alpha} n \ket{G_\alpha}
\end{align}
where the vector $\ket{G_\alpha}=\ket{E_\alpha^0}$ represents the ground state in the gauge $\alpha$ and where $a_\lambda({\bf k})$ is defined in Eq.~(\ref{phot}). At first glance it seems that the predicted photon number $N_\alpha$ is fundamentally gauge noninvariant, and that this is because $\ket{G_\alpha}$ depends on $\alpha$, but this is not the case. Rather, the operator $n$ represents the gauge invariant number of photons defined {\em relative} to the gauge $\alpha$. In the gauge $\alpha'$ the same observable is represented by $n'=R_{\alpha\alpha'}nR_{\alpha\alpha'}^\dagger$ and the physical ground state is represented by the vector $\ket{G_{\alpha'}}=R_{\alpha\alpha'}\ket{G_\alpha}$. Thus, $N_\alpha$ is gauge invariant; $N_\alpha = \bra{G_\alpha} n \ket{G_\alpha} = \bra{G_{\alpha'}} n' \ket{G_{\alpha'}}$. For each different fixed value of $\alpha$ the average $N_\alpha$ is that of a different physical observable and it is therefore a different gauge invariant prediction. The subscript $\alpha$ labels which particular gauge invariant definition of photon is being considered. A special case is the number of ${\bf E}_{\rm T}$-type photons given by $N_0=:N_{{\bf E}_{\rm T}}$, because ${\tilde {\bf \Pi}}({\bf k})=-{\tilde {\bf E}}_{\rm T}({\bf k})$ when $\alpha=0$. Another special case is the number of ${\bf D}_{\rm T}$-type photons, which is given by $N_1=:N_{{\bf D}_{\rm T}}$, because ${\tilde {\bf \Pi}}({\bf k})=-{\tilde {\bf D}}_{\rm T}({\bf k})$ when $\alpha=1$.

Let us consider a system prepared in the ground state before we suddenly switch-off the interaction. When the interaction vanishes photons are defined as in Eq.~(\ref{phot}), but this definition is now {\em unique}, because the non-interacting canonical momentum is unique; ${\bf \Pi}=-{\bf D}_{\rm T}=-{\bf E}_{\rm T}=-{\bf E}$. We can therefore ask how many of these unique photons are present for times $t>t_f$ if the interaction is suddenly switched-off at $t=t_f$? Modelling this situation using a time-dependent coupling in the gauge $\alpha$ gives the Hamiltonian
\begin{align}
H_\alpha(t) = H_m+H_{\rm ph}+\theta(t-t_f)V^\alpha(\eta)
\end{align}
where $\theta$ is the Heaviside step-function and $\eta$ is a coupling parameter such that $V^\alpha(0)=0$. These $H_\alpha(t)$ are clearly not equivalent to each other for different $\alpha$ \cite{stokes_ultrastrong_2021}. This is unsurprising because for $\alpha\neq \alpha'$, $H_\alpha(t)$ and $H_{\alpha'}(t)$ clearly model two different experiments in which $V^\alpha$ and $V^{\alpha'}$ are suddenly switched-off, respectively. For each $\alpha$ the evolution generated by $H_\alpha(t)$ from time $t=0$ consists of sequential evolutions; $U_\alpha(t) = e^{-i(H_m+H_{\rm ph})(t-t_f)}e^{-iH_\alpha t_f}$. It follows that the gauge invariant physical prediction $N_\alpha=:N_{{\bf E}_{\rm T}+\alpha{\bf P}_{\rm T}}$ gives the number of photons left over in an experiment realising a sudden switch-off of the $\alpha$-gauge interaction. In Ref.~\cite{settineri_gauge_2021} it is noted that the particular prediction $N_0=N_{{\bf E}_{\rm T}}$ is gauge invariant, but as we have shown more generally the same is true of any of the predictions $N_\alpha$.

There is a famous set of experiments for which it is well-known that the sudden switching condition appears ill-justified in the Coulomb gauge as compared with the multipolar gauge, these being the early experiments of Lamb \cite{lamb_fine_1952,power_coulomb_1959,fried_vector_1973,davidovich_theory_1980,milonni_natural_1989,woolley_gauge_2000,stokes_gauge_2013}. The natural lineshape prediction can be obtained by assuming the atom to be initially in a bare excited state with no photons. This amounts to a sudden switch-on of the interaction \cite{milonni_natural_1989}. Within the multipolar gauge the prediction is sufficiently close to the experimental result to rule out the corresponding Coulomb gauge prediction \cite{power_coulomb_1959,fried_vector_1973,davidovich_theory_1980,milonni_natural_1989,woolley_gauge_2000,stokes_gauge_2013}. Put differently, the multipolar gauge subsystems are more relevant for the description of this experiment. The natural lineshape of spontaneous emission is discussed in detail in Sec.~\ref{line}.

It should be clear that one can consider more general time-dependent interactions and the same considerations will apply. The generalisation can be achieved by letting
\begin{align}
H_\alpha(t) = H_m+H_{\rm ph}+V^\alpha(\eta\mu(t))
\end{align}
where $\mu(t)$ is an arbitrary coupling envelope that vanishes smoothly after some time $t_f$. Let us suppose, as in Ref.~\cite{stokes_ultrastrong_2021}, that $\mu(t)$ vanishes before some time $t_i$ so the system can be prepared at $t=0<t_i$ in the ground state represented by eigenvector $\ket{g}$ of $h$. The total number of photons at time $t>t_f$ is
\begin{align}\label{nat}
N_\alpha(t) = \bra{g}U_\alpha(t) n U_\alpha(t)^\dagger \ket{g}
\end{align}
where $U_\alpha(t)$ is the evolution operator generated by $H_\alpha(t)$. To prove the gauge invariance of $N_\alpha(t)$ one must of course take into account that gauge transformations are now time-dependent, because they depend on the coupling parameter; $R_{\alpha\alpha'}(\eta\mu(t))\equiv R_{\alpha\alpha'}(t)$. The vector $\ket{g_\alpha(t)}=U_\alpha(t)\ket{g}$ represents the Schr\"odinger-picture state at time $t$ in the gauge $\alpha$. The same physical state is represented in the gauge $\alpha'$ by the vector $\ket{g_\alpha^{\alpha'}(t)}=R_{\alpha\alpha'}(t)\ket{g_\alpha(t)}$. The physical observable represented by $n$ in the Schr\"odinger picture in the gauge $\alpha$, is represented by $n^{\alpha'}(t)=R_{\alpha\alpha'}(t)nR_{\alpha\alpha'}(t)^\dagger$ in the gauge $\alpha'$. We see therefore that $N_\alpha(t)$ is a gauge invariant prediction.

The two different vector representations $\ket{g_\alpha(t)}$ and $\ket{g_\alpha^{\alpha'}(t)}$ of the state at $t$, satisfy the Schr\"odinger equations $id\ket{g_\alpha(t)}/dt=H_\alpha(t)\ket{g_\alpha(t)}$ and $id \ket{g_\alpha^{\alpha'}(t)}/dt=H_\alpha^{\alpha'}(t)\ket{g_\alpha^{\alpha'}(t)}$. The Hamiltonians $H_\alpha^{\alpha'}(t)$ and $H_\alpha(t)$ are easily related via direct differentiation of the expression $\ket{g_\alpha^{\alpha'}(t)} = R_{\alpha\alpha'}(t)\ket{g_\alpha(t)}$, which implies
\begin{align}\label{hamtim}
H^{\alpha'}_\alpha(t) = R_{\alpha\alpha'}(t)H_\alpha(t)R_{\alpha\alpha'}(t)^\dagger + i{\dot R}_{\alpha\alpha'}(t)R_{\alpha\alpha'}(t)^\dagger.
\end{align}
It is a trivial matter to generate an equivalent model to any one of the $H_\alpha(t)$ by properly accounting for the time-dependence of gauge transformations. The Hamiltonian $H^{\alpha'}_\alpha(t)$ depends on two parameters $\alpha$ and $\alpha'$ which have different roles. The parameter $\alpha$ selects the gauge within which the time-dependent coupling assumption, $e\to e(t)$, has been made, whereas the parameter $\alpha'$ selects the choice of gauge used for calculations {\em after} this assumption has been made. The non-equivalence of the $H_\alpha(t)$ for different $\alpha$ shows that $e\to e(t)$ constitutes a different physical assumption in different gauges. In other words, gauge ambiguities arise because each $H_\alpha(t)$ generates its own equivalence class ${\cal S}_\alpha = \{H^{\alpha'}_\alpha(t):\alpha'\in {\mathbb R}\}$ and distinct classes describe different experiments. The particular prediction $N_\alpha(t)$ is {\em relevant} if the experimental protocol being modelled happens to realise a switch-on/off of the interaction $V^\alpha$. If, for example, the experimental arrangement considered is somehow capable of effectively manipulating the (gauge invariant) bare dipole moment $q{\bf r}$, then the multipolar gauge interaction might be controlled.

%%%%%%%%%%%%%%%%%%%%%%%%%%%%%%%%%%%%%%%%%%%%%%%%%%%%%%%%%%%%%%%%%%%%%%%%%%%%%%%%%%%%%%%%%%%%
%%
%%	F I G U R E S  S T A R T
%%
%%%%%%%%%%%%%%%%%%%%%%%%%%%%%%%%%%%%%%%%%%%%%%%%%%%%%%%%%%%%%%%%%%%%%%%%%%%%%%%%%%%%%%%%%%%
\begin{figure}[t]
\includegraphics[width=0.9\linewidth, height=7cm, keepaspectratio]{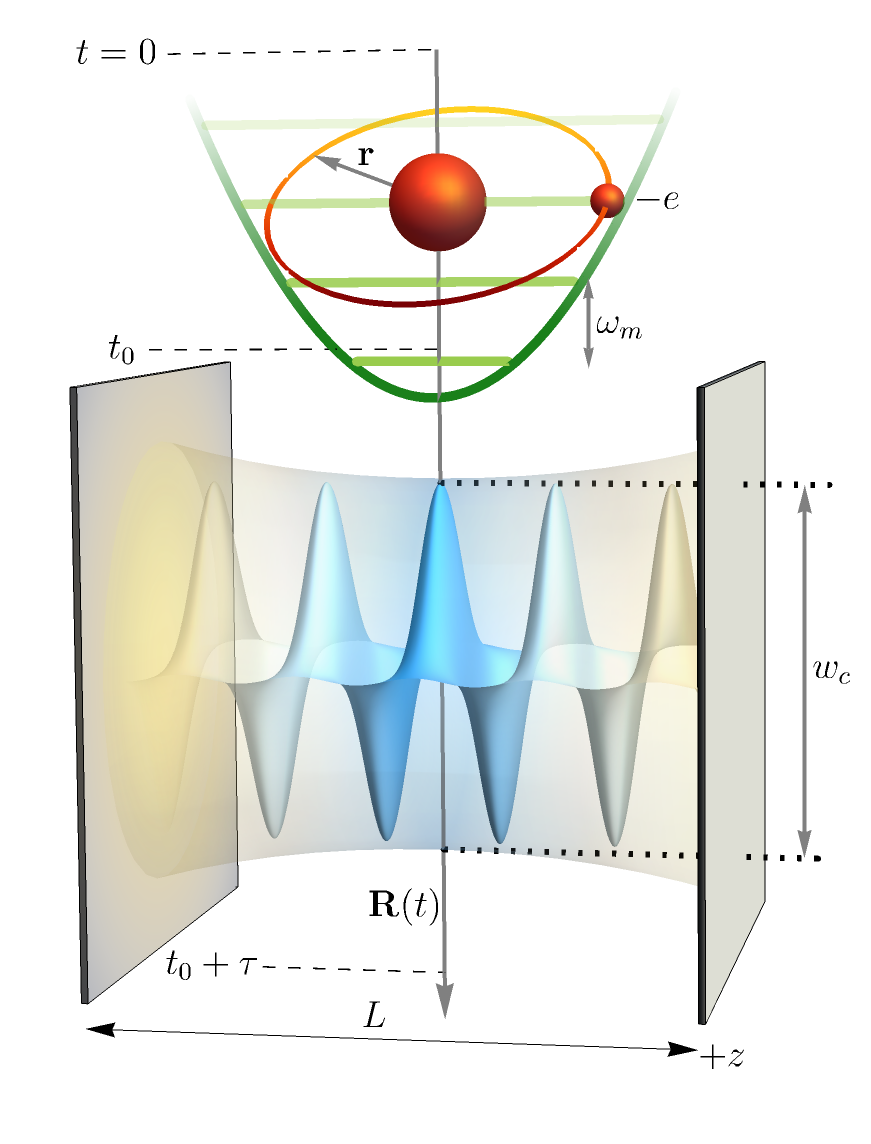}\vspace*{-4mm}
\caption{A cavity of length $L$ supporting standing waves in the $z$-direction and a Gaussian perpendicular mode profile with waist $w_c$ is depicted, along with a dipole $-er$ oscillating with frequency $\omega_m$. At $t=0$ the cavity and dipole are non-interacting. The dipole follows a classical trajectory ${\bf R}(t)$ through the cavity, entering the cavity at $t_0$ and exiting at $t_0+\tau$.}\label{pic2}
\end{figure}
%%%%%%%%%%%%%%%%%%%%%%%%%%%%%%%%%%%%%%%%%%%%%%%%%%%%%%%%%%%%%%%%%%%%%%%%%%%%%%%%%%%%%%%%%%%%
%%
%%	F I G U R E S  E N D
%%
%%%%%%%%%%%%%%%%%%%%%%%%%%%%%%%%%%%%%%%%%%%%%%%%%%%%%%%%%%%%%%%%%%%%%%%%%%%%%%%%%%%%%%%%%%%
These points are demonstrated directly in Ref.~\cite{stokes_ultrastrong_2021}, which considers the concrete setup of a dipole uniformly moving in and out of a Gaussian cavity mode, as depicted in Fig.~\ref{pic2}. This situation can be modelled using a Gaussian envelope $\mu(t)$. In addition to the non-equivalent models $H_\alpha(t)$, a more complete description ${\tilde H}_\alpha(t)$ is provided by retaining an explicit model for the control system, which in this example is the centre-of-mass motion of the dipole. Unlike the $H_\alpha(t)$ the more complete descriptions ${\tilde H}_\alpha(t)$ are equivalent to each other for different $\alpha$. In this way, the procedure of using a time-dependent coupling $\eta\mu(t)$ can be viewed as an approximation. The value of $\alpha$ such that ${\tilde H}_\alpha(t)=H_\alpha(t)$ is then the correct value to choose when describing the experiment using the result of this approximation, $H_\alpha(t)$.

It is shown in Ref. \cite{stokes_ultrastrong_2021} that if there exists a value $\alpha$ for which ${\tilde H}_\alpha(t)=H_\alpha(t)$, then the value depends strongly on the experimental protocol. The prediction $N_\alpha(t)$ obtained using $H_\alpha(t)$ is correct if and only if the dipole moment is aligned with the mode polarisation and these vectors make an angle $\theta$ with the direction of the centre-of-mass motion such that $\cos^2\theta=\alpha$. The result clearly demonstrates that in general, which prediction $N_\alpha(t)$ is the correct ({\em relevant}) one, depends strongly on the experimental context. It is certainly not the case that $N_0(t)$ is always the correct prediction. The result further illustrates why there are indeed gauge ambiguities. In order to find which of the predictions $N_\alpha(t)$ may be relevant for describing a concrete setup and experimental protocol, Ref. \cite{stokes_ultrastrong_2021} resorts to invoking an explicit model of the control system. The result obtained could not be anticipated without such a description, and yet such descriptions are only available in the simplest of cases whereby the control system accommodates tractable modelling.

\section{Measurements and virtual photons
}\label{pd}

We now turn to the topic of subsystem measurements. Their description when dealing with virtual processes within the weak-coupling regime was considered some time ago via simple models \cite{compagno_detection_1988,compagno_dressed_1988,compagno_dressed_1990,compagno_bare_1991,compagno_atom-field_1995}. The topic has recently been taken up when dealing with ultrastrong interactions \cite{di_stefano_photodetection_2018,settineri_gauge_2021}. We focus on a system consisting of a source and/or a detector within a single photonic environment. This situation is distinguished from the case of a source and a detector occupying different environments that are modelled separately, such as a source within a cavity with a detector external to the cavity. The outlook for this latter situation is discussed from Sec.~\ref{excavfs} onward.

The natural starting point for our considerations is Glauber's photodetection theory \cite{glauber_quantum_1963,glauber_quantum_2007}. We review aspects of photodetection that are important beyond the standard quantum optics paradigm including how photodetection divergences are related to virtual excitations. We consider the gauge relativity of the predicted natural lineshape of spontaneous emission   and determine the relation between subsystem gauge relativity, locality and dressing.

The important conclusion of this section is that outside of conventional weak-coupling and Markovian regimes there is necessarily a trade-off between defining material systems as localised objects versus avoiding virtual vacuum excitations. In the multipolar gauge material systems are most localised. We will see in Sec.~\ref{loc_caus} that if such a ``detector" is deemed accessible and is therefore prepared in its lowest energetic state, then under the influence of the interaction it will necessarily become excited even within the corresponding photonic vacuum, because this state is not the ground state of the interacting composite. These virtual excitations are not encountered if one instead defines physical subsystem excitations relative to the true ground state. This, however, constitutes defining the physical subsystems relative to an unconventional gauge (neither Coulomb nor multipolar). Material systems defined in this way, are necessarily delocalised to some extent. Thus, while in practice a detection process necessarily possesses finite extent in space and time, theoretically some degree of spatial localisation of a detector must be sacrificed if one wishes to eliminate the prediction of its virtual excitation.

\subsection{Conventional photodetection theory and its limitations}\label{glauber}

Glauber photodetection theory \cite{glauber_quantum_1963,glauber_quantum_2007} has been a major workhorse in weak-coupling quantum optics and constitutes a natural starting point. Here we briefly review this theory and its limitations. Photodetection in the context of ultrastrongly-coupled light-matter systems is discussed in sections~\ref{excavfs} and \ref{specs}.

\subsubsection{Real excitations}

Typical photodetectors work by photon-ionisation amplified to produce a macroscopic current. As such they are substantial objects consisting of photoconductive electrons over a cross-sectional area $S$, that is correlated with detection efficiency. As well as being big, such detectors are also typically slow to respond, at least, compared to the correlation times of the photonic reservoirs that they monitor. Thus, actual photon measurements are not restricted to individual points in spacetime and this fact is certainly relevant outside of weak-coupling regimes. However, as a model for dealing with weakly-coupled detectors we may consider a localised ``detector" dipole ${\bf d}=q{\bf r}$ fixed at the origin ${\bf 0}$. The charge $q$ is a suitable perturbation parameter [the fine-structure constant being $q^2/(4\pi)$].

In each gauge the unperturbed eigenvectors of $h =H_\alpha-V^\alpha$ represent different physical states. Photons are by definition quanta of the ``light" subsystem and a detector is a ``material" subsystem. A photo-detection process therefore involves an energetic change of the material system, usually accompanied by a change in the number of photons, i.e., it is a process between unperturbed states. In general these states do not coincide with well-defined states of energy of the light-matter composite and so they are not stationary.  Examining photo-detection probabilities in a particular gauge $\alpha$ provides insight into the physical natures of the ``light" and ``matter" subsystems defined relative to the gauge $\alpha$.

In conventional treatments (e.g \cite{glauber_quantum_2007}) a linear dipolar form of coupling is adopted as occurs in the multipolar gauge. This is often written $V^1=-{\bf d}\cdot {\bf E}_{\rm T}({\bf 0})$ or else $V^1=-{\bf d}\cdot {\bf E}({\bf 0})$. Neither expression is correct. As noted in Sec.~\ref{minc} the correct linear part of the multipolar interaction in the EDA is $V^1=-{\bf d}\cdot {\bf D}_{\rm T}({\bf 0})$. Two further common misconceptions are that the Coulomb gauge defines photons using the electric field, and that this is the basic field that first enters into Glauber's photodetection theory [e.g. Ref.~\cite{settineri_gauge_2021}]. In fact, the Coulomb gauge defines photons using ${\bf E}_{\rm T}\neq {\bf E}$ and %in conventional treatments that employ a dipolar coupling the relevant field entering the theory is correctly identified as ${\bf D}_{\rm T}({\bf 0})$. A
at the dipole's position ${\bf 0}$, the relevant field ${\bf D}_{\rm T}$ is also {\em infinitely different} to ${\bf E}({\bf 0})={\bf D}_{\rm T}({\bf 0})-q{\bf r}\delta({\bf 0})$. However, the infinite term ${\bf P}_1({\bf x})=q{\bf r}\delta({\bf 0})$ is a difference in the source components of the two fields, which are of at least $O(q)$. Since the detector's dipole moment is of order $q$ only the free (vacuum) component of ${\bf D}_{\rm T}({\bf 0})$ contributes to detection probabilities to order $q^2$  in an initially unperturbed state, and this may be taken to coincide with the free component of ${\bf E}({\bf 0})$.

We begin by following conventional treatments, which define the subsystems photons$_\alpha$ and detector$_\alpha$ relative to the multipolar gauge $\alpha=1$, and then employ perturbation theory to order $q^2$. The probability to find the detector$_1$ excited into the $n$'th level at time $t$, given the initial state $\ket{\epsilon^m,\psi_{\rm ph}}$ with a fixed number of photons$_1$ and with $m<n$, is
\begin{align}\label{gpd}
&{\cal P}_{\rm d}^{nm}(t)=d_{nm,i}d_{mn,j}\int_0^t ds \int_0^t ds' e^{i\omega_{nm}(s'-s)} G_{ij}(s,s')
\end{align}
where repeated indices are summed, and where
\begin{align}
G_{ij}(s,s') = \bra{\psi_{\rm ph}}E_{{\rm vac},i}(s,{\bf 0})E_{{\rm vac},j}(s',{\bf 0})\ket{\psi_{\rm ph}}
\end{align}
in which
\begin{align}\label{modeE}
{\bf E}_{\rm vac}(t,{\bf x})=&-i\int d^3 k \sum_\lambda \sqrt{\omega\over 2(2\pi)^3} {\bf e}_\lambda({\bf k}) \nonumber \\ & \times \left[a^\dagger_\lambda(0,{\bf k})e^{i\omega t-i{\bf k}\cdot {\bf x}}-a_\lambda(0,{\bf k})e^{-i\omega t+i{\bf k}\cdot {\bf x}}\right]
\end{align}
denotes the free component of ${\bf D}_{\rm T}(t,{\bf x})$. Since $\omega_{nm}>0$ the anti-normally ordered contribution in Eq.~(\ref{gpd}) is taken as rapidly oscillating and is neglected in a rotating-wave approximation (RWA), such that we may let
\begin{align}\label{gnon}
G_{ij}(s,s') = \bra{\psi_{\rm ph}}E^{(-)}_{{\rm vac},i}(s,{\bf 0})E^{(+)}_{{\rm vac},j}(s',{\bf 0})\ket{\psi_{\rm ph}}
\end{align}
where
\begin{align}
&{\bf E}^{(+)}_{\rm vac}(t,{\bf 0}) = i\int d^3 k \sum_\lambda \sqrt{\omega\over 2(2\pi)^3} {\bf e}_\lambda({\bf k})a_\lambda(0,{\bf k})e^{-i\omega t},\nonumber \\ & {\bf E}^{(-)}=({\bf E}^{(+)})^\dagger.
\end{align}
We see that normal-ordering occurs as an approximation based on the detector$_1$ excitation process having a supposedly dominant contribution coming from photon$_1$ absorption. The neglected contribution is virtual, i.e., number non-conserving, corresponding to detector$_1$ excitation with emission of a photon$_1$.

The detector$_1$ level $n$ typically belongs to the ionisation continuum and after excitation a number of physical processes must occur for a detection event to actually be registered. The description of these processes is subsumed into a classical epistemic probability ${\cal D}_n$ for a detection event given excitation to the level $n$. The total probability of detection is therefore
\begin{align}
{\mathscr P}_{\rm d}^{m}(t) =\sum_n {\cal D}_n {\cal P}_{\rm d}^{nm}(t)
\end{align}
where formally the summation over $n$ is understood to include integration over continuum levels. Defining the spectral density (sensitivity)
\begin{align}
{\cal S}_{ij}(\omega) =2\pi\sum_n d_{nm,i}d_{mn,j}{\cal D}_n \delta(\omega-\omega_{nm})
\end{align}
enables one to model different detection schemes by assuming different forms of ${\cal S}_{ij}(\omega)$. The photon$_1$ counting rate is
\begin{align}\label{rate}
{d {\cal P}_{\rm d}^{nm} \over dt} = 2{\rm Re}\int_{-\infty}^\infty {d\omega\over 2\pi} {\cal S}_{ij}(\omega) {\cal G}_{ij}(\omega,t)
\end{align}
where
\begin{align}\label{gnorm}
{\cal G}_{ij}(\omega,t)= \int_0^t ds\, e^{i\omega(t-s)} G_{ij}(s,t)
\end{align}
whose Fourier transform is
\begin{align}\label{gnorm2}
{\mathscr G}_{ij}(s,t)&= \int {d\omega\over 2\pi} \, e^{i\omega s} {\cal G}_{ij}(\omega,t)\nonumber \\
&=\theta(s)\theta(t-s)G_{ij}(t-s,t),
\end{align}
which vanishes unless $0\leq s\leq t$. Since photodetectors are slow, the measurement time $t$ is typically much longer than the reservoir correlation time $T_c = 1/\Delta\omega_G$ where $\Delta \omega_G$ is the bandwidth of the correlation function $G_{ij}$. Therefore, the $s$-width of ${\mathscr G}_{ij}(s,t)$ is approximately $T_c$.

Glauber defines an ideal broadband detector as one with a flat spectral density ${\cal S}_{ij}(\omega)={\cal S}_{ij}$ \cite{glauber_quantum_2007}. This requires that the width of the sensitivity function must be much larger than $\Delta\omega_G=1/T_c$, such that ${\cal G}_{ij}(\omega,t)$ is sharply peaked as a function of $\omega$ when compared with ${\cal S}_{ij}(\omega)$. The photon counting rate is then simply $S_{ij}{\cal G}_{ij}(t,t)$ such that if $S_{ij} \sim \delta_{ij}$ then the rate is proportional to the Glauber intensity
\begin{align}
I_G(t)=\langle {\bf E}_{\rm vac}^{(-)}(t,{\bf 0})\cdot {\bf E}_{\rm vac}^{(+)}(t,{\bf 0})\rangle.
\end{align}

\subsubsection{Virtual excitations}\label{virtual}

As in the textbook \cite{mandel_optical_1995} the actual field involved in photodetection theory can be left open by defining
\begin{align}\label{epm}
{\bf F}^{(+)}(t,{\bf x})=\int {d^3 k\over \sqrt{2(2\pi)^3}} \sum_\lambda {\bf e}_\lambda({\bf k}) \beta(\omega) a_\lambda(t,{\bf k})e^{i{\bf k}\cdot {\bf x}}
\end{align}
where a number of noteworthy choices of $\beta(\omega)$ can be made. For example, if $\beta(\omega)=i\sqrt{\omega}$ then ${\bf F}={\bf D}_{\rm T}$. If $\beta(\omega)=1/\sqrt{\omega}$ then ${\bf F}={\bf A}_{\rm T}$. If $\beta(\omega)=1$ then ${\bf F}^{(+)}$ defines a direct inverse Fourier transform of $\sum_\lambda {\bf e}_\lambda({\bf k})a_\lambda({\bf k})/\sqrt{2}$. This last choice of $\beta$ is noteworthy for the reason that although it is impossible to define a local number operator for relativistic quanta \cite{fulling_aspects_1989,mandel_optical_1995,haag_local_1996}, the operator ${\bf F}^{(-)}({\bf x})\cdot {\bf F}^{(+)}({\bf x})$ can be interpreted as a real-space number density of photons that are approximately localised on a scale much larger than the corresponding wavelengths \cite{mandel_optical_1995} (see also Supplementary Note~XII). We remark that being local in ${\bf k}$-space, the relation between fields corresponding to different $\beta(\omega)$ in Eq.~(\ref{epm}) is highly non-local in spacetime. This point is relevant to understanding the interplay between electromagnetic dressing and localisation and is discussed further in Supplementary Note~XIII

To understand the limitations of conventional photodetection theory we return to Eq.~(\ref{gpd}). If we assume the vacuum state $\ket{\psi_{\rm ph}}=\ket{0}$ and we allow the levels $m$ and $n$ to be arbitrary, then evaluating the polarisation summation and angular integrals gives
\begin{align}\label{Pvac}
{\cal P}_{\rm d,vac}^{nm}(t)= {|{\bf d}_{nm}|^2 \over 3\pi} \int_0^\infty d\omega\, \omega^3 {\sin^2 \left[(\omega_{mn}-\omega)t/2\right] \over \pi(\omega_{mn}-\omega)^2/2}.
\end{align}
If $m>n$ the process described is spontaneous emission. If $n>m$ then the process described is virtual. The dominant peak of the integrand then lies outside of the domain of integration and is oscillatory for positive frequencies. The amplitude of the oscillations in the integrand grows with $\omega$ due to the prefactor of $\omega^3$. This behaviour is only bounded by an ultra-violet cut-off $\omega_M$ and the integral is in fact quadratically divergent with $\omega_M$. The divergence is relatively severe, such that ${\cal P}_{\rm d,vac}^{nm}(t)$ is non-negligible even for realistic, yet modest values of $\omega_M$ that are consistent with, for example, the EDA and the nonrelativistic treatment \cite{drummond_unifying_1987,stokes_extending_2012}.

If we repeat the derivation of the detector excitation rate for a detector$_0$, i.e., for a detector defined relative to the Coulomb gauge, then the field entering into the theory is now ${\bf A}_{\rm T}({\bf 0})$, which amounts to letting $\beta(\omega)=1/\sqrt{\omega}$ in Eq.~(\ref{epm}). In place of Eq.~(\ref{Pvac}) we obtain
\begin{align}\label{Pvac2}
{\cal P}_{\rm d,vac}^{nm}(t)= {|{\bf d}_{nm}|^2 \over 3\pi} \int_0^\infty d\omega\, \omega\omega_{mn}^2 {\sin^2 \left[(\omega_{mn}-\omega)t/2\right] \over \pi(\omega_{mn}-\omega)^2/2}.
\end{align}
When $n>m$, the probability is in this case only logarithmically divergent. This is a direct consequence of the ${\bf k}$-space normalisation of the field ${\bf A}_{\rm T}$, which varies as $1/\sqrt{\omega}$. 

The probability ${\cal P}_{\rm d,vac}^{nm}(t)$ is generally non-zero because the initial unperturbed state consisting of no photons and $m$ excitations of the detector is not an eigenstate of the Hamiltonian and in particular it is not the ground state even if $m$ is the lowest dipolar level. If this final result is deemed unphysical then we must conclude that the assumed physical states are not operationally relevant in the description of photo-detection. In particular, if the physical detector is not the localised detector$_1$, then it must be delocalised to some extent. The interplay between localisation and dressing is discussed further from Sec.~\ref{loc_caus} onward.

The virtual detection probability ${\cal P}_{\rm d,vac}^{nm}(t)$ with $n>m$ was removed in the progression from Eq.~(\ref{gpd}) to Eq.~(\ref{rate}) using the RWA. The counting rate $d{\cal P}_{\rm d,vac}^{nm}/dt$ without the RWA can be found by direct differentiation of Eq.~(\ref{Pvac}) and can again be reduced to the gauge non-relative Fermi-golden-rule rate
\begin{align}\label{fgrr}
{d{\cal P}_{\rm d,vac}^{nm}\over dt}=\begin{cases}
\omega_{mn}^3|{\bf d}_{nm}|^2/(3\pi)=:\Gamma_{mn},~~~&n<m\\
0,~~~&n>m.
\end{cases}
\end{align}
in three different ways, all of which amount to imposing strict bare-energy conservation as in the $S$-matrix:\\

1) Differentiation of Eq.~(\ref{Pvac}) yields the frequency integrand $\omega^3 \sin[(\omega_{mn}-\omega)t]/(\omega_{mn}-\omega)$, which expresses a bare-energy-time uncertainty constraint. Taking the infinite-time limit $\lim_{t\to\infty} {\sin (\omega t)\over \pi \omega} = \delta(\omega)$ gives Eq.~(\ref{fgrr}).

2) Defining the counting rate as the difference quotient $({\cal P}_{\rm d,vac}^{nm}(t)-{\cal P}_{\rm d,vac}^{nm}(0))/t = {\cal P}_{\rm d,vac}^{nm}(t)/t$ yields via Eq.~(\ref{Pvac}) the frequency integrand $\omega^3 \sin^2[(\omega_{mn}-\omega)t]/([\omega_{mn}-\omega]^2t/2)$. In the limit $t\to \infty$ one obtains the right-hand-side of Eq.~(\ref{fgrr}) by using $\lim_{t\to\infty} {\sin^2 (\omega t/2)\over \pi \omega^2 t/2} = \delta(\omega)$. Meanwhile, the derivative $d{\cal P}_{\rm d,vac}^{nm}/dt$ on the left-hand-side of Eq.~(\ref{fgrr}) is  recovered in the limit $t\to 0$. This shows that the procedure for obtaining Eq.~(\ref{fgrr}) constitutes a form of Markov approximation that requires a clear separation of time scales as specified by the {\em Markovian regime} ${1\over \omega_{mn}}\ll t \ll {1\over \Gamma_{mn}}$. The final result is valid provided that matrix elements of the interaction Hamiltonian between initial and final unperturbed states are sufficiently small and slowly varying, as demonstrated by method 3).

3) Evaluating the prefactor $\omega^3$ in Eq.~(\ref{Pvac}) at resonance, $\omega=\omega_{mn}$, is valid if it can be considered sufficiently slowly varying compared with the peak in $\sin^2[(\omega_{mn}-\omega)t]/([\omega_{mn}-\omega]^2t/2)$ near to $\omega_{mn}$. One may then extend the lower integration limit to $-\infty$ by supposing that the integrand is dominated by this peak for sufficiently long times $\omega_Mt \gg 1$. This again yields Eq.~(\ref{fgrr}).
\\

It is not clear that any of the procedures 1), 2), or 3) can be justified for virtual excitation with $n>m$, because as already noted the dominant peak in $\sin^2[(\omega_{mn}-\omega)t]/([\omega_{mn}-\omega]^2t/2)$ then lies outside of the range of integration and the integral diverges quadratically with $\omega_M$. In this sense virtual contributions are non-Markovian.

Both of the predictions in Eqs.~(\ref{Pvac}) and (\ref{Pvac2}) are gauge invariant in the sense of Eq.~(\ref{eqt}), but without use of the Markovian approximation they are clearly different. This is an example of the gauge relativity expressed by inequality (\ref{ineqt}), which as noted in Sec.~\ref{adiabatic}, becomes important outside of Markovian regimes. We note that in any gauge, if the RWA is avoided and the broadband limit is taken then the photon counting rate is $S_{ij}G_{ij}(t,t)$ with $G_{ij}(t,t)$ given by Eq.~(\ref{gnon}) rather than Eq.~(\ref{gnorm}). Thus, a generally large virtual contribution occurs. However, the broadband limit is inapplicable to this contribution because the vacuum has infinite bandwidth. Thus, the significance of such contributions is in general dependent on the measurement schemes available.

In comparing the different predictions given by Eqs.~(\ref{Pvac}) and (\ref{Pvac2}), Power and Thirunamachandran noted that which one is the more accurate will depend on which set of {\em distinct physical states} represented by the same unperturbed vectors within the two gauges, are closer to the states actually realised in the considered experiment \cite{power_time_1999,power_time_1999-1}. Power and Thirunamachandran also noted that experiments could be used to determine which descriptions are most appropriate. Spectroscopic experimental signatures in particular, are discussed in Secs.~\ref{line} and \ref{specs}.

The elimination of divergent contributions requires ``renormalisation" of the ``bare" dipole by defining the ``physical" dipole relative to the appropriate gauge as recognised some time ago by Drummond \cite{drummond_unifying_1987}. One can use the elimination of virtual excitations as a criterion by which to select the most operationally relevant subsystem definitions, that is, to select the most appropriate gauge relative to which the dipole is to be defined in the context of photodetection. To this end let us consider a one-dimensional dipole harmonically quantised in the direction ${\hat {\bf u}}$ with canonical operators ${\bf r}= {\hat {\bf u}}(b^\dagger + b)/ \sqrt{2 m\omega_m}$ and ${\bf p} = i{\hat {\bf u}}(b^\dagger -b)\sqrt{m\omega_m / 2}$. From very early on purely bosonic models of this kind have been relevant to ultrastrong-coupling in polaritonic systems with quantum wells and microcavities \cite{todorov_ultrastrong_2010,ciuti_input-output_2006,ciuti_quantum_2005,bamba_dissipation_2012}.

We consider gauges of the form specified by Eq.~(\ref{Galph2}) while assuming that $\alpha({\bf k})=\alpha(\omega)$ is real and depends only on the magnitude of ${\bf k}$. We discretise the Fourier modes within a volume $v$ and combine wavevector and polarisation indices into a single mode label, writing $\alpha(\omega)=\alpha_k$. The polarisation self-energy term $\int d^3 x\, {\bf P}_{{\rm T}g}^2/2$ can be absorbed via new material modes such that ${\bf r}= {\hat {\bf u}}(d^\dagger + d)/\sqrt{2 m{\tilde \omega}_m}$ and ${\bf p} = i{\hat {\bf u}} \sqrt{m{\tilde \omega}_m}(d^\dagger -d)/\sqrt{2}$ where
\begin{align}
{\tilde \omega}_m^2=\omega_m^2+{q^2\over mv}\sum_k {({\bf e}_k\cdot {\hat {\bf u}})^2}\alpha_k^2.
\end{align}
Similarly, the order $q^2$ field self-energy term $q^2({\bf A}_g^{\rm EDA})^2/(2m)$ can be absorbed via radiative mode operators $c_k$ such that
\begin{align}
a_k &= \sum_j \left([\cosh\theta]_{kj} c_j + [\sinh\theta]_{kj} c^\dagger_j \right)\nonumber \\
&\approx c_k+\sum_j \theta_{kj}c_j^\dagger
\end{align}
where the approximate equality holds to order $q^2$ and
\begin{align}\label{theta}
\theta_{kj} = -{q^2\over 2m v}{{\bf e}_k\cdot {\bf e}_j (1-\alpha_k)(1-\alpha_j)\over \sqrt{\omega_k\omega_j}(\omega_k+\omega_j)}.
\end{align} 

The arbitrary-gauge Hamiltonian can now be written correct to order $q^2$ as
\begin{align}\label{hgak}
H_g =\, & {\tilde \omega}_m\left(d^\dagger d +{1\over 2}\right)+\sum_{k,j}\omega_{kj}\left(c_k^\dagger c_j+{\delta_{kj}\over 2}\right) \nonumber \\ &-{q \over m}{\bf p}\cdot {\tilde {\bf A}}_g({\bf 0}) + q{\bf r}\cdot {\tilde {\bf \Pi}}_g({\bf 0})
\end{align}
where $\omega_{kj} = \omega_k \delta_{kj}+(\omega_k+\omega_j)\theta_{kj}$ and
\begin{align}
&{\tilde {\bf A}}_g({\bf 0}) := \sum_{k,j} {{\bf e}_k \over \sqrt{2\omega_k v}}(1-\alpha_k)[e^\theta]_{kj}(c_j^\dagger +c_j),\label{atilg} \\ 
&{\tilde {\bf \Pi}}_g({\bf 0}) := i\sum_{k,j} {\bf e}_k \sqrt{\omega_k \over 2 v}\alpha_k[e^{-\theta}]_{kj}(c_j^\dagger -c_j).\label{pitilg}
\end{align}
Since the linear interaction components in Eq.~(\ref{hgak}) contain a prefactor of $q$ we may let $[e^\theta]_{kj} = \delta_{kj}$ in the mode expansions (\ref{atilg}) and (\ref{pitilg}) to obtain results correct to order $q^2$. This amounts to making the straightforward replacement $a_k\to c_k$ within the interaction Hamiltonian. Similarly, when used within the interaction Hamiltonian we may let ${\tilde \omega}_m = \omega_m$ within the expressions for ${\bf r}$ and ${\bf p}$ in terms of the material ladder operators, amounting to the replacement $b\to d$. We remark that the renormalisation of self-terms is consistent with an interpretation in which bare frequencies are not viewed as physical. The renormalisation does not affect the choice of gauge or the subsystem partition.

\begin{figure}[t]
\begin{minipage}{\columnwidth}
\begin{center}
\hspace*{-6mm}\includegraphics[scale=0.4
]{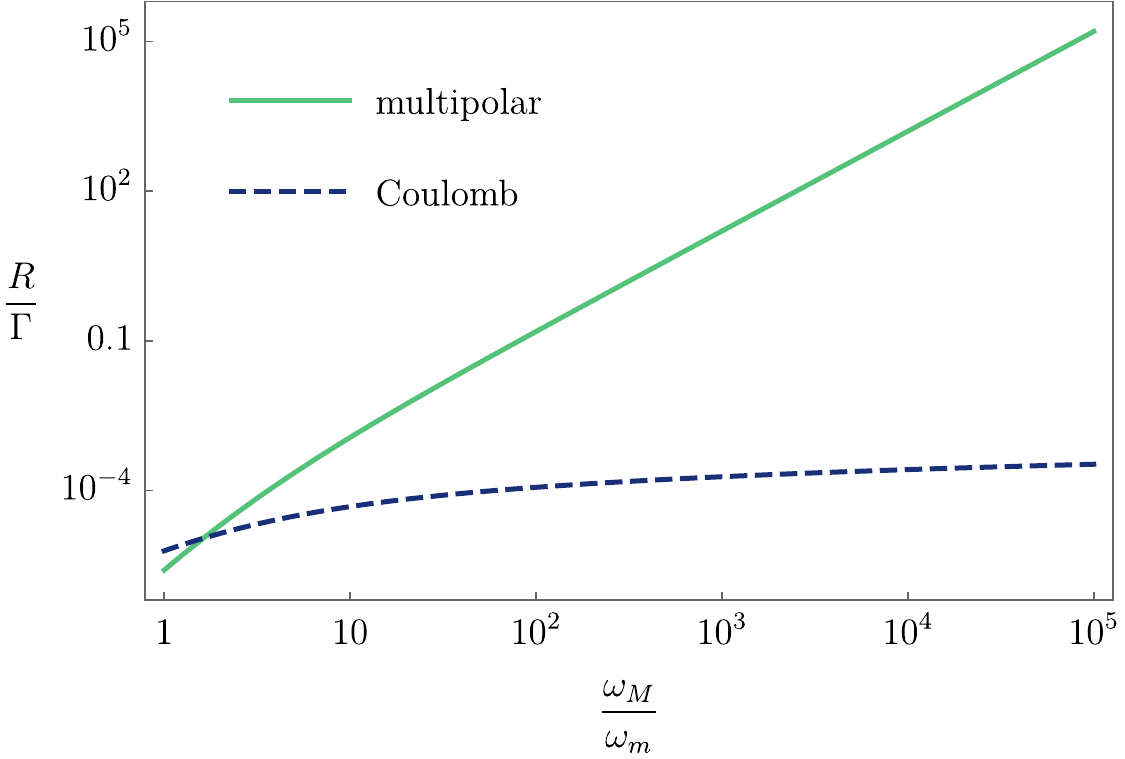}
\caption{The time-averaged detector excitation rate $R$ is plotted as a function of the cut-off $\omega_M/\omega_m$ in the Coulomb gauge and multipolar gauge, assuming $\omega_m T=10^4$. The multipolar rate in particular is severely divergent with $\omega_M$ whereas the Coulomb gauge rate is logarithmically divergent.}\label{rateR}
\end{center}
\end{minipage}
\end{figure}

Assuming the initial state $\ket{0_d,0_c}$ with no photons and no initial detector excitation we calculate the average detector population as
\begin{align}\label{n}
\langle d^\dagger(t)&d(t) \rangle_{0_d,0_c} \nonumber \\ &= {2\Gamma\over \pi} \int_0^\infty d\omega\,\left[{\omega u^+(\omega)  \sin \left[(\omega_m+\omega)t\right] \over \omega_m(\omega_m+\omega)}\right]^2
\end{align}
where $\Gamma = q^2 \omega_m^2/(6m\pi)$ is the total oscillator spontaneous emission rate into the ground state and where
\begin{align}\label{upm}
u^+(\omega) = \sqrt{\omega_m\over \omega}\left([1-\alpha(\omega)]-{\omega \over \omega_m}\alpha(\omega)\right).
\end{align}
The multipolar- and Coulomb gauge results are obtained by letting $\alpha(\omega)=1$ and $\alpha(\omega)=0$ respectively, and are consistent with Eqs.~(\ref{Pvac}) and (\ref{Pvac2}) respectively. The rate $d\langle d^\dagger(t)d(t) \rangle_{0_d,0_c}/dt$ is highly oscillatory. These oscillations can be removed by taking the time-average over an interval $T\gg1/\omega_m$ defined by
\begin{align}
R &= {1\over T}\int_0^T dt\, {d\over dt} \langle d^\dagger(t)d(t) \rangle_{0_d,0_c} = {1\over T} \langle d^\dagger(t)d(t) \rangle_{0_d,0_c} \nonumber  \\ &= {\Gamma\over \pi T }\int_0^\infty d\omega\,\left[{\omega u^+(\omega) \over \omega_m (\omega_m+\omega)}\right]^2
\end{align}
where we have replaced $\sin^2 \left[(\omega_m+\omega)T/2\right]$ in Eq.~(\ref{n}) by its average $1/2$ for $\omega_m T \gg 1$. The Coulomb and multipolar gauge time-averaged rates are plotted in Fig. \ref{rateR}. The multipolar rate in particular is quadratically divergent with $\omega_M$ and is clearly unphysical for values of $\omega_M$ consistent with the EDA. 
%%%%%%%%%%%%%%%%%%%%%%%%%%%%%%%%%%%%%%%%%%%%%%%%%%%%%%%%%
%%%%%%%%%%%%%%%%%%%%%%%%%%%%%%%%%%%%%%%%%%%%%%%%%%%%%%%%%
However, if we choose $\alpha=\omega_m/(\omega_m+\omega_k)$ [see Eq.~(\ref{alJC})] then we obtain the Hamiltonian [see Eq.~(\ref{nc})]
\begin{align}\label{nc2}
H_{\rm JC}=&{\tilde \omega}_m \left(d^\dagger d +{1\over 2}\right)+\sum_{k,j}\omega_{kj}\left( c_k^\dagger c_j +{\delta_{kj}\over 2}\right)\nonumber \\ &- iq\sum_k \sqrt{\omega_k \omega_m \over mv} {1\over \omega_m+\omega_k}(d^\dagger c_k -d c_k^\dagger).
\end{align}
In this gauge the ground state is represented by the vector $\ket{0_d,0_c}$ annihilated by $d$ and $c_k$. It is easy to verify that the ground energy eigenvalue of $H_{\rm JC}$ produces the expected order $q^2$ ground state Lamb shift \cite{drummond_unifying_1987}. In this gauge the detector excitation rate is identically zero because $u^+(\omega)\equiv 0$.\\
  
\subsubsection{Dressing transformation for an arbitrary multi-level dipole}\label{dressingmulti}

The idea of the JC-gauge can be extended beyond the simple systems considered above through a systematic approach to defining and understanding the concept of {\em dressing}. The task was undertaken relatively early on by Van Hove \cite{van_hove_energy_1955} whereby dressing is understood in terms of the Hamiltonian resolvent $G(z)=1/(z-H),~z\in{\mathbb C}$. Let $L(z)$ be the part of $G(z)$ that is diagonal in the eigenstates of $h=H-V$ and let subscript $_i$ refer to any state represented by an eigenvector of $h$. We express the eigenvalues of $L(z)$ in the form 
\begin{align}\label{Li}
L_i(z) = {1\over z-\omega_i-\Delta_i(z)+{i\over 2}\Gamma_i(z)}
\end{align}
where $\Delta_i(z)$ and $\Gamma_i(z)$ are real. One can characterise states in terms of these quantities \cite{van_hove_energy_1955,davidovich_theory_1980,cohen-tannoudji_atom-photon_2010}.

Following Ref.~\cite{davidovich_theory_1980} we assume that the equation $\omega-\omega_i-\Delta_i(\omega)=0$ has only one real root. There are then three further possible cases:
\begin{enumerate}[1)]
\item{$\Gamma_i(\omega_i)\neq 0$. }
\item{$\Gamma_i(\omega)= 0,~\forall \omega\in {\mathbb R}$.}
\item{$\Gamma_i(\omega_i)=0$ but $\Gamma_i(\omega)\neq 0$ for some $\omega\in {\mathbb R}$.}
\end{enumerate}
In case 1), $\omega_i$ lies on a cut of $L_i(z)$ and the state $i$ is said to be {\em dissipative}, because it will typically decay in the presence of the interaction. The quantities $\Gamma_i(\omega)$ and $\Delta_i(\omega)$ are the associated linewidth and level shift respectively (see Sec.~\ref{line}). An example is the state represented by the eigenvector $\ket{\epsilon^e,0}$ of $h$ in Eq.~(\ref{HAd}), in which the dipole is excited and there are no photons (this state will be considered as an initial state in Sec.~\ref{line}). In case 2), $L_i(\omega)$ has a simple pole at $\omega_i$. The state $i$ is said to be {\em asymptotically stationary}, because asymptotically it is unaffected by the interaction $V$. The interaction produces only transient effects and the $S$-matrix elements between such states are given by Eq.~(\ref{Sma}) without having to invoke the condition of adiabatic interaction switching. Case 3) lies inbetween cases 1) and 2). The state $i$ is not asymptotically stationary, but it is also distinguished from a dissipative state. In this case the interaction is said to give rise to {\em persistent perturbation effects}. Physically this can be thought of as {\em dressing} by a virtual ``cloud" of quanta. The ground state of $h$ represented by the vector $\ket{\epsilon^0,0}$ is an example.

The JC-gauge can be defined as a representation in which the ground state of $H$ is represented by the ground eigenvector of $h$, removing the affects of ``persistent perturbations". Physically, this means absorbing virtual dressing excitations, such that subsystem excitations are defined relative to the true ground state of the composite. To show how such a representation can be derived systematically, the authors of Ref.~\cite{davidovich_theory_1980} let $H=h+qV$ and $H'=e^{iS}[h+qV]e^{-iS}=h+qV'$ where $q$ is a small parameter. Writing $S=\sum_{n=1}^\infty q^n S_n$ and $V' = \sum_{n=1}^\infty q^n V_n'$ and equating coefficients in powers of $q$ gives
\begin{align}
&V_1'=V+i[S,h],\label{1cor}\\
&V_2'= i[S_1,V_1]+i[S_2,h]-{1\over 2}[S_1,[S_1,h]],\\
&V_3'=\dots \nonumber
\end{align}
One now chooses $S$ such that in $V'$ the component of $V$ that is responsible for persistent perturbation effects is cancelled out to the required order in $q$.

We consider the example of the dipole-field Hamiltonian $H_\alpha$ in Eq.~(\ref{HAd}). To illustrate the procedure we will eliminate the cause of persistent perturbations up to order $q$, which will often be sufficient for applications within the weak-coupling regime. We begin in the Coulomb gauge $\alpha=0$. The order $q$ part of the interaction Hamiltonian can be partitioned into rotating-wave and counter-rotating parts as $V_1^0 = V_{1,\rm rot}^0+V_{1,\rm counter}^0$ where $V_{1,\rm rot}^0:= \sum_{\substack{n,p \\ n>p}} r_{np} +{\rm H.c.}$ and $V_{1,\rm counter}^0:= \sum_{\substack{n,p \\ n<p}} r_{np} +{\rm H.c.}$ with $r_{np} := -\sum_{{\bf k}\lambda}i\omega_{np}g_k [{\bf d}_{np}\cdot {\bf e}_{{\bf k}\lambda}] a_{{\bf k}\lambda} \ket{\epsilon^n}\bra{\epsilon^p}$. Here ${\rm H.c.}$ stands for Hermitian conjugate. The term $V_{1,\rm rot}^0$ satisfies $V_{1,\rm rot}^0\ket{\epsilon^0,0}=0$ whereas the term $V_{1,\rm counter}^0$ is responsible for persistent perturbation effects. We define the generalised gauge fixing transformation \cite{stokes_gauge_2013}
\begin{align}\label{Rv}
&R_{0\{\alpha\}} :=e^{iS_{\{\alpha\}}},\\
&S_{\{\alpha\}}:=-\sum_{{\bf k}\lambda} \sum_{n,p} g_k ({\bf e}_{{\bf k}\lambda}\cdot{\bf d}_{np}) \alpha_{k,np} \ket{\epsilon^n}\bra{\epsilon^p} (a_{{\bf k}\lambda}^\dagger + a_{{\bf k}\lambda}),
\end{align}
which reduces to $R_{0\alpha}$ if $\alpha_{k,np}=\alpha$. It is easily verified that if
\begin{align}\label{alJC2}
\alpha_{k,np} = \begin{cases}
{\omega_{np}\over \omega_{np}+\omega_k},~~n>p\\
{\omega_{np}\over \omega_{np}-\omega_k},~~n<p
\end{cases} = {|\omega_{np}|\over |\omega_{np}|+\omega_k}
\end{align}
then the interaction within the transformed representation, $V^{\{\alpha\}} = R_{0\{\alpha\}}H_0R_{0\{\alpha\}}^\dagger-h$, satisfies $V^{\{\alpha\}}_1\ket{\epsilon^0,0} =0$ and so contains no persistent perturbation contributions of order $q$ \cite{stokes_gauge_2013}. 

This choice of $\alpha_{k,np}$ clearly generalises the JC-gauge defined by Eq.~(\ref{alJC}), which applies to a two-level or harmonic dipole. It must be borne in mind however, that in general, i.e., for an arbitrary anharmonic multi-level dipole, the particular choice in Eq.~(\ref{alJC2}) results in the sought cancellation only up to order $q$. The resulting interaction Hamiltonian correct to order $q$ is \cite{stokes_gauge_2013}
\begin{align}\label{Vv}
V_1^{\{\alpha_{\rm JC}\}} =& -i\sum_{{\bf k}\lambda } \sum_{\substack{n,p \\ n>p}}  \sqrt{\omega_k \over 2 v}
 {2 \omega_{np} \over \omega_{np} + \omega_k} {\bf d}_{np} \cdot {\bf  e}_{{\bf k}\lambda }\ket{\epsilon^n}\bra{\epsilon^p} a_{{\bf k}\lambda }\nonumber \\ &+ {\rm H.c.},
\end{align}
which allows photon annihilation (creation) if and only if the dipole transitions to a higher (lower) level. In particular, $H_{\{\alpha\}}=h+V_1^{\{\alpha\}}$ possesses the same ground state as $h$ and  the model possesses additional symmetry that allows the Hilbert space to be split into sectors.  These are of course the prototypical properties of the JC model.

\subsection{Natural lineshape}\label{line}

We now discuss the natural lineshape of spontaneous emission, which can be calculated using similar techniques to those reviewed above. Lamb noted in 1952 that two different expressions can be obtained for the natural lineshape, depending on whether the Coulomb gauge coupling or dipolar coupling is assumed. The prediction is a simple example of an experimentally testable signature of subsystem gauge relativity. 

\subsubsection{Gauge relativity of the prediction}

Excited atoms decay via spontaneous emission. The lineshape is defined as the frequency distribution (spectrum) of the emitted photons. Let the initial state of a dipole-field system be represented by $\ket{\epsilon^e,0}$, where $\ket{\epsilon^e}$ represents an excited dipolar level and $\ket{0}$ denotes the photonic vacuum. The average number of photons ${\bf k}\lambda$ at time $t$ is given by
\begin{align}\label{Nalphk}
N_\alpha({\bf k}\lambda, t) &= \langle a_{{\bf k}\lambda}^\dagger(t)a_{{\bf k}\lambda}(t)\rangle_{e0}\nonumber \\
&=\sum_m \sum_{n_{{\bf k}\lambda}} n_{{\bf k}\lambda}|\bra {\epsilon^m,n_{{\bf k}\lambda}}U_\alpha(t,0)\ket{\epsilon^e,0}|^2
\end{align}
where $\ket{n_{{\bf k}\lambda}}$ denotes the $n_{{\bf k}\lambda}$-photon Fock state. For each different $\alpha$, the quantity $N_\alpha({\bf k}\lambda, t)$ is gauge invariant and gauge relative (see Secs.~\ref{rel} and \ref{nonadiabatic}). The distinction between gauge invariance and gauge relativity [Eqs.~(\ref{eqt}) and (\ref{ineqt})] is important but there appears to have been a lack of recognition of this distinction within the literature on the natural lineshape, as discussed further in Sec.~\ref{Lamb}.

The lineshape may be defined in the mode continuum limit $\omega_k \to \omega$ by
\begin{align}\label{specph}
S_\alpha(\omega) ={v\over (2\pi)^3}\rho(\omega) \int d\Omega \sum_\lambda  \lim_{t\to\infty} N_\alpha({\bf k}\lambda, t) 
\end{align}
where $\omega = |{\bf k}|$ and $\rho(\omega)$ is the density of modes ($\rho(\omega)=\omega^2$ in free space), summation is over polarisations $\lambda=1,2$, and integration is over all directions for ${\bf k}$. We have assumed photonic modes confined to a volume $v$. 

In Ref.~\cite{power_time_1999-1} Coulomb and multipolar gauge photon number averages and atomic populations are calculated up to second order in the dipole moment, such that the gauge relativity of these quantities can be seen explicitly (see also Sec.~\ref{glauber} for the case of atomic populations). Damping is then described by adding explicit exponential temporal decay of the dipole moment operator, such that the gauge relativity of the spectrum is also confirmed through the attaining of different results for the $\alpha=0$ and $\alpha=1$ cases. It is remarked in Ref.~\cite{power_time_1999-1} that in principle such differences should be possible to test experimentally.

The multipolar prediction, which ignoring details of the dipole's excitation is $S_1(\omega)$, appeared to be in better agreement with the experiments of Lamb \cite{lamb_fine_1952} than the Coulomb gauge prediction \cite{power_coulomb_1959,milonni_natural_1989,fried_vector_1973,woolley_gauge_2000,stokes_gauge_2013}. Power and Zienau explained this by using what is now known as the PZW transformation to remove ``static precursor" contributions that occur in the Coulomb gauge \cite{power_coulomb_1959}. In other words, passage to the multipolar gauge removes the electrostatic field implicit within the definition of the Coulomb gauge dipole [see Secs.~\ref{share} and \ref{sourcefs}], which Power and Zienau deemed to be unphysical, at least within the context of the natural lineshape prediction.
This amounts to the stipulation that the subsystems defined relative to the multipolar gauge are more operationally relevant than the corresponding Coulomb gauge ones, as appears to have been borne out by the experiments. 

The prediction $S_1(\omega)$ is gauge invariant and it can therefore be calculated in any gauge [see~Eq.~(\ref{eqt})]. Milonni {\em et~al.} provide a derivation of the ``correct" lineshape $S_1(\omega)$ using the Coulomb gauge. This works by neglecting the difference between the source components of the Coulomb gauge and multipolar gauge photonic operators as follows \cite{milonni_natural_1989}. For a dipole at the origin, the integrated equation of motion for the $\alpha$-gauge annihilation operator is found using Eq.~(\ref{HAd}) and possesses the source term
\begin{align}\label{ibpa}
&a_{{\bf k}\lambda,{\rm s}}(t) \nonumber \\ &= i\int_0^t dt' \,{e^{-i\omega_k (t-t')}\over \sqrt{2\omega_k v}}{\bf e}_{{\bf k}\lambda} \cdot [i(1-\alpha){\dot {\bf d}}(t')+\alpha \omega_k {\bf d}(t')] \nonumber \\ &= {i(1-\alpha)\over \sqrt{2\omega_k v}} {\bf e}_{{\bf k}\lambda} \cdot [{\bf d}(0)e^{-i\omega_k t}-{\bf d}(t)] \nonumber \\ &+ \int_0^t dt' \sqrt{\omega_k \over 2 v}{\bf e}_{{\bf k}\lambda} \cdot {\bf d}(t') e^{-i\omega_k (t-t')}
\end{align}
where the second equality follows from an integration by parts. Neglecting the boundary term $\sim {\bf d}(0)e^{-i\omega_k t}-{\bf d}(t)$ gives the source part of the integrated equation of motion for the multipolar gauge annihilation operator, which in turn yields the ``correct" spectrum $S_1(\omega)$. Milonni {\em et~al.} argue that ignoring this term can be justified based on a sensible choice of boundary conditions. Specifically, exponential decay implies that the contribution from ${\bf d}(t)$ will vanish in the long-time limit, $t\gg 1/\Gamma_{e},~\Gamma_e = \sum_{m<e} \Gamma_{em}$, while the term depending on ${\bf d}(0)$ may be set to zero provided that the motion of the bare mechanical dipole (as defined relative to the multipolar gauge) is assumed to start {\em after} $t=0$. This is another way to understand the procedure of removing ``static precursor" contributions found in the Coulomb gauge, but more generally the argument can be applied for any $\alpha\neq 1$. Equivalently, it can be understood as a method of implementing the sudden switch-on of the multipolar gauge interaction within the $\alpha$-gauge. Again, these arguments essentially amount to the submission that the multipolar subsystems are the more physically relevant ones.

\subsubsection{Radiation damping}

There are different methods available to move beyond a second-order phenomenological calculation. These include Hamiltonian resolvent and projection operator techniques \cite{cohen-tannoudji_atom-photon_2010, davidovich_theory_1980}. An elegant and exact derivation of the lineshape is found using the formal theory of radiation damping, which goes back to the early work of Heitler \cite{heitler_quantum_2003}. Details are given in Supplementary Note~XV. To calculate the lineshape one assumes a dipole initially in an excited state with no photons present and one then calculates the long-time probability, $|b_{n{\bf k}\lambda,e0}(\infty)|^2$, that a transition has occurred into a state with the dipole in level $n$ and with one photon ${{\bf k}\lambda}$ present. The frequency spectrum is defined by
\begin{align}\label{spec0}
S(\omega) = {v\over (2\pi)^3} \omega^2 \int d\Omega \sum_\lambda  \lim_{t\to\infty}|b_{n{\bf k}\lambda,e0}(t)|^2.
\end{align}
We note that one could assume that the dipole is excited adiabatically starting in the distant past, and that the interaction is switched-off adiabatically, such that $b_{n{\bf k}\lambda,e0}(\infty)$ becomes an $S$-matrix element and therefore no longer gauge-relative [see Sec.~\ref{adiabatic} and the discussion below in Sec.~\ref{Lamb}]. This assumption may or may not be realistic when modelling an experiment.

For a Hamiltonian $H=h+V$, the long-time probability for the transition from initial ($t=0$) state $i$  represented by an eigenvactor of $h$, to final state $f$ represented by a different eigenvector of $h$, is given by
\begin{align}\label{probtext}
|b_f(\infty)|^2 = {| R_{fi}(\omega_f)|^2 \over [\omega_{fi}-\Delta_i(\omega_f)]^2+(\Gamma_i(\omega_f)/2)^2}
\end{align}
where
\begin{align}
&\Gamma_i(\omega) =2\pi\sum_{m\neq i} | R_{mi}(\omega)|^2\delta(\omega-\omega_m), \label{gamitext} \\
&\Delta_i(\omega) =V_{ii}+{\mathcal P} \sum_{m\neq i} {| R_{mi}(\omega_m)|^2 \over \omega-\omega_m}, \label{deltext}
\end{align}
and
\begin{align}\label{Rni2text}
R_{ni}(\omega) = V_{ni} + \sum_{m\neq i} V_{nm}R_{mi}(\omega)\zeta(\omega-\omega_m),~~~~~~n\neq i
\end{align}
in which
\begin{align}
\zeta(x):= {\mathcal P}{1\over x} -i\pi\delta(x).
\end{align}
Note that $b_f(\infty)$ can be written $R_{fi}(\omega_f)L_i(\omega_f)$ where $L_i(z)$ is defined in Eq.~(\ref{Li}).

Now consider the case that $V=V^\alpha=V^\alpha_1+V^\alpha_2$ is the $\alpha$-gauge interaction Hamiltonian for a dipole-field system in Eq.~(\ref{HAd}), while $\ket{i}=\ket{\epsilon^e,0}$ and $\ket{f}=\ket{\epsilon^n,{{\bf k}\lambda}}$. The matrix elements $R_{mi}(\omega_m)$ are in general gauge-relative ($\alpha$-dependent) as is $|b_f(\infty)|^2$. This gauge relativity can however be eliminated by invoking gauge non-relativistic approximations. Specifically, if bare energy conservation $\omega_f=\omega_i$ is imposed (from outside the theory) then the quantities in Eq.~(\ref{probtext}) are evaluated at $\omega_i$ (on-energy-shell) and they are then $\alpha$-independent \cite{woolley_gauge_2000}. This exemplifies the general result discussed in Sec.~\ref{adiabatic}, that strict bare-energy conservation is required to eliminate subsystem gauge relativity within probability amplitudes connecting bare states. 

To make contact with the Markovian approximations used in Sec.~\ref{glauber}, let us look more closely at the quantities in Eq.~(\ref{probtext}). To lowest order in $V$, Eq.~(\ref{Rni2text}) gives $R_{mi}(\omega) = V_{mi}$, which we can use to find $\Gamma_i(\omega_f)$ and $\Delta_i(f)$ as
\begin{align}
&\Gamma_i(\omega_f) = 2\pi \sum_m \sum_{{\bf k}'\lambda'} |[V^\alpha_1]_{m{\bf k}'\lambda',e0}|^2\delta(\omega_k+\omega_{nm}-\omega_{k'}), \\ 
&\Delta_i(\omega_f) = [V^\alpha_2]_{e0,e0}+ {\cal P}\sum_m \sum_{{\bf k}'\lambda'} {|[V^\alpha_1]_{m{{\bf k}'\lambda'},e0}|^2\over \omega_k +\omega_{nm}-\omega_{k'}}.
\end{align}
If we evaluate $[V^\alpha_1]_{m{{\bf k}\lambda},e0}$ at $\omega_f=\omega_i$, that is, at $\omega_k = \omega_{en}$, then $\Gamma_i(\omega_f)=\Gamma_e$ and $\Delta_i(\omega_f)=\Delta_e$ where $\Gamma_e = \sum_{m<e} \Gamma_{em}$ and $\Delta_e$ are the total spontaneous emission rate and on-energy-shell Lamb-shift associated with the dipole level $e$, as calculated using Fermi's golden rule and second order perturbation theory respectively. Both of these quantities are $\alpha$-independent. 

An on-energy-shell evaluation may be justified within the quantity $L_i(\omega)$ defined in Eq.~(\ref{Li}) and is known as the pole approximation, which is commonly employed in the calculation of Lorentzian spectra \cite{barnett_methods_1997}. Specifically, it is justified in the Markovian regime $\Gamma_{em} t \gg 1$ provided $\Gamma_i(\omega)$ and $\Delta_i(\omega)$ are sufficiently slowly varying near to $\omega_i$, because then $L_i(\omega)$ has a pole near to $\omega_i$ such that it may with sufficient accuracy be approximated by $L_i(\omega_i+i\eta)$ with $\eta\to 0^+$. In a similar fashion, Markovian approximations were used in Sec.~\ref{glauber} to derive the (gauge non-relative) rate $\Gamma$ from either of the gauge-relative expressions (\ref{Pvac}) or (\ref{Pvac2}). Applying the pole approximation to $L_i(\omega_f)$ within $b_f(\infty) = R_{fi}(\omega_f)L_i(\omega_f)$ yields the final result
\begin{align}\label{spec2}
S_\alpha(\omega) = {\Gamma_{en} \over 2\pi}{(\omega / \omega_{en}^3)[(1-\alpha)\omega_{en}+\alpha\omega]^2\over (\omega-{\tilde \omega}_{en})^2+\left(\Gamma_{e}/ 2\right)^2}
\end{align}
where ${\tilde \omega}_{en} = \omega_{en} + \Delta_{e}$. 
Further evaluating the numerator on-energy-shell implies that all remaining $\alpha$-dependence drops out and we obtain the pure Lorentzian
\begin{align}\label{spec2b}
S(\omega) =  {\Gamma_{en} \over 2\pi}{1\over (\omega-{\tilde \omega}_{en})^2+\left(\Gamma_{en}/ 2\right)^2}.
\end{align}
Note that away from resonance, $\omega=\omega_{en}$, this is significantly different from the ``correct" lineshape $S_1(\omega)$, but it should also be noted that a description of the dipole's excitation has not been included and this is important when describing, for example, the experiments of Lamb. Nevertheless, it is clear that even if an on-energy-shell evaluation of the lineshape denominator can be justified, the same procedure applied to the numerator may be difficult to justify because the numerator may not be sufficiently slowly varying compared with the denominator.

\subsubsection{The $2s_{1\over 2}\to 1s_{1\over 2}$ transition in hydrogen}\label{Lamb}

We now consider more closely the experiments of Lamb \cite{lamb_fine_1952} which probed the $2s_{1\over 2}\to1s_{1\over 2}$
transition in hydrogen. The atoms start in the meta-stable state $2s_{1\over 2}$. They are irradiated with a microwave frequency $\omega$ close to the frequency $\omega_0$ of the $2s_{1\over  2}\to2p_{1\over 2}$ transition. The microwave resonance $\omega-\omega_0$ is detected by photons spontaneously emitted in the $2p_{1\over 2}\to1s_{1\over 2}$ transition. 

A main goal of previous studies \cite{fried_vector_1973,bassani_choice_1977,davidovich_theory_1980,cohen-tannoudji_photons_1989,woolley_gauge_2000} has been to eliminate the $\alpha$-dependence of the lineshape prediction, which has been viewed as a paradoxical property (gauge noninvariance). However, we have provided a precise mathematical definition of gauge invariance  [Eq.~\ref{gi})] according to which the $\alpha$-dependence of the lineshape prediction does not constitute gauge noninvariance. Eq.~(\ref{probtext}) in particular is {\em exact}, but for each different value of $\alpha$, the labels $i$ and $f$ therein refer to different physical states. The result is therefore {\em gauge-relative}, which is not paradoxical and is simply an example of the expected inequality~(\ref{ineqt}). Each one of the predictions is certainly gauge invariant in the sense of Eq.~(\ref{eqt}). 

While the prediction is fundamentally gauge-relative, use of the $S$-matrix will circumvent this relativity [see Sec.~\ref{adiabatic}]. However, subsequent approximation of the $S$-matrix may in turn eliminate this special property. We must obviously distinguish the simplifying assumptions that eliminate $\alpha$-dependence by {\em defining} the $S$-matrix, from subsequent approximations {\em of} the $S$-matrix, that may then eliminate its $\alpha$-independence. Previous studies have identified which approximations of relevant $S$-matrix elements must be avoided in order that they are $\alpha$-independent (gauge non-relative). In particular, full sets of intermediate states must be retained in calculations \cite{cohen-tannoudji_photons_1989,bassani_choice_1977}, despite the apparent dominance of the intermediate state $2p_{1\over 2}$ \cite{power_coulomb_1959}. As we have already seen in Sec.~\ref{s1}, the significance of higher dipole levels is greater in the Coulomb gauge such that summation over these levels converges much more quickly in the multipolar gauge. In the context of calculating two-photon transition matrix elements this has been understood for some time \cite{cohen-tannoudji_photons_1989,bassani_choice_1977}.

The $S$-matrix is perturbative, essentially by definition [the $T$-matrix in Eq.~(\ref{Sma}) is expanded iteratively in powers of $V$], making it difficult to obtain an expression for the lineshape with finite-width corresponding to exponential decay. However, damping can of course be included by alteration of the two-photon on-energy-shell (Kramers-Heisenberg) transition matrix element describing the process \cite{fried_vector_1973,power_coulomb_1959,davidovich_theory_1980,stokes_gauge_2013}. Fried showed using a semi-classical treatment, \cite{fried_vector_1973}, that when damping is included in this way, and ``non-resonant background" terms that are present within the modified matrix element are not ignored, then the Coulomb gauge and multipolar gauge predictions can be brought into significantly closer agreement. The situation in which excitation occurs via a tunable microwave field as well as the alternative situation of a fixed microwave field with an applied magnetic field are both considered. This method was extended to a full quantum treatment in Ref.~\cite{davidovich_theory_1980}.

Nevertheless, it is clear that if an {\em exact} radiation damping treatment is adopted [Eq.~(\ref{probtext})], then the prediction will be $\alpha$-dependent (gauge-relative). As a somewhat extreme position, one might reject outright the validity of treating eigenvectors of $h$ as physically meaningful outside of scattering theory. One would be confined to the use of scattering theory or else alternative {\em physical} states would have to be identified. Over finite-times without adiabatic switching the eigenvectors of $H_\alpha=h+V^\alpha$ uniquely represent (fully-dressed) physical states. However, obviously these states are stationary, such that a rejection of initial and final states that are not eigenstates of $H$ would appear to preclude the possibility of studying non-trivial dynamics. One could instead consider the ground state of $H_\alpha$ as the initial state that is then subjected to a time-dependent external perturbation, but in this case a microscopic description would entail identifying a gauge relative to which the time-dependent interaction is to be defined, such that the prediction again becomes gauge-relative [see Sec.~\ref{nonadiabatic}].

A criterion by which the most meaningful physical states can be identified was suggested in Sec.~\ref{pd}. Specifically, one may attempt to define physical ``light" and ``matter" excitations relative to the ground state of $H_\alpha$. The virtual admixtures otherwise present in the ground state have therefore been absorbed into the subsystem definitions. This ``JC-gauge" for an arbitrary multi-level dipole was discussed in Sec.~\ref{dressingmulti} and has been discussed in the context of the lineshape in Refs.~\cite{davidovich_theory_1980,stokes_gauge_2013}. In Ref.~\cite{stokes_gauge_2013} a radiation damping treatment is adopted under the assumption that excitation of the dipole to the state $2s_{1\over 2}$ occurs through absorption of photons with a spectrum much sharper than the emitted spectrum. Excitation via continuous laser irradiation prior to emission is also considered. The fluorescence rates found using different gauges have Lorentzian forms with $L_i(\omega_f)$ evaluated on-energy-shell, but with differing numerators. The associated lineshapes are compared for the Coulomb gauge, multipolar gauge, and ``JC-gauge". The multipolar prediction is closest to a bare Lorentzian curve. As expected, the ``JC-gauge" curve interpolates between this curve and the Coulomb gauge result. Differences are increasingly conspicuous further away from resonance $\omega_f=\omega_i$. We note finally that in the weak-coupling regime differences between gauge-relative predictions such as emission lineshapes will typically be small. Spectroscopic experimental signatures outside of this regime are discussed in Sec.~\ref{specs}.

\subsection{Localisation and causality}\label{loc_caus}

\subsubsection{Electromagnetic source fields in an arbitrary gauge}\label{sourcefs}

To understand the balance between localisation and dressing it is necessary to determine the electromagnetic fields generated by a source in an arbitrary gauge. In particular, if we consider a system consisting of both a source ${\rm s}$ and detector ${\rm d}$ then the total electric field is a superposition of vacuum, source, and detector fields; 
\begin{align}\label{EL}
{\bf E}={\bf E}_{\rm vac}+{\bf E}_{\rm s}+{\bf E}_{\rm d}.
\end{align}
A full description of the source-detector-field system is postponed until Sec.~\ref{sdf}. First we note that due to subsystem gauge relativity the partitioning of a gauge invariant field into vacuum, source, and detector components, is gauge-relative \cite{power_time_1999,power_time_1999-1}. In other words, while the left-hand-side of Eq.~(\ref{EL}) is unique, the individual components on the right-hand-side represent different physical fields in different gauges. We therefore start by considering only one material system; a point dipole fixed at ${\bf 0}$ and with dipole moment $q{\bf r}$. For simplicity we again restrict our attention to the one-parameter $\alpha$-gauge framework.

Let us consider the canonical field ${\bf \Pi}$ at an arbitrary point ${\bf x}\neq {\bf 0}$, which can be partitioned as
\begin{align}
{\bf \Pi}(t,{\bf x})
&= -{\bf E}_{\rm T}(t,{\bf x})-\alpha{\bf E}_{\rm L}(t,{\bf x})\nonumber \\ 
&= {\bf \Pi}_{\rm vac}^\alpha(t,{\bf x})+{\bf \Pi}_{\rm s}^\alpha(t,{\bf x}).
\end{align} 
In the gauge $\alpha$ the vacuum and source components ${\bf \Pi}_{\rm vac}^\alpha(t,{\bf x})$ and ${\bf \Pi}_{\rm s}^\alpha(t,{\bf x})$ are defined as the components whose dynamics are generated by $H_{\rm ph}$ and $V^\alpha$ respectively. The vacuum field is defined by the right-hand-side of Eq.~(\ref{modeE}). Since the photons defined by $a_\lambda(0,{\bf k})$ are physically distinct for each $\alpha$ the vacuum field depends on $\alpha$. The source field also depends on $\alpha$ and the dynamics generated by $H_\alpha=H_m+H_{\rm ph}+V^\alpha$ yield
\begin{align}\label{pis}
{\bf \Pi}_{\rm s}^\alpha(t,{\bf x}) =& -\theta(t_r){\bf X}_{\rm T}(t_r,{\bf x})\nonumber \\&+(1-\alpha)\left[{\bf P}_{\rm T}(t,{\bf x})-\theta(-t_r){\bf P}_{\rm T}(0,{\bf x})\right]
\end{align}
where $t_r=t-x$ is the retarded time (in units with $c=1$) in which $x=|{\bf x}|$ is the distance from the dipole source at ${\bf 0}$ and where for ${\bf x\neq 0}$
\begin{align}\label{ptc}
X_{{\rm T},i}(t,{\bf x})= \left(-\partial^2 \delta_{ij} +\partial_i\partial_j\right){qr_j(t_r)\over 4\pi x}.
\end{align}
Note that the derivative operators in Eq.~(\ref{ptc}) act on $t_r$ as well as on $1/x$. Only the top line on the right-hand-side of Eq. (\ref{pis}) is causal, by which we mean vanishing for $t_r<0$, and the second line only vanishes for $\alpha=1$.

Using the fact that the $a_\lambda(0,{\bf k})$ of different gauges are related by $R_{\alpha\alpha'}$, one finds that the different vacuum components ${\bf \Pi}_{\rm vac}^\alpha$ are related by
\begin{align}\label{vac}
{\bf \Pi}_{\rm vac}^\alpha(t,{\bf x}) &= {\bf \Pi}^{\alpha'}_{\rm vac}(t,{\bf x})-(\alpha-\alpha')\theta(-t_r){\bf P}_{\rm T}(0,{\bf x}).
\end{align}
It follows that the combination ${\bf \Pi}^{\alpha}_{\rm vac}(t,{\bf x})+\alpha\theta(-t_r){\bf P}_{\rm T}(0,{\bf x})$ is actually $\alpha$-independent. We see also that for different $\alpha$ the vacuum components ${\bf \Pi}_{\rm vac}^\alpha$ differ by an $\alpha-\alpha'$ weighted factor of ${\bf P}_{\rm T}={\bf E}_{\rm L}$ evaluated at $t=0$, and that this contribution is restricted to the complement of the interior lightcone of the origin $(0,{\bf 0})$ of the dipole's rest-frame.

It is instructive to consider some specific physical fields. For example, ${\bf E}_{\rm T}=-{\bf \Pi}|_{\alpha=0}$ for which ${\bf E}^\alpha_{\rm Ts}=-{\bf \Pi}^\alpha_{\rm s}-\alpha{\bf P}_{\rm T}$ and ${\bf E}^\alpha_{\rm Tvac}=-{\bf \Pi}^\alpha_{\rm vac}$. Clearly the free and source components are different in different gauges, but their sum is
\begin{align}\label{et}
{\bf E}_{\rm T}(t,{\bf x})= &\theta(t_r){\bf X}_{\rm T}(t,{\bf x}) +\theta(-t_r){\bf P}_{\rm T}(0,{\bf x}) -{\bf P}_{\rm T}(t,{\bf x})\nonumber \\ & -{\bf \Pi}^\alpha_{\rm vac}(t,{\bf x})-\alpha \theta(-t_r){\bf P}_{\rm T}(0,{\bf x})
\end{align}
which upon taking into account Eq.~(\ref{vac}) is seen to be unique ($\alpha$-independent) as required. The total electric field is for ${\bf x}\neq {\bf 0}$ given by ${\bf E}={\bf D}_{\rm T}= -{\bf \Pi}|_{\alpha=1}={\bf E}_{\rm T}+{\bf P}_{\rm T}$, which can be read-off immediately from Eq. (\ref{et}) as
\begin{align}\label{e}
{\bf E}(t,{\bf x})={\bf D}_{\rm T}(t,{\bf x})= &\theta(t_r){\bf X}_{\rm T}(t,{\bf x}) +\theta(-t_r){\bf P}_{\rm T}(0,{\bf x}) \nonumber \\ & -{\bf \Pi}^\alpha_{\rm vac}(t,{\bf x})-\alpha \theta(-t_r){\bf P}_{\rm T}(0,{\bf x}).
\end{align}

Similarly to Sec.~\ref{share}, the results above demonstrate that what differs for different choices of $\alpha$ are the localisation properties of the source. For $t_r>0$, we have that ${\bf E}_{\rm s}(t,{\bf x})={\bf D}_{\rm Ts}(t,{\bf x})={\bf X}_{\rm T}(t,{\bf x})$ and ${\bf E}_{\rm Ts}(t,{\bf x}) = {\bf X}_{\rm T}(t,{\bf x})-{\bf P}_{\rm T}(t,{\bf x})$ {\em for all} $\alpha$. In words, at all points ${\bf x}$ that can be connected to the source's centre by a light signal emitted a time $x$ earlier, each physical field's source component is independent of the source's definition. In contrast, for $t_r<0$ the source-vacuum partitioning of a given physical field differs between different gauges $\alpha$. 

As explained in Sec.~\ref{share}, within the EDA the gauge controls the extent to which the instantaneous field ${\bf E}_{\rm L}({\bf x})={\bf P}_{\rm T}({\bf x})$ (where ${\bf x}\neq {\bf 0}$) is included within the source's definition. The gauges $\alpha=0$ and $\alpha=1$ are extremal cases whereby ${\bf E}_{\rm L}$ is fully included and completely absent respectively. For this reason, the source-component of the field ${\bf \Pi}=-{\bf E}_{\rm T}-\alpha'{\bf P}_{\rm T}$, when partitioned according to the gauge $\alpha$ to give ${\bf \Pi}_{\rm s}^\alpha=-{\bf E}_{\rm Ts}^\alpha-\alpha'{\bf P}_{\rm T}$, is causal (meaning vanishing for $t_r<0$), if and only if $\alpha=1$ and $\alpha'=1$. The latter equality $\alpha'=1$ specifies that the physical field being considered is ${\bf E}$, which is a local field, and the former equality $\alpha=1$ specifies that the source producing this field is defined relative to the multipolar gauge, and is therefore itself also local. It is easy to show that unlike ${\bf E}$ the magnetic field ${\bf B}={\bf B}_{\rm vac}+{\bf B}_{\rm s}$ has unique vacuum and source components and that ${\bf B}_{\rm s}$ is causal \cite{power_time_1999-1}.

These results generalise those of Ref. \cite{power_time_1999-1} by giving vacuum-source partitions of the physically arbitrary field ${\bf \Pi}$, using an arbitrary gauge $\alpha$. For any given physical field the relative magnitude of the non-local contributions occurring for $t_r<0$ vary with $\alpha$ and provide a measure of the delocalisation of the source, as will be elaborated further below and in   Supplementary Note~XIII.

\subsubsection{Source-detector-field system}\label{sdf}

Let us now consider the tripartite source-detector-field system. If we require the detector dipole to be fully localised at ${\bf x}$ and a source dipole to be fully localised at ${\bf 0}$, then ``matter" must be defined relative to the multipolar gauge. From the results of Sec.~\ref{sourcefs} it is also clear that the response of the detector$_1$ to the source$_1$ is causal as required \cite{cohen-tannoudji_photons_1989,biswas_virtual_1990,milonni_photodetection_1995,power_analysis_1997,power_time_1999,power_time_1999-1,sabin_fermi_2011,stokes_noncovariant_2012}. In any other gauge $\alpha\neq 1$ ``matter" is dressed by $\alpha{\bf E}_{\rm L}$ and so is not fully localised. However, questions regarding the causal nature of an interaction are only well-posed for separated localised objects. The instantaneous response of a delocalised detector to a delocalised source will vary with $\alpha$ and can be taken as a measure of the overlap of the source and detector, and hence as a measure of the delocalisation of ``matter" as defined within the gauge $\alpha$.

To make these statements concrete, let us consider a system of two identical dipoles labelled ${\rm s}$ (source) and ${\rm d}$ (detector) at positions ${\bf R}_{\rm s}={\bf 0}$ and ${\bf R}_{\rm d}$ respectively. To quantify the response of the detector to the source it suffices to consider the rate of change of the detector's energy. Excitation probabilities such as those considered in Sec.~\ref{glauber} are determined from the spectral projections of the detector's energy. The multipolar Hamiltonian can be partitioned as
\begin{align}\label{h1}
H_1
&= {\tilde H}_{\rm d}+{\tilde H}_{\rm s}+{\tilde V}_{\rm d}+{\tilde V}_{\rm s}+H_{\rm ph}
\end{align}
where $\mu={\rm s,\,d}$,
\begin{align}
{\tilde H}_\mu &= {{\bf p}_\mu^2\over 2m}+V({\bf r}_\mu)+S_\mu,~~~{\tilde V}_\mu = q{\bf r}_\mu \cdot {\bf \Pi}({\bf R}_\mu) 
\end{align}
in which the term $S_\mu:={1\over 2}\int d^3x\, {\bf P}_{{\rm T}\mu}^2$ with ${\bf P}_{\rm T\mu}({\bf x}):= q{\bf r}_\mu \cdot \delta^{\rm T}({\bf x}-{\bf R}_\mu)$ has not been placed in the interaction Hamiltonian. The rate of change of ${\tilde H}_{\rm d}$ is
\begin{align}\label{wd}
{\dot {\tilde H}}_{\rm d}(t) = -q{\dot {\bf r}}_{\rm d}(t)\cdot {\bf \Pi}(t,{\bf R}_{\rm d}) = q{\dot {\bf r}}_{\rm d}(t)\cdot {\bf D}_{\rm T}(t,{\bf R}_{\rm d}).
\end{align}
If one instead considers $H_{\rm d}={\tilde H}_{\rm d}-S_{\rm d}$, then the rate of change includes an additional self-term that depends only on the detector, which does not affect its response to the source. The total displacement field at ${\bf R}_{\rm d}$ can be partitioned as in Eq.~(\ref{EL}). We therefore obtain an expression of Poynting's theorem for the detector$_1$ in the presence of the external field ${\bf E}_{\rm s}^1(t,{\bf R}_{\rm d})$. Specifically, the rate at which work is done by ${\bf E}_{\rm s}^1$ on the detector$_1$ in the volume ${\cal V}$
is \cite{griffiths_introduction_2017,jackson_classical_1998}
\begin{align}
\int_{\cal V} d^3 x \, {\bf J}_{\rm d}(t,{\bf x}) \cdot {\bf E}_{\rm s}^1(t,{\bf x})=q{\dot {\bf r}}_{\rm d}(t)\cdot {\bf E}_{\rm s}^1(t,{\bf R}_{\rm d})=:{\dot {\tilde H}}_{\rm d,s}(t)
\end{align}
where ${\bf J}_{\rm d}(t,{\bf x}) = q{\dot {\bf r}}_{\rm d}(t)\delta({\bf x}-{\bf R}_{\rm d})$ is the detector current in the EDA. The detector$_1$ response rate ${\dot {\tilde H}}_{\rm d}(t)$ can be decomposed in its eigenbasis as ${\dot {\tilde H}}_{\rm d}(t)=\sum_n \epsilon^n_{\rm d} {\dot {\mathsf P}}_{\rm d}^n(t)$ where ${\mathsf P}_{\rm d}^n(t)$ is the projection onto the $n$'th level at time $t$. For a two-level detector$_1$ as is typically considered  \cite{fermi_quantum_1932,biswas_virtual_1990,milonni_photodetection_1995,power_analysis_1997,sabin_fermi_2011,stokes_noncovariant_2012} the rate of excitation into the excited state, ${\dot {\mathsf P}}^1_{\rm d}(t)$, is easily found as ${\dot {\mathsf P}}^1_{\rm d}(t) ={\dot {\tilde H}}_{\rm d}(t)/\omega_m$ where $\omega_m=\epsilon^1_{\rm d}-\epsilon^0_{\rm d}$ is the two-level detector$_1$ transition frequency. The source-dependent component is therefore ${\dot {\mathsf P}}^1_{\rm d,s}(t) = {\dot {\tilde H}}_{\rm d,s}(t)/\omega_m$.

\subsubsection{Discussion: Localisation and dressing}\label{discuss}

For fully localised and hence bare multipolar dipoles the detector's response to the source is causal because ${\bf E}_{\rm s}^1(t,{\bf R}_{\rm d})={\bf 0}$ for $t<t_r$ where $t_r=t-R_{\rm d}$. It follows that each of the spectral projections ${\mathsf P}_{\rm d}^n(t)$ must also depend causally on ${\rm s}$, and therefore, that
the probability to find the bare detector$_1$ in an excited state depends causally on ${\rm s}$. Crucially, there is also a non-zero component of ${\dot {\tilde H}}_{\rm d}(t)$ that is independent of the source$_1$, namely, ${\dot H}_{\rm d,0}(t)={\dot {\tilde H}}_{\rm d}(t)-{\dot {\tilde H}}_{\rm d,s}(t)$. In fact, such a contribution must exist if the response of the detector$_1$ to the source$_1$ is to be causal. This follows from Hegerfeldt's theorem, which is a general mathematical result that assumes {\em i)} the energy is bounded from below, {\em ii)} the source and detector are initially localised in disjoint regions, {\em iii)} the initial state consists of the source excited and the detector in its ground state with no photons present \cite{hegerfeldt_causality_1994}. Hegerfeldt showed that under these assumptions, the total probability of excitation of the detector, ${\cal P}_{\rm d}^e(t)= {\cal P}_{\rm d,0}^e(t)+{\cal P}_{\rm d,s}^e(t)$, is either necessarily non-zero for times $t_r<0$, or that it is identically zero for all times. It follows that for an initial state represented by the vector $\ket{\epsilon_{\rm s}^n,\epsilon_{\rm d}^0,0}$ in the multipolar gauge, if ${\cal P}_{\rm d,0}^e(t)$ were to vanish, then ${\cal P}_{\rm d,s}^e(t)$ would be non-zero for $t_r<0$ and this would violate Einstein causality, because the multipolar gauge dipoles are localised and spacelike separated.

  Hegerfeldt concludes that the two-atom system (source and detector) engenders a conflict with Einstein causality, modulo some ways out that he lists \cite{hegerfeldt_causality_1994}. The claimed violation was contested in, for example, Refs.~\cite{milonni_photodetection_1995,power_analysis_1997,buchholz_there_1994}, it being recognised that the possible ways out listed by Hegerfeldt are not mere technicalities and do have to be taken seriously. In particular, removing the virtual excitations of a localised material system means absorbing the ``cloud" of virtual particles around it \cite{hegerfeldt_causality_1994,buchholz_there_1994}, such that states in which there are no such excitations are not ones in which the atoms are strictly localised. States in which the atoms are localised in disjoint regions (and which therefore permit well-posed questions regarding signal propagation) will contain (virtual) photons \cite{milonni_photodetection_1995,buchholz_there_1994}.  

By assuming the initial state $\ket{\epsilon_{\rm s}^n,\epsilon_{\rm d}^0,0}$ in the multipolar gauge one is assuming that the bare multipolar gauge dipoles are those that are operationally relevant at the preparation stage, but since $\ket{\epsilon_{\rm d}^0,0}$ is not the ground state of the detector-field system this leads to the immediate virtual excitation of the detector for $t>0$. We have seen that this virtual excitation is actually necessary to preserve Einstein causality. However, like a violation of Einstein causality, such virtual (spontaneous) excitations are themselves conceptually problematic and are essentially what one seeks to eliminate within a successfully renormalised theory. Indeed, we saw in Sec.~\ref{glauber} that the multipolar dipole's virtual excitation was particularly unphysical and we identified a different gauge within which such excitations were eliminated. In any such theory the detector responds to the source for times $t_r<0$. To avoid a conflict with Einstein causality one must interpret the renormalised source and detector as objects that are delocalised around their centres at ${\bf 0}$ and ${\bf R}_{\rm d}$ respectively.

The representation in which virtual excitations are removed is one in which the initial state of the detector and field subsystems coincides with the detector-field ground state, which might be considered a more realistic initial state [see Sec.~\ref{dressingmulti}], but this state is not one that specifies definite energy of a localised detector. Since preparation and measurement procedures necessarily possess finite extent in spacetime, there is clearly a balance to be struck between dressing and localisation. The parameter $\alpha$ affects this balance by controlling the extent to which bare matter is dressed by ${\bf E}_{\rm L}$, which in turn effects the value of ${\dot {\cal P}}_{\rm d,0}$ resulting from the ground state virtual photons surrounding the bare detector$_1$. It therefore seems sensible to conclude that the value of $\alpha$ that specifies the most relevant subsystems will depend on coupling strengths, as well as on the experimental protocols for preparation and measurement, including their spatial and temporal properties.

These questions can essentially be ignored within the traditional quantum optical regime because as shown in Sec.~\ref{master} the reduced description of the detector is independent of the gauge relative to which it is defined and its stationary state is $\ket{\epsilon^0_{\rm d}}$. This is also the regime in which the fields ${{\bf E}^1}^{(\pm)}_{\rm s}$ are approximately causal \cite{milonni_photodetection_1995,stokes_vacuum_2018}. Thus, in this regime it is possible to define the detector dipole as a localised system, while also retaining a fully causal response to the source, but {\em without} spontaneous vacuum excitation. This combination of properties is forbidden by Hegerfeldt's theorem and must therefore be the culmination of weak-coupling approximations. In sufficiently strong-coupling regimes one or more of these properties must be sacrificed. The gauge $\alpha$ relative to which the detector is defined will affect which properties of its weak-coupling counterpart it continues to possess. The multipolar gauge continues to define localised dipoles with causal interactions, but with ${\cal P}_{\rm d,0}(t)\not\approx 0$. On the other hand values $\alpha\neq 1$ define dipoles that are delocalised to some extent, but  which may retain the property ${\cal P}_{\rm d,0}(t)\approx 0$ even outside of the weak-coupling regime.   
In   Supplementary Note~XIII we review concrete demonstrations of this by considering the average electromagnetic energy-momentum in the vicinity of a dipole.
  
\section{Measurements and cavity QED beyond weak-coupling approximations}\label{cavers}
 
We now turn our attention to understanding photonic fields confined to a cavity where weak-coupling theory is generally inapplicable and subsystem gauge relativity is expected to be important.    We first provide a simple but arbitrary-gauge description of the field inside a cavity containing a two-level dipole. This extends the results of Ref.~\cite{sanchez_munoz_resolution_2018} that identify the field bound to the dipole$_1$, as distinguished from the propagating field. The results can also be thought of as a simplified extension of the results for a dipole in free space presented in Supplementary Note~XIII. An early attempt to relate the dressing of a two-level dipole with (weak) measurement protocols through the explicit modelling of a pointer system, is detailed in Supplementary Note~XVI. We discuss the topic of ground state photon-condensation in cavity QED systems, which is of considerable current interest but also strongly gauge-relative. Finally we discuss extra-cavity fields, including reviewing simple models describing associated measurement signals.

\subsection{Simple model of intra-cavity fields}\label{intracav}

We first consider a simple analysis of intra-cavity fields produced by a dipole at the cavity centre. This closely mirrors the analysis in   Supplementary Note~XIII for free space. An early step towards evaluating the Glauber intensity within a cavity in the ultrastrong-coupling regime has been given in \cite{sanchez_munoz_resolution_2018}. Therein emphasis was placed upon the need for a multi-mode theory in accommodating the requisite spatio-temporal structure to elicit signal propagation. We will consider a similar analysis in an arbitrary-gauge.

We model the cavity as a one-dimensional field in the $x$-direction with periodic boundary conditions at $x=\pm L/2$ where $L$ is the cavity length. The allowed wavenumbers are $k=2\pi n/L$, $n\in {\mathbb Z}$. The canonical fields are assumed to point in the $z$-direction and have bosonic mode expansions
\begin{align}
&A(t,{\bf x}) = \sum_k {1\over \sqrt{2\omega_k v}}\left[a^\dagger_k(t)e^{-ikx}+a_k(t)e^{ikx}\right], \\
&\Pi(t,{\bf x}) = i\sum_k  \sqrt{\omega_k\over 2v}\left[a^\dagger_k(t)e^{-ikx}-a_k(t)e^{ikx}\right]\label{1dpi}
\end{align}
where $v$ is the cavity volume. The cross-sectional area is therefore $v/L$. As usual, we have $[a_k,a^\dagger_{k'}]=\delta_{kk'}$ and $\omega_k=|k|$. To be consistent with the assumed expressions for ${\bf A}_{\rm T}$ and ${\bf \Pi}$ the transverse polarisation ${\bf P}_{\rm T\alpha}$ is also assumed to point in the $z$-direction. 

We assume that the dipole within the cavity is sufficiently anharmonic that we can expect a two-level truncation in the multipolar gauge to be generally robust (see for example Fig.~\ref{fcomp} in Sec.~\ref{opt}). It should be borne in mind that for a less anharmonic dipole truncation may remain accurate for predicting the low-energy properties that we consider below, but the optimal gauge for truncation may no longer be the multipolar gauge. It is also important to note that while the procedure of two-level truncation is performed in the multipolar gauge, this certainly does not restrict our attention to subsystems defined relative to the multipolar gauge. It is straightforward to identify the observables that define the $\alpha$-gauge subsystems within the multipolar gauge wherein the truncation of these observables may then be performed. %as noted in Sec.~\ref{s5},
In particular, we are free to consider the canonical field $\Pi$ defined relative to an arbitrary gauge $\alpha$. The physical observable represented by the momentum $\Pi$ in the gauge $\alpha$ will be denoted $O_\alpha$. The notation $\Pi$ will be reserved for the multipolar gauge canonical momentum $-D_{\rm T}$, therefore $O_\alpha=\Pi+P_{\rm T1}-P_{\rm T\alpha}$.  Here the $\alpha$-gauge polarisation is within the multipolar gauge truncation given by
\begin{align}\label{PTcav}
P_{\rm T\alpha}(t,x)=\sum_{k} {d\over v}\sigma^x(t) \alpha \cos[kx]
\end{align}
where $d={\hat {\bf z}}\cdot {\bf d}$ is the two-level transition dipole moment in the $z$-direction. In fact, since ${\bf P}_{\rm T\alpha}$ commutes with gauge fixing transformations, Eq.~(\ref{PTcav}) is an example of a truncated expression that is actually independent of the gauge within which truncation is performed  .

To obtain the Hamiltonian, truncation within the multipolar gauge gives the multipolar gauge multi-mode QRM
\begin{align}\label{h1mult}
H_1^2 =\,& {\tilde \omega}_m\sigma^+\sigma^- + \sum_k \omega_k \left(a_k^\dagger a_k+{1\over 2}\right)\nonumber \\ &+i\sum_k g_k(a_k^\dagger -a_k)\sigma^x 
\end{align}
where $g_k = d\sqrt{\omega_k /2v}$ and where we have absorbed the multipolar gauge polarisation self-energy term into a renormalisation of the two-level transition frequency denoted ${\tilde \omega}_m$. In Ref.~\cite{sanchez_munoz_resolution_2018} [see also \cite{casanova_deep_2010}] it was demonstrated via comparison with numerical results utilising matrix product states that for sufficiently large coupling strengths and numbers of modes the two-level system frequency ${\tilde \omega}_m$ may be neglected in Eq.~(\ref{h1mult}) resulting in an independent-boson model;
\begin{align}\label{h1mult2}
H_1^2 \approx \sum_k \omega_k \left(a_k^\dagger a_k+{1\over 2}\right)+i\sum_{k} g_k(a_k^\dagger -a_k)\sigma^x.
\end{align}
Since $\sigma^x$ is now a symmetry, the Hamiltonian is easily diagonalised using a polaron transformation
\begin{align}\label{polmm}
{\cal T}_{10} = \exp\left[i\sum_{k} {g_k\over \omega_k}(a_k^\dagger +a_k)\sigma^x\right].
\end{align}
This is the same type of transformation as was encountered in Sec.~\ref{s1}. Although it is not in general a gauge transformation   [as defined by Eqs.~(\ref{min1c}) and (\ref{min2c})] , we noted in Sec.~\ref{notgt} that when acting on the canonical momentum $\Pi$ this transformation does have the same effect as the projected PZW gauge fixing transformation $PR_{10}$.

The dynamics of the observables $O_\alpha$ closely mirror those found for free space in Sec.~\ref{sourcefs}. Using Eq.~(\ref{h1mult2}) we obtain
\begin{align}\label{akt}
a_k(t) &= a_ke^{-i\omega_k t}+g_k \int_0^t ds\, e^{-i\omega_k(t-s)}\sigma^x(s) \nonumber \\ &\equiv a_{k,\rm vac}(t)+a_{k,\rm s}(t).
\end{align}
We note that the above vacuum-source partitioning is that given by the multipolar gauge. This is the most convenient partitioning if we wish to determine averages when assuming an initial bare state in the multipolar gauge, which corresponds to assuming a well-defined state of energy of a fully localised dipole. The operator $\sigma^x(s) = \sigma^x(0)=\sigma^x$ is time-independent because the two-level dipole energy has been neglected. As a result the temporal integral in Eq.~(\ref{akt}) can be evaluated immediately and since $\sigma^x$ is stationary, so too is the electrostatic field $P_{\rm T\alpha}(t,x)=P_{\rm T\alpha}(0,x)$ defined in Eq.~(\ref{PTcav}). The negative frequency fields are found to be $O_{\alpha}^{(-)}(t,x) =\Pi_{\rm vac}^{(-)}(t,x) +O_{\alpha,\rm s}^{(-)}(t,x)$, $O_{\alpha,\rm s}^{(-)}(t,x)=\Pi_{\rm s}^{(-)}(t,x)+P_{\rm T1}^{(-)}(t,x)-P_{\rm T\alpha}^{(-)}(t,x)={d}\sigma^x(1-\alpha)/(2v)+\sum_{k>0}^N {d}\sigma^x(e^{i\omega_k t}-\alpha)\cos[kx]/v$, and $P_{\rm T\alpha}^{(-)}:=\sum_{k}{d}\sigma^x\alpha e^{-ikx}/(2v) = P_{\rm T\alpha}/2$, where the integer $N$ sets the total number of modes retained within the model. Positive frequency components are obtained by Hermitian conjugation and the sum of positive and negative frequency parts of a field gives the total field. By construction these expressions yield $O_\alpha= O_{\alpha}^{(-)}+O_{\alpha}^{(+)}$ for any $\alpha$. Choosing $\alpha=1$ gives the particular case $O_1=\Pi = -D_{\rm T}=-E_{\rm T}-P_{\rm T1}$.
 
%%%%%%%%%%%%%%%%%%%%%%%%%%%%%%%%%%%%%%%%%%%%%%%%%%%%%%%%%
\begin{figure}[t]
\begin{minipage}{\columnwidth}
\begin{center}
(\textbf{a})\hspace*{-6mm}
\includegraphics[scale=0.39
]{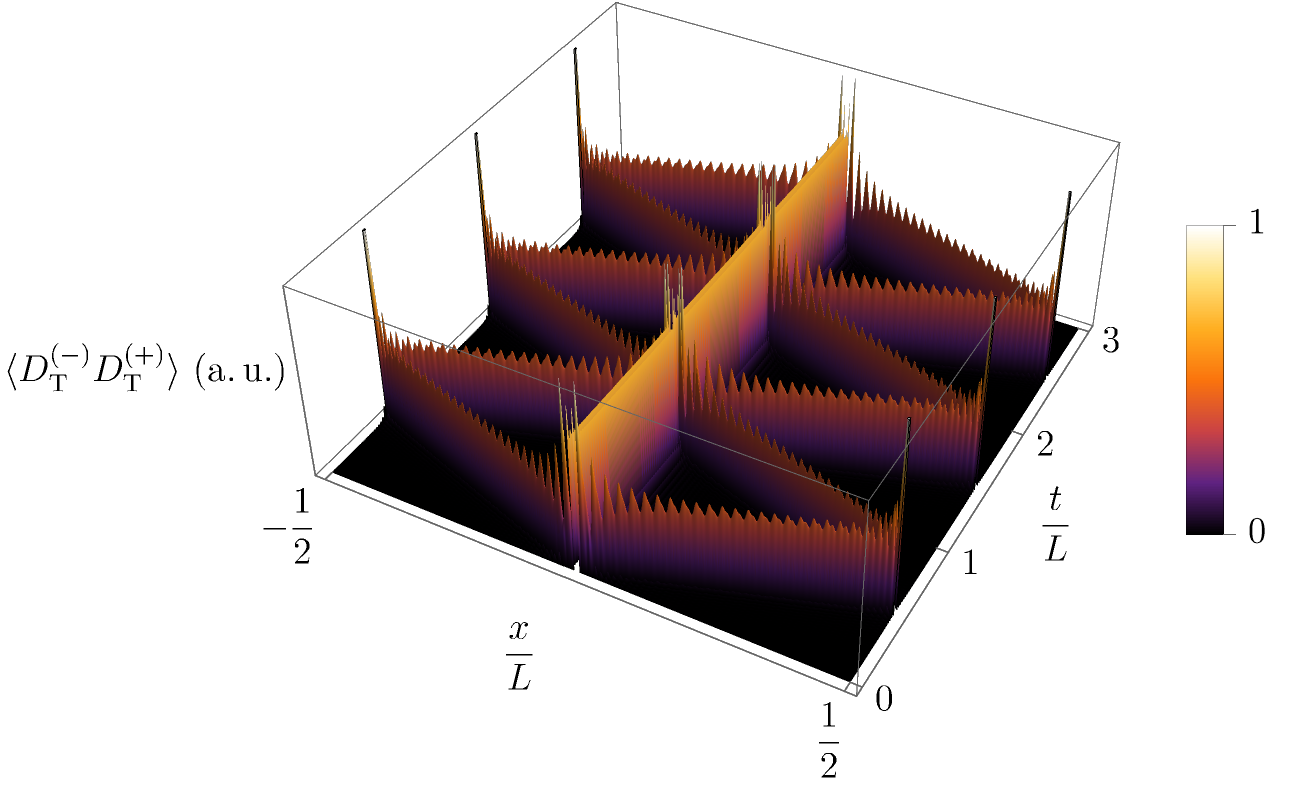}
(\textbf{b})\hspace*{-6mm}
\includegraphics[scale=0.39
]{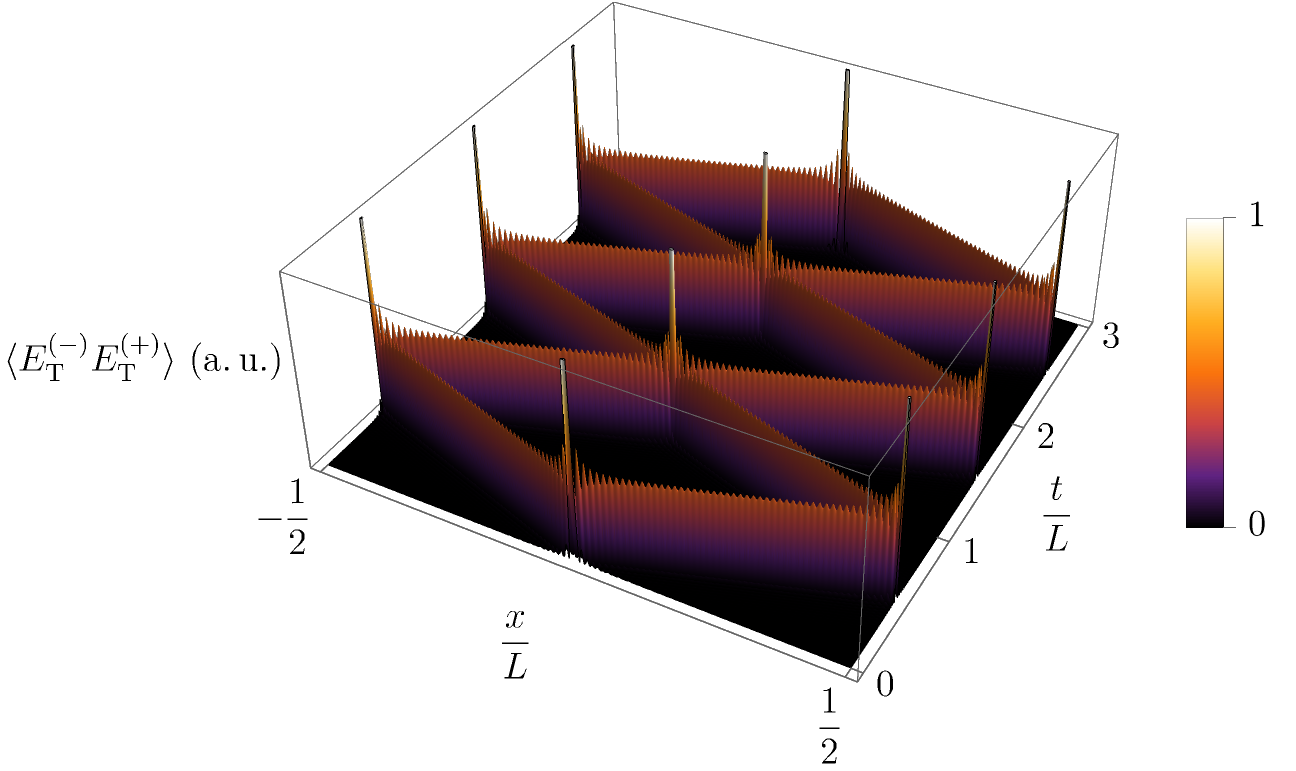}
\caption{The averages $\langle D_{\rm T}^{(-)}(t,x)D_{\rm T}^{(+)}(t,x)\rangle$ (\textbf{a}) and $\langle E_{\rm T}^{(-)}(t,x)E_{\rm T}^{(+)}(t,x)\rangle$ (\textbf{b}) are plotted with space and time, showing the presence and absence of a bound-field around the multipolar and Coulomb gauge dipoles respectively. Essentially the same propagating field is obtained in both cases. We have assumed $N=50$ and normalised both densities via the maximum value attained when the propagating field is coincident with the dipole; $(t,x)=(nL,0),~n\in {\mathbb Z}$.}\label{bound}
\end{center}
\end{minipage}
\end{figure}
%%%%%%%%%%%%%%%%%%%%%%%%%%%%%%%%%%%%%%%%%%%%%%%%%%%%%%%%%
It is now possible to evaluate the average of arbitrary functions of $O_\alpha$, $O_{\alpha}^{(-)}$ and $O_{\alpha}^{(+)}$ using any initial state. We use both the initial multipolar bare state $\ket{\epsilon^1,0}$ and the ground state, which is represented by the vector $\ket{\epsilon^0,0}$ in the polaron frame. Since we have neglected the dipole energy and since the polaron transformation coincides with the projected PZW transformation when acting on $\Pi$, for the purpose of finding the dynamics of $O_\alpha$ the polaron-frame is nothing but the Coulomb gauge. Specifically, we have ${\cal T}_{10}\Pi{\cal T}_{10}^\dagger = \Pi-P_{\rm T1}$, and ${\cal T}_{10}O_\alpha {\cal T}_{10}^\dagger = \Pi - P_{\rm T\alpha}$. Since the operator ${\cal T}_{10}\Pi{\cal T}_{10}^\dagger$ represents the observable $-D_{\rm T}$ in the polaron frame, the operator $\Pi$ represents the observable $-D_{\rm T}+P_{\rm T1}=-E_{\rm T}$, as in the Coulomb gauge. In this gauge the electrostatic field is absorbed into the definition of the dipole. Further still, within the approximations made the Coulomb gauge coincides with the JC-gauge; $\alpha_{\rm JC}={\tilde \omega}_m/({\tilde \omega}_m+\omega_k)\approx 0$. Thus, the very simple treatment in which the free dipole Hamiltonian has been neglected, is unable to distinguish between electrostatic and virtual-photonic bound-fields. In Supplementary Note~XIII it is seen that this distinction is also obscured when considering the near-field limit of the ground-state energy density in free space whereby the total electric energy density becomes approximately purely electrostatic, as shown by Eq.~(208) in Supplementary Note~XIV. We emphasize that the coincidence of the Coulomb gauge, the JC-gauge, and the polaron-frame for calculating averages of functions of $\Pi$ does not occur without the simplifications made. In general, these representations are distinct.

We now calculate various quadratic energy densities as in   Supplementary Note~XIII. For the initial state $\ket{\epsilon^1,0}$ we obtain
\begin{align}
&\langle O_\alpha(t,x)^2 \rangle -E_{\rm vac} \nonumber \\ &= \left[\sum_{k} {d\over v} (\cos[kx-\omega_k t]-\alpha \cos[kx])\right]^2 \\
&\langle O^{(-)}_\alpha(t,x)O^{(+)}_\alpha(t,x)\rangle=\nonumber \\ &\left|{d\over 2v}(1-\alpha)+\sum_{k>0}^N {d\over v}(e^{i\omega_k t}-\alpha)\cos[kx]\right|^2\label{oalphno}
\end{align} 
where $E_{\rm vac} = \sum_k \omega_k/(2v)$ is an energy density of the vacuum. For $\alpha=1$ (multipolar gauge), Eq.~(\ref{oalphno}) reduces to the result obtained in Ref.~\cite{sanchez_munoz_resolution_2018}. Ground-state averages are obtained using the polaron frame and are found to be $\langle O_\alpha(t,x)^2 \rangle_G - E_{\rm vac}=  \langle P_{\rm T\alpha}(t,x)^2 \rangle_G$, and $\langle O^{(-)}_\alpha(t,x)O^{(+)}_\alpha(t,x)\rangle_G ={1\over 4} \langle P_{\rm T\alpha}(t,x)^2 \rangle_G$ where
\begin{align}
 \langle P_{\rm T\alpha}(t,x)^2 \rangle_G = \left[\sum_{k\neq 0} {d\over v} \alpha \cos[kx]\right]^2. \label{gsoalph}
\end{align} 
This confirms that, within the approximations made, the bound-field tied to the $\alpha$-gauge dipole is nothing but the electrostatic field $P_{\rm T\alpha}$. In the Coulomb gauge this field is fully included within the definition of the dipole, so $\langle O_0(t,x)^2 \rangle_G - E_{\rm vac} = 0$. Fig.~\ref{bound} shows $\langle O^{(-)}_\alpha(t,x)O^{(+)}_\alpha(t,x)\rangle$ given in Eq.~(\ref{oalphno}) for the Coulomb and multipolar gauges $\alpha=0$ and $\alpha=1$ respectively. It can be seen clearly that all gauges possess essentially the same propagating fields. In contrast the ground-state bound-field energy has weight $\alpha^2$ within the gauge $\alpha$ and is evidently highly localised at the position of the dipole within the one-dimensional model employed.

Ref.~\cite{sanchez_munoz_resolution_2018} proposes that the initial multipolar bare-state $\ket{\epsilon^1,0}$ could be prepared by controlling the interaction. However, given the level of localisation of the bound field, it is far from clear that the latter could ever be separated from the dipole allowing the corresponding interaction to be controlled. A possible exception may be to move the dipole in and out of the cavity very quickly. As already described in Sec.~\ref{nonadiabatic} in this case the relevant gauge for modelling the interaction using a time-dependent coupling will depend strongly on the microscopic details of the system.

We remark that the treatment of this section is highly idealised. The cavity is taken as one-dimensional, the two-level truncation has also been made, and the dipole moment dynamics have been taken as approximately stationary. The extension of these results using more realistic treatments warrants further investigation, including a more physical model for the cavity and a more sophisticated method of solution, for example, via a variational polaron ansatz \cite{diaz-camacho_dynamical_2016}.

Evidently, the physical nature of the internal cavity field depends strongly on the gauge relative to which it is defined. As we have emphasized, gauge ambiguities arise because it is not always clear which subsystems should be considered operationally addressable. The interaction between the system of interest and apparatus used in preparation and measurement must be defined relative to a choice of gauge. Simple gedanken experiments for the weak measurement of intra-cavity subsystems within the weak-coupling regime were introduced some-time ago in Refs.~\cite{compagno_detection_1988,compagno_dressed_1988,compagno_dressed_1990,compagno_bare_1991,compagno_atom-field_1995}. We discuss these models in Supplementary Note~XVI.

\subsection{Ground state superradiance}\label{super}

Here we exemplify the importance of the preceding discussions concerning intra-cavity fields and subsystem gauge relativity by very briefly reviewing the phenomenon of ground state superradiance (also called photon condensation in Refs.~\cite{andolina_cavity_2019,andolina_theory_2020}).

\subsubsection{Dicke models}

Ground state superradiance was first predicted in the Dicke model \cite{dicke_coherence_1954,hepp_superradiant_1973,wang_phase_1973}. There is now extensive literature on this topic including extended Dicke models \cite{carmichael_higher_1973,hioe_phase_1973,pimentel_phase_1975,emeljanov_appearance_1976,sung_phase_1979}, connections with quantum chaos \cite{holstein_field_1940,emary_chaos_2003,emary_quantum_2003}, driven and open systems \cite{grimsmo_dissipative_2013,klinder_dynamical_2015,gegg_superradiant_2018,kirton_superradiant_2018,peng_unified_2019}, and artificial systems \cite{yamanoi_polariton_1979,haug_quantum_1994,lee_first-order_2004,nataf_no-go_2010,viehmann_superradiant_2011,todorov_intersubband_2012,leib_synchronized_2014,bamba_stability_2014,bamba_superradiant_2016,bamba_circuit_2017,jaako_ultrastrong-coupling_2016,de_bernardis_cavity_2018,rouse_theory_2022}. The topic has received renewed interest in light of rapid progress in magnonic systems and in controlling correlated electron systems inside cavities \cite{mazza_superradiant_2019,andolina_cavity_2019,nataf_rashba_2019,guerci_superradiant_2020,andolina_theory_2020,bamba_magnonic_2022}.

Despite this, whether or not a phase transition does indeed occur and its precise nature have remained open questions. This is due to the existence of so-called ``no-go theorems", which prohibit a superradiant phase and which are proved in the Coulomb gauge  \cite{rzazewski_phase_1975}. Further variants of this theorem have been both refuted and confirmed subsequently \cite{kudenko_interatomic_1975,rzazewski_remark_1976,emeljanov_appearance_1976,knight_are_1978,bialynicki-birula_no-go_1979,yamanoi_polariton_1979,sung_phase_1979,rzazewski_stability_1991,keeling_coulomb_2007,nataf_no-go_2010,vukics_adequacy_2012,vukics_elimination_2014,bamba_stability_2014,tufarelli_signatures_2015,grieser_depolarization_2016,bamba_circuit_2017,andolina_cavity_2019,andolina_theory_2020}.

Keeling noted that since the radiation modes are distinct in the Coulomb and multipolar gauges, a ground state phase transition may possess different characterisations, and showed that a ferroelectric phase transition occurs within the Coulomb gauge at the same point in parameter space as the superradiant phase transition of the conventional Dicke model \cite{keeling_coulomb_2007}. More recently, the present authors have shown \cite{stokes_uniqueness_2020} that a unique (gauge invariant) phase transition can be supported within cavity QED systems, by using the one-parameter $\alpha$-gauge framework. It was shown further that the macroscopic manifestation is gauge invariant, but that the classification of the phase transition depends on the gauge relative to which the quantum subsystems are defined.

For a cavity containing $N$ dipoles labelled by $\mu=1,...,N$, with dipole moments ${\bf d}_\mu$ and fixed positions ${\bf R}_\mu$, the $\alpha$-dependent canonical momenta are found to be \cite{stokes_uniqueness_2020}
\begin{align}
{\bf p}_{\mu} &= m{\dot {\bf r}}_\mu - e(1-\alpha){\bf A}({\bf R}_\mu),\label{mom1}\\
{\bf \Pi}({\bf x}) &= -{\bf E}_{\rm T}({\bf x}) - {\bf P}_{{\rm T}\alpha}({\bf x}),\label{mom2}
\end{align}
The Hamiltonian is the total energy \cite{stokes_uniqueness_2020} wherein the total electrostatic energy can be split into an atomic binding energy for each dipole, $V$, and an inter-dipole electrostatic coupling, $V_{\rm dip}$ (dipole-dipole interaction).

Assuming that the dipole moments $d ={\bf d}\cdot {\bf e}$ point in the direction of the cavity polarisation ${\bf e}$, the single-mode approximation is performed in such a way as to preserve gauge invariance (see Secs.~\ref{eda} and \ref{modalres}). This eliminates the need to regularise ${\bf P}_{\rm T}$ \cite{vukics_fundamental_2015}, and ensures that the transverse commutation relation for the canonical fields is preserved. The fundamental kinematic relations given by Eqs.~(\ref{mom1}) and (\ref{mom2}) are therefore also preserved. In order to obtain a Dicke Hamiltonian the limit of closely spaced dipoles around the origin; ${\bf R}_\mu \approx {\bf 0}$ is taken, and the dipoles are approximated as two-level systems. Collective operators are then introduced; $J_\alpha^i = \sum_{\mu=1}^N \sigma^i_{\mu\alpha}, i=\pm,z$, where $\sigma^\pm_{\mu\alpha}$ are the raising and lowering operators of the $\mu$'th two-level dipole and $\sigma^z_{\mu\alpha} = [\sigma^+_{\mu\alpha},\sigma^-_{\mu\alpha}]/2$. 

Although the non-truncated Hamiltonian $H$ is unique, we now have a continuous infinity of Dicke Hamiltonians $H^{\alpha,2}$ such that $H^{\alpha,2}$ and $H^{\alpha',2}$ are not equal when $\alpha\neq \alpha'$ \cite{de_bernardis_breakdown_2018,stokes_gauge_2019,stefano_resolution_2019,roth_optimal_2019}. The breaking of gauge invariance due to truncation turns out not to be a barrier in eliminating all ambiguities regarding the occurrence and nature of a quantum phase transition.

The thermodynamic limit is defined by $N\to \infty, V\to \infty$ with $\rho=N/V$ constant. In this limit the Holstein-Primakoff map defined by $J_\alpha^z = b_\alpha^\dagger b_\alpha -{N/ 2}$, $J_\alpha^+ = b_\alpha^\dagger \sqrt{N-b_\alpha^\dagger b_\alpha}$, and $J_\alpha^-=(J_\alpha^+)^\dagger$, where $[b_\alpha,b_\alpha^\dagger]=1$, is used \cite{holstein_field_1940,emary_chaos_2003,emary_quantum_2003}. The Hamiltonian obtained by substituting these expressions into $H^{\alpha,2}$ is denoted $H_{\rm th}^{\alpha,2}$. 

The Hamiltonian is found to support two distinct phases and reads \cite{stokes_uniqueness_2020}
\begin{align}\label{norm}
H_{\rm th}^{\alpha,2,\rm i} =& \,E^{\rm i}_{\alpha+} {f^{\rm i}_\alpha}^\dagger f^{\rm i}_\alpha + E^{\rm i}_{\alpha-}  {c^{\rm i}_\alpha}^\dagger c^{\rm i}_\alpha +{1\over 2}( E^{\rm i}_{\alpha+} +E^{\rm i}_{\alpha-} )+C^{\rm i}
\end{align}
where the superscript $\cdot^{\rm i}$ is either ${\rm i}={\rm n}$ for normal-phase, or ${\rm i}={\rm a}$ for abnormal-phase. The polariton operators $f^{\rm i}_{\alpha},\,c^{\rm i}_\alpha$ are bosonic satisfying $[f^{\rm i}_{\alpha},{f^{\rm i}_{\alpha}}^\dagger]=1=[c^{\rm i}_{\alpha},{c^{\rm i}_{\alpha}}^\dagger]$ with all other commutators vanishing. The polariton energies ${E^{\rm i}_{\alpha\pm}}$ and constant $C^{\rm i}$ are known functions of the couplings and frequencies appearing in the Hamiltonian $H^{\alpha,2}$. It can be shown that the lower polariton energy $E^{\rm n}_-$ is real provided that
\begin{align}
\tau:={\omega_m \over 2\rho d^2}\geq1
\end{align}
while the lower polariton energy $E^{\rm a}_-$ is real provided that
\begin{align}\label{pt2}
\tau \leq1.
\end{align}
It can also be shown that $H_{\rm th}^{\alpha,2,\rm n} =H_{\rm th}^{\alpha,2,\rm a} $ for $\tau=1$. As $\rho d^2$ is increased, a unique phase transition is predicted to occur at the critical point $\tau=1$ in parameter space, beyond which the normal phase Hamiltonian, $H_{\rm th}^{\alpha,2,\rm n}$ breaks down and the abnormal phase Hamiltonian, $H_{\rm th}^{\alpha,2,\rm a}$, takes over. This prediction is gauge invariant.

It remains only to determine the nature of the unique phase transition predicted. To demonstrate equivalence between all gauges the $\alpha$-gauge transverse polarisation $P_{\rm T\alpha}=\alpha {\bf e}\cdot{\bf P}_{\rm T}=\alpha(\Pi_0-\Pi_1)$ is calculated. In the normal phase the thermodynamic limit of this quantity, denoted $P_{\rm T\alpha, th}$, vanishes, whereas in the abnormal phase it is found to be $P_{\rm T\alpha, th}^{\rm a}  =- \alpha \rho d\sqrt{1-\tau^2}$. It can be further shown that in the thermodynamic limit one obtains $\Pi_{\rm th}^{\rm a}=-P_{\rm T\alpha, th}^{\rm a}$, such that choosing $\alpha=0$ we have $-E_{\rm T, th}^{\rm a}=\Pi_{\rm th}^{\rm a}=0$, verifying the fundamental kinematic relation (\ref{mom2}). This establishes consistency between all gauges. The onset of the abnormal phase manifests in the form of a macroscopic value of the gauge invariant field ${\bf P}_{\rm T}$;
\begin{align}\label{pabt}
P_{\rm T,th}^{\rm a}=P_{\rm T1,th}^{\rm a}= - \rho d\sqrt{1-\tau^2}.
\end{align}

Previous no-go and counter no-go results can be reconciled by noting that radiation is gauge-relative. In the Coulomb gauge radiation is defined by ${\bf \Pi}=-{\bf E}_{\rm T}$, such that the phase transition does not appear superradiant in character and only the ``material" subsystem acquires a macroscopic population. This constitutes a ``no-go theorem" for superradiance defined relative to the Coulomb gauge. In the multipolar gauge radiation is defined by ${\bf \Pi}=-{\bf E}_{\rm T}-{\bf P}_{\rm T}$ such that both the material and radiative subsystems acquire macroscopic population in the abnormal phase. This constitutes a ``counter no-go theorem" for superradiance defined relative to the multipolar gauge. Clearly these results are not in contradiction, because they are referring to different definitions of radiation. Indeed, the results above demonstrate that they are in fact equivalent \cite{stokes_uniqueness_2020}. More generally, since ${\bf \Pi} = -{\bf E}_{\rm T}-{\bf P}_{\rm T \alpha}$, the degree to which the unique phase transition is classed as superradiant is directly determined by the value of $\alpha$.

As we have seen, $\alpha$ controls the balance between localization and dressing in defining the quantum subsystem called matter. In Sec.~\ref{intracav} we observed that the field ${\bf P}_{\rm T}$ is highly localised at the position of the dipole within the approximations made and the one-dimensional model adopted. As discussed in Sec.~\ref{intracav} and Supplementary Note~XVI, which predictions are most relevant depends on which observables are accessible via the available preparation and measurement protocols.

\subsubsection{Condensed matter systems in the Coulomb gauge}

The superradiant phase transition has predominantly been discussed in the context of Dicke-type models. As reviewed above, the gauge invariance of the predicted instability, and the gauge invariance of its manifestation, are now established. However, there remains a question of whether such simplified models can realistically describe actual physical systems (see Ref.~\cite{grieser_depolarization_2016} for a discussion). Recent work in Refs.~\cite{mazza_superradiant_2019,andolina_cavity_2019,nataf_rashba_2019,guerci_superradiant_2020,andolina_theory_2020,bamba_magnonic_2022,rouse_theory_2022} moves beyond simplified Dicke model type treatments. Strongly-correlated electron systems of the type encountered in condensed matter theory are considered, rather than a gas of dipoles as in the Dicke model. 

Ref.~\cite{andolina_cavity_2019} shows that ground state photon condensation cannot occur in strongly-correlated electron systems, including an arbitrary electron-electron interaction potential, but considering only a single cavity mode and only photons defined relative to the Coulomb gauge. Ref.~\cite{andolina_theory_2020} progresses these findings by considering a three-dimensional electron system (3DES) in an inhomogeneous cavity field, i.e., one that varies in space. Again, only photons defined relative to the Coulomb gauge are considered, but in this case it is found that photon condensation can occur if
\begin{align}\label{chiorb}
\chi_{\rm orb}(k)>{1\over 4\pi}
\end{align}
where $\chi_{\rm orb}(k)$ is the ${\bf k}$-space non-local orbital magnetic susceptibility of the 3DES \cite{giuliani_quantum_2005}. If the model is extended to include the spin of electrons then this condition becomes $\chi_{\rm orb}(k)+\chi_{\rm spin}(k)>{1\over 4\pi}$ where $\chi_{\rm spin}(k)$ is the spin magnetic susceptibility.

This transition to photon condensation possesses a simple interpretation as a magnetic instability \cite{andolina_theory_2020}. Specifically, Ref.~\cite{andolina_theory_2020} defines the magnetic energy of a material subject to a magnetic field ${\bf B}$ as $E_M = \int d^3 x\, {\bf H}\cdot {\bf B}$ where ${\bf H}={\bf B}-{\bf M}$ and ${\bf M}$ is the (orbital) magnetisation of the material, which is traditionally interpreted as describing the response of the material to the applied field. Then, assuming linear response theory in which ${\bf M}$ is a linear functional of ${\bf B}$ and $\chi_{\rm orb}$, one finds that $E_M$ can be written \cite{andolina_theory_2020}
\begin{align}\label{magenfun}
&E_{\rm M}= \nonumber \\ &-2\pi \int d^3 x \int d^3x' \delta({\bf x}-{\bf x}')\chi_{\rm orb}(|{\bf x}-{\bf x}'|){\bf B}({\bf x})\cdot {\bf B}({\bf x}').
\end{align}
An instability occurs if $E_M<0$. Upon Fourier transforming $E_{\rm M}$ in Eq.~(\ref{magenfun}), this inequality gives inequality (\ref{chiorb}), which is the condition for photon condensation. We note that relative to gauge $\alpha$ photons are defined by ${\bf \Pi}=-{\bf E}_{\rm T}-\alpha{\bf P}_{\rm T}$ and so upon noting the traditional interpretation of ${\bf P}$ as describing the response of a material to an electric field, one might expect condensation of photons$_\alpha$ to be related to electric instability for any $\alpha\neq 0$. This was confirmed by \cite{rouse_theory_2022} for the case of a jellium source within a cavity.

Ref.~\cite{mazza_superradiant_2019} considers strongly correlated electrons coupled to a single cavity mode in the Coulomb gauge and affirms the no-go theorem for condensation of these photons. However, it is reported that the situation changes when electronic interactions {\em and} delocalisation are taken into account. It is found that in a two-band model of interacting electrons a phase supporting condensation of excitons and photons can occur, even while considering only one cavity mode. 

Ref.~\cite{guerci_superradiant_2020} considers one- and two-dimensional strongly-correlated electron systems coupled to a cavity field in the Coulomb gauge. The no-go theorem is again affirmed for the case of a single-mode homogeneous field while photon condensation is found to be possible for a non-uniform field. Ref.~\cite{nataf_rashba_2019} also considers an inhomogeneous cavity field coupled to a two-dimensional electron system in the Coulomb gauge, including spin-orbit coupling and a perpendicular applied magnetic field. It is found that a superradiant phase transition can occur. We conclude this section by remarking that the investigation of strongly-correlated electron cavity QED systems beyond a restriction to the Coulomb gauge, as undertaken initially by \cite{rouse_theory_2022}, warrants further study. 

\subsection{Extra-cavity fields: Overview}\label{excavfs}

The description of external coupling to the cavity has received considerable attention. We provide an overview here before discussing specific simple models in subsequent sections. We are again faced with two problems outside of traditional regimes. The first concerns the determination of which approximations might be applied and when, and the second concerns the determination of which physical states and observables are relevant in preparation and measurement.

Although the two problems are not unrelated let us consider the first problem first. For weakly coupled subsystems dissipation and decoherence can be modelled via separate loss mechanisms as though the subsystems are uncoupled. This constitutes the so-called local approach to deriving a master equation for the matter-cavity system. For example, the stationary state of a qubit in a cavity described by the local master equation
\begin{align}\label{local}
{\dot \rho} = -i[H,\rho] &+ {\Gamma\over 2}(2\sigma^-\rho\sigma^+ - \{\sigma^+\sigma^-,\rho\}) \nonumber \\
&+ {\kappa\over 2}(2a\rho a^\dagger - \{a^\dagger a,\rho\})
\end{align}
is simply $\ket{\epsilon^g,0}$. Here $\sigma^+ = \ket{\epsilon^e}\bra{\epsilon^g}$ is the qubit raising operator, $\sigma^-=(\sigma^+)^\dagger$, and $a$ is the annihilation operator for the cavity. Dissipation is described via separate Lindblad tails corresponding to the qubit and mode. In the so-called global approach dissipation is instead described in the dressed basis of the light-matter system.

The difference between local and global approaches has been discussed extensively and in various contexts \cite{walls_higher_1970,schwendimann_interference_1972,carmichael_master_1973,scala_microscopic_2007,scala_cavity_2007,santos_master_2014,stokes_master_2018,gonzalez_testing_2017,mitchison_non-additive_2018,maguire_environmental_2019,chiara_reconciliation_2018,joshi_markovian_2014,manrique_nonequilibrium_2015,purkayastha_out--equilibrium_2016,santos_microscopic_2016,decordi_two_2017,stockburger_thermodynamic_2017,hewgill_quantum_2018,naseem_thermodynamic_2018,seah_refrigeration_2018,hamedani_raja_thermodynamic_2018,cattaneo_local_2019}. Cresser noted early on that the local master equation could apparently break down when describing a lossy Jaynes-Cummings model \cite{cresser_thermal_1992}. Hoffer et al. found by comparison with exact predictions that the local equation may perform better in the weak-coupling regime while the global master equation is generally better in the strong-coupling regime \cite{hofer_markovian_2017}. However, the relative validity of the two approaches depends on the form of secular approximation used. Cattaneo et al. have shown that the global master equation with partial secular approximation is always most accurate when Born-Markov approximations are also valid \cite{cattaneo_local_2019}. The local approach is often claimed to fail \cite{santos_master_2014,manrique_nonequilibrium_2015,decordi_two_2017}, but it has been shown to be thermodynamically consistent for fairly large ranges of coupling strengths \cite{hofer_markovian_2017,gonzalez_testing_2017}.

Here we note that since the gauge-parameter $\alpha$ selects the form of the interaction, one would not expect the relative applicability of local versus global master equations to be independent of $\alpha$. In general, losses of a light-matter system will depend on how it couples to the external system or environment \cite{bamba_recipe_2014,bamba_system-environment_2013}. For example, Ref. \cite{ciuti_input-output_2006} applys input-output theory to quantum wells within a microcavity, such that the cavity couples to external photonic modes via a number-conserving interaction while the electronic system couples to another bosonic environment similarly. With this treatment it is predicted that ground state ``virtual" cavity and electronic excitations cannot leak out of the cavity. In contrast, Ref. \cite{de_liberato_extracavity_2009} used a form of non-Markovian master equation to describe a two-level system coupled to radiation while assuming fast modulation of the vacuum Rabi frequency. It was predicted that extra-cavity quantum vacuum radiation would occur for state-of-the-art circuit cavity QED systems. 

Predictions such as those in Refs.~\cite{ciuti_input-output_2006,de_liberato_extracavity_2009} are in general specific to the forms of coupling adopted, i.e., they are specific to the physical subsystems considered. Indeed, as we have noted the second task that we are faced with is identifying which states and observables are relevant. If counter-rotating terms are non-negligible in the interaction of a light-matter system then the local master equation description of its losses will result in photon generation in the environmental vacuum \cite{werlang_rabi_2008}. This would typically be taken as indicating the onset of the regime in which the bare states are no longer meaningful, such that one should switch to a global description in which dissipation is described holistically using the dressed states of the full light-matter Hamiltonian \cite{bamba_system-environment_2013,bamba_recipe_2014,beaudoin_dissipation_2011,boite_theoretical_2020}. Similarly, a coarse-grained projection onto the vacuum state, as in the Born approximation, will induce apparently paradoxical spontaneous excitations in polaritonic systems. The paradox is resolved by accounting for correlations between the dressed ground  
state of the system and the environmental vacuum within the reservoir correlation functions of the master equation \cite{bamba_dissipation_2012}.

If we are interested in determining measurement signals from a source then the generic problem consists of two multi-level systems, a source and a detector, coupled to a common reservoir as was considered in Sec.~\ref{sdf}. However, the multi-level source need not be elementary. In particular, it could be an ultrastrongly coupled light-matter composite. In a ``global approach", the light-matter composite is diagonalised and then weakly-coupled to whatever is external \cite{di_stefano_photodetection_2018,boite_theoretical_2020,bamba_recipe_2014,bamba_system-environment_2013,salmon_gauge-independent_2022}.   In particular, Ref.~\cite{di_stefano_photodetection_2018} adopts precisely this strategy as a means by which to apply Glauber photodetection theory when dealing with an ultrastrongly coupled light-matter composite that is weakly coupled to a photon absorber. The same method is applied in Ref.~\cite{salmon_gauge-independent_2022} to understand cavity leakage using a simple semi-phenomenological approach, which is reviewed below in Sec.~\ref{specs}  . In this case all weak-coupling results for loss and detection are recovered with the only difference being that the eigenstates of the source are the dressed states of a composite. As previously discussed, in this context there is obviously a balance to be struck between electromagnetic dressing and localisation in spacetime. This balance is affected by the choice of gauge. 

In Supplementary Note~XVII we review microscopic descriptions of cavity QED systems, including a perfect cavity containing matter and an imperfect empty cavity. The problem of describing leakage from an imperfect cavity containing matter is more involved. A phenomenological approach consists of matter coupled linearly to the cavity, which in turn couples linearly to an environment, with reasonable coupling functions being chosen. This is the approach employed in Ref. \cite{ciuti_input-output_2006} for example.

A promising means by which to provide a description from first principles is to use the theory of QED within absorbing and dispersing media, as reviewed in Supplementary Note~VII. Ref.~\cite{bamba_system-environment_2013}  (see also Ref.~\cite{bamba_dissipation_2012}) uses this theory in conjunction with Maxwell boundary conditions to describe dissipation from a cavity containing bosonic matter (a polaritonic system), while considering the good cavity limit. It is found that external modes couple linearly to polaritonic raising and lowering operators via number conserving form. Since these operators are linear combinations of the cavity and matter subsystem raising and lowering operators it is noted that this (global) description differs from a phenomenological (local) description via a Gardiner-Collett model \cite{gardiner_input_1985}. 

In Ref.~\cite{bamba_recipe_2014}, the same authors consider the coupling to external modes of an ultrastrongly-coupled light-matter system. Both cavity and circuit QED implementations are considered. It is again noted that the phenomenological Gardiner-Collett Hamiltonian will break down. It is also emphasized that  in this situation the form of the system-environment interaction Hamiltonian will become significant, as was noted in Sec.~\ref{excavfs}. Two forms of interaction Hamiltonian are considered. One in which the cavity couples to external modes via the position quadrature $\sim a^\dagger +a$ (this is referred to as magnetic or inductive coupling) and one in which the coupling instead occurs through the momentum quadrature $\sim i(a^\dagger-a)$ (this is referred to as electric or capacitive coupling). Both coupling forms can be derived from an underlying Lagrangian. It is noted that in the absence of a dissipative transmission line the inductive and capacitively coupled light-matter system Hamiltonians are unitarily equivalent, but this is no longer the case for the full Hamiltonians that include coupling of the system to a transmission line. This is similar to the situation encountered in Sec.~\ref{excavfs} wherein coupling to external modes was defined relative to different gauges, which resulted in different reduced descriptions that corresponded to physically distinct reduced systems of interest. It is noted in Ref.~\cite{bamba_recipe_2014} that the difference in results obtained from different coupling forms can be ignored in sufficiently weak-coupling regimes, as well as in the good-cavity limit, which is effectively defined by the applicability of certain Markovian approximations.

Ref.~\cite{khanbekyan_qed_2005} also employs the theory of absorbing and dispersing dielectrics in Supplementary Note.~VII. Choosing the multipolar gauge, the authors consider leakage from a one-dimensional high-$Q$ cavity consisting of one perfect and one imperfect mirror, and containing a dipole, in both the weak- and strong-coupling regimes. It is found that on time scales large compared with the inverse separation of neighbouring cavity resonances, the internal cavity field may be expressed in terms of internal bosonic mode operators that obey quantum Langevin equations. Radiative input-output coupling and absorption losses can then be viewed as independent, with each possessing a damping rate and corresponding Langevin noise force. Thus, in the regime considered, the phenomenological Gardiner-Collett approach \cite{gardiner_input_1985} is valid inasmuch that the description of absorption losses requires only that the model is supplemented with bilinear interaction Hamiltonians between the cavity modes and appropriately chosen bosonic loss channels.

Ref.~\cite{franke_quantization_2019} similarly uses the dielectric theory of Supplementary Note~VII applied to a single dipole within the multipolar gauge and weak-coupling regime. The approach of these authors is to approximate the Green's function defined by Eq.~(89) in Supplementary Note~VII.~1 by an expansion in mode-functions corresponding to only a few resonant modes that are assumed to be dominant; so-called quasinormal modes (QNMs). The internal field to which the dipole couples is expressed in terms of the QNM functions and global bosonic mode operators while the external field is described similarly but with the QNM functions replaced by regularised counterparts. The use of only one or two QNMs has been found to be accurate within weak-coupling regimes (e.g. \cite{kamandar_dezfouli_modal_2017}). The approach enables dissipative QNM-Jaynes-Cummings models to be constructed for arbitrary dissipative structures.

The extension of the descriptions in, for example, Refs.~\cite{khanbekyan_qed_2005,franke_quantization_2019} to ultrastrongly-coupled light-matter systems within an arbitrary gauge warrants further study. We remark however that  a plausible physical model for the description of a lossy cavity containing atomic systems can already be proposed by combining insights from the case of a perfect cavity containing atomic systems (Supplementary Note~XVII. 1) with insights from the case of an imperfect but empty cavity (Supplementary Note~XVII. 2). Specifically, in Supplementary Note~XVII. 2 it is shown that for a high-$Q$ cavity a linear-coupling model between the cavity and external modes can be justified, while in Supplementary Note~XVII. 1 it is shown that a localised polarisation which vanishes at the cavity boundary implies that the light-matter interaction is mediated entirely by the local cavity field evaluated at the positions of the atoms. Such local light-matter coupling away from the boundary should not affect the form of the coupling between the cavity and external modes at the boundary. Thus, within a gauge in which the atomic systems are highly localised, such as the multipolar gauge, a model in which atomic dipoles couple linearly to a cavity field that in turn couples linearly to external modes, would appear to be physically reasonable. In the following section we review a spectroscopic signature of gauge relativity \cite{salmon_gauge-independent_2022} that uses such a model and leads to a final master equation given in Eq.~(\ref{dg}).

\subsection{Spectroscopic signatures of gauge relativity via a simple model}\label{specs}

An early attempt at modelling cavity leakage from an ultrastrongly-coupled dipole-cavity system using different gauges has recently been given in Ref.~\cite{salmon_gauge-independent_2022}. The authors consider the simplest toy model system of a two-level dipole coupled to a single cavity mode, volume $v$, frequency $\omega$, in one spatial dimension, described by the multipolar gauge quantum Rabi model (QRM) $H_1^2$ defined by the $\alpha=1$ case of Eq.~(\ref{2lmst}). Up to a constant this model reads
\begin{align}\label{rabm}
H_1^2=\omega_m\sigma^+\sigma^-+\omega a^\dagger a + ig(a^\dagger -a)(\sigma^++\sigma^-)
\end{align}
where $\sigma^{\pm}$ are the raising and lowering operators for the two-level dipole with transition frequency $\omega_m$, $a$ and $a^\dagger$ are the cavity annihilation and creation operators for photons defined relative to the multipolar gauge, and $g=d\sqrt{\omega\over 2v}$ is the coupling strength. A dimensionless coupling strength is defined by $\eta=g/\omega$. Recall that the above multipolar gauge two-level truncation of the dipole is expected to be accurate for a sufficiently anharmonic dipole (see Sec.~\ref{opt}). Leakage at rate $\kappa$ to external environmental modes $k$ described by bosonic operators $b_k,\,b_k^\dagger$ can be described in the gauge $\alpha$ using a linear weak-coupling Hamiltonian $V_{\rm cav-ext}^\alpha = \pi \otimes \sum_k g_k (b_k+b_k^\dagger)$ where $\pi:= \sqrt{2v\over \omega}\Pi$ is the cavity canonical momentum quadrature. This operator represents a different physical observable in
 each different gauge $\alpha$.
 
Since the light-matter system is ultrastrongly-coupled, Ref.~\cite{salmon_gauge-independent_2022} assumes a global approach in which dissipation is described using the dressed states of the light-matter composite. If one applies the standard derivation of the Lindblad master equation (see Sec.~\ref{master}) with the reduced system of interest being the dipole-cavity system described by the dressed states of the QRM $H_1^2$ and with coupling to the bath $V_{\rm cav-ext}^\alpha$, then one obtains
\begin{align}\label{mast}
&{\dot \rho} = i[\rho,H_1^2]+{\cal L}(\rho,x),\\
&{\cal L}(\rho,x) = \kappa\left(x\rho x^\dagger - {1\over 2}\{x^\dagger x,\rho\}\right)
\end{align}
where $\rho$ is the density operator describing the dipole-cavity system, $H_1^2$ is the multipolar gauge QRM, and $x$ is a Lindblad operator obtained by expressing the $\alpha$-gauge canonical momentum quadrature $\pi$ in the eigenbasis $\{\ket{i}\}$ of the QRM $H_1^2$ as
\begin{align}
&\pi = x+x^\dagger, \\
&x = \sum_{\substack{i,j \\ i<j}} \bra{i}\pi\ket{j}\ket{i}\bra{j}.
\end{align} 

Eq.~(\ref{mast}) constitutes a different physical description of cavity leakage for each different physical definition of $\pi = \sqrt{2v\over \omega}\Pi$. In the multipolar gauge itself we have $\Pi=-D_{\rm T} = i\sqrt{\omega\over 2v}(a^\dagger -a)$ where $a$ and $a^\dagger$ are the same (multipolar gauge) operators as appear in Eq.~(\ref{rabm}). Thus, for $\alpha=1$ we have $\pi=\pi_1$ where $\pi_1:=i(a^\dagger -a)$, yielding the corresponding master equation
\begin{align}\label{dg}
&{\dot \rho} = i[\rho,H_1^2]+{\cal L}_1(\rho,x),\\
&{\cal L}_1(\rho,x) = \kappa\left(x\rho x^\dagger - {1\over 2}\{x^\dagger x,\rho\}\right),\\
&x = \sum_{\substack{i,j \\ i<j}} \bra{i}\pi_1 \ket{j}\ket{i}\bra{j}.
\end{align}
Ref.~\cite{salmon_gauge-independent_2022} refers to this result as the ``dipole-gauge" (DG) master equation.

However, Ref.~\cite{salmon_gauge-independent_2022} assumes that the ``correct" description is provided when the cavity couples to external modes via the transverse electric field, $E_{\rm T}$, which equals $-\Pi$ in the Coulomb gauge. The multipolar gauge two-level truncation of the observable $\sqrt{2v\over \omega}E_{\rm T}$ is $-\pi_1 - 2\eta\sigma^x$ where $\pi_1:=i(a^\dagger -a)$. The resulting master equation is therefore
\begin{align}\label{dgf}
&{\dot \rho} = i[\rho,H_1^2]+{\cal L}_0(\rho,x),\\
&{\cal L}_0(\rho,x) = \kappa\left(x\rho x^\dagger - {1\over 2}\{x^\dagger x,\rho\}\right),\\
&x = \sum_{\substack{i,j \\ i<j}} \bra{i}(\pi_1 + 2\eta\sigma^x)\ket{j}\ket{i}\bra{j}.
\end{align}
Ref.~\cite{salmon_gauge-independent_2022} refers to this master equation as the ``dipole gauge-fixed" (DGF) master equation, which is clearly different from Eq.~(\ref{dg}). Note that this master equation is obtained by assuming a coupling between the cavity and external modes using the Coulomb gauge cavity canonical momentum, $E_{\rm T}$, but the two-level truncation of the dipole has been performed within the multipolar gauge where, unlike in the Coulomb gauge, it is expected to be generally accurate for an anharmonic dipole (see Sec.~\ref{opt}). The observable $E_{\rm T}$ has therefore been expressed in terms of the multipolar gauge operators $\sigma^x$ and $a,\,a^\dagger$.

More generally, in the gauge $\alpha$ we have $\Pi = -E_{\rm T}-\alpha P_{\rm T}$ where $P_{\rm T} = d\sigma^x/v$. The multipolar gauge two-level truncation of the observable $-\sqrt{2v\over \omega}(E_{\rm T}+\alpha P_{\rm T})$ is represented by $\pi_1+2(1-\alpha)\eta\sigma^x$, which results in the master equation
\begin{align}\label{dgalph}
&{\dot \rho} = i[\rho,H_1^2]+{\cal L}_\alpha(\rho,x),\\
&{\cal L}_\alpha(\rho,x) = \kappa\left(x\rho x^\dagger - {1\over 2}\{x^\dagger x,\rho\}\right),\\
&x = \sum_{\substack{i,j \\ i<j}} \bra{i}(\pi_1 + 2(1-\alpha)\eta\sigma^x)\ket{j}\ket{i}\bra{j}\label{xalph}.
\end{align}
Eqs.~(\ref{dg}) and (\ref{dgf}) are the particular cases given by $\alpha=1$ and $\alpha=0$ respectively. For each different $\alpha$ the general master equation (\ref{dgalph}) constitutes a different physical model of cavity leakage in which the cavity is assumed to couple to external modes linearly through its canonical momentum $\Pi$, which represents the observable $-E_{\rm T}-\alpha P_{\rm T}$. In other words, cavity leakage is described relative to a choice of gauge.

\begin{figure}[t]
\begin{minipage}{\columnwidth}
\begin{center}
\vspace*{-1cm}
\hspace*{-1mm}\includegraphics[scale=0.48]{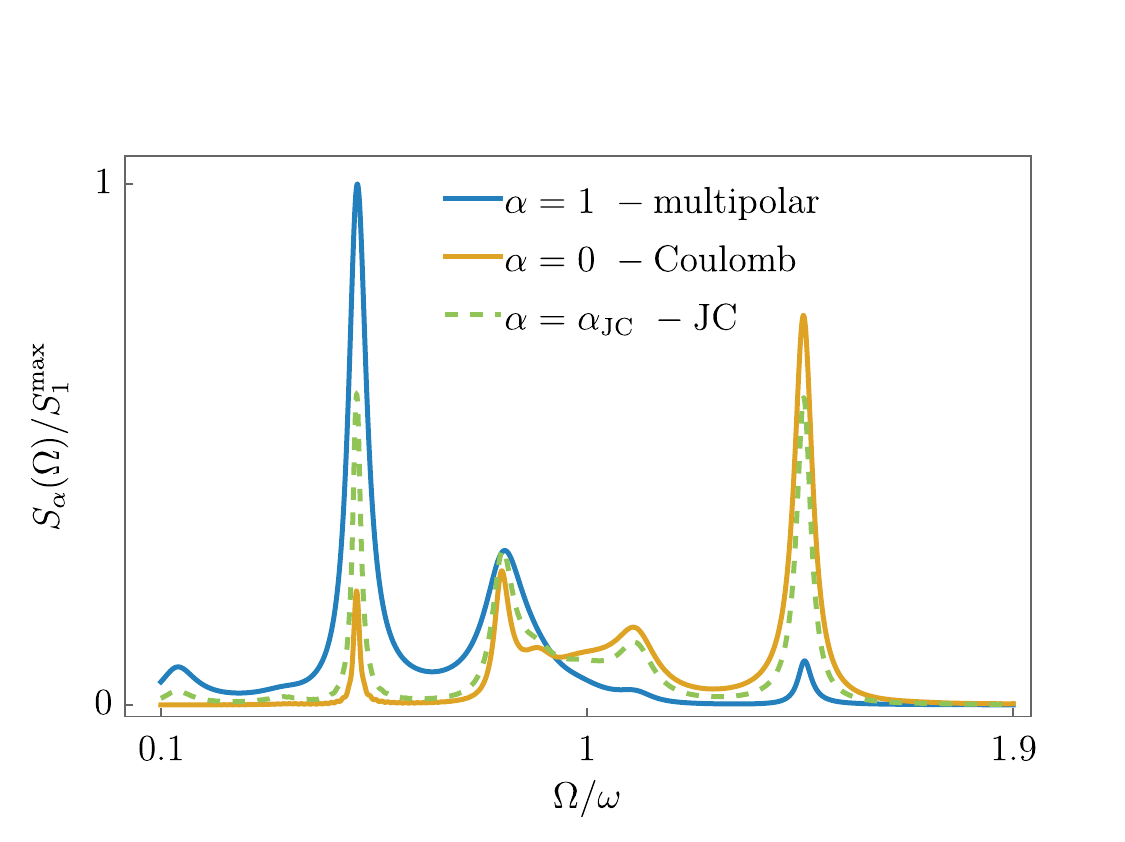}
\caption{Cavity emission spectra using the multipolar ($\alpha=1$, darker curve), Coulomb ($\alpha=0$, lighter curve) and Jaynes-Cummings gauge ($\alpha=\alpha_{\rm JC}=0.335115$, dashed curve) definitions of $x$, for ultrastrong light-matter coupling $\eta=g/\omega=0.5$ and weak incoherent pumping $P_{\rm inc}=0.01g$. The spectra are normalised to the multipolar maximum and $\alpha_{\rm JC}$ is determined for the same highly anharmonic double-well dipole as considered in Sec.~\ref{opt}. Other parameters are $\kappa=0.05g$ and $\delta=\omega/\omega_m=1$.}\label{spectraalph}
\end{center}
\end{minipage}
\end{figure}

The cavity emission spectrum is defined as the spectrum of the average external mode number operator, $\langle b_k^\dagger b_k \rangle$, and is given using $V_{\rm cav-ext}^\alpha$ by \cite{salmon_gauge-independent_2022} 
\begin{align}
S_{\alpha}(\Omega)\propto{\rm Re}\left[\int_0^{\infty}d\tau e^{i\Omega\tau}\langle x^{\dagger}(0)x(\tau)\rangle_{ss}\right],
\end{align}
where the conventional weak-coupling approximations have been applied in the dressed basis of the QRM and we consider the long-time limit. Like the master equation, the physical meaning of the spectrum is determined by the value of $\alpha$, which specifies the physical observable in terms of which $x$ is defined in Eq.~(\ref{xalph}). Ref.~\cite{salmon_gauge-independent_2022} considers the cases $\alpha=0$ and $\alpha=1$, which define $x$ in terms of $E_{\rm T}$ and $D_{\rm T}$ respectively. Incoherent excitation of the dipole and coherent excitation under semi-classical driving are both considered. For incoherent driving Ref.~\cite{salmon_gauge-independent_2022} considers a phenomenological pump term, for which the master equation~(\ref{dgalph}) becomes
\begin{align}\label{dgalphpump}
&{\dot \rho} = i[\rho,H_1^2]+{\cal L}_\alpha(\rho,x)+{\cal L}^{\rm inc}_\alpha(\rho,x),\\
&{\cal L}^{\rm inc}_\alpha(\rho,x) = P_{\rm inc}\left(x^\dagger\rho x - {1\over 2}\{x x^\dagger,\rho\}\right).
\end{align}
Example results for ultrastrong light-matter coupling and weak incoherent pumping are plotted in Fig.~\ref{spectraalph}. They are clearly markedly different for the two different gauges $\alpha=0$ and $\alpha=1$ for ultrastrong light-matter coupling, as well as for the JC gauge, which lies between the two. We note that Ref.~\cite{salmon_gauge-independent_2022} considers a larger value of the cavity leakage rate, $\kappa = g/4$, for which one again sees clear qualitative differences between the spectra corresponding to different $\alpha$, and so one obtains the same qualitative conclusions.

Ref.~\cite{salmon_gauge-independent_2022} assumes that the Coulomb gauge model in Eq.~(\ref{dgf}) and associated spectrum is ``correct". Accordingly the multipolar gauge model in Eq.~(\ref{dg}) and associated spectrum is deemed to ``fail". It is noted that the ``correct" result can be transformed using the $x_P$-phase transformation ${\cal T}_{10}$ to give an equivalent expression
\begin{align}\label{dgf2}
&{\dot \rho} = i[\rho,h_1^2(0)]+{\cal L}_0(\rho,x),\\
&{\cal L}_0(\rho,x) = \kappa\left(x\rho x^\dagger - {1\over 2}\{x^\dagger x,\rho\}\right),\\
&x = \sum_{\substack{i,j \\ i<j}} \bra{i}\pi_1 \ket{j}\ket{i}\bra{j}.
\end{align}
Here $h_1^2(0)={\cal T}_{10}H_1^2 {\cal T}_{01}$ is the two-level model encountered in Sec.~\ref{ss3}, the $\{\ket{i}\}$ denote its eigenvectors, and $\rho$ denotes the density operator in the rotated frame. More generally, {\em any} two-level model unitary operator $U^2$ can be applied to {\em any} one of the master equations (\ref{dgalph}) corresponding to a fixed value of $\alpha$, and this will of course result in an equivalent expression of the given master equation. Note that in all of these equations two-level truncation has been performed in the multipolar gauge and so provided this truncation is accurate each one of the equations is an accurate approximation of a gauge invariant equation. Importantly however, the master equations (\ref{dgalph}) corresponding to different $\alpha$ are not equivalent, because each one constitutes a different physical model of cavity leakage whereby the cavity couples to external modes via a different physical field.

The ``correct" master equation can only be determined through the provision of a physical argument to prefer one of the results over another. As already noted, Ref.~\cite{salmon_gauge-independent_2022} assumes that the $\alpha=0$ result is correct based on the assumption that the cavity should couple to external modes through the transverse electric field $E_{\rm T}$, referred to simply as the ``electric field" in Ref.~\cite{salmon_gauge-independent_2022}. The emission spectrum is then found using the same interaction Hamiltonian and is therefore given in terms of the same physical field. However, as described in Sec.~\ref{pd}, conventional photodetection theory uses the {\em total} electric field, which is equal to the field $D_{\rm T}$ at all points away from the source dipole itself. Moreover, boundary conditions defining an electromagnetic cavity are typically specified in terms of the {\em local total} electric field. A perfect conductor, for example, satisfies ${\hat {\bf n}}\times {\bf E}({\bf x})={\bf 0}$ for ${\bf x}$ on the boundary with unit normal vector ${\hat {\bf n}}$. Since ${\bf x}$ is not a point inside the source, we have ${\bf E}({\bf x})={\bf D}_{\rm T}({\bf x})$. Thus, in the above simplified toy model, leakage to external modes through the field $D_{\rm T}$ would seem to offer a more physically sensible description than leakage through $E_{\rm T}$. According to these arguments, the specification in Ref.~\cite{salmon_gauge-independent_2022}  of which result is ``correct" and of which result ``fails" should actually be reversed. 

Regardless, the results above demonstrate that the prediction of cavity leakage is strongly gauge-relative, because coupling of the cavity to external modes can only be defined relative to a choice of gauge. The relativity clearly becomes significant for sufficiently large values of the light-matter coupling strength, even though the coupling $V_{\rm cav-ext}^\alpha$ is weak. 

\section{Conclusions}\label{conc}

In this article we have focussed on the implications of gauge freedom for QED beyond conventional weak-coupling and Markovian regimes. We have shown that subsystems in QED are fundamentally gauge-relative meaning that in each different gauge they are defined in terms of {\em different} physical observables. The fundamental condition known as gauge invariance states that the predictions for any physical observable must always be the same when found in different gauges. This is guaranteed by the unitarity of gauge fixing transformations. However, if we compare predictions coming from different gauges of quantum subsystem properties such as ``photon" number or ``light"-``matter" entanglement, then we are comparing predictions for different physical observables which are, of course, different. This is not a violation of gauge invariance. It is analogous to the fact that an interval in space or time between two events possesses a different value in different inertial frames, even though the same labels ``space" and ``time" are used in every inertial frame.

Subsystem gauge relativity can be ignored within the idealised setting of scattering theory, beyond which it can only be eliminated using various weak-coupling and Markovian approximations. It is therefore an important fundamental feature whenever such approximations cannot be employed, i.e., outside of {\em gauge nonrelativistic regimes}. We have provided descriptions of a number of simple systems, showing that subsystem gauge relativity is significant in the description of so-called ``virtual" processes. It thereby affects the balance between localisation and electromagnetic dressing. This has non-trivial implications for modelling controllable interactions, for photodetection theory, and for cavity QED. In all instances, the quantum subsystems, including reservoirs and measurement devices, can only be defined relative to a choice of gauge. Beyond conventional weak-coupling and Markovian regimes the physical predictions for subsystems defined relative to different gauges can be markedly different.

\section*{SUPPLEMENTARY MATERIAL}

\setcounter{section}{0}

\section{Local $U(1)$-phase invariance}\label{u1ph}

The connection between gauge invariance and local $U(1)$-phase invariance of a material wave function was first recognised in the context of nonrelativistic wave mechanics by Fock \cite{fock_uber_1926} and was firmly established by Weyl \cite{weyl_elektron_1929}. This important connection now forms the basis for the modern development of gauge-field theories, as will be seen in Supplementary Note~\ref{symm}. Much less well-known however, is the application of the local phase invariance principle to the wave functional of the free electromagnetic field and it's connection to gauge redundancy in auxiliary material potentials.

Recently, attempts have been made to establish gauge invariant approximate models within the ultrastrong-coupling regime based on variants of the local phase invariance principle \cite{stefano_resolution_2019,taylor_resolution_2020}. This motivates our brief consideration now of the connection between local phase invariance and gauge freedom. We consider the case of matter interacting with an external electromagnetic field and then we consider an electromagnetic field interacting with external matter. Altogether, this enables us to understand the gauge freedoms in {\em both} ${\bf A}_{\rm L}$ {\em and} ${\bf P}_{\rm T}$ in terms of local $U(1)$-phases.

\subsubsection{Material wave function}

The nonrelativistic energy of charge $q$ with mass $m$ is $H_m={\bf p}^2/(2m)$ where ${\bf p}$ admits the representation $-i\nabla$ when acting on a position-space wave function $\psi(t,{\bf r})$. Predictions are invariant under a phase transformation $\psi\to e^{iq\chi}\psi$ where $\chi\in {\mathbb C}$ is arbitrary. However, the Schr\"odinger equation is not invariant under a {\em local}-phase transformation 
\begin{align}
&\psi(t,{\bf r})\to e^{iq\chi(t,{\bf r})}\psi(t,{\bf r}) \equiv R_\chi(t)\psi(t,{\bf r}),\\
&R_\chi(t)=\exp\left(i\int d^3 x \chi(t,{\bf x})\rho({\bf x})\right).\label{rchi1}
\end{align}
Physical invariance of the theory under such a transformation requires the introduction of an external potential $A$ with components $(A_\mu)=(A_0,-{\bf A})$ such that $A$ is physically equivalent to $A'$ with components $A'_\mu=A_\mu-\partial_\mu \chi$. This is precisely the property held by an electromagnetic potential. The Hamiltonian for the charge in the presence of the external electromagnetic field is
\begin{align}\label{enex}
H_m(A) = {1\over 2m}[{\bf p}-q{\bf A}(t,{\bf r})]^2+qA_0(t,{\bf r}).
\end{align}
Since we have already dealt with the coupling of matter to the quantised field in Sec.~II~A of the main text, the minimal coupling replacement ${\bf p}\to {\bf p}-q{\bf A}({\bf r})$ in Eq.~(\ref{enex}) is familiar. The additional scalar potential interaction $qA_0$ is necessary when the electromagnetic field is external. An example is the electrostatic potential energy $V=qA_0$ due to an external nucleus. It is easily verified that in the Heisenberg picture one obtains the Lorentz force law $m{\ddot {\bf r}} = q[{\bf E}(t,{\bf r})+\{{\dot {\bf r}} \times {\bf B}(t,{\bf r})-{\bf B}(t,{\bf r})\times {\dot {\bf r}}\}/2]$ where ${\bf E}=-\nabla A_0-\partial_t{\bf A}$ and ${\bf B}=\nabla\times {\bf A}$.

If we now define a local phase-transformed wave function, $\psi'(t,{\bf r})=R_\chi(t)\psi(t,{\bf r})= e^{iq\chi(t,{\bf r})}\psi(t,{\bf r})$, then we see that
\begin{align}
i{\dot \psi}(t,{\bf r}) = H_m(A)\psi(t,{\bf r}) 
\end{align}
if and only if
\begin{align}
i{\dot \psi}'(t,{\bf r}) = H_m(A')\psi'(t,{\bf r})
\end{align}
where $A'_\mu= A_\mu-\partial_\mu\chi$. Thus, requiring local phase invariance implies the existence of the electromagnetic gauge field and dictates how it couples to the charge $q$, such that the correct Heisenberg equation of motion for the quantum charge coupled to the external field is obtained. Moreover, this is ensured in any gauge because the Hamiltonians $H_m(A)$ and $H_m(A')$ are unitarily equivalent;
\begin{align}
H_m(A') = R_\chi(t) H_m(A) R_\chi(t)^\dagger + i{\dot R}_\chi(t)R_\chi(t)^\dagger.
\end{align}

\subsubsection{Electromagnetic wave functional}\label{polsup}

A gauge freedom occurs when expressing the physical material charge and current densities $\rho$ and ${\bf J}$ in terms of the auxiliary polarisation ${\bf P}$ and magnetisation ${\bf M}$. These fields are defined by the inhomogeneous Maxwell equations;
\begin{align}
\rho = -\nabla\cdot {\bf P},\qquad {\bf J}={\partial_t {\bf P}}+\nabla \times {\bf M}.
\end{align}
In the absence of any accompanying homogenous Maxwell equations the fields ${\bf P}$ and ${\bf M}$ are not unique. Specifically, the physical charge and current densities are invariant under a transformation by pseudo-magnetic and pseudo-electric fields as
\begin{align}\label{pmt}
{\bf P}\to {\bf P}+\nabla \times {\bf U},~~{\bf M} \to {\bf M}-\nabla U_0 -{\partial_t {\bf U}}
\end{align}
where $U$ is an arbitrary pseudo-four-potential. The polarisation and magnetisation fields are in turn invariant under a gauge transformation $U_\mu\to U_\mu -\partial_\mu \chi$ where $\chi$ is arbitrary. Since ${\bf M}_{\rm L}$ does not contribute to either $\rho$ or ${\bf J}$, only the transverse freedom in ${\bf P}$ and ${\bf M}$ is non-trivial. Defining ${\bf X}_{\rm T}=\nabla \times {\bf M}$ we see that $\rho$ and ${\bf J}$ are invariant under the transformations ${\bf P}_{\rm T}\to {\bf P}_{\rm T} +{\bf U}_{\rm T}$ and ${\bf X}_{\rm T}\to {\bf X}_{\rm T}-{\partial_t {\bf U}}_{\rm T}$ where ${\bf U}_{\rm T}$ is arbitrary.

In the same way that the gauge freedom in an external potential $A$ can be related to the local phase of the material wave function, it is possible to relate the freedom in external material potentials ${\bf P}_{\rm T}$ and ${\bf X}_{\rm T}$ to the local phase of the electromagnetic wave functional. The functional Schr\"odinger picture of quantum field theory \cite{jackiw_diverse_1994} is much less often employed than the Heisenberg picture, but it has the advantage of revealing useful structural analogies with wave-mechanics, as will be seen in the following.

Consider the free electromagnetic field. The electric and magnetic fields are transverse, and are fully specified in terms of the gauge invariant transverse potential ${\bf A}_{\rm T}$ by ${\bf E}={\bf E}_{\rm T} = -{\partial_t {\bf A}}_{\rm T}$ and ${\bf B}=\nabla \times {\bf A}_{\rm T}$. The energy of the field is
\begin{align}
H_{\rm ph} = {1\over 2}\int d^3 x \left({\bf \Pi}^2+[\nabla\times {\bf A}_{\rm T}]^2\right)
\end{align}
where the momentum ${\bf \Pi}=-{\bf E}_{\rm T}$ can be taken to admit the representation ${\bf \Pi}=-i\delta/\delta{\bf A}_{\rm T}$ when acting on configuration-space wave functionals $\psi[t,{\bf A}_{\rm T}]$. Predictions are invariant under a phase transformation $\psi\to e^{iq\chi}\psi$ where $\chi\in {\mathbb C}$ is arbitrary. However, the Schr\"odinger equation is not invariant under a {\em local}-phase transformation 
\begin{align}\label{rchi}
\psi[t,{\bf A}_{\rm T}]\to &\exp\left(i\int d^3 x\, \chi[t,{\bf x},{\bf A}_{\rm T}]\rho(t,{\bf x})\right)\psi[t,{\bf A}_{\rm T}] \nonumber \\ &\equiv R_\chi(t)\psi[t,{\bf A}_{\rm T}]
\end{align}
where $\rho(t,{\bf x})$ is an external charge density and $\chi$ is an arbitrary functional. Physical invariance of the theory under such a transformation when $\chi$ is linear  in ${\bf A}_{\rm T}$, can be ensured via the introduction of transverse potentials $({\bf P}_{\rm T},{\bf X}_{\rm T})$ such that ${\bf J}_{\rm T}(t,{\bf x})={\partial_t {\bf P}}_{\rm T}(t,{\bf x})+{\bf X}_{\rm T}(t,{\bf x})$. These potentials are physically equivalent to $({\bf P}'_{\rm T},{\bf X}'_{\rm T})$ where
\begin{align}
{\bf P}'_{\rm T}(t,{\bf x}) &= {\bf P}_{\rm T}(t,{\bf x}) +{\bf U}_{\rm T}(t,{\bf x}) ,\label{ptgt}\\
{\bf X}'_{\rm T}(t,{\bf x}) &= {\bf X}_{\rm T}(t,{\bf x}) -{\partial_t {\bf U}}_{\rm T}(t,{\bf x}) \label{xtgt},
\end{align}
in which
\begin{align}
{\bf U}_{\rm T}(t,{\bf x}):=-\int d^3 x' {\delta \chi[t,{\bf x}',{\bf A}_{\rm T}] \over \delta{\bf A}_{\rm T}({\bf x})}\rho(t,{\bf x}')
\end{align}
is determined by the functional $\chi$. Note that the transformation of ${\bf P}_{\rm T}$ in Eq.~(\ref{ptgt}) can be written
\begin{align}\label{polana}
{\bf P}'_{\rm T}(t,{\bf x}) = {\bf P}_{\rm T}(t,{\bf x}) +i{\delta F_\chi(t) \over \delta {\bf A}_{\rm T}({\bf x})}
\end{align}
where $F_\chi(t)$ is defined by $R_\chi(t)=e^{F_\chi(t)}$. This is analogous to the gauge transformation
\begin{align}\label{Aana}
-q{\bf A}'_{\rm L}(t,{\bf r}) = -q{\bf A}_{\rm L}(t,{\bf r}) +i{\partial G_\chi(t) \over \partial {\bf r}}
\end{align}
where $G_\chi(t)$ is defined by $R_\chi(t)=e^{G_\chi(t)}$ with $R_\chi(t)$ given in Eq.~(\ref{rchi1}). The freely choosable polarisation ${\bf P}_{\rm T}$ and coordinate ${\bf A}_{\rm T}$ in Eq.~(\ref{polana}) are respectively analogous to the freely choosable potential $-q{\bf A}_{\rm L}$ and coordinate ${\bf r}$ in Eq.~(\ref{Aana}). We note however, that in order for ${\bf J}_{\rm T}(t,{\bf x})$ to be an external current, i.e., to not explicitly depend on ${\bf A}_{\rm T}$, it must be the case that the potentials do not depend on ${\bf A}_{\rm T}$. Therefore $\chi$ must be a {\em linear} functional of ${\bf A}_{\rm T}$;
\begin{align}
 \chi[t,{\bf x}',{\bf A}_{\rm T}] = \int d^3 x \, {\bm \chi}(t,  {\bf x}',{\bf x})\cdot {\bf A}_{\rm T}({\bf x})
\end{align}
where ${\bm \chi}(t,{\bf x}',{\bf x}) = {\delta \chi[t,{\bf x}',{\bf A}_{\rm T}] /\delta {\bf A}_{\rm T}({\bf x})}$
is independent of ${\bf A}_{\rm T}$ but otherwise arbitrary.

If the Hamiltonian for the electromagnetic field in the presence of external matter is defined as
\begin{align}\label{enphex}
H_{\rm ph}({\bf P}_{\rm T},{\bf X}_{\rm T}) =  &{1\over 2}\int d^3 x \left([{\bf \Pi}+{\bf P}_{\rm T}(t)]^2+[\nabla\times {\bf A}_{\rm T}]^2\right)\nonumber \\ & - \int d^3 x\, {\bf X}_{\rm T}(t)\cdot {\bf A}_{\rm T}
\end{align}
then it is easily verified that in the Heisenberg-picture we obtain the expected transverse component of the Maxwell-Ampere Law; ${\partial_t {\bf E}}_{\rm T} = \nabla\times {\bf B}-{\bf J}_{\rm T}$. Since we have already dealt with the interaction of the electromagnetic field with quantised matter in Sec.~II~A of the main text, the coupling via the replacement ${\bf \Pi}\to {\bf \Pi}+{\bf P}_{\rm T}$ in Eq.~(\ref{enphex}) is familiar. The additional magnetic interaction $- \int d^3 x\, {\bf X}_{\rm T}\cdot {\bf A}_{\rm T}=-\int d^3x \, {\bf M}\cdot {\bf B}$ is necessary when the material field is external.

If we now define a local phase-transformed wave functional, $\psi'[t,{\bf A}_{\rm T}]=R_\chi(t)\psi[t,{\bf A}_{\rm T}]$ where $R_\chi$ is defined in Eq.~(\ref{rchi}) and $\chi$ is a linear functional of ${\bf A}_{\rm T}$, then we see that in the Schr\"odinger picture 
\begin{align}
i{\dot \psi}[t,{\bf A}_{\rm T}] = H_{\rm ph}({\bf P}_{\rm T},{\bf X}_{\rm T})\psi[t,{\bf A}_{\rm T}]
\end{align}
if and only if
\begin{align}
i{\dot \psi}'[t,{\bf A}_{\rm T}]  = H_{\rm ph}({\bf P}'_{\rm T},{\bf X}'_{\rm T})\psi'[t,{\bf A}_{\rm T}] 
\end{align}
where $({\bf P}_{\rm T},{\bf X}_{\rm T})$ and $({\bf P}'_{\rm T},{\bf X}'_{\rm T})$ are related as in Eqs.~(\ref{ptgt}) and (\ref{xtgt}). Thus, requiring local (linear) phase invariance implies the existence of the external material gauge field and dictates how it couples to the electromagnetic field, such that the correct Heisenberg equation of motion for the quantum field coupled to the external matter is obtained. Moreover, this is ensured in any gauge because the Hamiltonians $H_{\rm ph}({\bf P}_{\rm T},{\bf X}_{\rm T})$ and $H_{\rm ph}({\bf P}'_{\rm T},{\bf X}'_{\rm T})$ are unitarily equivalent;
\begin{align}
&H_{\rm ph}({\bf P}'_{\rm T},{\bf X}'_{\rm T}) \nonumber \\ &= R_\chi(t) H_{\rm ph}({\bf P}_{\rm T},{\bf X}_{\rm T}) R_\chi(t)^\dagger + i{\dot R}_\chi(t)R_\chi(t)^\dagger.
\end{align}
We have therefore shown that the gauge freedom in the material potentials can be related to $U(1)$-phase invariance in a way that closely resembles the relation between gauge freedom in the electromagnetic potentials and $U(1)$-phase invariance.

\section{The gauge principle and gauge freedom}\label{symm}

We now provide a rigorous derivation of arbitrary gauge nonrelativistic QED using the principles of modern gauge-field theory. Our purpose is to show that the implications of gauge freedom discussed in the main text are a fundamental feature of QED, and not in any way an artefact of approximations or simplifications. This derivation also shows that gauge freedom is much more general than the one-parameter freedom introduced in Sec.~II A of the main text.

We derive the theory of an atom within the quantised electromagnetic field. The Lagrangian is defined over Minkowski spacetime $E^{1,3}$. With respect to the atomic rest frame a vector $v\in E^{1,3}$ has components $v^\mu=(v_0,{\bf v})$. We assume a nonrelativistic (Schr\"odinger) matter-field $\psi$ with charge $q$ and without spin. The formalism is easily extended to include spin and is equally applicable to the relativistic Dirac-field \cite{stokes_noncovariant_2012}. The four-current density $j$ has components $j^\mu=(\rho,{\bf J})$ where $\rho = q\psi^\dagger \psi$ and for a free material field ${\bf J}=-iq(\psi^\dagger \nabla \psi - (\nabla\psi^\dagger) \psi)/2m$ with $m$ the electronic mass.

Let ${\cal G}$ be a (Lie) group called the gauge group and let $g:E^{1,3}\to {\cal G}$. The {\em gauge principle} asserts that:
\begin{itemize}
\item{The form of electromagnetic and other interactions should be invariant under the local action of ${\cal G}$ on the matter field $\psi$, written $\psi'(x)=g(x)\cdot \psi(x)$. In QED ${\cal G}=U(1)$ and $\psi'(x) = e^{iq\chi(x)}\psi(x)$ where $\chi$ is arbitrary.}
\end{itemize}
The definition of group action is textbook group theory \cite{hassani_mathematical_2013}. In electrodynamics the above requirement is fulfilled if the matter-field interacts with a gauge field via the replacement $-i\partial_\mu \to -i\partial_\mu +qA_\mu$ made for each $\mu$ within the material Lagrangian. Here $A_\mu = (A_0,-{\bf A})$ are the components of the gauge-potential $A$ and any two potentials $A$ and $A'$ such that
\begin{align}
iqA'(x)&=g(x)iqA(x)g(x)^{-1}+g(x)dg(x)^{-1}\nonumber  \\ &= iq[A(x) - d\chi(x)]
\end{align}
are physically equivalent, where $d$ denotes the exterior derivative \cite{fecko_differential_2006,frankel_geometry_2004}.

Mathematically, the classical field $\psi:E^{1,3}\to {\mathbb C}$ is a section of the trivial bundle $E=E^{1,3}\times {\mathbb C}$ and the phases $e^{iq\chi(x)}$ are identifiable as transition maps on the intersections of open regions in $E^{1,3}$ meaning that the gauge group ${\cal G}$ is the structure group of the bundle \cite{fecko_differential_2006,frankel_geometry_2004}. The gauge-potential $A$ is an $E^{1,3}$-valued connection one-form mapping from the tangent bundle $TE^{1,3}$ \cite{frankel_geometry_2004}. The potentials $A$ and $A'$ are said to be related by a gauge symmetry transformation. The spatial replacement $-i\nabla\to -i\nabla -q{\bf A}$ is called the {\em minimal coupling replacement} and the gauge principle asserts that within this replacement the longitudinal potential ${\bf A}_{\rm L}$ can be {\em freely} chosen. In this sense ${\bf A}_{\rm L}$ is superfluous, i.e., {\em redundant}. A choice of ${\bf A}_{\rm L}$ fixes the gauge.

A suitable Lagrangian-density is \cite{cohen-tannoudji_photons_1989}
\begin{align}
{\mathscr L}=&{i\over 2}\left(\psi^\dagger{\dot \psi}-{\dot \psi}^\dagger \psi\right)- (U+qA_0)\psi^\dagger \psi +{1\over 2}({\bf E}^2-{\bf B}^2)\nonumber \\&-{1\over 2m}\left[(-i\nabla-q{\bf A})\psi^\dagger\right] \cdot \left[(-i\nabla-q{\bf A})\psi \right]
\end{align}
where ${\bf E}$ and ${\bf B}$ are the electric and magnetic fields and where $U$ is an external potential due for example to nuclei. We note that the results in this section continue to hold if an external charge density $\rho_{\rm ext}({\bf x})$ is included, as the source of the potential $U$, within the definition of $\rho({\bf x})$. In the case of a positive nucleus at ${\bf 0}$ with charge $-qZ$ the external charge density is $\rho_{\rm ext}({\bf x}) = -qZ\delta({\bf x})$.

The Lagrangian is degenerate due to gauge-redundancy, which is implied by the occurrence of non-dynamical constraints $\{C\}$ \cite{dirac_lectures_2003,muller-kirsten_introduction_2006}. The Hamiltonian description must therefore be obtained by Dirac's method \cite{dirac_lectures_2003}. The naive Hamiltonian acts within a space ${\cal H}$ containing the physical state space $\mathcal{H}_p$ as a proper subspace comprised of vectors $\ket{\psi}$ such that $C\ket{\psi}=0$. The momentum $\Pi_0$ conjugate to $A_0$ vanishes identically, while the momentum conjugate to ${\bf A}$ is ${\bf \Pi} =-{\bf E}$, where ${\bf E}={\bf E}_{\rm T}+{\bf E}_{\rm L}$ is the {\em total} electric field.

Altogether there are three constraints, which are $\{\Pi_0$,\, $\rho +\nabla\cdot {\bf \Pi},\,{\cal F}(A)\}$ where ${\cal F}(A)$ is a gauge fixing constraint. The naive equal-time canonical brackets are \cite{woolley_r._g._charged_1999,muller-kirsten_introduction_2006,dirac_lectures_2003,chernyak_gauge_1995}
\begin{align}
&\{\psi({\bf x}),\psi^\dagger({\bf x}')\}=\delta({\bf x}-{\bf x}'),\label{comnapsi}\\
&[A_\mu({\bf x}),\Pi_\nu({\bf x}')]=i\delta_{\mu\nu}\delta({\bf x}-{\bf x}')\label{comna}.
\end{align}
The constraints $\Pi_0$ and $G=\nabla\cdot {\bf \Pi}+\rho$ generate transformations between the redundant degrees of freedom. More specifically, the infinitesimal generator ${\mathfrak G}[\chi]$ of a $U(1)$ gauge symmetry transformation $S_\chi = e^{-i{\mathfrak G}[\chi]}$ is given by
\begin{align}\label{gsymm}
{\mathfrak G}[\chi] = \int d^3 x \, \left[\Pi_0 {\dot \chi} + (\nabla\cdot {\bf \Pi}+\rho)\chi\right].
\end{align}
As is easily verified using Eqs.~(\ref{comnapsi}) and (\ref{comna}), $S_\chi$ transforms the matter-field as
\begin{align}
S_\chi\psi(x)S_\chi^\dagger = e^{iq\chi(x)}\psi(x)
\end{align}
and the gauge-potential as
\begin{align}
S_\chi A(x)S_\chi^\dagger = A(x)-d\chi(x).
\end{align}

The naive Hamiltonian defined using ${\mathscr L}$ is found to be $H =\int d^3 x \left[{\mathscr H} + A_0G\right]$ where ${\mathscr H}$ is defined below in Eq.~(\ref{hamna}). The (gauge) term $\int d^3 x A_0G$ shows that the scalar potential acts as a Lagrange multiplier for the constraint $G$ (Gauss' law) \cite{muller-kirsten_introduction_2006}. The time evolution of $A_0$ is completely arbitrary and is restricted to the non-physical subspace, so $A_0$ can be removed immediately. The constrained degrees of freedom ${\bf A}_{\rm L}$ and $A_0$ will later be seen to emerge in terms of gauge invariant quantities within the final unconstrained theory. The Hamiltonian-density is therefore
\begin{align}\label{hamna}
{\mathscr H}=&{1\over 2m}\left[(i\nabla-q{\bf A})\psi^\dagger\right] \cdot \left[(-i\nabla-q{\bf A})\psi \right] +\psi^\dagger U\psi \nonumber \\&+{1\over 2}:\left[{\bf \Pi}^2+(\nabla \times {\bf A})^2\right]:
\end{align}
where colons denote the normal-ordering required to eliminate divergent vacuum terms, and gauge symmetry transformations are given by \cite{lenz_quantum_1994}
\begin{align}\label{gs}
S_\chi = \exp\left[-i\int d^3 x\, G\chi\right].
\end{align}
Since $G$ commutes with the Hamiltonian the subspace defined by $G$ is dynamically invariant. %The naive theory does not yield the correct Maxwell-Schr\"odinger system of equations, because the remaining constraints have not yet been imposed.
The procedure for obtaining the unconstrained theory is now given.

A realization of the algebra of the canonical Maxwell operators ${\bf A}$ and ${\bm \Pi}$ is given on ${\mathcal H}$ using the representations
\begin{align}\label{oprep}
({\hat{\bf A}}\varphi)[{\bf A}] = {\bf A}\varphi[{\bf A}],~~~~~~({\hat {\bm \Pi}}\varphi)[{\bf A}] = -i{\delta \varphi[{\bf A}] \over \delta {\bf A}} 
\end{align}
where $\varphi$ is a wave functional of ${\bf A}$ and where we have introduced hats to distinguish between operators and classical vector fields. Letting $ {\bf A}_{\rm L}=\nabla \chi$, we can vary the wave functional $\varphi$ with respect to $\chi$ and make use of Eq.~(\ref{oprep}) to obtain
\begin{align}\label{st}
i{\delta \varphi \over \delta \chi} = -i\nabla \cdot {\delta \varphi \over \delta \nabla\chi} =-i\nabla \cdot {\delta \varphi \over \delta {\bf A}_{\rm L}} = \nabla \cdot {\hat {\bm \Pi}_{\rm L}}\varphi =\nabla \cdot {\hat {\bm \Pi}}\varphi.
\end{align}
Using the constraint $G$ a physical state $\varphi_p$ is therefore seen to be such that
\begin{align}\label{physst}
i{\delta \varphi_p \over \delta \chi} = -\rho\varphi_p.
\end{align}
Solving this equation gives the general form of a physical state $\varphi_p$ \cite{lenz_quantum_1994,chernyak_gauge_1995,stokes_noncovariant_2012};
\begin{align}\label{physst2}
\varphi_p[{\bf A}]= \exp\left(i\int d^3x \, \chi({\bf x})\rho({\bf x}) \right)\varphi_p[{\bf A}_{\rm T}].
\end{align}
We note that in a similar fashion, had we initially employed the representation $\Pi_0=-i\delta/\delta A_0$ we would have immediately found that $\varphi_p$ does not depend on $A_0$ by solving the equation $\Pi_0\varphi_p=0$ \cite{chernyak_gauge_1995}.

In Ref.~\cite{lenz_quantum_1994} a unitary gauge fixing transformation yielding the Coulomb gauge theory is given as
\begin{align}\label{ucoul}
U \equiv \exp\left(-i\int d^3x \, {\hat \chi}({\bf x})\rho({\bf x}) \right)
\end{align}
where $({\hat \chi}\varphi)[{\bf A}] = \chi\varphi[{\bf A}]$ for all $\varphi[{\bf A}]$. In the present context we see clearly that $U$ eliminates the dependence of the physical state on ${\bf A}_{\rm L}$;
\begin{align}\label{ucoul2}
(U\varphi_p)[{\bf A}] = \varphi_p[{\bf A}_{\rm T}].
\end{align}
This corresponds to choosing the constraint ${\cal F}({\bf A}_{\rm L})={\bf A}_{\rm L}$ for which the physical subspace is such that ${\bf A}_{\rm L}\ket{\psi}={\bf 0}$. More generally, we can use the transverse vector potential to specify any other vector potential \cite{chernyak_gauge_1995,stokes_noncovariant_2012}. This results from employing the gauge fixing constraint ${\cal F}({\bf A}_{\rm L})={\bf A}_{\rm L}({\bf x})-\nabla \chi_g({\bf x},{\bf A}_{\rm T})$ such that on the physical subspace;
\begin{align}\label{a3}
{\bf A}_{\rm L}= \nabla \chi_g({\bf x},{\bf A}_{\rm T})
\end{align}
where we could, for example, follow Ref.~\cite{woolley_r._g._charged_1999} by setting
\begin{align}\label{chi}
\chi_g({\bf x},[{\bf A}_{\rm T}]) = \int d^3x' \,{\bf g}({\bf x}',{\bf x})\cdot{\bf A}_{\rm T}({\bf x}')
\end{align}
in which ${\bf g}$ is the Green's function for the divergence operator; $\nabla\cdot {\bf g}({\bf x},{\bf x}')=\delta({\bf x}-{\bf x}')$. The gauge is now set by a choice of transverse Green's function, which beyond a requirement of sensible mathematical behaviour, is completely arbitrary, but is also non-dynamical and classical. We refer to a specific choice of ${\bf g}_{\rm T}$ as selecting the gauge $g$. The above form of gauge function $\chi_g$ is sufficiently general to yield the standard Coulomb and multipolar gauge descriptions of nonrelativistic QED as special cases \cite{woolley_r._g._charged_1999}.

A general unitary gauge fixing transformation $U_g$ is defined by \cite{lenz_quantum_1994,stokes_noncovariant_2012}
\begin{align}\label{ug}
U_g :=\exp\bigg(-i\int d^3x \, \big[{\hat \chi}({\bf x})-\chi_g({\bf x},{\hat {\bf A}}_{\rm T})\big]\rho({\bf x}) \bigg).
\end{align}
The physical subspace can be realised as any of the isomorphic spaces ${\cal H}_g=\{\ket{\psi} \in U_g{\cal H}: U_gGU_g^\dagger \ket{\psi}=0\}$ labelled by the gauge $g$. Evidently ${\cal H}_g$ is dynamically invariant. A generic element of ${\cal H}_g$ is
\begin{align}\label{ug2}
(U_g\varphi_p)[{\bf A}] &= \exp\bigg(i\int d^3x\,\chi_g({\bf x},{\bf A}_{\rm T})\rho({\bf x})\bigg)\varphi_p[{\bf A}_{\rm T}] \nonumber \\ &= \varphi_p[{\bf A}_{\rm T}+\nabla\chi_g] =:\varphi_g[{\bf A}_{\rm T}] \in {\mathcal H}_g.
\end{align}
The vector potential operator in the gauge $g$ is defined by ${\hat {\bf A}}_g({\bf x}) := {\hat {\bf A}}_{\rm T}({\bf x}) + \nabla\chi_g({\bf x},{\hat {\bf A}}_{\rm T})$ such that
\begin{align}\label{Ag}
({\hat {\bf A}}_g\varphi_g)[{\bf A}_{\rm T}] = ({\bf A}_{\rm T}+\nabla\chi_g)\varphi_g[{\bf A}_{\rm T}].
\end{align}
The unitary transformation from the fixed gauge $g$ to the fixed gauge $g'$ is \cite{chernyak_gauge_1995,stokes_noncovariant_2012}
\begin{align}\label{ugg'}
U_{gg'}&:= \exp\left(-i\int d^3x \, \big[\chi_g({\bf x},{\hat {\bf A}}_{\rm T})-\chi_{g'}({\bf x},{\hat {\bf A}}_{\rm T})\big]\rho({\bf x}) \right),
\nonumber \\
&=\exp\left(i\int d^3x \, \big[{\bf P}_g({\bf x})-{\bf P}_{g'}({\bf x})\big]\cdot {\bf A}_{\rm T}({\bf x})\right)
\end{align}
an example of which is the well known Power-Zienau-Woolley transformation. These transformations are clearly distinct from the gauge symmetry transformations $S_\chi$ of the original (constrained) theory in that they do not directly transform ${\bf A}_g$, with which they commute. It is therefore evident that within Hamiltonian QED the single label ``gauge transformation" is semantically inadequate, because transforming to a new gauge requires us to use different mathematical generators depending on the stage of development of the theory. Before any constraints are imposed a gauge symmetry transformation $S_\chi$ is required whereas in the final unconstrained theory a gauge fixing transformation $U_{gg'}$ is required. The significance of this distinction is discussed in further detail in Sec.~II~F of the main text.

\section{Hamiltonian in gauge $g$}\label{haming}

To obtain the Hamiltonian in the gauge $g$ we simply need to determine the effect of $U_g$ on the remaining operators of the theory, namely $\psi,\, \psi^\dagger$ and ${\bm \Pi}$. In  so doing we will resume denoting operators without hats. To find the transformation of ${\bf \Pi}$ it is convenient to define the polarisation field ${\bf P}_g$ such that $ -\nabla \cdot {\bf P}_g=\rho$. As was seen in Sec.~\ref{polsup}, the longitudinal part of ${\bf P}_g$ is unique being given by ${\bf P}_{\rm L} =\nabla\phi$ where $\phi$ is the Coulomb potential defined in Eq.~(8) of the main text, whereas the transverse part ${\bf P}_{{\rm T}g}$ is completely arbitrary and is defined by
\begin{align}\label{PTgen}
{\bf P}_{{\rm T}g}({\bf x}) = -\int d^3 x' {\delta \chi_g({\bf x'},[{\bf A}_{\rm T}]) \over \delta {\bf A}_{\rm T}({\bf x})}\rho({\bf x}').
\end{align}
Expressing $\chi_g$ as in Eq.~(\ref{chi}) gives
\begin{align}\label{pg}
{\bf P}_g({\bf x})=-\int d^3x'\, {\bf g}({\bf x},{\bf x}')\rho({\bf x}')
\end{align}
where ${\bf g}={\bf g}_{\rm L}+{\bf g}_{\rm T}$ with ${\bf g}_{\rm L}({\bf x},{\bf x}')=-\nabla(1/4\pi|{\bf x}-{\bf x}'|)$ and ${\bf g}_{\rm T}({\bf x},{\bf x}')$ arbitrary. Using Eq.~(\ref{chi}), Eq.~(\ref{pg}), and ${\bf A}_{\rm L}=\nabla\chi$ within Eq.~(\ref{ug}), we obtain using integration by parts
\begin{align}
U_g= \exp \left(-i\int d^3x \,{\bf P}_g({\bf x})\cdot {\bf A}({\bf x})\right)
\end{align}
where ${\bf A}={\bf A}_{\rm T}+\nabla\chi$. We therefore obtain
\begin{align}\label{t2}
U_g{\bm \Pi} U_g^\dagger = {\bm \Pi} + {\bf P}_g.
\end{align}
The constraint $G$ and the residual gauge transformation $S_\chi$ therefore transform as;
\begin{align}
U_gGU_g^\dagger &= \nabla\cdot{\bm \Pi},\\
U_g S_\chi U_g^\dagger &= \exp\left(i\int d^3x\, {\bf \Pi}\cdot\nabla \chi \right),
\end{align}
which are both independent of $g$. The constraint $U_gGU_g^\dagger\ket{\psi}=0$ implies that the longitudinal canonical momentum ${\bf \Pi}_{\rm L}$ vanishes on $\mathcal{H}_g$, i.e., that ${\bf \Pi}={\bf \Pi}_{\rm T}$ such that ${\bf \Pi}$ admits the representation ${\bf \Pi}=-i\delta/\delta{\bf A}_{\rm T}$. It also follows that $S_\chi$ is the identity on $\mathcal{H}_g$. Thus, all gauge-redundancy within the state space has been eliminated.

Before transformation by $U_g$ the operator ${\bf \Pi}$ represented the field $-{\bf E}$, which implies that in the gauge $g$ the operator $U_g{\bf \Pi} U_g^\dagger = {\bf \Pi}+{\bf P}_g$ represents $-{\bf E}$. Since on ${\cal H}_g$ we have ${\bf \Pi}={\bf \Pi}_{\rm T}$, it follows that 
\begin{itemize}
\item{In the gauge $g$ the operator ${\bm \Pi}$ represents the field $-{\bf E}-{\bf P}_g=-{\bf E}_{\rm T}-{\bf P}_{{\rm T}g}$.}
\end{itemize}
In applications this is an especially important feature of the theory. Hereafter we use subscripts to denote contravariant indices. The commutator $[A_{{\rm T},i}({\bf x}),\Pi_j({\bf x}')]$ follows from the naive commutator in Eq.~(\ref{comna});
\begin{align}\label{com3}
&[{\rm A}_{{\rm T},i}({\bf x}),{\rm \Pi}_j({\bf x}')]\nonumber \\ 
&= \int d^3 y \, \delta_{ik}^{\rm T}({\bf x}-{\bf y}) \left[{\rm A}_k({\bf y}),\Pi_j({\bf x}')\right] = i\delta_{ij}^{\rm T}({\bf x}-{\bf x}').
\end{align}
Finally, the transformation of $\psi$ by $U_g$ is easily found to be
\begin{align}\label{t1}
U_g\psi U_g^\dagger = e^{iq(\chi - \chi_g)}\psi.
\end{align}
Like ${\bf \Pi}$ the fermionic operator $\psi$ is implicitly different in each gauge $g$.

Having determined all operators in the gauge $g$ we can now write the Hamiltonian density ${\mathscr H}$ in the gauge $g$ as
\begin{align}\label{h2b}
\mathscr{H}_g=&{1\over 2m}\left[(i\nabla-q{\bf A}_g)\psi^\dagger\right] \cdot \left[(-i\nabla-q{\bf A}_g)\psi \right] +\psi^\dagger U\psi\nonumber \\&+{1\over 2}:\left[({\bm \Pi} + {\bf P}_{g})^2+(\nabla \times {\bf A}_g)^2\right]: ~= U_g{\mathscr H}U_g^\dagger
\end{align}
where it is understood that ${\mathscr H}_g$ is defined over ${\cal H}_g$. Colons again indicate normal-ordering, which includes that of the material operators $\psi$ within the quadratic ${\bf P}_g^2$-term. The ordering can be implemented using the anti-commutation relations for $\psi$ and is seen to eliminate an infinite self-term. It should be borne in mind however, that this term is ${\bf g}_{\rm T}$-dependent, such that manipulations of it may generally need to be tracked when verifying gauge invariance. Similarly, once photonic operators are defined in terms of ${\bf A}_{\rm T}$ and ${\bf \Pi}$, their normal-order within the free photonic Hamiltonian is implemented using their commutation relations and is seen to eliminate the infinite and ${\bf g}_{\rm T}$-independent vacuum energy. We note that if $\rho$ includes an external component $\rho_{\rm ext}$ as the source of $U$, then the $U$-dependent term in Eq.~(\ref{h2b}) is included in the ${\bf P}_g^2$-term.

Both the longitudinal part of ${\bf A}_g$ and the transverse part of ${\bf P}_g$ are arbitrary. Within the Hamiltonian a gauge transformation of either one of these quantities using $U_{gg'}$ necessarily incurs an accompanying gauge transformation of the other. The transformations are implemented via the canonical momenta $\psi^\dagger(-i\nabla)\psi$ and ${\bf \Pi}$ as
\begin{align}
&U_{gg'}\psi^\dagger(-i\nabla-q{\bf A}_g)\psi U_{gg'}^\dagger = \psi^\dagger(-i\nabla-q{\bf A}_{g'})\psi,\label{mingena}\\
&U_{gg'}({\bf \Pi}+{\bf P}_{{\rm T}g}) U_{gg'}^\dagger = {\bf \Pi}+{\bf P}_{{\rm T}g'}\label{mingenb},
\end{align}
which generalise Eqs. (27) and (28) of the main text respectively. The Hamiltonians of different gauges are related by
\begin{align}\label{hgg}
H_{g'}=U_{gg'}H_g U_{gg'}^\dagger.
\end{align}

The Hamiltonian $H_g$ can be partitioned in a number of illuminating ways. Noting that $m{\dot {\bf x}}:=-i\nabla-q{\bf A}_g$ is the single-particle mechanical momentum operator (gauge covariant derivative), the first term on the top line of Eq.~(\ref{h2b}) is the material kinetic energy density ${\mathscr E}_{\rm KE}$, while the second term ${\mathscr E}_U:=\psi^\dagger U\psi$ is the potential energy density due to the external potential $U$. Since on ${\cal H}_g$ we have ${\bf E}=-{\bf \Pi}-{\bf P}_g$, the term on the second line in Eq.~(\ref{h2b}) is the electromagnetic energy density ${\mathscr E}_{\rm EM}:=\,:({\bf E}^2+{\bf B}^2):/2$. The Hamiltonian therefore represents the total energy in {\em any} gauge \cite{stokes_noncovariant_2012};
\begin{align}\label{EN}
H_g =E = E_{\rm KE}+ E_U+E_{\rm EM}.
\end{align}
Furthermore, on the space ${\cal H}_g$ we have ${\bf \Pi} =-{\bf E}-{\bf P}_g$ and ${\bf \Pi}_{\rm L}={\bf 0}$, so the longitudinal field ${\bf E}_{\rm L}=-{\bf P}_{\rm L}$ is uniquely specified as a function of $\rho$. Thus, the electromagnetic energy $E_{\rm EM}$ can be  partitioned into transverse and Coulomb components as
\begin{align}\label{ENEM}
E_{\rm EM} = V_{\rm Coul} + E_{\rm TEM}
\end{align}
where $V_{\rm Coul} =\int d^3 x :{\bf E}_{\rm L}^2:/2$ is the Coulomb energy density and where ${\mathscr E}_{\rm TEM}:=\, :({\bf E}_{\rm T}^2+{\bf B}^2):/2$. 

\section{Re-emergence of the scalar potential}\label{reem}

Since we have now fixed ${\bf A}_{\rm L}$ as ${\bf A}_{\rm L} =\nabla\chi_g$ and we have also identified that the electric field is ${\bf E}=-{\bf \Pi}-{\bf P}_g$ we can identify, up to a constant, the scalar potential $\phi_g$ within the gauge $g$ from its fundamental definition $\nabla\phi_g=-{\bf E}-{\partial_t {\bf A}}_g$. We use this equality and the definition of ${\bf A}_g$ together with ${\bf E}= -{\bf \Pi}-{\bf P}_g$ and ${\bf P}_{\rm L}=-{\bf E}_{\rm L}=\nabla \phi$ where $\phi$ is the Coulomb gauge scalar potential (Coulomb potential) given in Eq.~(8) of the main text, to obtain
\begin{align}\label{sc1}
\nabla \phi_g = \nabla(\phi-\partial_t \chi_g)-{\partial_t {\bf A}}_{\rm T} +{\bf \Pi}+{\bf P}_{{\rm T}g}.
\end{align}
Thus, we see that $\phi_g$ is fully determined in terms of the transverse canonical operators and the matter field. Moreover, from the Hamiltonian $H_g$ we easily find that
\begin{align}\label{Adot}
{\partial_t {\bf A}}_{\rm T} =-i[{\bf A}_{\rm T},H_g]={\bf \Pi}+{\bf P}_{{\rm T}g} = -{\bf E}_{\rm T}
\end{align}
as expected, and using this result together with Eq.~(\ref{sc1}) we find that up to a constant
\begin{align}
\phi_g = \phi-\partial_t\chi_g,
\end{align}
which is the expected result for the scalar potential corresponding to the vector potential ${\bf A}_g={\bf A}_{\rm T}+\nabla\chi_g$.

It is instructive to calculate in the arbitrary gauge $g$, the equation of motion for the Schr\"odinger operator $\psi$, which should be the Schr\"odinger equation in the presence of the Maxwell field and the external potential $U$. A straightforward calculation yields the correct result
\begin{align}\label{schro}
i{\dot \psi} = [\psi,H_g]= \left[{1\over 2m}(-i\nabla-q{\bf A}_g)^2 + U+q\phi_g \right]\psi.
\end{align}
Under the local phase transformation
\begin{align}
& \psi \to  e^{-iq(\chi_{g'} - \chi_{g})}\psi,\label{psig}
\end{align}
the Schr\"odinger equation is unchanged in form but as required by the gauge principle the potentials therein are replaced with the gauge-transformed potentials
\begin{align}
 & \phi_{g'} = \phi_{g} - \partial_t(\chi_{g'} -\chi_{g}),\label{phit} \\ & {\bf A}_{g'} = {\bf A}_g +\nabla(\chi_{g'}-\chi_{g}).\label{agt}
\end{align}
Eq.~(\ref{schro}) reproduces as two special cases the separately derived Coulomb gauge and multipolar gauge Schr\"odinger equations given in Ref.~\cite{power_quantum_1983}, which were not expressed in terms of potentials. We have shown that these Schr\"odinger equations are particular fixed-gauge cases of the expected general result that must be obtained according to the gauge principle, and that they are related by a gauge transformation.

\section{Relation to the particle-based description}\label{red}

In the nonrelativistic setting where matter is described by a Schr\"odinger field rather than a Dirac field there is no anti-matter, so the total material number operator is a conserved quantity \cite{cohen-tannoudji_photons_1989}. One can therefore employ an equivalent description to the field-theoretic description derived above, whereby each electron is described using single-particle canonical position and momentum operators ${\bf r}$ and ${\bf p}$ such that $[r_i,p_j]=i\delta_{ij}$. For a given number of electrons the descriptions are strictly equivalent, but the particle-based description may be less cumbersome when dealing with simple systems.

The field density $\rho=q\psi^\dagger\psi$ corresponds to the single-electron density $q\delta({\bf x}-{\bf r})$. In Sec.~II~A of the main text we considered a single-electron atom with nucleus fixed at the origin such that the charge density is $\rho({\bf x}) = -q\delta({\bf x})+q\delta({\bf x}-{\bf r})$. The nuclear potential $U({\bf x})/q=-q/4\pi |{\bf x}|$ is included in the longitudinal electric field energy along with the infinite self-energies $V_{\rm self}$ as
\begin{align}
{1\over 2}\int d^3 x \,{\bf E}_{\rm L}^2 = {1\over 2}\int d^3 x\, {\bf P}_{\rm L}^2 = U({\bf r})+V_{\rm self}.
\end{align}
The transverse field ${\bf P}_{\rm T}$ is unaffected. The Hamiltonian $H_g$ with density in Eq.~(\ref{h2b}) can now be written
\begin{align}\label{hggg}
H_g=&{1\over 2m}\left[{\bf p}-q{\bf A}_g({\bf r})\right]^2 + U({\bf r}) + V_{\rm self} \nonumber \\ &+{1\over 2}\int d^3 x \left[({\bf \Pi}+{\bf P}_{\rm Tg})^2+(\nabla\times {\bf A}_{\rm T})^2\right]
\end{align}
where, assuming $\chi_g$ as in Eq.~(\ref{chi}), we have
\begin{align}
&{\bf A}_g({\bf x}) = {\bf A}_{\rm T}({\bf x})+\nabla\int d^3x' {\bf g}({\bf x}',{\bf x})\cdot{\bf A}_{\rm T}({\bf x}'), \\
&{\bf P}_{{\rm T}g}({\bf x}) = -\int d^3 x' {\bf g}_{\rm T}({\bf x},{\bf x}')\rho({\bf x}').\label{PTg}
\end{align}

The theory is simplified further by restricting ${\bf g}_{\rm T}$ via Eq.~(12) of the main text in terms of the gauge-parameter $\alpha$. These simplifications are not approximations, so the theory remains exact and it becomes the theory presented in Sec.~II~A of the main text. Therein gauge freedom is the freedom to choose the parameter $\alpha$ which specifies ${\bf P}_{\rm T\alpha}$ and ${\bf A}_\alpha$ as in Eqs.~(14) and (9) of the main text respectively \cite{stokes_gauge_2019,stokes_ultrastrong_2021,stokes_uniqueness_2020}. The Hamiltonian $H_g$ in Eq. (\ref{hggg}) becomes $H_\alpha$ given in Eq.~(22) of the main text and the gauge fixing transformation $U_{gg'}$ in Eq.~(\ref{ugg'}) becomes $R_{\alpha\alpha'}$ in Eq.~(26) of the main text. Hamiltonians belonging to different gauges are unitarily related as in Eq.~(25) of the main text.

\section{Generalisation to many charges}\label{chd}

\subsubsection{Charge distributions referred to fixed centres}

In nonrelativistic QED it is useful to partition the collection of charges into certain groups called atoms and molecules. In Sec.~II~A of the main text, we describe a single hydrogen atom with positive charge $-q$ assumed fixed (non-dynamical). This is equivalent to describing the system using relative and centre-of-mass coordinates instead of the charge coordinates themselves, and assuming that the centre-of-mass is fixed, all centre-of-mass couplings being ignored. The atom is then described using the single coordinate ${\bf r}$, which is the position of charge $q$ relative to charge $-q$. We now provide the extension to arbitrary charge distributions in the vicinity of fixed molecular centres. The use of the same formalism to describe electrons in crystal lattices is given in Supplementary Note \ref{lattice}.

A molecule can be described by grouping arbitrary charges $\{q_\mu\}$ with positions ${\bf r}_\mu$ in the vicinity of a single fixed point ${\bf R}$. Often this point is assumed to coincide with a fixed molecular centre-of-mass \cite{craig_molecular_1998}. A given subset of positive charges may be assumed to be coincident at a fixed point and thereby define an atomic nucleus within the molecule \cite{craig_molecular_1998}. If relative and centre-of-mass coordinates are introduced rigorously in terms of the charge coordinates, then the centre-of-mass is an independent dynamical variable and it is necessary to introduce equations of constraint in order to preserve the number of degrees of freedom \cite{baxter_canonical_1993}. The theory can be developed along these general lines allowing centre-of-mass motion and also accommodating non-neutral charge distributions \cite{baxter_canonical_1993}. Here however, we will confine our attention to neutral charge-distributions in the vicinity of non-dynamical fixed points in space. A formulation in which no such fixed centres occur is outlined in Supplementary Note \ref{removal}.

\begin{figure}[t]
\begin{minipage}{\columnwidth}
\begin{center}
\hspace*{-3.5mm}\includegraphics[scale=0.132]{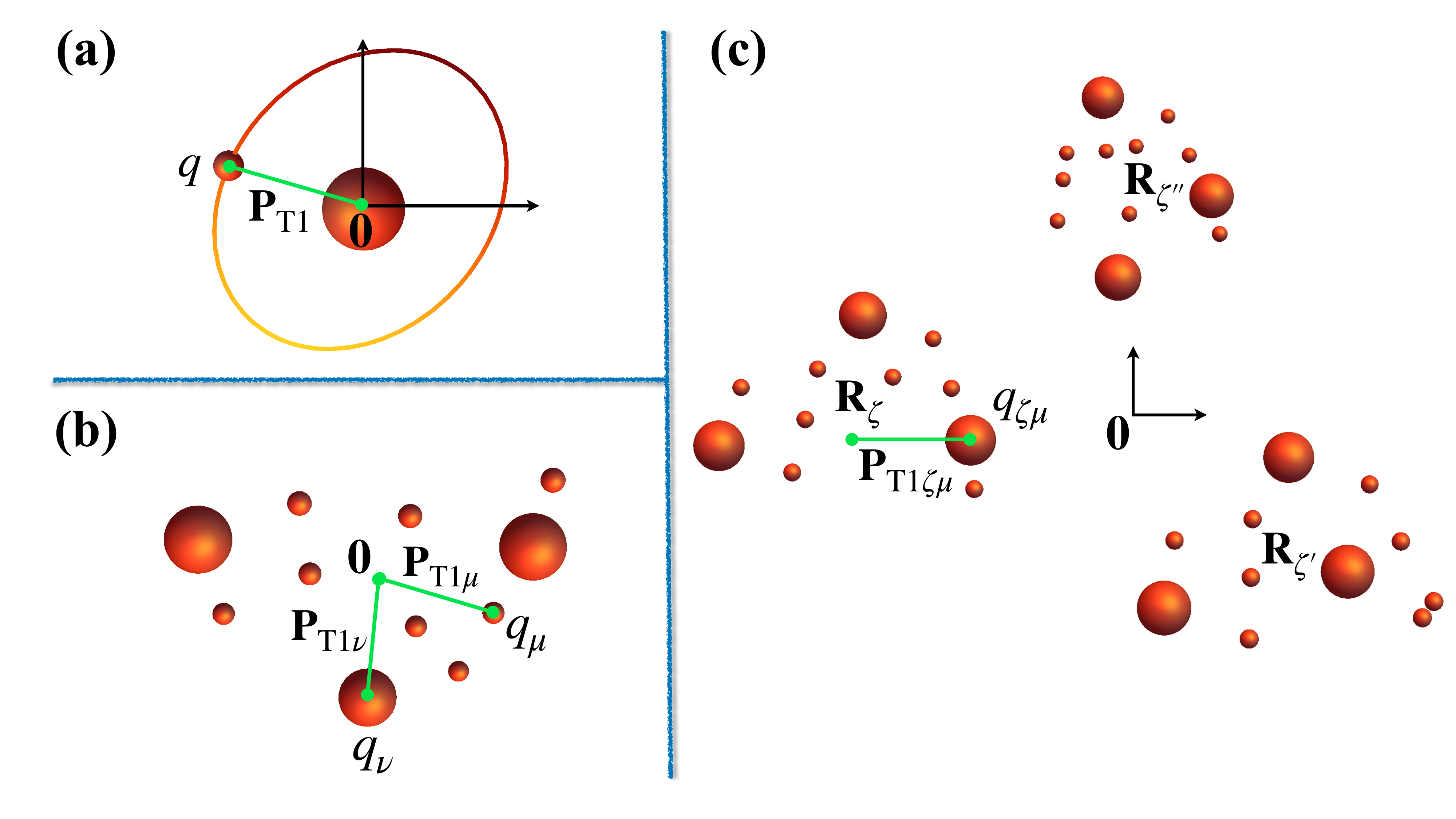}
\vspace*{-4mm}\caption{\textbf{(a)} A single-electron atom is described by the theory of Sec.~II~A of the main text. The multipolar polarisation ${\bf P}_{\rm T1}$ refers the dynamical charge  $q$ to the fixed centre ${\bf 0}$. \textbf{(b)} A single molecule consisting of charges with values $\pm q$. The multipolar polarisation refers each charge to a fixed centre at ${\bf 0}$. The system can be described using a single Poincar\'e gauge fixing condition. \textbf{(c)} A collection of molecules consisting of charges $\pm q$ in the vicinity of several fixed points ${\bf R}_{\zeta},\,{\bf R}_{\zeta'},...$ defining distinct molecules. The multipolar polarisation refers each charge to one of these fixed points. Each centre ${\bf R}_\zeta$ now corresponds to a different $\zeta$-Poincar\'e gauge fixing condition.}\label{moles}
\vspace*{-2mm}
\end{center}
\end{minipage}
\end{figure}

In its full generality, multipolar electrodynamics is designed to describe an arbitrary number of  molecular charge distributions each localised in the vicinity of a different fixed-point ${\bf R}_\zeta$ where $\zeta=1,...,N$ with $N$ the total number of molecules (Fig.~\ref{moles}). Whether or not the multipolar framework defines a choice of gauge for $N>1$ has been the subject of discussion and is related to controversy surrounding the nature and validity of the multipolar framework in general \cite{rousseau_quantum-optics_2017,vukics_gauge-invariant_2021,andrews_perspective_2018,rousseau_reply_2018}. The relation of the multipolar framework to the Poincar\'e gauge will now be clarified (see also Supplementary Note \ref{app1}).

In the above development of the theory the multipolar formalism is obtained from the Poincar\'e gauge choice ${\bf x}\cdot {\bf A}({\bf x})=0$. This condition can obviously also be written $({\bf x}-{\bf 0})\cdot {\bf A}({\bf x})=0$, an expression intended to signify the importance of the fixed distribution centre ${\bf 0}$ on the left-hand-side. More generally, we may specify a $\zeta$-dependent gauge fixing condition $({\bf x}-{\bf R}_\zeta)\cdot {\bf A}({\bf x})=0$, which we may call the $\zeta$-Poincar\'e gauge condition. The fixed potentials obtained from these conditions are different for different $\zeta$. In general, i.e., for $N>1$, multipolar electrodynamics constitutes a framework in which the gauge of the potential to which the distribution $\zeta$ couples within the Hamiltonian, is the $\zeta$-Poincar\'e gauge, as will be shown below.

We consider a total of $\sum_{\zeta=1}^N Z_\zeta$ charges with each $\zeta=1,...,N$ labelling a neutral molecule comprised of $Z_\zeta$ charges $q_{\zeta\mu}$,~$\mu=1,...,Z_\zeta$. The charge and current densities are as above but with summations over all charges now entailing a partition into separate molecules;
\begin{align}
\rho({\bf x})&= \sum_{\zeta=1}^N \rho_\zeta({\bf x})  = \sum_{\zeta=1}^N\sum_{\mu=1}^{Z_\zeta} \rho_{\zeta\mu}({\bf x})\label{rhomany}\\
{\bf J}({\bf x})&= \sum_{\zeta=1}^N {\bf J}_\zeta({\bf x})  = \sum_{\zeta=1}^N\sum_{\mu=1}^{Z_\zeta} {\bf J}_{\zeta\mu}({\bf x})
\end{align}
with $\rho_{\zeta\mu}({\bf x}) := q_{\zeta\mu} \delta({\bf x}-{\bf r}_{\zeta\mu})$ and ${\bf J}_{\zeta\mu}({\bf x}):=q_{\zeta\mu}[{\dot {\bf r}}_{\zeta\mu} \delta({\bf x}-{\bf r}_{\zeta\mu}) + \delta({\bf x}-{\bf r}_{\zeta\mu}){\dot {\bf r}}_{\zeta\mu}]/2$. The Coulomb energy is
\begin{align}
V&= {1\over 2}\int d^3 x\, {\bf E}_{\rm L}({\bf x})^2= \sum_{\zeta,\xi=1}^N\sum_{\mu=1}^{Z_\zeta} \sum_{\nu=1}^{Z_\xi} {q_{\zeta\mu} q_{\xi\nu} \over 8\pi |{\bf r}_{\zeta\mu}-{\bf r}_{\xi\nu}|}.
\end{align}
Using Eqs.~(36) and (41) of the main text, to define the multipolar polarisation associated with $\rho_{\zeta\mu}$, we assume the straight line path from the origin ${\bf o}={\bf R}_\zeta$ giving
\begin{align}
{\bf P}_{\zeta\mu}&({\bf x}) = -q_{\zeta\mu}{\bf g}_{\rm L}({\bf x},{\bf R}_\zeta) + q_{\zeta\mu} \int_{C({\bf R}_\zeta,{\bf r}_{\zeta\mu})}d{\bf z}\,\delta({\bf z}-{\bf x})\nonumber \\
 =& -q_{\zeta\mu}{\bf g}_{\rm L}({\bf x},{\bf R}_\zeta) \nonumber \\ &+ q_{\zeta\mu} \int_0^1 d\lambda\, ({\bf r}_{\zeta\mu}-{\bf R}_\zeta)\delta({\bf x}-{\bf R}_\zeta -\lambda[{\bf r}_{\zeta\mu}-{\bf R}_\zeta]).
\end{align}
For a neutral molecule $\zeta$ the first term $-q_{\zeta\mu}{\bf g}_{\rm L}({\bf x},{\bf R}_\zeta) $ does not contribute to ${\bf P}_\zeta({\bf x}) =\sum_{\mu=1}^{Z_\zeta} {\bf P}_{\zeta\mu}({\bf x})$. More generally, the arbitrary $\alpha$-gauge transverse polarisation may be defined as
\begin{align}
&{\bf P}_{\rm T\alpha}({\bf x})= \sum_{\zeta=1}^N \sum_{\mu=1}^{Z_\zeta} {\bf P}_{\rm T\alpha\zeta\mu}({\bf x}), \nonumber \\ &{\bf P}_{\rm T\alpha\zeta\mu}({\bf x})\nonumber \\ &=\alpha q_{\zeta\mu} \int_0^1 d\lambda\, ({\bf r}_{\zeta\mu}-{\bf R}_\zeta)\cdot \delta^{\rm T}({\bf x}-{\bf R}_\zeta -\lambda[{\bf r}_{\zeta\mu}-{\bf R}_\zeta]). \label{poltotZ2}
\end{align}

The total energy is
\begin{align}\label{enH2}
E =&\sum_{\zeta=1}^N \sum_{\mu=1}^{Z_\zeta} {1\over 2}m_{\zeta\mu}{\dot {\bf r}}_{\zeta\mu}^2 +{1\over 2}\int d^3 x \left({\bf E}^2+{\bf B}^2\right)\nonumber \\
=&\sum_{\zeta=1}^N \sum_{\mu=1}^{Z_\zeta} {1\over 2}m_{\zeta\mu}{\dot {\bf r}}_{\zeta\mu}^2+ V +{1\over 2}\int d^3 x \left({\bf E}_{\rm T}^2+{\bf B}^2\right).
\end{align}
Canonical momenta are defined as before by $m{\dot {\bf r}}_{\zeta\mu} = {\bf p}_{\zeta\mu}-q_{\zeta\mu}{\bf A}_\alpha({\bf r}_{\zeta\mu})$ and ${\partial_t {\bf A}}_{\rm T} = {\bf \Pi}+{\bf P}_{\rm T\alpha}$ with the material canonical commutation relation now being $[r_{\zeta\mu,i},p_{\xi\nu,j}]=i\delta_{\zeta\xi}\delta_{\mu\nu}\delta_{ij}$. The gauge fixing transformation $R_{\alpha\alpha'}$ is again given by Eq.~(26) of the main text but with ${\bf P}_{{\rm T}\alpha}$ defined in Eq.~(\ref{poltotZ2}). The transformation of the material canonical momenta in Eq.~(23) of the main text now holds for each charge and Eq.~(24) of the main text continues to hold for ${\bf \Pi}$. The $\alpha$-gauge Hamiltonian is again the energy expressed in terms of canonical operators; $H_\alpha({\bf y})=E$. Hamiltonians belonging to different gauges are unitarily related as in Eq.~(25) of the main text.

The $\alpha$-generalised Power-Zienau-Woolley transformation is
\begin{align}\label{pzw2}
R_{0\alpha} &:= \exp \left(-i\int d^3 x \,{\bf P}_{\rm T\alpha}({\bf x})\cdot {\bf A}_{\rm T}({\bf x})\right).
\end{align}
It is instructive to write the generator in terms of the molecular charge densities as
\begin{align}
\int d^3 x \,{\bf P}_{\rm T\alpha}({\bf x})\cdot {\bf A}_{\rm T}({\bf x}) = -\int d^3 x \sum_{\zeta=1}^N \rho_\zeta({\bf x})\chi_{\alpha\zeta}({\bf x})
\end{align}
where
\begin{align}
\chi_{\alpha\zeta}({\bf x}) = -\alpha \int_0^1d\lambda \, ({\bf x}-{\bf R}_\zeta)\cdot {\bf A}_{\rm T}({\bf R}_\zeta+\lambda[{\bf x}-{\bf R}_\zeta]) .
\end{align}
We see therefore, that $R_{0\alpha}$ in Eq.~(\ref{pzw2}) is a product of distinct local gauge transformations acting on each distribution $\zeta$ separately, rather than producing a global gauge transformation of all distributions by the same gauge function. The $\zeta\mu$'th material momentum $m_{\zeta\mu}{\dot {\bf r}}_{\zeta\mu}={\bf p}_{\zeta\mu}-q_{\zeta\mu}{\bf A}_{\rm T}({\bf r}_{\zeta\mu})$ of the Coulomb gauge transforms as
\begin{align}
&R_{0\alpha}\left[{\bf p}_{\zeta\mu}-q_{\zeta\mu}{\bf A}_{\rm T}({\bf r}_{\zeta\mu})\right]R_{0\alpha}^\dagger = {\bf p}_{\zeta\mu}-q_{\zeta\mu}{\bf A}_{\alpha\zeta}({\bf r}_{\zeta\mu})\label{mina2}
\end{align}
where
\begin{align}
{\bf A}_{\alpha\zeta}({\bf x}) = {\bf A}_{\rm T}({\bf x})+\nabla \chi_{\alpha\zeta}({\bf x}).
\end{align}
The transformation of the canonical momentum ${\bf \Pi}$ continues to be given by Eq.~(24) of the main text. It is easily verified that the $\alpha=1$ potentials satisfy the $\zeta$-Poincar\'e gauge conditions $({\bf x}-{\bf R}_\zeta)\cdot {\bf A}_{1\zeta}({\bf x}) = 0$. More generally, whenever $\alpha\neq 0$ the $\alpha$-gauge coupling involves potentials specified by $N$ distinct gauge fixing conditions. The single standard Poincar\'e gauge choice is obtained when $\alpha=1$ and $N=1$ and the standard Coulomb gauge choice is obtained for any $N$ when $\alpha=0$.    In Supplementary Note \ref{lattice}, we review a similar use of multipolar theory in deriving a description of electrons within a crystal lattice via the so-called Peierls substitution. In this case lattice vectors ${\bf R}_l$ play the role of the molecular centres ${\bf R}_\zeta$. 

The multipolar framework is obviously an equivalent formulation to the Coulomb gauge theory, but in the general case of arbitrary $N$, this equivalence is often viewed as distinct to equivalence under gauge transformations \cite{vukics_gauge-invariant_2021,andrews_perspective_2018}. Clearly gauge fixing transformations comprise only a subgroup of the unitary group, but evidently the potentials ${\bf A}_{\zeta\alpha}$ are all gauge transformations of ${\bf A}_{\rm T}$ and, therefore, of one another. Thus, the freedom to transform from the Coulomb gauge to the equivalent multipolar framework can be viewed as gauge freedom, but without requiring that every charge's interaction is transformed by the same gauge function [cf. Eq.~(\ref{pzw2})].

Recognition that the multipolar framework results from gauge transformations possesses the advantage of making clear that the Coulomb gauge and multipolar frameworks differ only in how they eliminate inherent mathematical redundancy that occurs within the formalism through ${\bf A}_{\rm L}$ and ${\bf P}_{\rm T}$. We view gauge freedom and gauge fixing in the generalised sense of being nothing less than the occurrence and elimination of such mathematical redundancy. Conversely, the physical differences between the Coulomb gauge and multipolar canonical momenta are well-known and these differences therefore immediately exemplify the impact of gauge freedom. This is discussed in detail from Sec.~II~D of the main text onward.

\subsubsection{Removal of arbitrary fixed centres}\label{removal}

For a globally neutral system the charge density can be partitioned in such a way that the polarisation field does not depend on arbitrary molecular centres ${\bf R}_\zeta$ \cite{woolley_power-zienau-woolley_2020}. Considering $N$ nuclei labelled by $\zeta=1,...,N$. The nuclei $\zeta$ has $Z^{\rm n}_\zeta$ positive charges, $-q$, that are located at ${\bf r}_\zeta$. For each positive charge there is a negative charge $q$, such that a total of $Z_\zeta= 2Z_\zeta^{\rm n}$ charges can be associated with each index $\zeta$. %, half of which are positive and half of which are negative. 
The charge density in Eq.~(\ref{rhomany}) can then be written \cite{woolley_power-zienau-woolley_2020}
\begin{align}
\rho({\bf x}) &= \sum_{\zeta=1}^N \sum_{\mu=1}^{Z^{\rm n}_\zeta} \rho_{\zeta\mu}({\bf x})
\end{align}
where now $\rho_{\zeta\mu}({\bf x}):=-q\left[\delta({\bf x}-{\bf r}_{\zeta})-\delta({\bf x}-{\bf r}_{\zeta\mu})\right]$. The charge density in Eq.~(1) of the main text can be understood as the special case with $N=1$ and $Z_{\zeta}^{\rm n}=1$, consisting of one nucleus with total charge $-q$ fixed at ${\bf r}_1 = {\bf 0}$, and a single charge $q$ with position ${\bf r}_{11}={\bf r}$. We can define the polarisation field associated with $\rho_{\zeta\mu}$ using Eqs.~(36) and (41) of the main text as
\begin{align}
{\bf P}_{\zeta\mu}({\bf x}) = q\int_{C({\bf r}_\zeta,{\bf r}_{\zeta\mu})}d{\bf z}\, \delta({\bf x}-{\bf z}).
\end{align}
In this formulation the polarisation field is localised along paths between charges rather than on paths between the charges and arbitrary fixed centres. The total polarisation is defined as \cite{woolley_power-zienau-woolley_2020}
\begin{align}
{\bf P}({\bf x}) = \sum_{\zeta=1}^N {\bf P}_{\zeta}({\bf x})=\sum_{\zeta=1}^N \sum_{\mu=1}^{Z^{\rm n}_\zeta}{\bf P}_{\zeta\mu}({\bf x}) 
\end{align}
where ${\bf P}_{\zeta}({\bf x})$ is the polarisation of the $N$'th atom.

\section{Generalisation to dispersing and absorbing media}\label{absm}

The arbitrary-gauge formalism is easily adapted to the description of linear dispersing and absorbing dielectric media, which is a valuable tool in describing cavity QED systems \cite{knoll_resonators_1991,gruner_green-function_1996,knoll_QED_2003,khanbekyan_qed_2005,viviescas_field_2003} (see also~Supplementary Note~\ref{cavabs}). More comprehensive details of the medium assisted quantum Maxwell theory reviewed below can be found in Refs.~\cite{gruner_green-function_1996,dung_three-dimensional_1998,viviescas_field_2003,knoll_QED_2003}. We provide the extension of the formalism to arbitrary gauges. Concerning further extensions, we note that Ref.~\cite{wei_quantization_2009} considers an anisotropic medium, and Ref.~\cite{ judge_canonical_2013} considers a linear magnetoelectric medium.

\subsubsection{QED of linear dielectrics}

Consider a medium, M, that responds linearly and locally to changes in the occupying electric field, such that it can be characterised by a polarisation of the form \cite{knoll_QED_2003}
\begin{align}\label{medpol1}
{\bf P}_{\rm M}(t,{\bf x}) = {\bf P}_{\rm n}(t,{\bf x}) + \int_0^\infty d\tau \chi(\tau,{\bf x}){\bf E}_{\rm M}(t-\tau,{\bf x})
\end{align}
where $\chi$ is the dielectric susceptibility and ${\bf P}_{\rm n}$ is a noise term describing losses via absorption. The corresponding noise charge and current densities are defined by $\rho_{\rm n} = -\nabla \cdot {\bf P}_{\rm n}$ and ${\bf J}_n ={\dot {\bf P}}_{\rm n}$, such that continuity equation ${\dot \rho}_{\rm n}=-\nabla\cdot {\bf J}_{\rm n}$ is satisfied identically.

We now define the Fourier transformation of any Hermitian operator-valued function $f(t)$ by
\begin{align}
f(t) = \int_0^\infty d\omega f(\omega)e^{-i\omega t}+{\rm H.c.}
\end{align}
We rely upon the argument, $t$ or $\omega$, to distinguish an operator from its Fourier transform. We may now define the Fourier transforms of the polarisation and its accompanying displacement field as
\begin{align}
{\bf P}_{\rm M}(\omega,{\bf x}) &= {\bf P}_{\rm n}(\omega ,{\bf x}) +[\epsilon(\omega,{\bf x})-1]{\bf E}_{\rm M}(\omega,{\bf x}),\\
{\bf D}_{\rm M} (\omega,{\bf x}) &:= {\bf E}_{\rm M}(\omega,{\bf x}) +{\bf P}_{\rm M}(\omega,{\bf x}) \nonumber \\&= \epsilon(\omega,{\bf x}){\bf E}_{\rm M}(\omega,{\bf x})+{\bf P}_{\rm n}(\omega ,{\bf x})
\end{align}
where
\begin{align}
\epsilon(\omega,{\bf x}) = 1+\int_0^\infty d\tau \,\chi(\tau,{\bf x})e^{i\omega \tau}
\end{align}
is the dielectric permittivity. The permittivity satisfies $\epsilon(z,{\bf x})^* = \epsilon(-z^*,{\bf x})$, is analytic in the upper-half complex plane, and is such that $\epsilon(\omega,{\bf x}) -1$ vanishes sufficiently fast as $\omega \to \infty$ that the real and imaginary parts satisfy the Kramers-Kronig relations
\begin{align}
&\epsilon_{\rm R}(\omega,{\bf x})-1 = {{\cal P}\over \pi} \int d\omega' {\epsilon_{\rm I}(\omega',{\bf x}) \over \omega' - \omega},\\
&\epsilon_{\rm I}(\omega,{\bf x}) = -{{\cal P}\over \pi} \int d\omega' {\epsilon_{\rm R}(\omega',{\bf x})-1 \over \omega' - \omega }.
\end{align}
These real and imaginary parts describe dispersion and absorption respectively.

Maxwell's equations can be written in Fourier space in terms of the medium's electric and magnetic fields, and the noise and charge currents; ${\bf E}_{\rm M},\, {\bf B},\, \rho_{\rm n},\, {\bf J}_{\rm n}$ \cite{knoll_QED_2003}. It follows that the electric field satisfies $\nabla\times\nabla\times {\bf E}_{\rm M}(\omega,{\bf x}) - \omega^2\epsilon(\omega,{\bf x}){\bf E}_{\rm M}(\omega,{\bf x}) = i\omega {\bf J}_{\rm n}(\omega,{\bf x})$, 
which possesses solution
\begin{align}\label{elm}
{\bf E}_{\rm M}(\omega,{\bf x}) = i\omega \int d^3 x'\, {\bf G}(\omega,{\bf x},{\bf x}')\cdot {\bf J}_{\rm n}(\omega,{\bf x}')
\end{align}
via the dyadic Green's function ${\bf G}$ defined by
\begin{align}\label{grndyad}
&([\partial^x_i \partial^x_j - \delta_{ij} \partial^2] -\omega^2 \epsilon(\omega,{\bf x})\delta_{ij})G_{jk}(\omega,{\bf x},{\bf x}') \nonumber \\ &= \delta_{ik}\delta({\bf x}-{\bf x}').
\end{align}
The Green's function possesses the following properties \cite{knoll_QED_2003}
\begin{align}
&G_{ij}(z,{\bf x},{\bf x}')^*=G_{ij}(-z^*,{\bf x},{\bf x}'),\label{idg1}\\
&G_{ij}(\omega,{\bf x},{\bf x}')=G_{ji}(\omega,{\bf x}',{\bf x}),\\
&{\rm Im}G_{ik}(\omega,{\bf x},{\bf x}')\nonumber \\ &=\int d^3y \, \omega^2\epsilon_{\rm I}(\omega,{\bf y})G_{ij}(\omega,{\bf x},{\bf y})G_{kj}(\omega,{\bf x}',{\bf y})^*,\\
&\pi\delta_{ik}^{\rm T}({\bf x}-{\bf x}')\nonumber \\ &=2\int_0^\infty d\omega\, \omega\,{\rm Im} \int d^3 y\, G_{ij}(\omega,{\bf x},{\bf y})\delta^{\rm T}_{jk}({\bf y}-{\bf x}').\label{idg4}
\end{align}
For further details we refer the reader to Ref.~\cite{knoll_QED_2003}.

The Maxwell equations can be derived from the Hamiltonian \cite{knoll_QED_2003}
\begin{align}
H_{\rm M} = \int d^3 x \int_0^\infty d\omega\,  \omega {\bf f}^\dagger(\omega,{\bf x})\cdot {\bf f}(\omega,{\bf x})
\end{align}
where the noise operators are understood as comprising a bosonic system defined by
\begin{align}
{\bf P}_{\rm n}(\omega,{\bf x}) = i\sqrt{\epsilon_{\rm I}(\omega,{\bf x})\over \pi }{\bf f}(\omega,{\bf x})
\end{align}
with
\begin{align}
&[f_i(\omega,{\bf x}),f_j(\omega',{\bf x}')]=0,\\
&[f_i(\omega,{\bf x}),f^\dagger_j(\omega',{\bf x}')]=i\delta_{ij}\delta(\omega-\omega')\delta({\bf x}-{\bf x}').
\end{align}
Given a Fourier-space field, ${\bf F}_{\rm M}(\omega,{\bf x})$, the corresponding Schr\"odinger picture field in real space is given by
\begin{align}\label{elm2}
&{\bf F}({\bf x}) = \int_0^\infty d\omega \, {\bf F}(\omega,{\bf x}) +{\rm H.c.} \nonumber \\ 
\end{align}
Examples are ${\bf E}_{\rm M},\, {\bf B},\, {\bf A}_{\rm T}$, and ${\bm \Pi}$ where
\begin{align}
&{\bf E}_{\rm M}(\omega,{\bf x}) = i\omega^2\int d^3 x'\,\sqrt{\epsilon_{\rm I}(\omega,{\bf x}')\over 
\pi} {\bf G}(\omega,{\bf x},{\bf x}')\cdot {\bf f}(\omega,{\bf x}'),\label{modeexm}\\
&{\bf B}(\omega,{\bf x}) = -{i\over \omega}\nabla \times {\bf E}_{\rm M}(\omega,{\bf x})=-{i\over \omega}\nabla \times {\bf E}_{\rm MT}(\omega,{\bf x}),\\
&{\bf A}_{\rm T}(\omega,{\bf x}) = -{i\over \omega}{\bf E}_{\rm MT}(\omega,{\bf x}),\\
&{\bm \Pi}(\omega,{\bf x}) = -i\omega {\bf A}_{\rm T}(\omega,{\bf x}) = -{\bf E}_{\rm MT}(\omega,{\bf x}).\label{last}
\end{align}
Using Eqs.~(\ref{modeexm})-(\ref{last}), it is clear that each of ${\bf E}_{\rm M},\, {\bf B},\, {\bf A}_{\rm T}$, and ${\bm \Pi}$, admits a mode expansion in terms the bosonic operators ${\bf f}(\omega,{\bf x})$ and ${\bf f}^\dagger(\omega,{\bf x})$.

The longitudinal electric field defines the Coulomb gauge scalar potential as
\begin{align}\label{elml}
{\bf E}_{\rm ML}({\bf x}) = {\bf E}_{\rm M}({\bf x}) -{\bf E}_{\rm MT}({\bf x})= -\nabla\phi_{\rm M}({\bf x}).
\end{align}
The non-zero commutation relations among the above fields are as in vacuum QED \cite{knoll_QED_2003};
\begin{align}
&[E_{{\rm M}i}({\bf x}),B_j({\bf x}')]=i\varepsilon_{ijk}\partial_k^x\delta({\bf x}-{\bf x}'),\label{comm1}\\
&[A_{{\rm T}i}({\bf x}),\Pi_j({\bf x}')]=i\delta_{ij}^{\rm T}({\bf x}-{\bf x}')\label{comm2}
\end{align}
where $\varepsilon_{ijk}$ is the Levi-Civita symbol.

\subsubsection{Arbitrary gauge coupling to guest charges}

Coupling to atomic systems within the dielectric proceeds as in the case of the vacuum, but with the addition of direct electrostatic interactions between the guest charges and medium. The Coulomb gauge Hamiltonian for the medium assisted Maxwell field coupled to a system of $N$ guest charges $q_\mu$, that is neutral ($\sum_{\mu =1}^N q_\mu =0$) and that has charge and current densities $\rho_{\rm A}({\bf x})=\sum_{\mu=1}^N q_\mu \delta({\bf x}-{\bf r}_\mu)$ and ${\bf J}_{\rm A}({\bf x}) = \sum_{\mu=1}^Nq_\mu[{\dot {\bf r}}_\mu \delta({\bf x}-{\bf r}_\mu)+\delta({\bf x}-{\bf r}_\mu) {\dot {\bf r}}_\mu]/2$, is \cite{knoll_QED_2003}
\begin{align}
H_0 =& \sum_{\mu=1}^N {1\over 2m_\mu}\left[{\bf p}_\mu -q_\mu{\bf A}_{\rm T}({\bf r}_\mu)\right]^2 + V_{\rm A} +V_{\rm AM}+H_{\rm M}
\end{align}
where $[r_{\mu i},p_{\nu j}]=i\delta_{ij}\delta_{\mu\nu}$, and
\begin{align}
V_{\rm A}& = {1\over 2}\int d^3 x\, \rho_{\rm A}({\bf x})\phi_{\rm A}({\bf x}) ={1\over 2}\int d^3 x\, {\bf P}_{\rm AL}({\bf x})^2 =\nonumber \\ &=  {1\over 2}\int d^3 x\, {\bf E}_{\rm AL}({\bf x})^2 = {1\over 2}\int d^3 x d^3x' { \rho_{\rm A}({\bf x})\rho_{\rm A}({\bf x}')\over 4\pi|{\bf x}-{\bf x}'|},\label{Amcoul}\\
V_{\rm AM} &= \int d^3 x\, \rho_{\rm A}({\bf x})\phi_{\rm M}({\bf x}) =  \int d^3 x\, {\bf E}_{\rm AL}({\bf x}) \cdot  {\bf E}_{\rm ML}({\bf x}) \nonumber \\ &=  \int d^3 x\, {\bf P}_{\rm AL}({\bf x}) \cdot  {\bf P}_{\rm ML}({\bf x}) \label{Amedsc}
\end{align}
are respectively the Coulomb energy of the guest charges and the electrostatic interaction energy of the guest charges with the medium. The longitudinal atomic polarisation ${\bf P}_{\rm AL}$ is minus the electrostatic field ${\bf E}_{\rm AL}$ associated with the guest charges.

 The commutation relations in Eqs.~(\ref{comm1}) and (\ref{comm2}) continue to hold. The Hamiltonian $H_0$ in conjunction with the commutation relations yields the Maxwell-Lorentz equations \cite{knoll_QED_2003}. For the present case of a non-magnetic medium the magnetic flux and magnetic fields coincide; ${\bf B}_{\rm M}={\bf H}_{\rm M}$ (the medium magnetisation vanishes), and the total electric field is
\begin{align}\label{emtot}
{\bf E} = {\bf E}_{\rm M} + {\bf E}_{\rm AL},
\end{align}
which includes both the transverse field ${\bf E}_{\rm MT}$ and the total electrostatic field ${\bf E}_{\rm ML}+{\bf E}_{\rm AL}$.

The $g$-gauge guest charge polarisation is defined as before by
\begin{align}
{\bf P}_{{\rm A}g}({\bf x}) = -\int d^3 x' \, {\bf g}({\bf x},{\bf x}')\rho_{\rm A}({\bf x}')
\end{align}
such that $\nabla \cdot {\bf P}_{{\rm A}g} = -\rho_{\rm A}$, and the unitary gauge fixing transformation from the Coulomb gauge to the gauge ${\bf g}_{\rm T}$ is the generalised PZW transformation given by
\begin{align}
U_{0g} = \exp\left[-i\int d^3 x \,{\bf P}_{{\rm A}g}({\bf x})\cdot {\bf A}_{\rm T}({\bf x})\right].
\end{align}
The $g$-gauge Hamiltonian is obtained by finding the expressions for the transformed operators $U_{0g}{\bf p}_\mu U_{0g}^\dagger$ and $U_{0g}{\bf f}(\omega,{\bf x})U_{0g}^\dagger$, and by making use of the identities (\ref{idg1})-(\ref{idg4}), with the final result
\begin{align}\label{Hgmed}
H_g =&U_{0g}H_0U_{0g}^\dagger \nonumber \\=& \sum_{\mu=1}^N {1\over 2m_\mu}\left[{\bf p}_\mu -q_\mu{\bf A}_g({\bf r}_\mu)\right]^2 + V_{\rm A}+V_{\rm AM} +H_{\rm M} \nonumber \\  & +{1\over 2}\int d^3 x\, {\bf P}_{{\rm A}g{\rm T}}({\bf x})^2 + \int d^3 x\, {\bf P}_{{\rm A}g{\rm T}}({\bf x})\cdot {\bf \Pi}({\bf x}).
\end{align}
As in vacuum (no medium) QED we have
\begin{align}
{\bf A}_g({\bf x}) = {\bf A}_{\rm T}({\bf x}) + \nabla \int d^3 x'\, {\bf g}({\bf x}',{\bf x})\cdot {\bf A}_{\rm T}({\bf x}'),
\end{align}
and the momentum ${\bf \Pi}$ conjugate to ${\bf A}_{\rm T}$ represents in gauge ${\bf g}_{\rm T}$, (minus) the generalised transverse atomic displacement field, ${\bf D}_{{\rm A}g{\rm T}}$, defined by
\begin{align}
{\bf D}_{{\rm A}g{\rm T}} = {\bf E}_{\rm MT}+{\bf P}_{{\rm A}g{\rm T}}.
\end{align}
The Coulomb and multipolar gauge theories found in Ref.~\cite{knoll_QED_2003} are special cases of the $g$-gauge formalism obtained by making the Coulomb gauge and multipolar gauge choices of ${\bf g}_{\rm T}$ respectively. Using Eqs.~(\ref{Amcoul}) and (\ref{Amedsc}) we may also write $H_g$ more compactly in terms of total polarisation fields rather than transverse and longitudinal parts, as 
\begin{align}\label{Hgmed2}
H_g =& H_{\rm KE} +H_{\rm M} +{1\over 2} \int d^3 x\, {\bf P}_{{\rm A}g}({\bf x})^2 \nonumber \\ &+ \int d^3 x {\bf P}_{{\rm A}g}({\bf x}) \cdot [{\bf P}_{\rm M}({\bf x}) - {\bf D}_{g{\rm T}}({\bf x})]
\end{align}
where $H_{\rm KE}:= \sum_{\mu=1}^N {1\over 2m_\mu}\left[{\bf p}_\mu -q_\mu{\bf A}_g({\bf r}_\mu)\right]^2$, and where the {\em total} $g$-gauge polarisation ${\bf P}_g$ and total $g$-gauge transverse displacement field ${\bf D}_{g{\rm T}}$ are defined by
\begin{align}
&{\bf P}_g := {\bf P}_{{\rm A}g}+{\bf P}_{\rm M},\\
&{\bf D}_{g{\rm T}} := {\bf E}_{\rm MT}+ {\bf P}_{g{\rm T}}.
\end{align}

In summary, in a way that closely mirrors the case of vacuum QED, we have provided a Hamiltonian for charges coupled to the Maxwell field within a linear dielectric medium. The gauge is arbitrary and is determined by the transverse function ${\bf g}_{\rm T}$. For simplicity, we focus primarily on QED in vacuum as described in Sec.~II of the main text. However, it should be borne in mind that  the same general conclusions that we will draw also apply to the present theory of linear dielectric media, wherein unitary gauge fixing transformations have precisely the same form as in the vacuum theory, and essentially the same effect. The dielectric theory reviewed above has important applications in cavity QED, which is discussed in more detail in Sec.~VII of the main text.

\section{Lattice systems}\label{lattice}

We briefly review here the description of electrons within a crystal lattice in terms of the so-called Peierls substitution \cite{peierls_zur_1933,luttinger_effect_1951,graf_electromagnetic_1995}. The formalism closely resembles the multipolar framework for $N$ separate charge distributions in which   lattice vectors replace the molecular centres ${\bf R}_\zeta$. The thermodynamic phases of strongly correlated electron systems and their description via the Peierls substitution is a topic of current interest in cavity QED \cite{mazza_superradiant_2019,andolina_cavity_2019,nataf_rashba_2019,guerci_superradiant_2020,andolina_theory_2020,bamba_magnonic_2022}. The significance of gauge freedom when describing the thermodynamic limits of cavity QED systems has been the subject of perennial debate, as is briefly reviewed in Sec.~VII~B of the main text.

For simplicity we consider a single electron confined within an $N$-site lattice and we will restrict the electronic excitations to a single-band. Each lattice site is labelled by a position ${\bf R}_l~,l=1,..,N$. The free material Hamiltonian is \cite{luttinger_effect_1951}
\begin{align}
H_m={{\bf p}^2\over 2m} +V({\bf r})
\end{align}
where $V({\bf r})$ is the periodic potential provided by the lattice. The orthonormal electronic energy eigenfunctions are Bloch functions labelled by a single reciprocal lattice index ${\bf k}$;
\begin{align}
H_m \psi_{\bf k}({\bf r}) = E_{\bf k} \psi_{\bf k}({\bf r}).
\end{align}
A localised Wannier function can be defined for each lattice site as
\begin{align}
w_l=w({\bf r}-{\bf R}_l) = {1\over \sqrt{N}} \sum_{\bf k}  \psi_{\bf k}({\bf r}) e^{-i{\bf k}\cdot {\bf R}_l}.
\end{align}
These functions are orthonormal in the sense that $\braket{w_l|w_{l'}}=\delta_{ll'}$ where $\braket{\cdot|\cdot }$ denotes the usual inner-product on $L^2({\mathbb R}^3)$. The matrix representation of $H_m$ in the site basis is denoted $-t$, viz.,
\begin{align}
H_m = -\sum_{l,l'} t_{ll'}\ket{w_l}\bra{w_{l'}}, \qquad t_{ll'} = -\bra{w_l}H_m\ket{w_{l'}}.
\end{align}
Introducing coupling to the transverse vector potential via ${\bf p} \to {\bf p}-q{\bf A}_{\rm T}({\bf r})$ in $H_m$ gives the Coulomb gauge mechanical energy
\begin{align}
{\cal H}_m[{\bf A}_{\rm T}]={1\over 2m}\left[{\bf p}-q{\bf A}_{\rm T}({\bf r})\right]^2 +V({\bf r}).
\end{align}

The Peierls substitution is a matrix transformation of $t$ that gives an approximation of the corresponding matrix for ${\cal H}_m[{\bf A}_{\rm T}]$. It makes use of the existence of alternative choices of free basis states $\ket{{\bar w}_l}$ and $\ket{w_l}$ that are defined to be related by a PZW transformation in which the vector ${\bf R}_l$ acts as a multipole centre \cite{luttinger_effect_1951};
\begin{align}
\ket{{\bar w}_l} &= e^{-iq\chi_{1l}({\bf r})} \ket{w_l}\\
q\chi_{1l}({\bf r}) &= -q\int_{{\bf R}_l}^{\bf r} d{\bf s}\cdot {\bf A}_{\rm T}({\bf s}) =-\int d^3 x\,  {\bf P}_{{\rm T}1l}({\bf x})\cdot {\bf A}_{\rm T}({\bf x})
\end{align}
where
\begin{align}
{\bf P}_{{\rm T}1l}({\bf x}) = q\int_0^1 d\lambda \, ({\bf r}-{\bf R}_l)\cdot \delta_{\rm T}({\bf x}-{\bf R}_l-\lambda[{\bf r}-{\bf R}_l])
\end{align}
is the transverse multipolar polarisation connecting the $l$'th lattice site vector ${\bf R}_l$ to the charge position ${\bf r}$.

We can represent the Hamiltonian ${\cal H}_m[{\bf A}_{\rm T}]$ in the basis $\ket{{\bar w}_l}$ as
\begin{align}\label{coullat}
{\cal H}_m[{\bf A}_{\rm T}] = \sum_{l,l'} \bra{{\bar w}_l}{\cal H}_m[{\bf A}_{\rm T}]\ket{{\bar w}_{l'}}\ket{{\bar w}_l}\bra{{\bar w}_{l'}}.
\end{align}
The matrix elements are computed as
\begin{align}\label{matel}
&\bra{{\bar w}_l}{\cal H}_m[{\bf A}_{\rm T}]\ket{{\bar w}_{l'}} =\nonumber \\& \int d^3 r\, e^{iq\chi_{1l}({\bf r})}w^*({\bf r}-{\bf R}_l){\cal H}_m[{\bf A}_{\rm T}]e^{-iq\chi_{1l'}({\bf r})}w({\bf r}-{\bf R}_{l'}) \nonumber \\ &= \int d^3 r\, e^{iq[\chi_{1l}({\bf r})-\chi_{1l'}({\bf r})]}w^*({\bf r}-{\bf R}_l){\cal H}_m[{\bf A}_{1l'}]w({\bf r}-{\bf R}_{l'})
\end{align}
where
\begin{align}
{\bf A}_{1l'}({\bf r})&={\bf A}_{\rm T}({\bf r})+\nabla \chi_{1l'}({\bf r}) \nonumber \\
&= -\int_0^1 d\lambda \, \lambda({\bf r}-{\bf R}_{l'})\times {\bf B}({\bf R}_{l'}+\lambda[{\bf r}-{\bf R}_{l'}])
\end{align}
is the multipolar potential referred to the $l'$'th site. Neglecting this potential is an electric dipole approximation that assumes ${\bf A}_{\rm T}$ does not vary appreciably over the extent of $w({\bf r}-{\bf R}_{l'})$. We then have ${\cal H}_m[A_{1l'}] =H_m$, i.e., the multipolar gauge mechanical energy is the bare material energy, and Eq.~(\ref{matel}) becomes
\begin{align}\label{matel2}
&\bra{{\bar w}_l}{\cal H}_m[{\bf A}_{\rm T}]\ket{{\bar w}_{l'}} \nonumber \\&= \int d^3 r\, e^{iq[\chi_{1l}({\bf r})-\chi_{1l'}({\bf r})]}w^*({\bf r}-{\bf R}_l)H_mw({\bf r}-{\bf R}_{l'}).
\end{align}
The phase $e^{iq[\chi_{1l}({\bf r})-\chi_{1l'}({\bf r})]}$ can be simplified by noting that the magnetic flux threading the loop $C = {\bf r}\to {\bf R}_l\to {\bf R}_{l'} \to {\bf r}$, is negligible over the extent of the Wannier functions, which are localised at the lattice sites. We therefore have
\begin{align}
 0 = \oint_C d{\bf s} \cdot {\bf A}_{\rm T}({\bf s}) = \chi_{1l}({\bf r}) - \chi_{1l'}({\bf r}) - \int_{{\bf R}_{l'}}^{{\bf R}_l} d{\bf s} \cdot {\bf A}_{\rm T}({\bf s})
\end{align}
and so Eq.~(\ref{matel2}) can be written
\begin{align}\label{matel3}
&\bra{{\bar w}_l}{\cal H}_m[{\bf A}_{\rm T}]\ket{{\bar w}_{l'}} = e^{iq\int_{{\bf R}_{l'}}^{{\bf R}_l} d{\bf s} \cdot {\bf A}_{\rm T}({\bf s})} t_{ll'}.
\end{align}
Substituted into Eq.~(\ref{coullat}) this yields
\begin{align}\label{coullat2}
{\cal H}_m[{\bf A}_{\rm T}] = -\sum_{l,l'} e^{iq\int_{{\bf R}_{l'}}^{{\bf R}_l} d{\bf s} \cdot {\bf A}_{\rm T}({\bf s})} t_{ll'}\ket{{\bar w}_l}\bra{{\bar w}_{l'}}.
\end{align}
Thus, we see that the Coulomb gauge Hamiltonian can be obtained by making the Peierls substitution $t_{ll'} \to e^{iq\int_{{\bf R}_{l'}}^{{\bf R}_l} d{\bf s} \cdot {\bf A}_{\rm T}({\bf s})} t_{ll'}$ within the free Hamiltonian $H_m = -\sum_{l,l'} t_{ll'}\ket{{\bar w}_l}\bra{{\bar w}_{l'}}$. The substitution has been derived via the PZW transformation and the EDA. It closely resembles the means by which lattice gauge-field theories are defined and understood in terms of Wilson's parallel transport operator \cite{wilson_confinement_1974,wiese_ultracold_2013}.

\section{Proof that multipolar and Poincar\'e gauge QED are identical; resolution of controversy}\label{app1}

The procedure we follow closely resembles that in Supplementary Note~\ref{symm}. The mechanical momentum of the dynamical charge of the system with charge and current densities in Eqs.~(1) and (2) of the main text, is given in terms of the canonical momentum ${\bf p}$ via minimal coupling; $m{\dot {\bf r}}={\bf p}-q{\bf A}({\bf r})$. As in Refs.~\cite{rousseau_quantum-optics_2017,rousseau_reply_2018} and as in Supplementary Note~\ref{symm}, the momentum ${\tilde {\bf \Pi}}$ conjugate to ${\bf A}$ is $-{\bf E}$, noting that here we will use the notation ${\tilde \Pi}_\mu$ for the field canonical momentum conjugate to $A_\mu$. The naive Hamiltonian is
  
\begin{align}\label{Hamtot}
H ={1\over 2m}\left[{\bf p}-q{\bf A}({\bf r})\right]^2 + {1\over 2} \int d^3 x\,\left[ {\tilde {\bf \Pi}}^2+{\bf B}^2\right] + {\mathfrak G}[A_0]
\end{align}
where the infinitesimal generator of gauge transformations ${\mathfrak G}[\chi]$ is defined in Eq.~(\ref{gsymm}).  The Poisson brackets of the naive theory are 
\begin{align}
\{r_i,p_j\}=&\delta_{ij}, \\ \{A_\mu({\bf x}),{\tilde \Pi}_\nu({\bf x}')\}=&\delta_{\mu\nu}\delta({\bf x}-{\bf x}')\label{npois},
\end{align}
which can be used to determine the naive time evolution of any classical observable written as function of canonical variables. There are three constraints; $C_0={\tilde \Pi}_0,\,C_1=\rho+\nabla\cdot{\tilde {\bf \Pi}}$, and a gauge fixing constraint $C_2={\mathcal F}(A)$.   The infinitesimal generator of gauge transformations, ${\mathfrak G}[\chi] = \int d^3x [C_0{\partial_t \chi}+C_1\chi]$,  [Eq.~(\ref{gsymm})] is such that $\{{\mathfrak G}[\chi],{\bf A}\}=\nabla\chi$ and $\{{\mathfrak G}[\chi],A_0\}=-\partial_t \chi$. Clearly, the final term in Eq.~(\ref{Hamtot}) vanishes on the physical subspace. Following Woolley, we take the general gauge fixing constraint \cite{woolley_r._g._charged_1999}
\begin{align}
C_2:=\int d^3 x' {\bf g}({\bf x}',{\bf x})\cdot {\bf A}({\bf x'})
\end{align}
in which ${\bf g}$ is the Green's function for the divergence operator; $\nabla\cdot {\bf g}({\bf x},{\bf x}')=\delta({\bf x}-{\bf x}')$.

Since $\{C_0,C_1\}=0$ and $\{C_0,C_2\}=0$, the constraint $C_0=0$ can be imposed immediately, which removes $A_0$ and ${\tilde \Pi}_0$ from the formalism completely. The Poisson brackets $C_{ij}({\bf x},{\bf x}'):=\{C_i({\bf x}),C_j({\bf x}')\}$ of the remaining two constraints form a matrix with inverse
\begin{align}
C^{-1}({\bf x},{\bf x}')=\delta({\bf x}-{\bf x}')\left( {\begin{array}{cc}
 0 & 1 \\
-1 & 0 
 \end{array} } \right).
\end{align}
The equal-time {\em Dirac bracket} is defined by
\begin{align}\label{dbrack}
&\{\cdot,\cdot\}_D := \{\cdot,\cdot\} \nonumber \\ &- \int d^3 x \int d^3 x' \, \{\cdot ,C_i({\bf x})\}C_{ij}^{-1}({\bf x},{\bf x}')\{C_j({\bf x}'),\cdot\}.
\end{align}
Like the Poisson bracket the Dirac bracket is a Lie bracket, but unlike the Poisson bracket, it will yield the correct equations of motion when used in conjunction with the Hamiltonian, even once the constraints $C_i=0$ have been imposed.

Hereafter we denote contravariant indices with subscripts. The nonzero Dirac brackets between the dynamical variables are \cite{woolley_r._g._charged_1999}
\begin{align}
\{r_i,p_j\}_D &=\delta_{ij}, \\  \{A_i({\bf x}),{\tilde \Pi}_j({\bf x}')\}_D &=\delta_{ij}\delta({\bf x} - {\bf x}') + \nabla^{\bf x}_i{\rm g}_j({\bf x}',{\bf x})\label{dbapi}, \\  \{p_i,{\tilde \Pi}_j({\bf x})\}_D &= q \nabla^{\bf r}_i{\rm g}_j ({\bf x},{\bf r})=-\partial_i^{\bf r} P_{g,j}({\bf x})\label{pPi}
\end{align}
where ${\bf P}_g$ is defined by Eq.~(\ref{pg}). These Dirac brackets are consistent with those given in Ref.~\cite{rousseau_quantum-optics_2017}.
Quantisation of the theory may now be carried out via the replacement $\{\cdot,\cdot\}_D\to-i[\cdot,\cdot]$. The construction of the quantum theory is complete. However, so far only the Dirac brackets of ${\bf A}$ and ${\tilde {\bf \Pi}}$ have been determined and as operators these fields provide an inconvenient expression of the quantum theory, because of Eq.~(\ref{pPi}). This feature is noted in Ref.~\cite{vukics_gauge-invariant_2021} and its response Ref.~\cite{rousseau_reply_2018}. The ensuing lack of commutativity between ${\bf p}$ and ${\tilde {\bf \Pi}}$ within the final quantum theory, implies that the canonical pairs $({\bf r},{\bf p})$ and $({\bf A},{\tilde {\bf \Pi}})$ do not define separate (``matter" and ``light") quantum subsystems therein.

It is straightforward to construct canonical operator pairs that define quantum subsystems by imposing the constraints. The constraint $C_1=0$ implies that ${\tilde {\bf \Pi}}_{\rm L} = -{\bf E}_{\rm L} = {\bf P}_{\rm L}$ is fully determined by ${\bf r}$, while $C_2=0$ implies that ${\bf A}$ can be written \cite{woolley_r._g._charged_1999}
\begin{align}
{\bf A}({\bf x})={\bf A}_{\rm T}({\bf x})+\nabla\int d^3 x' {\bf g}({\bf x'},{\bf x})\cdot {\bf A}_{\rm T}({\bf x}')
\end{align}
and so it is fully determined by ${\bf A}_{\rm T}$ and ${\bf g}_{\rm T}$. We define the field ${\bf \Pi}$ by
\begin{align}\label{pitilde}
{\bf \Pi} &= {\tilde {\bm \Pi}} - {\bf P}_g = {\tilde {\bm \Pi}}_{\rm T} - {\bf P}_{{\rm T}g}\nonumber \\
&=-{\bf E} - {\bf P}_g=-{\bf E}_{\rm T} - {\bf P}_{{\rm T}g},
\end{align}
where ${\bf P}_{g}$ is defined by Eq.~(\ref{pg}) and where the second, third, and fourth equalities hold for $C_1=0$. Since immediately we have $\{p_i,P_{g,j}({\bf x})\}_D = -\partial_i^{\bf r} P_{g,j}({\bf x})$, it follows from Eq.~(\ref{pPi}) that
\begin{align}
\{p_i,\Pi_j({\bf x})\}_D  =\{p_i,{\tilde \Pi}_j({\bf x})\}_D -  \{p_i,P_{g,j}({\bf x})\}_D = 0.\label{dbppicgt}
\end{align}
Thus, the only non-zero Dirac brackets of the canonical pairs $({\bf r},{\bf p})$ and $({\bf A}_{\rm T},{\bf \Pi})$ are
\begin{align}
\{r_i,p_j\}_D & = \delta_{ij},\\
\{A_{{\rm T},i}({\bf x}),\Pi_j({\bf x}')\}_D &=\delta_{ij}^{\rm T}({\bf x} - {\bf x}'),\label{trd}
\end{align}
where the second bracket follows immediately from Eq.~(\ref{dbapi}). The theory can be expressed entirely in terms of these canonical pairs, which respectively define matter and light quantum subsystems upon quantisation. If we let ${\bf g}_{\rm T}({\bf x},{\bf x}')=-\alpha\int_0^1 d\lambda \, {\bf x}'\cdot \delta^{\rm T}({\bf x}-\lambda{\bf x}')$ then the theory expressed in this way, coincides with the $\alpha$-gauge theory derived in Sec.~II~A of the main text. In the Poincar\'e gauge ($\alpha=1$) in particular, we have ${\bf \Pi}=-{\bf D}_{\rm T}:=-{\bf E}_{\rm T}-{\bf P}_{\rm T1}$, which is the well-known momentum conjugate to ${\bf A}_{\rm T}$ within multipolar QED \cite{cohen-tannoudji_photons_1989,craig_molecular_1998}.

Refs.~\cite{rousseau_quantum-optics_2017,rousseau_reply_2018} conclude that when written in terms of ${\bf A}_{\rm T}$ and ${\tilde {\bf \Pi}}=-{\bf E}_{\rm T}-{\bf E}_{\rm L}$, the Poincar\'e gauge Hamiltonian is not the multipolar Hamiltonian, because ${\tilde {\bf \Pi}}_{\rm T}$ equals $-{\bf E}_{\rm T}$ rather than $-{\bf D}_{\rm T}$ and so the %Poincar\'e gauge
momentum ${\tilde {\bf \Pi}}_{\rm T}$ is {\em not} the well-known canonical momentum encountered in textbook multipolar theory. However, what is required in order that the two theories coincide is that ${\bf \Pi}_{\rm T}=-{\bf D}_{\rm T}$, and this is the case. Indeed, as we have shown, this equality is {\em implied} by the equality ${\tilde {\bf \Pi}}_{\rm T}=-{\bf E}_{\rm T}$, which therefore proves that the two theories are identical  rather than disparate.

The misunderstanding stems from a one-to-two usage of the name ``canonical momentum". In multipolar QED we call ${\bf \Pi}=-{\bf D}_{\rm T}$ the canonical momentum, because in the final unconstrained theory it is conjugate to ${\bf A}_{\rm T}$ [in the sense of Eq.~(\ref{trd})] and it commutes with ${\bf r}$ and ${\bf p}$. On the other hand, when we follow Dirac's method of quantisation (as is done in Refs.~\cite{rousseau_quantum-optics_2017,rousseau_reply_2018}) the object termed ``canonical momentum" is ${\tilde {\bf \Pi}}=-{\bf E}$, because in the starting naive (constrained) theory this momentum is conjugate to ${\bf A}$ [in the sense of Eq.~(\ref{npois})] and it commutes with ${\bf r}$ and ${\bf p}$. Thus, the same name ``canonical momentum" has been used for distinct fields that are not equal but that are instead related by Eq.~(\ref{pitilde}). Both of these nomenclatures are reasonable, but misunderstanding results from attempting to adopt them simultaneously. We must recognise that neither ${\tilde {\bf \Pi}}$ nor ${\tilde {\bf \Pi}}_{\rm T}$ equals ${\bf \Pi}$ in general.

Refs.~\cite{rousseau_quantum-optics_2017,rousseau_reply_2018} express the Poincar\'e gauge theory in terms of the Poincar\'e gauge potential ${\bf A}_1$ and the momentum ${\tilde {\bf \Pi}}$ (see for example Eq.~(12) of Ref.~\cite{rousseau_reply_2018}). The multipolar framework is the same theory expressed in terms of different fields ${\bf A}_{\rm T}$ and ${\bf \Pi}$, which are more convenient for use within the quantum theory. We have now verified this via three separate derivations in Sec.~II~A of the main text, Supplementary Note~\ref{haming}, and again above via the construction of Dirac brackets. Before now this latter demonstration had not been clearly provided within the literature. Indeed, as well as being unrecognised in Ref.~\cite{rousseau_quantum-optics_2017}, the distinction between ${\tilde {\bf \Pi}}$ and ${\bf \Pi}$ is perhaps also obfuscated elsewhere. For example, the constraint $C_2$ used above was first employed by Woolley in Ref.~\cite{woolley_r._g._charged_1999}, who then also constructs the Dirac brackets for the theory, but chooses the notation ${\bf E}^\perp$ for $-{\bf \Pi}$, despite that $-{\bf \Pi}$ does not represent the transverse electric field except when ${\bf g}_{\rm T}={\bf 0}$ (Coulomb gauge). We emphasize that the distinction between Coulomb gauge and multipolar QED is no more or less than a distinction between gauge choices.

\section{QED in terms of dual coordinates}\label{app2}

We briefly review the use of dual coordinates in QED, as for example are used in Ref.~\cite{rouse_avoiding_2021}. The dual-potential ${\bf C}_{\rm T}$ is such that
\begin{align}\label{invcurl}
{\bf C}_{\rm T}({\bf x})= -(\nabla \times)^{-1}{\bf \Pi}({\bf x}) = -\int d^3 x' {\nabla'\times {\bf \Pi}({\bf x}') \over 4\pi|{\bf x}-{\bf x}'|}.
\end{align}
In the same way that ${\bf A}_{\rm T}$ is conjugate to ${\bf \Pi}$ the potential ${\bf C}_{\rm T}$ can be viewed as a coordinate conjugate to the magnetic field ${\bf B}$, because as is easily verified
\begin{align}
[C_{{\rm T},i}({\bf x}),B_j({\bf x}')]=i\delta_{ij}^{\rm T}({\bf x}-{\bf x}').
\end{align}
Due to the non-existence of magnetic charge, as specified by $\nabla \cdot {\bf B}=0$, the magnetic quantities ${\bf A}_{\rm T}$ and ${\bf B}=\nabla \times {\bf A}_{\rm T}$ are physically unique. In contrast, due to Gauss' law $\nabla \cdot {\bf E}=\rho$, which generates gauge symmetry transformations [Eq.~(\ref{gsymm})], the electric quantities ${\bf C}_{\rm T}$ and ${\bf \Pi}=-\nabla \times {\bf C}_{\rm T}$ represent different observables in different gauges. The field canonical subsystem is defined using $({\bf A}_{\rm T},{\bf \Pi})$ or equivalently using $({\bf C}_{\rm T},{\bf B})$. Since the curl operator, $\nabla\times$, is invertible on the space of transverse fields [cf. Eq.~(\ref{invcurl})], any function of $({\bf A}_{\rm T},{\bf \Pi})$ can instead be written as a function of $({\bf C}_{\rm T},{\bf B})$ and vice versa. The $\alpha$-gauge Hamiltonian given in Eq.~(22) in the main text can be written in terms of ${\bf C}_{\rm T}$ and ${\bf B}$ using ${\bf \Pi}=-\nabla \times {\bf C}_{\rm T}$ and Eq.~(57) of the main text. Choosing the multipolar gauge, $\alpha=1$, then gives the result of Ref. \cite{rouse_avoiding_2021}.

Ref.~\cite{rouse_avoiding_2021} refers to gauge freedom as a freedom to choose ${\bf C}_{\rm L}$. This freedom has no non-trivial consequences in the absence of magnetic charge, and it is independent of the gauge freedom in ${\bf A}_{\rm L}$. The latter freedom is highly non-trivial and it is necessarily present as a fundamental feature of QED. An expression of the theory in terms of $({\bf C}_{\rm T},{\bf B})$ is always possible, but this cannot circumvent gauge freedom in ${\bf A}_{\rm L}$.  However, when written in terms of dual coordinates the dependence of the theory on ${\bf A}_{\rm L}$ is no longer explicit. The freedom within the theory is understood in terms of the ``polarisation" ${\bf P}$ and the accompanying ``magnetisation" ${\bf M}$ as defined in Sec.~\ref{polsup}. 

Ref.~\cite{rouse_avoiding_2021} argues that the potentials ${\bf P}$ and ${\bf M}$ may offer a more intuitive way to understand the relativity within the light-matter subsystem decomposition. It must however be noted that ${\bf P}_{\rm T}$ is completely arbitrary and once fixed determines ${\bf M}$, in the same way that ${\bf A}_{\rm L}$ is completely arbitrary and once fixed determines $A_0$. As noted in Supplementary Note~\ref{haming}, a gauge transformation of ${\bf P}_{\rm T}$ as defined in Eq.~(\ref{mingenb}) is necessarily accompanied by a gauge transformation in ${\bf A}_{\rm L}$ [Eq.~(\ref{mingena})].

Ref.~\cite{rouse_avoiding_2021} concludes that approximations within the multipolar gauge, $\alpha=1$, will typically most accurately represent the physics of small, bound dipoles interacting with a single mode. A wide range of system types is considered along with the effects of both material truncation and the EDA. However, as noted above, it has been found elsewhere that while the multipolar gauge may often be optimal (or very close to optimal) for performing material level truncations, this is not always the case when considering low energy properties involving more than one radiation mode or less anharmonic material dipoles \cite{stokes_gauge_2019,roth_optimal_2019}.

Most importantly, as noted in the preceding section, it is essential to recognise that gauge ambiguities are much broader than the gauge noninvariance resulting from approximations, which are always avoidable in principle. As we have shown the canonical dipole defined by $({\bf r},{\bf p})$ possesses a continuously varying level of localisation directly controlled by the gauge. The strict multipolar dipole is unphysical due to its singular nature. The interplay between localisation and dressing is directly relevant in determining measurable properties. In particular, the distinction between real and virtual photons is important and is intimately related to the choice of gauge. These points are discussed in detail in the context of time-dependent interactions in Sec.~V and photodetection theory in Sec.~VI A. It is shown that the multipolar gauge may yield especially unphysical results in photodetection theory.

\section{Derivation of the quantum optical master equation in an arbitrary gauge}\label{meapp}

We use the EDA of the arbitrary gauge Hamiltonian in Eq.~(22) of the main text, which is
\begin{align}\label{HAd}
H&=h+V^\alpha_1+V^\alpha_2\\
h&= \sum_n \epsilon^n\ket{\epsilon^n}\bra{\epsilon^n} + \int d^3 k \sum_\lambda \omega \left(a^\dagger_\lambda({\bf k})a_\lambda({\bf k})+{1\over 2}\right),\\
V^\alpha_1 &= -(1-\alpha){q\over m}{\bf p}\cdot {\bf A}_{\rm T}({\bf 0}) + \alpha q{\bf r}\cdot {\bf \Pi}({\bf 0}),\\
V^\alpha_2 &=(1-\alpha)^2{q^2\over 2m}{\bf A}_{\rm T}({\bf 0})^2 + {\alpha^2q^2\over 2}{\bf r}\cdot \delta^{\rm T}({\bf 0})\cdot {\bf r}.
\end{align}
The terms $h$, $V^\alpha_1$, and $V^\alpha_2$ are zeroth, first, and second order in $q$ respectively ($q^2/(4\pi)$ is the fine structure constant serving as a dimensionless small parameter). Despite the EDA, the theory remains gauge invariant because $R_{\alpha\alpha'}$ remains unitary.

We will view $h$ as the unperturbed Hamiltonian, whose definition we have made sure to keep independent of $\alpha$, because this is essential in order that the $S$-matrix is $\alpha$-independent (cf. Sec~V~B of the main text). In approximation 1 we assume that the system's density matrix can be written $\rho(t)\otimes \ket{0}\bra{0}$ where $\ket{0}$ is the photonic vacuum and $\rho(t)$ is the dipole state in the interaction picture with respect to $h$. In approximation 2 the Von-Neumann equation for the density matrix is integrated and iterated up to second order in $q$ to give \cite{breuer_theory_2007}
\begin{align}
{\dot\rho}(t) =& i[\rho(0),\Delta^\alpha_2(t)] \nonumber \\ &- \int_0^t ds\, {\rm tr}_{\rm ph}\left[V^\alpha_1(t),[V^\alpha_1(t-s),\rho(s)\otimes \ket{0}\bra{0}]\right]
\end{align}
where $\Delta^\alpha_2(t) := \bra{0}V^\alpha_2(t)\ket{0}$. In approximation 3 the density matrix $\rho(s)$ is approximated as $\rho(s)\approx\rho(t)$ for all $s\in[0,t]$ resulting in the time-local equation
\begin{align}
{\dot \rho}(t) =& i[\rho(t),\Delta^\alpha_2(t)] \nonumber \\ &- \int_0^t ds\, {\rm tr}_{\rm ph}\left[V^\alpha_1(t),[V^\alpha_1(t-s),\rho(t)\otimes \ket{0}\bra{0}]\right].
\end{align}
In principle, all terms can now be calculated as known functions of $t$ that are second order in $q$. In approximation 4 the limit of integration is extended; $t\to \infty$, which gives the Markovian equation
\begin{align}\label{mark}
&{\dot \rho}(t)= \nonumber \\ & i[\rho(t),\Delta^\alpha_2(t)]- \bigg[\int_0^\infty ds \bra{0}V^\alpha_1(t)V^\alpha_1(t-s)\ket{0}\rho(t)  \nonumber \\ &
-{\rm tr}_{\rm ph}\left[V^\alpha_1(t)\ket{0}\rho(t)\bra{0}V^\alpha_1(t-s)\right] +{\rm H.c.}\bigg]
\end{align}
where H.c. stands for Hermitian conjugate. Typically the $s$-integral will not converge and must be regularised. In the Schr\"odinger-picture, all master equation coefficients are now time-independent. Having used approximations 1-4 the complete positivity of the reduced evolution is not guaranteed. Complete positivity requires approximation 5 \cite{breuer_theory_2007}.

We will first deal with the unitary part of the master equation, which is given by
\begin{align}\label{rhou}
&{\dot \rho}(t)|_{\rm u}= i[\rho(t),\Delta^\alpha(t)]
\end{align}
where $\Delta^\alpha(t)=\Delta^\alpha_2(t)+ \Delta^\alpha_1(t)$ in which $\Delta^\alpha_1(t)$ comes from partitioning the coefficient of $\rho(t)$ in the second term in Eq.~(\ref{mark}) as
\begin{align}\label{0vv02}
\int_0^\infty ds \bra{0}V^\alpha_1(t)V^\alpha_1(t-s)\ket{0} = \gamma^\alpha(t)+i\Delta_1^\alpha(t).
\end{align}
The dipole operators $\gamma^\alpha(t)$ and $\Delta_1^\alpha(t)$ will be seen in the end to be separately Hermitian. We will now show that within approximation 5 we obtain
\begin{align}
\Delta^\alpha(t) = \sum_n \Delta^n\ket{\epsilon^n}\bra{\epsilon^n}
\end{align}
where $\Delta^n$ is the $\alpha$-independent on-energy-shell second order $T$-matrix element for the vacuum shift of the dipole's $n$'th energy level;
\begin{align}\label{tmd}
\Delta^n = \bra{\epsilon^n,0}V^\alpha_2\ket{\epsilon^n,0} + \sum_{m\neq n}{|\bra{e^m}V^\alpha_1\ket{\epsilon^n,0}|^2\over \epsilon_n-e_m}.
\end{align}
Here the summation is over all unperturbed states $\ket{e^m}\neq \ket{\epsilon^n,0}$. Direct calculation yields
\begin{align}\label{0vv0}
& \int_0^\infty ds \bra{0}V^\alpha_1(t)V^\alpha_1(t-s)\ket{0}=\int d^3 k\sum_\lambda \sum_{n,m,q} \nonumber \\
&~\times g_{nm\lambda} g_{mq\lambda} u_{nm\alpha}^+u_{mq\alpha}^-
\ket{\epsilon^n}\bra{\epsilon^q}e^{i\omega_{nq}t}\int_0^\infty ds\, e^{i(\omega_{qm}-\omega)s}
\end{align}
where $\omega_{nm}:=\epsilon^n-\epsilon^m$ and
\begin{align}
&g_{nm\lambda} : = {q{\bf e}_\lambda({\bf k})\cdot {\bf r}_{nm}\over \sqrt{2\omega(2\pi)^3}},\\
&u_{nm\alpha}^\pm : = \alpha\omega \pm (1-\alpha)\omega_{nm}.
\end{align}
Forcing the $s$-integral in Eq.~(\ref{0vv0}) to converge by adding damping $e^{-\eta s},~\eta \to0_+$, and using the identity
\begin{align}\label{idd}
\lim_{\eta\to 0_+}\int_0^\infty ds \,e^{is\epsilon}e^{-\eta s} = \pi\delta(\epsilon)+i{{\rm p.v.}\over \epsilon},
\end{align}
the quantity $\Delta^\alpha_1(t)$ is identified as the component of Eq.~(\ref{0vv0}) coming from the principal value (p.v.) term;
\begin{align}
&\Delta^\alpha_1(t) \nonumber \\
&= \int d^3 k\sum_\lambda \sum_{n,m,q} g_{nm\lambda} g_{mq\lambda} {u_{nm\alpha}^+u_{mq\alpha}^-\over \omega_{qm}-\omega}
\ket{\epsilon^n}\bra{\epsilon^q}e^{i\omega_{nq}t}
\end{align}
where the ${\bf k}$-integral takes its principal value. For the term $\Delta_2^\alpha(t)$ direct calculation yields
\begin{align}
\Delta_2^\alpha(t) =\int d^3 k &\sum_\lambda \bigg({(1-\alpha)^2q^2|{\bf e}_\lambda({\bf k})|^2\over 4m\omega(2\pi)^3} \nonumber \\ & +\sum_{n,m,q}\alpha^2 \omega g_{nm\lambda}g_{mq\lambda}\ket{\epsilon^n}\bra{\epsilon^q}e^{i\omega_{nq}t}\bigg)
\end{align}
and using the TRK identity
\begin{align}
{1\over 2m}\delta_{ij} = \sum_{m} \omega_{mn}r_{nm,i}r_{mn,j}
\end{align}
we obtain
\begin{align}
\Delta_2^\alpha(t) =\int d^3 k &\sum_\lambda \sum_{n,m} \bigg((1-\alpha)^2 \omega_{mn}|g_{nm\lambda}|^2\ket{\epsilon^n}\bra{\epsilon^n} \nonumber \\ & +\sum_{q}\alpha^2 \omega g_{nm\lambda}g_{mq\lambda}\ket{\epsilon^n}\bra{\epsilon^q}e^{i\omega_{nq}t}\bigg).
\end{align}
In approximation 5 the off-diagonal terms in $\Delta^\alpha_1(t)$ and $\Delta^\alpha_2(t)$ for which $q\neq n$ are assumed to be rapidly oscillating and are ignored. We thereby obtain
\begin{align} 
\Delta_2^\alpha = \int d^3 k \sum_\lambda \sum_{n,m} &|g_{nm\lambda}|^2\left[(1-\alpha)^2 \omega_{mn} +\alpha^2 \omega\right] \nonumber \\&\times \ket{\epsilon^n}\bra{\epsilon^n}\\ 
\Delta^\alpha_1 
= \int d^3 k\sum_\lambda \sum_{n,m}& |g_{nm\lambda}|^2 {[(1-\alpha) \omega_{nm} +\alpha \omega]^2\over \omega_{nm}-\omega}\nonumber \\
&\times \ket{\epsilon^n}\bra{\epsilon^n}
\end{align}
These terms give respectively the contributions of $V^\alpha_2$ and $V_1^\alpha$ to the right hand-side of Eq.~(\ref{tmd}). Their sum is therefore $\alpha$-independent and is found to be
\begin{align}
\Delta^\alpha=\Delta =&\sum_{n,m} \int d^3 k\sum_\lambda |g_{nm\lambda}|^2 {\omega \omega_{nm}\over \omega_{nm}-\omega}\ket{\epsilon^n}\bra{\epsilon^n}\nonumber \\ 
=:&\sum_n \Delta^n\ket{\epsilon^n}\bra{\epsilon^n}
\end{align}
Let us recap how this result has been obtained. Approximation 1 ensured that $\Delta^\alpha(t)$ could be calculated using the photonic vacuum at any time $t$. Approximation 2 ensured that it was of second order in $q$. Approximation 3 ensured it could be calculated independent of $\rho$. Approximation 4 ensured that the expected energy denominators were obtained as in the $T$-matrix, and approximation 5 ensured that the $T$-matrix element was evaluated on-energy-shell. It follows that the unitary part of the master equation [Eq.~(\ref{rhou})] is $\alpha$-independent. 

We now consider the dissipative part. We first calculate $\gamma^\alpha(t)$ defined by Eq.~(\ref{0vv02}), which is the remaining component of Eq.~(\ref{0vv0}) that comes from the delta-function term of Eq.~(\ref{idd});
\begin{align}
\gamma^\alpha(t)  =\int d^3 k\sum_\lambda \sum_{n,m,q} &\pi g_{nm\lambda} g_{mq\lambda} u_{nm\alpha}^+u_{mq\alpha}^-e^{i\omega_{nq}t}\nonumber \\&\times \delta(\omega_{qm}-\omega)\ket{\epsilon^n}\bra{\epsilon^q}.
\end{align}
We see immediately that approximation 4 has resulted in an evaluation of the photonic frequencies on resonance with dipolar transitions. Invoking the approximation 5 of neglecting terms for which $q\neq n$ we obtain
\begin{align}
\gamma^\alpha(t)=\gamma =\int d^3 k\sum_\lambda \sum_{n,m} &\pi |g_{nm\lambda}|^2 [(1-\alpha) \omega_{nm} +\alpha \omega]^2 \nonumber \\&\times \delta(\omega_{nm}-\omega)\ket{\epsilon^n}\bra{\epsilon^n} \nonumber \\ = \sum_{n,m}\int d^3 k\sum_\lambda \pi  |g_{nm\lambda}&|^2 \omega_{nm}^2\delta(\omega_{nm}-\omega) \ket{\epsilon^n}\bra{\epsilon^n}
\end{align}
where in the final equality all $\alpha$-dependence has dropped out due to the delta function. Within the approximations 1-5 the coefficient in the summand over dipole levels is half the rate of emission into the photonic continuum via a downward transition $\ket{\epsilon^n}\to \ket{\epsilon^m}$, the latter being exactly as is found using the corresponding $S$-matrix element. This calculation is also commonly called Fermi's golden rule \cite{craig_molecular_1998}. Evaluating the ${\bf k}$-integral and polarisation summation gives
\begin{align}
\gamma =\sum_{\substack{n,m \\ n>m}}{\Gamma_{nm}\over 2}\ket{\epsilon^n}\bra{\epsilon^n},~~~ \Gamma_{nm} = {q^2\omega_{nm}^3 |{\bf r}_{nm}|^2 \over 3\pi}.
\end{align}

The remaining part of the master equation is another dissipative part coming from the second line in Eq.~(\ref{mark}). This can be calculated in a similar fashion using Eq.~(\ref{idd}) and approximation 5. The final coefficients are again found to be the $\alpha$-independent $\Gamma_{nm}$. Collecting these results we obtain the quantum optical master equation at zero temperature \cite{breuer_theory_2007}, which is given by Eq.~(148) of the main text.

\section{Non-local connections between free photonic fields}\label{nlphot}

Relativistic quanta such as photons do not possess a position operator and cannot be localised \cite{fulling_aspects_1989,mandel_optical_1995,haag_local_1996}. Quadratic functions of ${\bf E}$ and ${\bf B}$ such as the energy density and Poynting vector, are local, but have the dimensions of energy density rather than number density. We noted in Sec.~VI of the main text that distinct gauge invariant fields ${\bf A}_{\rm T}$ and ${\bf D}_{\rm T}$ were relevant in photodetection for subsystems defined relative to the Coulomb and multipolar gauges respectively, and we noted that both of these fields are special cases of Eq.~(165) of the main text. Before calculating average values of local energy densities we briefly review the connection between some examples of fields defined by Eq.~(169) of the main text for different $\beta(\omega)$. We first consider free electrodynamics (no charges).

There are several commonly encountered operations performed on the local fields ${\bf E}$ and ${\bf B}$, which are local in ${\bf k}$-space and therefore non-local in spacetime. Specifically, $i)$ the longitudinal and transverse projection of a local field is non-local; $ii)$ the projection of a local field onto it's positive and negative frequency components is non-local in time. Moreover, causal wave propagation requires both signs of the frequency \cite{fulling_aspects_1989}, so the positive and negative frequency components of a causal field are only themselves causal within the Markovian approximation of extending frequency integrals over the whole real line \cite{milonni_photodetection_1995,stokes_vacuum_2018}; $iii)$ the (arbitrary-gauge) fields defined by Eq.~(169) of the main text corresponding to different choices of $\beta(\omega)$ are non-locally connected.

To exemplify point $iii)$ note that the Glauber intensity is non-locally connected to the naive ``photon number density" ${\bf V}^{(-)}\cdot {\bf V}^{(+)}$ defined by \cite{mandel_optical_1995}
\begin{align}
{\bf V}^{(+)}(t,{\bf x}) &= \int {d^3 k\over \sqrt{(2\pi)^3}} \sum_\lambda {\bf e}_\lambda({\bf k}) a_\lambda({\bf k})e^{i{\bf k}\cdot {\bf x}-i\omega t} \\ &=
 \int d^3 x' K({\bf x}-{\bf x}'){\bf E}^{(+)}(t,{\bf x}),\\ K({\bf x}) &= \int {d^3k\over (2\pi)^3}\sqrt{\omega} e^{i{\bf k}\cdot {\bf x}}= {3\over 8\sqrt{2\pi^3 x^7}}.
\end{align}
Because of this, if a single photon were to be considered localised around ${\bf 0}$, then its energy would be less localised, falling off as $x^{-7}$ \cite{mandel_optical_1995}. Similarly, the fields ${\bf A}_{\rm T}^{(+)}$ and ${\bf E}^{(+)}=-{\partial_t {\bf A}}^{(+)}_{\rm T}$ are related in ${\bf k}$-space by a factor of $\omega$, so the relevant integral kernel is 
\begin{align}
K({\bf x}) =  \int {d^3k\over (2\pi)^3} \omega e^{i{\bf k}\cdot {\bf x}}= -{1\over \pi^2 x^4}.
\end{align}
Given the non-local connections between free photonic fields corresponding to different $\beta(\omega)$ in Eq.~(165) of the main text it is unsurprising that the inclusion of virtual photons within the definition of a source requires non-local operations in spacetime. Further understanding is gained by analysing local energy densities, as reviewed below.

\section{Local densities}\label{locden}

The connections between free photonic fields corresponding to different $\beta(\omega)$ in Eq.~(169) of the main text are non-local, as detailed above. Given this fact, it is unsurprising that the inclusion of virtual photons within the definition of a source requires non-local operations in spacetime. Below we determine the average electromagnetic energy-momentum density in the vicinity of dipoles defined relative to different gauges.

In order to understand the interplay between local-fields, virtual processes, and subsystem gauge relativity, we now consider various energy densities in the vicinity of a dipole \cite{power_quantum_1983,power_quantum_1983-1,power_quantum_1983-2,power_quantum_1992,power_quantum_1993,power_time_1999-1,power_time_1999,passante_cloud_1985,persico_time_1987,salam_molecular_2008,salam_molecular_2009,stokes_quantum_2016,stokes_vacuum_2018}. In finding energy densities different methods are available and are suitable for different purposes. For a two-level dipole, energy densities can be found without resorting to perturbative expansion of the electromagnetic fields by instead using the rotating-wave and Markov approximations. The RWA imposes number conservation and thereby restricts to processes for which the Markov approximation $\sigma^\pm(s)\approx \sigma^\pm(t) e^{\pm i\omega_m (s-t)},~0\leq s\leq t$, can subsequently be applied when calculating canonical photonic fields. Combined with the extension of frequency integrals over the whole real line the overall result of these approximations is to enforce strict bare-energy conservation. This description captures the exponential decay of excited states, but causes virtual contributions to vanish identically.

To describe virtual contributions we evaluate expressions perturbatively, which does not imply a lower bound on the time-scales described.  In this case, in all expressions that are second order in $q$ one approximates the photonic and dipolar operators at times $s\in [0,t]$ as freely evolving within the interaction picture \cite{power_quantum_1992,salam_molecular_2008}. This does not capture the exponential decay of excited states, but it can be seen to be consistent with the non-perturbative approach applicable to a two-level dipole. Specifically, if negative frequencies are included as a form of Markov approximation, then only real contributions remain and for a two-level dipole the results obtained coincide with the short-time limit of the non-perturbative results found using the rotating-wave and Markov approximations. Below we summarise the main results, while detailed expressions are given in Supplementary Note \ref{app3}.

We begin by considering the multipolar gauge interaction $V^1={\bf d}\cdot {\bf \Pi}({\bf 0})$,~${\bf d}=q{\bf r}$. The second order interaction $\int d^3 x {\bf P}_{\rm T}({\bf x})^2/2$ can be ignored for predictions up to order $q^2$. The electric and magnetic fields are expanded up to order $q^2$ for ${\bf x}\neq {\bf 0}$;
\begin{align}
&{\bf E}={\bf D}_{\rm T} = {\bf E}_{\rm vac} + {\bf E}_{1}+{\bf E}_{2}, \label{d2}\\
&{\bf B} = {\bf B}_{\rm vac} + {\bf B}_1+{\bf B}_2.  \label{b2}
\end{align}
Recall that, as discussed in Sec.~VI~C~1 of the main text, the partitioning of the electric field into vacuum and source fields is gauge-relative, so the components on the right-hand-side of Eq.~(\ref{d2}) must be understood as being specific to the multipolar gauge (they are generated by interaction $V^1$).

We first calculate the Glauber intensity in the state $\ket{\epsilon^p,0}$ which was used in Sec.~VI~A of the main text. Only the first-order field ${\bf E}_1$ contributes, because $I_G$ is normal-ordered. The radiation component, which varies as $x^{-2}$ is, within the rotating-wave and Markov approximations, given by
\begin{align}\label{glau1}
I^{\rm rad}_G(t,{\bf x}) = \left({1\over 4\pi x}\right)^2\sum_{l<p} \omega_{pl}^4 {\bf d}_{pl}\cdot \theta \cdot {\bf d}_{lp}
\end{align}
in which strict bare-energy conservation is observed. Integrating this expression over a sphere surrounding the dipole gives
\begin{align}\label{glaureal}
\int d\Omega \, x^2 I_G(t,{\bf x})  = {1\over 2}\sum_{l<p} \omega_{pl}\Gamma_{pl}
\end{align}
which is half of the expected radiated energy-flux. The total energy flux is found using the Poynting vector \cite{power_quantum_1992,power_quantum_1993,salam_molecular_2008,salam_molecular_2009};
\begin{align}\label{Sen}
{\bf S}(t,{\bf x}) := &{1\over 2}[{\bf E}(t,{\bf x})\times {\bf B}(t,{\bf x}) -{\bf B}(t,{\bf x})\times {\bf E}(t,{\bf x})]\nonumber\\=\,&
{\bf S}^{\rm vac}(t,{\bf x}) +{\bf S}^{\rm real}(t,{\bf x}) +{\bf S}^{\rm virt}(t,{\bf x}).
\end{align}
Note that, as was found in Sec~VI~C~1, the vacuum-source partitioning of a given physical field is gauge-relative and so again the individual components on the right-hand-side of Eq.~(\ref{Sen}) differ in different gauges, each of which defines the corresponding physical ``sources" and ``vacuum" differently \cite{power_time_1999-1}. We are presently using the multipolar gauge.

The vacuum component is defined as the part that depends on the vacuum fields alone and so we will focus on the remaining source part. The real and virtual components will be defined below. Using Eqs.~(\ref{d2}) and (\ref{b2}) we see that in addition to a normally-ordered combination of first order fields as occurs in the Glauber intensity, there is also an anti-normally ordered contribution from the first order fields, and there are also correlations between the vacuum and second-order fields. The contribution of first order fields to the Poynting vector is \cite{power_quantum_1992}
\begin{align}\label{S1}
&{1\over 2}\langle{\bf E}_1(t,{\bf x})\times {\bf B}_1(t,{\bf x}) -{\bf B}_1(t,{\bf x})\times {\bf E}_1(t,{\bf x})\rangle \nonumber \\ &
=  {{\hat {\bf x}}\over (4\pi x)^2} \sum_l \omega_{pl}^4 {\bf d}_{pl}\cdot \theta \cdot {\bf d}_{lp},
\end{align}
which unlike the radiative part of $I_G$ involves summation over all dipole levels. The contribution from the vacuum source-field correlations is the sum of time-independent and time-dependent terms \cite{power_quantum_1992};
\begin{align}\label{Sv2}
&{1\over 2}\langle{\bf E}_{\rm vac}(t,{\bf x})\times {\bf B}_2(t,{\bf x}) -{\bf B}_{\rm vac}(t,{\bf x})\times {\bf E}_2(t,{\bf x})\rangle \nonumber \\ &
= {{\hat {\bf x}}\over (4\pi x)^2}\sum_l {\rm sgn}(\omega_{pl}) \omega_{pl}^4 {\bf d}_{pl}\cdot \theta \cdot {\bf d}_{lp} +\langle {\bf S}^{\rm virt}(t,{\bf x})\rangle.
\end{align}
The time-dependent term $\langle {\bf S}^{\rm virt}(t,{\bf x})\rangle$ is the contribution from virtual processes as will now be shown. The contribution from real photons $\langle {\bf S}^{\rm real}(t,{\bf x})\rangle$ is time-independent, being defined as the sum of Eq.~(\ref{S1}) and the the first term in Eq.~(\ref{Sv2}), which is \cite{power_quantum_1992}
\begin{align}\label{Sr}
\langle{\bf S}^{\rm real}(t,{\bf x})\rangle&:= {{\hat {\bf x}}\over 8\pi^2 x^2} \sum_{l<p} \omega_{pl}^4 {\bf d}_{pl}\cdot \theta \cdot {\bf d}_{lp} =2{\hat {\bf x}}I^{\rm rad}_G(t,{\bf x}).
\end{align}
The partition into real and virtual parts is justified by integrating ${\hat {\bf x}} \cdot \langle {\bf S}^{\rm real}(t,{\bf x})\rangle$ over a sphere surrounding the dipole to give
\begin{align}\label{poyntv}
\int d\Omega \, x^2 {\hat {\bf x}} \cdot \langle{\bf S}^{\rm real}(t,{\bf x})\rangle  = \sum_{l<p} \omega_{pl}\Gamma_{pl} =: P^{\rm real}
\end{align}
which is clearly the expected total radiated energy-flux (power) due to real photon emission. The time-dependent virtual component is specified in Supplementary Note \ref{app3}. It is transient and rapidly decaying in the sense that, for fixed $x$, it vanishes both when $t\gg x$, and when an infinite time-average is taken \cite{power_quantum_1992}. It also vanishes if the Markov approximation is performed. However, we saw in Sec.~VI~A of the main text that virtual contributions are not necessarily small when averaged over a finite time for sensible ultra-violet cut-offs.

To gain further insight one can calculate the average of the electromagnetic energy density, which is the sum of electric and magnetic components
\begin{align}\label{EMenden}
{\mathscr E}_{\rm EM}(t,{\bf x}) &:=  {1\over 2}\left[{\bf E}(t,{\bf x})^2+{\bf B}(t,{\bf x})^2\right] \nonumber \\ &={\mathscr E}^{\rm vac}(t,{\bf x})+{\mathscr E}_{\rm E}(t,{\bf x})+{\mathscr E}_{\rm M}(t,{\bf x})
\end{align}
where we have separated-off pure-vacuum contributions into the term ${\mathscr E}^{\rm vac}(t,{\bf x})$. Again we note that the individual components on the right-hand-side of Eq.~(\ref{EMenden}) differ in different gauges, which each define the corresponding physical ``sources" and ``vacuum" differently \cite{power_time_1999-1}. We are presently using the multipolar gauge.

Both the electric and magnetic source energy densities ${\mathscr E}_{\rm E}(t,{\bf x})$ and ${\mathscr E}_{\rm M}(t,{\bf x})$ respectively, receive contributions from the first order fields as well as from vacuum source-field correlations. Concurrently, both densities can be partitioned into a time-independent component plus a time-dependent component; ${\mathscr E}_{\rm X}(t,{\bf x}) = {\mathscr E}_{\rm X}^{\bcancel t}({\bf x})+{\mathscr E}_{\rm X}^t(t,{\bf x}),~{\rm X}={\rm E,\,M}$. The time-dependent parts are again purely virtual, but in contrast to the Poynting vector, the time-independent parts also have a virtual component. The time-independent electric energy density itself is made up of two distinct terms for which different limits are given in Supplementary Note \ref{app3}. For an excited state $p>0$ the first term dominates in the far-field, $x \gg 1/\omega_{p0}$, and corresponds to real photon emission, decaying as $1/x^2$. In contrast, for a ground state bare dipole, $p=0$, this same term vanishes, so only the second term remains. In the near-field $x\ll 1/\omega_{l0}$ this term is essentially the electrostatic energy of the dipole, while the far-field limit possesses the characteristic (Casimir-Polder) $x^{-7}$ decay. 

We are now in a position to understand how the the vacuum and source components of the electric energy density would be different if we had instead assumed a Coulomb gauge dipole prepared in the state $\ket{\epsilon^p,0}$. In the Coulomb gauge the electrostatic field is included within the definition of the dipole. Thus, the electric energy density for $t_r>0$ would be identical to that above whereas for $t_r<0$, ${\mathscr E}_{\rm E}(t,{\bf x})$ would coincide the electrostatic energy [Eq.~(\ref{tindepnear}), Supplementary Note \ref{app3}] \cite{power_time_1999-1}. Since the multipolar gauge and Coulomb gauge vacuum densities also differ by the same amount the sum ${\mathscr E}^{\rm vac}+{\mathscr E}_{\rm E}$ is unique and the same in both gauges. This is consistent with the results of Sec~VI~C~1 whereby the vacuum-source partitioning of the electric field itself differs between the Coulomb and multipolar gauges in precisely this way [Eq.~(205) of the main text].

Similar results are obtained for the time-independent part of the magnetic energy density $\langle {\mathscr E}_{\rm M}^{\bcancel t}({\bf x})\rangle$ [Supplementary Note \ref{app3}]. It possesses a non-vanishing real-photonic part for $p>0$ which dominates in the far-field via $x^{-2}$ decay, and which vanishes for $p=0$. For $p=0$, the remaining time-independent part exhibits different behaviour in the near and far zone limits similar to the electric energy density. The near-field limit varies as $x^{-5}$ rather than $x^{-6}$, while the far-field limit again decays as $x^{-7}$. Note that unlike the electric field the vacuum-source partitioning of the magnetic-field is not gauge-relative so these results for the magnetic energy density are identical in every gauge \cite{power_time_1999-1}.

The remaining components not yet discussed are the time-dependent components, $\langle{\mathscr E}_{\rm X}^t(t,{\bf x})\rangle$, for which expressions are given in Supplementary Note \ref{app3}. These parts are purely virtual and for ${\bf x}\neq {\bf 0}$, they comprise the only non-trivial contributions within the local continuity equation for energy. Poynting's theorem reads;
\begin{align}\label{poyntt0}
{d\over dt} \langle{\mathscr E}_{\rm EM}(t,{\bf x})+{\mathscr E}_{\rm d}(t,{\bf x})\rangle = -\nabla \cdot \langle{\bf S}(t,{\bf x})\rangle
\end{align}
where ${\mathscr E}_{\rm d}(t,{\bf x})$ is the energy density of the bare dipole localised at ${\bf 0}$. For ${\bf x}\neq {\bf 0}$, this becomes
\begin{align}\label{poyntt}
{d\over dt} \langle{\mathscr E}_{\rm EM}(t,{\bf x})\rangle = -\nabla \cdot \langle{\bf S}(t,{\bf x})\rangle.
\end{align}
It is noteworthy that the vacuum, real and virtual components of ${\bf S}$ separately satisfy local energy conservation as can be directly verified. For the vacuum parts, which are space and time-independent this is immediate. For the time-independent energy density such that $d\langle{\mathscr E}_{\rm EM}^{\bcancel t}({\bf x})\rangle/dt\equiv 0$ the corresponding Poynting vector is $\langle{\bf S}^{\rm real}(t,{\bf x})\rangle$, which is such that $\nabla \cdot \langle {\bf S}^{\rm real}(t,{\bf x})\rangle = 0$ for ${\bf x}\neq {\bf 0}$. Therefore, Eq.~(\ref{poyntt}) is also trivially satisfied for the real part of ${\bf S}$. The integral of the divergence of the real Poynting vector over a sphere ${\mathscr S}$ containing the dipole has already been calculated and is given by Eq.~(\ref{poyntv}). Finally, it can be verified using the expressions in Supplementary Note \ref{app3} that
\begin{align}\label{poyntvirt}
{d\over dt} \langle{\mathscr E}_{\rm E}^t(t,{\bf x}) + {\mathscr E}_{\rm M}^t(t,{\bf x})\rangle = -\nabla \cdot \langle{\bf S}^{\rm virt}(t,{\bf x})\rangle.
\end{align}

Virtual contributions violate bare-energy conservation by definition whereas global energy conservation is fundamental and is automatically satisfied; $[H,H]\equiv 0$. The stronger condition of local energy conservation, namely Eq.~(\ref{poyntt0}), is also fundamental, yet its explicit verification is more involved. The calculation above shows that virtual processes do satisfy this fundamental requirement and in this sense they are not unphysical, indicating again that the term virtual is a misnomer.

\begin{figure}[t]
\vspace*{5mm}
{
	\begin{minipage}[c][0.6\width]{
	   0.2\textwidth}
	   \centering
	   \includegraphics[width=1.10\textwidth]{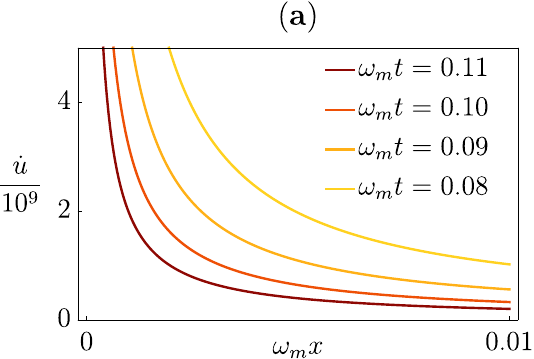}
	\end{minipage}}
 \hfill 	
{
	\begin{minipage}[c][0.5\width]{
	   0.2\textwidth}
	   \centering
	   \hspace*{-13mm}\includegraphics[width=1.14\textwidth]{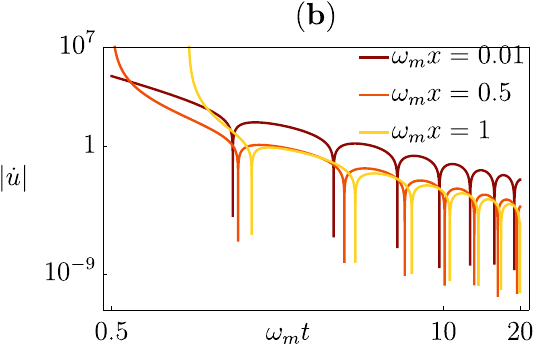}
	\end{minipage}}
 \hfill	
{
	\begin{minipage}[c][1.2\width]{
	   0.3\textwidth}
	 \centering 
	 \hspace*{-26mm}\includegraphics[width=1.95\textwidth]{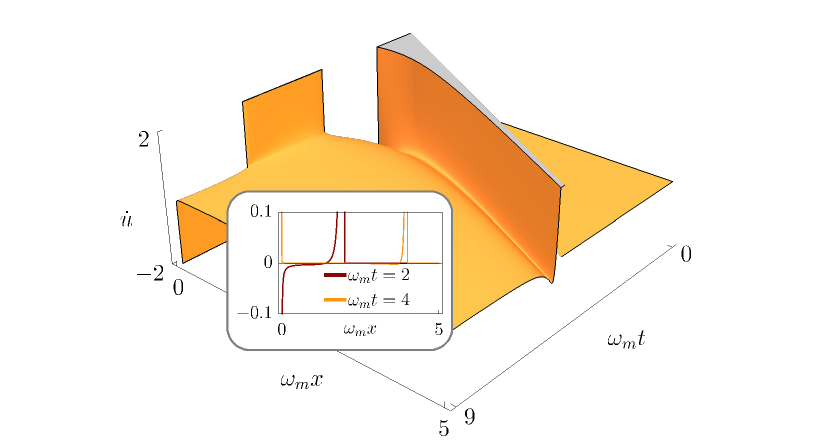} {\scriptsize(\textbf{c})} 
	\end{minipage}}
\caption{(\textbf{a}) ${\dot u}(t,x)$ is plotted for slightly increasing values of $t$ within one bare dipole cycle of the initial time $t=0$, indicating a rapidly localised virtual field around $x=0$. (\textbf{b}) Logarithmic plot of ${\dot u}(t,x)$ with time for three near-field values of $x$, showing the rapidly increasing localisation of the corresponding bound field in the near-zone $\omega_m x < 1$ as well as its oscillations in time. (\textbf{c}) Plot of both the oscillating bound field and a causally propagating outgoing pulse from the origin $(t,{\bf x})=(0,{\bf 0})$, which is localised on the light-cone. The inset shows cross-sections in the $xz$-plane corresponding to distinct time-slices separated by 3 bare dipole cycles.}\label{edotg}
\end{figure}

For simplicity we now restrict our attention to the lowest two-dipole levels with energy difference $\omega_m$, and calculate the variations in the time-dependent part of the energy density on the surface of the sphere with radius $x<t$ surrounding the bare dipole in its ground state. This is found to be
\begin{align}
{\dot u}(t, x):=&{2\pi \over \omega_m^3 P^{\rm real}} \int d\Omega {d\over dt} \langle{\mathscr E}_{\rm E}^t(t,{\bf x}) \rangle \nonumber \\  =& {8\theta(t_r) \over \omega_m x q_rq_a}\left[2q_a\cos q_r+(q_a^2-2)\sin q_r\right]
\end{align}
where we have chosen a spacetime-independent normalisation $\omega^3_m P^{\rm real}$ to obtain a dimensionless measure, and where $q_r=\omega_m(t-x)$ and $q_a=\omega_m(t+x)$. The quantity $u(t,x)$ is a normalised electromagnetic energy density associated with time-dependent processes at a distance $x$ from the source, that has been averaged over all directions. The variations in $u(t,x)$ are plotted in Fig.~\ref{edotg}. There is a causally propagating pulse localised on the light-cone consistent with the assumption of an initial unperturbed state, which is not an energy eigenstate. There is also a highly oscillatory component that is highly localised at the position of the dipole. This is consistent with an interpretation of the bare dipole as undergoing rapid virtual emission and absorption processes. The extreme localisation and oscillations of the virtual bound field suggest that it be interpreted as an inseparable component of the physical dipole.

The gauge choice affects the extent to which bound virtual energy is included within the definition of the dipole. This can be seen by noting that time-dependent virtual contributions are not present if one instead considers a full energy eigenstate. The (unnormalised) eigenvector $\ket{E_1^p}$ of the full multipolar Hamiltonian corresponding to the unperturbed state $\ket{\epsilon^p,0}$ is found using second order perturbation theory and is given in Supplementary Note \ref{app3}. The average Poynting vector in the state $\ket{E_1^p}$ is found to be \cite{power_quantum_1993}
\begin{align}
\bra{E_1^p} {\bf S}(t,{\bf x})\ket{E_1^p} = \langle {\bf S}^{\rm real}(t,{\bf x}) \rangle 
\end{align}
where the right-hand-side is defined as in Eq.~(\ref{Sr}). This vanishes for $p=0$ showing that there can be no energy loss in the ground state \cite{power_quantum_1993}. On the other hand the electric energy density in the vicinity of the dipole does not vanish for the ground state and is given in Supplementary Note \ref{app3}. Thus, the differences between the ground state predictions found using $\ket{E_1^0}$ and those found using the bare ground state $\ket{\epsilon^0,0}$ are the time-dependent components $\langle {\bf S}^{\rm virt}\rangle$ and $\langle{\mathscr E}_{\rm EM}^t\rangle$.

If the system is prepared in the state $\ket{\epsilon^0,0}$ in the multipolar gauge, then $\langle{\mathscr E}^{\bcancel t}_{\rm EM}\rangle$ represents the only part of the full average $\langle {\mathscr E}_{\rm EM}\rangle$ that is not typically neglected, that is, the purely virtual part $\langle{\mathscr E}^t_{\rm EM}\rangle$ is often ignored \cite{power_quantum_1992}. Within this approximation we have $\langle {\mathscr E}_{\rm EM}\rangle = \langle {\mathscr E}_{\rm EM}\rangle_0 = \langle {\mathscr E}^{\bcancel t}_{\rm EM}\rangle$ consistent with the results of Sec.~V~C~2 in the main text, in which the dipole's stationary state was found within the conventional weak-coupling approximations to be the bare ground state.

Let us now again consider the example of the harmonic dipole as in Sec.~VI~A of the main text. In this case the full ground state is represented by the vacuum $\ket{0_d,0_c}$ of light and matter modes $c$ and $d$ defined relative to the JC-gauge (see Sec.~VI~A~2 of the main text). In the JC-gauge the canonical momentum ${\bf \Pi}({\bf x})$ represents the physical observable $O_{\rm JC}({\bf x}) = -{\bf D}_{\rm T}({\bf x})+{\bf P}_{\rm T}({\bf x})-{\bf P}_{\rm TJC}({\bf x})$ where to order $q^2$
\begin{align}\label{ptiljc}
{\bf P}_{\rm T}({\bf x})-{\bf P}_{\rm TJC}({\bf x}) = \sum_{{\bf k}\lambda} {{\bf e}_{{\bf k}\lambda}\over v} {({\bf e}_{{\bf k}\lambda}\cdot {\bf d}) \omega\over \omega+\omega_m}e^{i{\bf k}\cdot {\bf x}}.
\end{align}
The electric field is therefore given for ${\bf x}\neq {\bf 0}$ by ${\bf E} = -{\bf \Pi}+{\bf P}_{\rm T}-{\bf P}_{\rm TJC}$. Writing ${\bf \Pi}$ in terms of the modes $c_k$ we find the average using the vector $\ket{0_d,0_c}$ to be
\begin{align}
\langle {\bf \Pi}({\bf x})\rangle_0 =\sum_{k,j} q^2 {({\bf e}_k \cdot {\bf e}_j) \over 2m\omega_mv^2} {\omega_m \omega_k \omega_j e^{i({\bf k}+{\bf j}) \cdot {\bf x}}\over (\omega_k+\omega_j)(\omega_m+\omega_k)(\omega_m+\omega_j)}
\end{align}
where we have labelled the modes using a single index as in Sec.~VI~A~2 of the main text. Using Eq.~(\ref{ptiljc}), one can show that, as required, $\langle {\mathscr E}_{\rm E}\rangle_0 = \langle {\bf \Pi}^2 +[{\bf P}_{\rm T}-{\bf P}_{\rm TJC}]^2\rangle_0/2$ coincides with the electric energy density in Eq.~(\ref{enden}) when assuming a harmonic dipole. We see therefore that the average $\langle {\mathscr E}_{\rm EM}\rangle$ is found to coincide with $\langle{\mathscr E}^{\bcancel t}_{\rm EM}\rangle =\langle {\mathscr E}_{\rm EM}\rangle_0$, without neglecting (as a weak-coupling or Markov approximation) any time-dependent component $\langle{\mathscr E}^t_{\rm EM}\rangle$, provided that the subsystems are defined relative to the JC-gauge. The prepared state then coincides with the true ground state.

Above we have compared the same physical observable, namely the electromagnetic (EM) energy density, while assuming different initial physical states. These different states are ones of well-defined energy of different physical subsystems. When choosing the multipolar gauge, the vector $\ket{\epsilon^0,0}$ specifies a localised ``bare" dipole$_1$ in its own lowest energy state and with no accompanying photons$_1$. In this case the EM energy density possesses a virtual time-dependent component $\langle{\mathscr E}^t_{\rm EM}\rangle$. The same vector $\ket{\epsilon^0,0}$ in the Coulomb gauge specifies an electrostatically dressed dipole$_0$ with no accompanying photons$_0$. Thus, the same EM energy density is obtained as in the multipolar gauge, with the exception that for $t_r<0$ we obtain a non-vanishing electrostatic energy density given by Eq.~(\ref{tindepnear}). In the case of a harmonic dipole, the vector $\ket{0_d,0_c}$ specifies a state of well-defined energy of the subsystems defined relative to the JC-gauge and this coincides with the ground state. In this case the EM energy density is again the same but possesses no time-dependent virtual component. This is consistent with an interpretation of the JC-gauge subsystems as having subsumed the virtual ground state quanta that exist in conventional gauges.

The most {\em relevant} physical predictions will depend on which of these physical states is closest to that which has been prepared in the experiment considered. This, in turn, will depend on the extent to which the purely virtual field that results in the energy density $\langle{\mathscr E}^t_{\rm EM}\rangle$, is separate from the addressable dipole. The transient and highly localised nature of this field suggest that it should be considered part of the dipole on the accessible time and length scales. However, this may depend on the available preparation and measurement procedures. Similarly, in stronger-coupling regimes, whether or not the ground state is entangled and contains a large number of photons depends on the gauge-relative to which the subsystems are defined, i.e., on the relative extent to which the virtual bound-field is taken as separate from the physical (measurable) dipole.

\section{Second order energy predictions for a free dipole}\label{app3}

To supplement the analysis above we provide explicit expressions for averages obtained using the first and second order electric and magnetic (source) fields for a free dipole in the multipolar gauge. The electric source fields are
\begin{align}
&E_{1,i}(t,{\bf x}) = {\theta(t_r)\over 4\pi}\sum_{l,n} \ket{\epsilon^n}\bra{\epsilon^l}d_{nl}^j\omega_{nl}^3 f_{ij}(\omega_{nl}x)e^{-i\omega_{nl}t_r} \\
&E_{2,i}^{(+)}(t,{\bf x}) = {i\theta(t_r)\over 4\pi}\sum_{{\bf k}\lambda}\sqrt{\omega \over 2v}\sum_{l,n,p} \ket{\epsilon^l}\bra{\epsilon^p} a_{{\bf k}\lambda} e_{{\bf k}\lambda}^s\Bigg[ \nonumber \\ & \left( {d^j_{ln}d^s_{np}\over \omega_{np}-\omega}+{d^s_{ln}d^j_{np}\over \omega_{nl}+\omega}\right)\omega_{np}^3f_{ij}([\omega_{lp}+\omega]x)e^{i(\omega_{lp}-\omega)t_r} \nonumber \\ 
&- {d^j_{ln}d^s_{np}\over \omega_{np}-\omega}\omega_{nl}^3 f_{ij}(\omega_{nl}x)e^{i\omega_{ln}t_r} \nonumber \\ &-{d^s_{ln}d^j_{np}\over \omega_{nl}+\omega}\omega_{pn}^3 f_{ij}(\omega_{pn}x)e^{i\omega_{np}t_r} \Bigg]
\end{align}
where the modes have been discretised in a volume $v$. The magnetic counterparts are given by the same expressions with $f_{ij}$ replaced by $g_{ij}$. These tensor components are defined by
\begin{align}
&f_{ij}(\omega x) : = -{\theta_{ij}\over \omega x} + \phi_{ij}\left(-{i\over (\omega x)^2} -+{1\over (\omega x)^3}\right), \\
&g_{ij}(\omega x): = \varphi_{ij}\left({1\over \omega x} + {i\over (\omega x)^2}\right)
\end{align}
where for convenience we have defined
\begin{align}
&\theta_{ij} = \delta_{ij}-{\hat x}_i{\hat x}_j ,\\ 
&\phi_{ij} = \delta_{ij}-3{\hat x}_i{\hat x}_j ,\\
&\varphi_{ij} = -\epsilon_{ijk}{\hat x}_k.
\end{align}
We remark that due to the causality constraint imposed by the function $\theta(t_r)$ within the source fields, all results that follow are to be understood as holding for $t_r>0$.

We provide averages in the multipolar bare state $\ket{\epsilon^p,0}$ with no photons$_1$ and the dipole$_1$ in the state $\ket{\epsilon^p}$ where $p$ is arbitrary. The time-dependent virtual component of the Poynting vector is
\begin{align}
\langle S_s^{\rm virt}(t,{\bf x})\rangle = &-{i\epsilon_{sij}\over 4(2\pi)^3} \int_0^\infty d\omega\,\omega^3 \nonumber \\
&\times \big[{\rm Re} (g_{jq}(\omega x)e^{i\omega x}) A^{iq}_p(t,x,\omega)e^{i\omega t} \nonumber \\ &- {\rm Im}(f_{iq}(\omega x)e^{i\omega x})C^{jq}_p(t,x,\omega)e^{-i\omega t}\big]\nonumber \\&+{\rm c.c.}
\end{align}
where c.c. stands for complex conjugate and
\begin{align}
&A^{iq}_p(t,x,\omega) :=\sum_n \Bigg[{d^l_{pn} d^q_{np} \over \omega_{np}-\omega}\omega_{np}^3 f_{il}(\omega_{np}x) e^{i\omega_{pn}t_r} \nonumber \\ &\qquad\qquad\qquad  +{d^q_{pn} d^l_{np} \over \omega_{np}+\omega}\omega_{pn}^3 f_{il}(\omega_{pn}x)e^{-i\omega_{pn}t_r}\bigg], \\
&C^{jq}_p(t,x,\omega) :=\sum_n \Bigg[{d^q_{pn} d^l_{np} \over \omega_{np}-\omega}\omega_{np}^3 g^*_{jl}(\omega_{np}x)e^{-i\omega_{pn}t_r}\nonumber \\ &\qquad\qquad\qquad  +{d^l_{pn} d^q_{np} \over \omega_{np}+\omega}\omega_{pn}^3 g^*_{jl}(\omega_{pn}x)e^{i\omega_{pn}t_r}\bigg]. 
\end{align}
The time-independent component of the electric energy density component is \cite{power_quantum_1992,power_quantum_1993}
\begin{align}\label{tindepee}
\langle {\mathscr E}_{\rm E}^{\bcancel t}&({\bf x})\rangle =
{1\over 16\pi^2}\sum_{l<p} d^j_{pl}d^q_{lp} \omega_{pl}^6 f_{ij}^*(\omega_{pn}x)f_{iq}(\omega_{pn}x)\nonumber \\+&{1\over 16\pi^3}\sum_l \omega_{pl} d^j_{pl}d^q_{lp}\int_0^\infty du\, {u^6 e^{-2 ux} \over u^2+\omega_{pl}^2}f_{ij}(iux)f_{iq}(iux)
\end{align}
while the time-dependent component is
\begin{align}\label{teed}
\langle {\mathscr E}^t_{\rm E}(t,{\bf x})\rangle=&{i\over 4(2\pi)^3} \int_0^\infty d\omega\,\omega^3 \nonumber \\
&\times {\rm Im}(f_{iq}(\omega x)e^{i\omega x})A^{iq}_p(t,x,\omega)e^{i\omega t} +{\rm c.c.}
\end{align}

For $p>0$ in the far-field the first term dominates and gives
\begin{align}
&\langle {\mathscr E}^{\bcancel t}_{\rm E}({\bf x})\rangle \nonumber \\ &= {1\over 16 \pi^2} \sum_{l<p} \omega_{pl}^6\Bigg[{\bf d}_{pl} \cdot \theta \cdot {\bf d}_{lp}\left({1\over (\omega_{pl}x)^2}-{2\over (\omega_{pl}x)^4}\right) \nonumber \\ & \qquad \qquad+{\bf d}_{pl} \cdot \varphi \cdot {\bf d}_{lp}\left({1\over (\omega_{pl}x)^4}+{1\over (\omega_{pl}x)^6}\right)\Bigg] \nonumber \\
&\approx  {1\over 16 \pi^2 x^2} \sum_{l<p} \omega_{pl}^4{\bf d}_{pl} \cdot \theta \cdot {\bf d}_{lp},\qquad \omega_{p0}x\gg 1.
\end{align}
corresponding to real photon emission. For $p=0$ the first term vanishes and the second term in the near-field limit is
\begin{align}\label{tindepnear}
&\langle {\mathscr E}^{\bcancel t}_{\rm E}({\bf x})\rangle_0= {1\over 32\pi^2 x^6}\sum_l {\bf d}_{0l}\cdot \phi^2 \cdot {\bf d}_{l0},\qquad \omega_{l0} x\ll 1
\end{align}
which is the electrostatic energy. The far-field limit is
\begin{align}\label{tindepeefar}
&\langle {\mathscr E}^{\bcancel t}({\bf x})\rangle_0={1\over 64\pi^3 x^7}\sum_{l\neq 0} {d_{l0}^id_{0l}^j \over \omega_{l0}}(13\delta_{ij}+7{\hat x}_i{\hat x}_j) ,~~ \omega_{l0}x\gg 1
\end{align}
possessing the Casimir-Polder $x^{-7}$-decay.

Similarly to the above, for the magnetic energy density we find
\begin{align}
\langle {\mathscr E}_{\rm M}^{\bcancel t}&({\bf x})\rangle =
{1\over 16\pi^2}\sum_{l<p} d^j_{pl}d^q_{lp} \omega_{pl}^6 g_{ij}^*(\omega_{pn}x)g_{iq}(\omega_{pn}x)\nonumber \\+&{1\over 16\pi^3}\sum_l \omega_{pl}d^j_{pl}d^q_{lp}\int_0^\infty du\, {u^6 e^{-2 ux}\over u^2+\omega_{pl}^2}g_{ij}(iux)g_{iq}(iux)
\end{align}
whose behaviour with $x$ was described in Supplementary Note~\ref{locden}, and
\begin{align}\label{tmed}
\langle{\mathscr E}^t_{\rm M}(t,{\bf x})\rangle=&-{i\over 4(2\pi)^3} \int_0^\infty d\omega\,\omega^3 \nonumber \\
&\times {\rm Re}(g_{iq}(\omega x)e^{i\omega x})C^{iq}_p(t,x,\omega)e^{-i\omega t} +{\rm c.c.}
\end{align}

To provide further understanding consider the stationary case for which the composite system is in a global Hamiltonian eigenstate. The (unnormalised) eigenvector $\ket{E_1^p}$ of the full multipolar Hamiltonian corresponding to the unperturbed state $\ket{\epsilon^p,0}$ is found using second order perturbation theory as
\begin{align}\label{sos}
&\ket{E_1^p} =T\ket{\epsilon^p,0}= (1+T_1+T_2)\ket{\epsilon^p,0},\\ 
&T_1\ket{\epsilon^n,0}  = \sum_{i\neq p,{\bf k}\lambda}\ket{\epsilon^i,{\bf k}\lambda} {|\bra{\epsilon^i,{\bf k}\lambda}V^1\ket{\epsilon^p,0}|^2 \over \omega+ \omega_{ip}}, \\ 
&T_2\ket{\epsilon^p,0} =  \sum_{\substack{i,j\neq p,\\ {\bf k}\lambda,{\bf k}'\lambda'}}\ket{\epsilon^i,{\bf k}\lambda,{\bf k}'\lambda'} \nonumber \\ &~~\times {\bra{\epsilon^i,{\bf k}\lambda,{\bf k}'\lambda'}V^1\ket{\epsilon^j,{\bf k}\lambda}\bra{\epsilon^j,{\bf k}\lambda}V^1\ket{\epsilon^i,{\bf k}\lambda,{\bf k}'\lambda'}\over (\omega+\omega_{jp})(\omega+\omega'+\omega_{ip})}.
\end{align}

The electric energy density in the vicinity of the dipole does not vanish for the ground state and is found to be \cite{power_quantum_1993}
\begin{align}\label{enden}
&\bra{E_1^0}{\mathscr E}_{\rm E}\ket{E_1^0}=\langle {\mathscr E}_{\rm E}\rangle_0 = \langle {\mathscr E}_{\rm E}^{\bcancel t}({\bf x})\rangle |_{p=0} \nonumber \\ 
&={1\over 16\pi^3}\sum_l \omega_{0l} d^j_{0l}d^q_{l0}\int_0^\infty du\, {u^6 e^{-2 ux} \over u^2+\omega_{0l}^2}f_{ij}(iux)f_{iq}(iux)
\end{align}
where $\langle {\mathscr E}_{\rm E}^{\bcancel t}({\bf x})\rangle$ is defined in Eq.~(\ref{tindepee}). Thus, as discussed in Supplementary Note~\ref{locden}, the differences between the ground state predictions found using $\ket{E_1^0}$ and those found using the bare ground state $\ket{\epsilon^0,0}$ are the time-dependent components $\langle {\bf S}^{\rm virt}\rangle$ and $\langle{\mathscr E}_{\rm EM}^t\rangle$.

\section{Radiation damping}\label{app4}

Consider a Hamiltonian $H=h+V$ and let eigenvalues and eigenvectors of $h$ be denoted $\omega_n$ and $\ket{n}$ respectively. We introduce the Fourier transform
\begin{align}\label{G}
b_f(t) &:=\bra{f}U(t,0)\ket{i}e^{i\omega_f t}  \nonumber \\ &=-{1\over 2\pi i}\int d\omega \, G_{fi}(\omega)e^{i(\omega_f-\omega)t},
\end{align}
where $U(t,0)=e^{-iHt}$ and $G(\omega)={\cal G}(\omega+i\eta)$ in which ${\cal G}(z)=1/(z-H)$ is the Hamiltonian resolvent. The limit $\eta \to 0^+$ is understood in Eq.~(\ref{G}) and all subsequent equations. The Schr\"odinger equation yields the amplitude equation
\begin{align}\label{bfdot}
i{\dot b}_f(t) =\sum_m V_{fm} b_m(t)e^{i\omega_{fm}t}
\end{align}
to which we seek a solution for $t>0$ subject to the initial conditions $\lim_{t\to 0^+}b_m(t) =\delta_{mi}$. Equivalently we require a solution to the equation
\begin{align}\label{bfdot2}
i{\dot b}_f(t) = \sum_m V_{fm} b_m(t)e^{i\omega_{fm}t} + i\delta_{fi}\delta(t)
\end{align}
for all $t\in {\mathbb R}$, subject to the condition that the amplitudes be normalised to zero for negative times; $\forall m,~b_m(t) = 0$ whenever $t<0$. This corresponds to imposing a sudden jump from $0$ to $1$ of the amplitude $b_i$ at time $t=0$.

Substituting the representation
\begin{align}
i\delta(t) = -{1\over 2\pi i}\int d\omega\, e^{i(\omega_i - \omega)t}
\end{align}
and Eq.~(\ref{G}) into Eq.~(\ref{bfdot2}) one finds that
\begin{align}\label{Grel1}
(\omega-\omega_n)G_{ni}(\omega) = \delta_{ni} +\sum_m V_{nm}G_{mi}(\omega),
\end{align}
which implies that
\begin{align}\label{Grel2}
G_{ni}(\omega) = \zeta(\omega-\omega_n)\left[\delta_{ni} +\sum_m V_{nm}G_{mi}(\omega)\right]
\end{align}
where $\zeta$ is a distribution defined by
\begin{align}
\zeta(x):=\lim_{\eta \to 0^+} {1\over x +i\eta} = {\mathcal P}{1\over x} -i\pi\delta(x)
\end{align}
in which ${\mathcal P}$ indicates the principal value. Next we define for all $n\neq i$ the matrix elements $R_{ni}(\omega)$ through the relation
\begin{align}\label{Rni}
G_{ni}(\omega) = R_{ni}(\omega)L_i(\omega)\zeta(\omega-\omega_n).
\end{align}
Substituting Eq.~(\ref{Rni}) into the right-hand-side of Eq.~(\ref{Grel2}), imposing the condition $n\neq i$, and finally equating the resulting expression with the right-hand-side of Eq.~(\ref{Rni}), gives
\begin{align}\label{Rni2}
R_{ni}(\omega) = V_{ni} + \sum_{m\neq i} V_{nm}R_{mi}(\omega)\zeta(\omega-\omega_m),~~~~~~n\neq i.
\end{align}
Meanwhile, Eq.~(\ref{Grel1}) with $n=i$, and Eq.~(\ref{Rni}) imply that
\begin{align}
&(\omega -\omega_i)L_i(\omega) \nonumber \\ &= 1 + V_{ii}L_i(\omega) + \sum_{m\neq i} V_{im} R_{mi}(\omega)L_i(\omega)\zeta(\omega -\omega_m)
\end{align}
from which it follows that
\begin{align}\label{Gi}
L_i(\omega) = {1\over \omega -\omega_i - R_i(\omega)}
\end{align}
where
\begin{align}
R_i(\omega):= V_{ii} + \sum_{m\neq i} V_{im}R_{mi}(\omega)\zeta(\omega-\omega_m).
\end{align}
We split this quantity into real and imaginary parts as
\begin{align}\label{reim}
R_i(\omega) = \Delta_i(\omega)-{i\over 2}\Gamma_i(\omega).
\end{align}
The imaginary part is found from
\begin{align}
&\Gamma_i(\omega) = i[R_i(\omega)-R_i^*(\omega)] \nonumber \\ 
&= i\sum_{m\neq i}[V_{im}R_{mi}(\omega)\zeta(\omega-\omega_m) +V_{mi}R_{mi}^*(\omega)\zeta(\omega_m-\omega)].
\end{align}
Noting that $\zeta^*(x)=-\zeta(-x)$ we can add zero to the right-hand-side of this equation in the form
\begin{align}
0= \sum_{m,p\neq i} \big[&V_{mp}R_{pi}(\omega)R_{mi}^*(\omega)\zeta(\omega-\omega_p)\zeta(\omega_m-\omega) \nonumber \\ &+ V^*_{mp}R^*_{pi}(\omega)R_{mi}(\omega)\zeta^*(\omega-\omega_p)\zeta(\omega-\omega_m) \big],
\end{align}
such that upon using $\zeta(x)+\zeta(-x)=-2\pi i \delta(x)$ and Eq.~(\ref{Rni2}) we obtain
\begin{align}\label{gami}
\Gamma_i(\omega) =2\pi\sum_{m\neq i} | R_{mi}(\omega)|^2\delta(\omega-\omega_m).
\end{align}
The real part $\Delta_i(\omega)$ is given by
\begin{align}\label{delt}
&\Delta_i(\omega) =V_{ii} \nonumber \\ &+ {1\over 2}\sum_{m\neq i}[V_{im}R_{mi}(\omega)\zeta(\omega-\omega_m) -V_{mi}R_{mi}^*(\omega)\zeta(\omega_m-\omega)].
\end{align}
Alternatively, we can define $f(z) = f_1(z)+if_2(z)$ with $f_1(z)=\Delta_i(z)-V_{ii}$ and $f_2(z)=-\Gamma_i(z)/2$ viewed as functions of complex variable $z$. If we assume that $f(z)$ is analytic in the upper-half plane and vanishes sufficiently fast as $|z|\to \infty$, then we can apply the Kramers-Kronig relation
\begin{align}
f_1(z) = {1\over \pi}{\cal P}\int_{-\infty}^\infty d\omega' {f_2(\omega')\over \omega'-z},
\end{align}
such that by making use of Eq.~(\ref{gami}) we obtain
\begin{align}\label{deltaom}
\Delta_i(\omega) =V_{ii}+{\mathcal P} \sum_{m\neq i} {| R_{mi}(\omega_m)|^2 \over \omega-\omega_m}.
\end{align}
Thus, $R_i(\omega)$ can be expressed entirely in terms of the $R_{mi}(\omega)$ in Eq.~(\ref{Rni2}).

We are now in a position to obtain an expression for the amplitude $b_f(\infty)$ purely in terms of the $R_{mi}(\omega)$. The amplitude can therefore be expressed to arbitrary order in the interaction $V$ by applying Eq.~(\ref{Rni2}) iteratively. We begin by substituting Eq.~(\ref{Rni}) into Eq.~(\ref{G}), which yields for $f\neq i$
\begin{align}\label{bf1}
b_f(t) = -{1\over 2\pi i} \int d\omega \, R_{fi}(\omega)L_i(\omega)\zeta(\omega-\omega_f)e^{i(\omega_f-\omega)t}.
\end{align}
Since we require $b_f(0)=0$ for $f\neq i$ we require that
\begin{align}\label{rfi0}
{1\over 2\pi i}\int d\omega \, R_{fi}(\omega)L_i(\omega)\zeta(\omega-\omega_f) = 0.
\end{align}
We can therefore add the left-hand-side of Eq.~(\ref{rfi0}) to the right-hand-side of Eq.~(\ref{bf1}), which gives
\begin{align}\label{bf2}
&b_f(t) \nonumber \\ &=-{1\over 2\pi i} \int d\omega \,R_{fi}(\omega)L_i(\omega)\zeta(\omega-\omega_f)\left[e^{i(\omega_f-\omega)t}-1\right].
\end{align}
The delta-function contribution from the $\zeta$-distribution in Eq.~(\ref{bf2}) is zero, because the complex exponential term in square brackets vanishes at $\omega=\omega_f$. Furthermore,
\begin{align}
\lim_{t\to\infty} {e^{ixt}-1 \over x} = \zeta(x)
\end{align}
and so
\begin{align}
b_f(\infty) = -{1\over 2\pi i}{\mathcal P}\int d\omega \,R_{fi}(\omega)L_i(\omega)\zeta(\omega_f-\omega).
\end{align}
Subtracting Eq.~(\ref{rfi0}) from this expression gives
\begin{align}
&b_f(\infty) \nonumber \\ &= -{1\over 2\pi i}{\mathcal P}\int d\omega \,R_{fi}(\omega)L_i(\omega)\left[\zeta(\omega_f-\omega) + \zeta(\omega-\omega_f)\right],
\end{align}
and since $\zeta(x)+\zeta(-x)=-2\pi i \delta(x)$ we find using Eq.~(\ref{Gi}) that
\begin{align}\label{bff}
b_f(\infty) = R_{fi}(\omega_f)L_i(\omega_f) = {R_{fi}(\omega_f) \over \omega_{fi} - R_i(\omega_f)}.
\end{align}
This is the expression for $b_f(\infty)$ in terms of the $R_{mi}$ that we sought. Finally using Eq.~(\ref{reim}) the associated probability can be written
\begin{align}\label{prob}
| b_f(\infty)|^2 = {| R_{fi}(\omega_f)|^2 \over [\omega_{fi}-\Delta_i(\omega_f)]^2+(\Gamma_i(\omega_f)/2)^2}.
\end{align}
In Sec.~VI~B of the main text we make use of this equation together with Eqs.~(\ref{deltaom}), (\ref{gami}) and (\ref{Rni2}), which define $\Delta_i$, $\Gamma_i$ and $R_{fi}$ respectively.

\section{Gedanken experiment for weak measurements in cavity QED}\label{weakm}

The explicit modelling of measurements of light and matter subsystems via a pointer system was considered in the form of simple gedanken experiments by Compagno et al. \cite{compagno_detection_1988,compagno_dressed_1988,compagno_dressed_1990,compagno_bare_1991,compagno_atom-field_1995}. Such models indicate how measurement procedures might be related to subsystem dressing. To review the weak measurement concept we consider first a bare two-level system coupled to a ``macroscopic" pointer with large mass $M$ and position and momentum $r$ and $p$ with $[r,p]=i$ \cite{peres_quantum_2002}. The position of the pointer is assumed to provide information about the energy of the two-level system. Hence, the Hamiltonian is taken to be
\begin{align}\label{smint}
H = \omega_m \sigma^z + {p^2\over 2M} + \eta(t)p\sigma^z
\end{align}
where $\sigma^z = [\sigma^+,\sigma^-]/2$ and where $\eta(t)$ is a dimensionless system-pointer coupling envelope determining the speed and duration of the interaction, which is assumed to vanish at the initial and final times. We take an initially uncorrelated state of the system and pointer with a sharp Gaussian distribution of pointer positions, with standard deviation $\sigma$ and centre at $r=0$.Compagno et al. assume an instantaneous switching function \cite{compagno_atom-field_1995} $\eta(t) = { {\mathfrak r} \over t_P} [\theta(t)-\theta(t-t_P)]$ where $t_P$ is the measurement duration after which the pointer's position is observed. The parameter ${\mathfrak r}$ has the dimensions of $r$. The evolution operator is $U(t) = e^{-iht}e^{-i{\mathfrak r} \sigma^z p}$ where $h=p^2/(2M)+\omega_m\sigma^z$ generates free evolution. The diagonal matrix elements in the position basis of the corresponding reduced pointer state at time $t$ are
\begin{align}\label{pointerP}
&{\mathsf P}_t(r):=\bra{r}\rho_I(t)\ket{r} \nonumber \\ &= {1\over \sqrt{2\pi}\sigma}\left(p_0 e^{-(r+{\mathfrak r}/2)^2/(2\sigma^2)}+p_1 e^{-(r-{\mathfrak r}/2)^2/(2\sigma^2)}\right)
\end{align}
where $p_1 = |c_1|^2$ and $p_0=1-p_1 = |c_0|^2$ are the excited and ground state probabilities. Thus, the system pointer coupling splits the initial single Gaussian peak into two Gaussian peaks at $\pm {\mathfrak r}/2$ with relative heights that give the probabilities to find the two-level system excited or not excited. In this sense, the pointer measures the energetic state of the two-level system. 

Ignoring the free evolution, the average $\langle r(t)\rangle$ and variance $\llangle r(t) \rrangle:= \langle r(t)^2\rangle-\langle r(t)\rangle^2$ of the pointer position at time $t$ are easily found
to be
\begin{align}
\langle r(t) \rangle &= {\mathfrak r}\left(p_1-{1\over 2}\right) =  {\mathfrak r}\langle\sigma^z\rangle_0,\label{pointpos0} \\
\llangle r(t) \rrangle &= {\mathfrak r}^2p_1(1-p_1) +\sigma^2= {\mathfrak r}^2\left({1\over 4}-\langle\sigma^z\rangle_0^2\right)+\sigma^2\label{varP}
\end{align}
where on the right-hand-sides $\langle\cdot\rangle_0$ denotes averaging in the initial state and $\sigma^z\equiv \sigma^z(0)$. We may assume that the initial Gaussian state is sharp with vanishingly small variance, $\sigma \to 0$, such that the final term $\sigma^2$ in Eq.~(\ref{varP}) can be ignored.

When $p_1=0$ (ground state dipole) there is a peak in the distribution of pointer positions at $\langle r(t) \rangle = -{\mathfrak r}/2$ and there are no other peaks, consistent with $\llangle r(t) \rrangle=0$. Similarly, when $p_1=1$ (excited dipole) there is a peak at $\langle r(t) \rangle = +{\mathfrak r}/2$ and there are again no other peaks; $\llangle r(t) \rrangle=0$. For $p_1=1/2$ we have $\langle r(t) \rangle = 0$ and $\llangle r(t) \rrangle={\mathfrak r}^2/4$, corresponding to symmetric peaks at $\pm {\mathfrak r}/2$, which indicate equal probabilities that the detector will register the dipole in either of its two states.

Compagno et al. considered the same dipole-pointer interaction and the same initial pointer state in the case of a two-level dipole$_1$ coupled to a single radiation$_1$-mode with polarisation ${\bf e}$ and frequency $\omega$ in volume $v$, starting in the ground state of the dipole-mode system \cite{compagno_atom-field_1995}. More generally, we can consider light and matter subsystems defined relative to an arbitrary-gauge specified by $\alpha(\omega)$. To order $q^2$ we obtain
\begin{align}
\langle r(t) \rangle &= {\mathfrak r} \langle \sigma^z\rangle, \label{pointpos}\\
\llangle r(t) \rrangle &= {\mathfrak r}^2 \left({1\over 4}-\langle\sigma^z\rangle^2\right){\rm sinc}^2 \left[{1\over 2}(\omega_m+\omega)t_P\right],
\end{align}
where ${\rm sinc} \,x : = \sin(x)/x$ and
\begin{align}
\langle \sigma^z\rangle \equiv \langle \sigma^z(0)\rangle =-{1\over 2}+ {|{\bf e}\cdot {\bf d}|^2 \over 2v} {\omega_m u^+(\omega)^2 \over (\omega_m + \omega)^2}\label{sigzpoint}
\end{align}
in which $u^+(\omega)$ is the coefficient defined in Eq.~(180) of the main text for the counter-rotating terms within the bilinear $\alpha(\omega)$-gauge interaction Hamiltonian. 

We see that the choice of gauge determines the physical model for the pointer, which is implicitly assumed to couple to the dipole quantum subsystem defined relative to the $\alpha(\omega)$-gauge. In particular, the gauge choice determines the extent to which the pointer is defined as being able to register ground state virtual photons, which arise from counter-rotating terms. In the JC-gauge the ground state of the dipole-mode system simply comprises a ground state dipole$_{\rm JC}$ and no photons$_{\rm JC}$ such that  $u^+(\omega)\equiv 0$ and therefore $\langle r(t) \rangle =- {\mathfrak r} /2$ and $\llangle r(t) \rrangle =0$ for all times. These are identical to the previous results for an uncoupled ground-state dipole [the $p_1=0$ cases of Eqs~(\ref{pointpos0}) and (\ref{varP})]. Thus, the relative strength of the counter-rotating terms within the interaction, as specified by $u^+(\omega)$, determines the relative deviation from the case of a ground state dipole$_{\rm JC}$, which looks to the pointer exactly the same as an uncoupled ground state dipole. In this sense $\alpha(\omega)$ directly controls the degree of virtual dressing explicitly registered by the pointer.

For $u^+(\omega)\neq 0$ the pointer position's variance is time-dependent due to the addition of the ${\rm sinc}$-function, which represents the (bare) energy-time uncertainty relation as encountered in Sec.~VI~A of the main text. This means that as well as the dipole's definition, the dressing registered by the pointer also depends on the measurement duration compared with bare cycle times. For short measurements compared with a bare cycle,  $t_P (\omega_m+\omega)\ll 1$, the average and variance again reduce to the uncoupled dipole result, such that the dipole$_{\alpha(\omega)}$ is perceived as bare by the pointer. For long measurements $t_P (\omega_m+\omega)\gg 1$ the variance vanishes, indicating a single peak in the distribution of pointer positions, but not one located at $-{\mathfrak r}/2$ as for an uncoupled dipole, instead the peak's position is determined by Eqs.~(\ref{pointpos}) and (\ref{sigzpoint}). This will be the same as the uncoupled dipole case only for gauges sufficiently close to the JC-gauge. In this way, the extent to which the dipole appears to the pointer as being the same as an uncoupled bare dipole is controlled by the balance between $u^+(\omega)$ and the measurement duration $t_P$. For a given $u^+(\omega)$ longer measurements result in increasing deviation from the uncoupled dipole case, while for fixed $t_P$ a larger $u^+(\omega)$ similarly results in increased deviation.

The weak measurement formalism is general in that it is obviously not restricted to any particular subsystem or observable. It can be used to model the measurement of arbitrary light or matter subsystem observables. However, a generic feature of weak measurements in QED is that a gauge must be selected relative to which the subsystem that the pointer couples is defined. Each gauge then provides a description of a different physical measurement process. Furthermore, since the system-pointer coupling is by assumption controllable, it is modelled via an explicitly time-dependent coupling, and so the considerations of Sec.~V~D of the main text apply. Specifically, the assumption that the system-pointer coupling is time-dependent is not a gauge invariant assumption. Distinct models resulting when this assumptions is made in distinct gauges describe different experimental protocols and will yield different predictions even for the pointer's measurement of the same physical observable. We end by remarking that the extension of the simple framework presented in this section to ultrastrong-coupling regimes and specific experimental contexts warrants further study.

\section{Microscopic descriptions of cavity interactions}\label{cavabs}

\subsubsection{Perfect cavity}\label{perfect}

Before considering leakage from an imperfect cavity it will be useful to briefly review the idealised case of a cavity comprised of perfectly conducting surfaces. We follow the approach in Ref.~\cite{power_quantum_1982}. For an empty cavity, with volume $v$ the transverse vector potential ${\bf A}_{\rm T}$ and its conjugate momentum ${\bf \Pi}=-{\bf E}_{\rm T}={\dot {\bf A}}_{\rm T}$ possess the mode expansions
\begin{align}
&{\bf A}_{\rm T}({\bf x}) = {1\over \sqrt{v}}\sum_{{\bf k}\lambda}q_{{\bf k}\lambda}{\bf f}_{{\bf k}\lambda}({\bf x}),\label{atcav}\\
&{\bf \Pi}({\bf x}) = {1\over \sqrt{v}}\sum_{{\bf k}\lambda}p_{{\bf k}\lambda}{\bf f}_{{\bf k}\lambda}({\bf x}),\label{picav}
\end{align}
where $q_{{\bf k}\lambda}$ and $p_{{\bf k}\lambda}$ are canonical mode quadratures satisfying $[q_{{\bf k}\lambda},p_{{\bf k}'\lambda'}]=i\delta_{\lambda\lambda'}\delta_{{\bf kk}'}$. The mode functions ${\bf f}_{{\bf k}\lambda}$ are normalised
\begin{align}
{1\over v}\int_v d^3 x\, {\bf f}_{{\bf k}\lambda}({\bf x})\cdot {\bf f}_{{\bf k}'\lambda'}({\bf x}) = \delta_{\lambda\lambda'}\delta_{{\bf kk}'}.
\end{align}
Maxwells equations and the boundary conditions for a perfect conductor imply \cite{power_quantum_1982}
\begin{align}
&\nabla \cdot {\bf f}_{{\bf k}\lambda}({\bf x})=0,\\
&\nabla \times [\nabla \times {\bf f}_{{\bf k}\lambda}({\bf x})]+|{\bf k}|^2 {\bf f}_{{\bf k}\lambda}({\bf x}) = {\bf 0},\\
& {\hat {\bf n}}\times {\bf f}_{{\bf k}\lambda}({\bf x})|_{\cal B} = {\bf 0}.\label{boundary}
\end{align}
where ${\hat {\bf n}}$ is normal to the boundary ${\cal B}$. The canonical commutation relation reads \cite{power_quantum_1982}
\begin{align}
[A_{{\rm T},i}({\bf x}),\Pi_j({\bf x}')] &= {i\over v}\sum_{{\bf k}\lambda}f_{{\bf k}\lambda,i}({\bf x})f_{{\bf k}\lambda,j}({\bf x}')\nonumber \\ &=i\left[\delta_{ij}^{\rm T}({\bf x}-{\bf x}')+\Delta_{ij}({\bf x},{\bf x}')\right].\label{cavcom}
\end{align}
where
\begin{align}
\Delta_{ij}({\bf x},{\bf x}') = -\partial_i\partial'_j\int d^3  z {\theta({\bf z},{\bf x})\over 4\pi|{\bf x}'-{\bf z}|}.\label{Del}
\end{align}
The steps leading to the second equality in Eq.~(\ref{cavcom}) can be found in Ref.~\cite{power_quantum_1982}. The term dependent on $\Delta({\bf x},{\bf x}')$ is the contribution of the boundary. The function $\theta$ is zero for ${\bf z}$ inside the cavity and is also such that
\begin{align}
\chi({\bf x},{\bf x}') = {1\over 4\pi|{\bf x}-{\bf x}'|} + \int_{\bar v} d^3  z {\theta({\bf z},{\bf x}')\over 4\pi|{\bf x}-{\bf z}|}
\end{align}
vanishes for ${\bf x}\in {\cal B}$. We have used the notation ${\bar v}$ to indicate integration over all points outside the cavity. Note that the commutator in Eq.~(\ref{cavcom}) is not purely transverse because $\Delta({\bf x},{\bf x}')$ is the gradient of a function with respect to both ${\bf x}$ and ${\bf x}'$.

As an illustrative example consider a rectangular parallelepiped and consider a point ${\bf x}$ near the reflecting wall located at $z=0$. The reflection in the wall of point ${\bf x}'$ is $\sigma{\bf x}'$ where $\sigma = {\rm diag}(1,1,-1)$. The function $\theta$ may be defined as 
\begin{align}\label{the1}
\theta({\bf z},{\bf x}') = -\delta({\bf z}-\sigma{\bf x}')
\end{align}
such that
\begin{align}\label{chicav}
\chi({\bf x},{\bf x}') = {1\over 4\pi|{\bf x}-{\bf x}'|} - {1\over 4\pi |{\bf x}-\sigma{\bf x}'|}.
\end{align}
and
\begin{align}
[A_{{\rm T},i}({\bf x}),\Pi_i({\bf x}')] = i\delta_{ij}^{\rm T}({\bf x}-{\bf x}') -i\sigma_{jk}\delta_{ki}^{\rm T}({\bf x}-\sigma{\bf x}').
\end{align}
Essentially the same result can be derived by using the explicit forms of the mode functions for a rectangular box. 

Let us now consider an atom comprised of a charge $-q$ at fixed position ${\bf R}$ and charge $q$ at position ${\bf r}_-$ inside the cavity. The charge density is $\rho({\bf x}) = -q\delta({\bf x}-{\bf R})+q\delta({\bf x}-{\bf r}_-)$. The Hamiltonian is the total energy and in the Coulomb gauge reads
\begin{align}
H_0 = &{1\over 2m}\left[{\bf p}-q{\bf A}_{\rm T}({\bf r}_-)\right]^2 + V + H_c
\end{align}
where ${\bf A}_{\rm T}$ and ${\bf \Pi}=-{\bf E}_{\rm T}$ are given by Eqs.~(\ref{atcav}) and (\ref{picav}), the cavity Hamiltonian is
\begin{align}
H_c={1\over 2}\int_v  d^3 x \left[{\bf \Pi}^2+{\bf B}^2\right] ={1\over 2v} \sum_{{\bf k}\lambda} \left[p^2_{{\bf k}\lambda}+\omega^2 q^2_{{\bf k}\lambda}\right],
\end{align}
and the Coulomb energy is given by
\begin{align}\label{vfull}
V&={1\over 2}\int_v d^3 x \,{\bf E}_{\rm L}^2 =- {1\over 2}\int_v d^3 x\, \phi \nabla^2 \phi \nonumber \\ &= \int_v d^3 x \int_v d^3 x' {\rho({\bf x})\rho({\bf x}')\over 8\pi|{\bf x}-{\bf x}'|} + \int_v d^3 x \int_{\bar v} d^3 z {\rho({\bf x}){\bar \rho}({\bf z})\over 8\pi|{\bf x}-{\bf z}|}\nonumber \\
&=V_{\rm charges}+V_{\rm image}.
\end{align}
Here ${\bf E}_{\rm L}=-\nabla\phi$, $-\nabla^2\phi =\rho + {\bar \rho}$ and
\begin{align}
&\phi({\bf x}) = \int_v d^3 x' {\rho({\bf x})\over 4\pi|{\bf x}-{\bf x}'|} + \int_{\bar v} d^3 z {{\bar \rho}({\bf z})\over 4\pi|{\bf x}-{\bf z}|},\label{phifull}\\
&{\bar \rho}({\bf z}) = \int_v d^3 x \, \theta({\bf z},{\bf x})\rho({\bf x}).
\end{align}
The first term in Eq.~(\ref{phifull}) is the usual Coulomb potential of the charges while the second term  arises from the image distribution ${\bar \rho}$. This term is required to imply that the Coulomb potential vanishes at the boundary. For a conducting wall at $z=0$ the function $\theta$ is given in Eq.~(\ref{the1}) yielding the image charge density ${\bar \rho}({\bf z}) = q\delta({\bf z}-\sigma{\bf R})-q\delta({\bf z}-\sigma{\bf r}_-)$. This density consists of charges opposite to those in $\rho$ and located at the corresponding reflected positions. It is clear that the associated electrostatic field will cancel the electrostatic field produced by $\rho$ at all points on the reflecting boundary. The Coulomb potential and Coulomb energies are then given by
\begin{align}
&4\pi\phi({\bf x}) =\nonumber  \\&{q\over |{\bf x}-{\bf r}_-|}- {q\over |{\bf x}-{\bf R}|}+ {q\over |{\bf x}-\sigma{\bf R}|}- {q\over |{\bf x}-\sigma{\bf r}_-|},\label{coulpcav}\\
&V_{\rm charges} = V_{\rm self} - {q^2\over 4\pi|{\bf r}_--{\bf R}|},\\
&8\pi V_{\rm image} =\nonumber \\ & \left[{q^2\over |{\bf r}_--\sigma{\bf R}|}+{q^2\over |{\bf R}-\sigma{\bf r}_-|}-{q^2\over |{\bf R}-\sigma{\bf R}|}-{q^2\over |{\bf r}_--\sigma{\bf r}_-|}\right]\label{image1}.
\end{align}
It is easily verified that ${\hat {\bf z}}\times \nabla\phi({\bf x})|_{\cal B}=0$ and since ${\hat {\bf z}}\times {\bf E}_{\rm T}({\bf x})|_{\cal B}=0$ is satisfied due to Eq.~(\ref{boundary}), the required Maxwell boundary condition ${\hat {\bf z}}\times {\bf E}({\bf x})|_{\cal B}=0$ is satisfied.

For parallel conducting walls at $z=0$ and $z=L$ the situation is more complicated, because the images corresponding to the reflection of charges in one of the walls will result in an imbalance of charges on either side of the other wall. A proliferation of images is required in order that the electric field vanishes at both boundaries, such that the potential must be expressed as an infinite series
\begin{align}
&4\pi\phi({\bf x}) = \sum_{n=-\infty}^\infty \nonumber  \\&\left[{q\over |{\bf x}-{\bf r}_{-n}^+|}- {q\over |{\bf x}-{\bf R}_n^+|}+ {q\over |{\bf x}-{\bf R}_n^-|}- {q\over |{\bf x}-{\bf r}_{-n}^-|}\right]
\end{align}
where ${\bf y}_n^\pm :={\bf y}\pm 2nL{\hat {\bf z}}$ with ${\bf y}={\bf r}_-,\, {\bf R}$. The previous results in Eqs.~(\ref{coulpcav})-(\ref{image1}) are recovered if the effects of the wall at $z=L$ can be neglected, for example, if the atom is far from this wall and near to the wall at $z=0$.

To define the Hamiltonian in gauge $g$ we use the generalised PZW transformation
\begin{align}
U_{0g} = \exp\left[-i\int_v d^3x\, {\bf P}_g({\bf x})\cdot {\bf A}_{\rm T}({\bf x})\right]
\end{align}
where ${\bf P}_g$ is defined in Eq.~(36) of the main text. Note that the transformation involves the total polarisation and the integration is taken over the cavity volume. This is significant because as already noted the commutator in Eq.~(\ref{cavcom}) is not purely transverse. The momentum ${\bf p}$ transforms such that 
\begin{align}
&U_{0g}[{\bf p}-q{\bf A}_{\rm T}({\bf r}_-)]U_{0g}^\dagger = {\bf p}-q{\bf A}_g({\bf r}_-),
\end{align}
where
\begin{align}
{\bf A}_g({\bf x}) = {\bf A}_{\rm T}({\bf x}) +\nabla \int_v d^3 x' {\bf g}({\bf x}',{\bf x})\cdot {\bf A}_{\rm T}({\bf x}').
\end{align}
Using $[A_{{\rm T},i}({\bf x}),H_c] = i\Pi_i({\bf x})$ and Eq.~(\ref{cavcom}) the cavity Hamiltonian is found to transform as
\begin{align}
&U_{0g}H_c U_{0g}^\dagger = H_c - i \int_v d^3 x\, P_{g,i}({\bf x})[A_{{\rm T},i}({\bf x}),H_c] \nonumber \\ &- {1\over 2} \int_v d^3x \,d^3x'\, P_{g,i}({\bf x})P_{g,j}({\bf x}')[A_{{\rm T},j}({\bf x}'),[A_{{\rm T},i}({\bf x}),H_c]] \nonumber \\
&\qquad \qquad ~=H_c +H_{\rm pol} + H_{{\bf P},\rm self}+H_{\Delta}\label{cavtrans}
\end{align}
where
\begin{align}
H_{\rm pol}&=\int_v d^3x\, {\bf P}_g({\bf x})\cdot {\bf \Pi}({\bf x}),\\
H_{{\bf P},\rm self} &= {1\over 2}\int_v d^3x\, d^3 x'\, {\bf P}_g({\bf x})\cdot \delta^{\rm T}({\bf x}-{\bf x}')\cdot {\bf P}_g({\bf x}'),\\
H_\Delta &= {1\over 2}\int_v d^3x\, d^3 x'\, {\bf P}_g({\bf x})\cdot \Delta({\bf x},{\bf x}')\cdot {\bf P}_g({\bf x}').
\end{align}
The terms $H_{\rm pol}$ and $H_{{\bf P},\rm self}$ are the cavity analogs of the usual ``${\bf P}\cdot {\bf \Pi}$" light-matter interaction term and transverse polarisation material self-energy term respectively. The final term $H_\Delta$ is a contribution due to the boundary. For a localised polarisation field ${\bf P}_g$ that vanishes at the cavity boundary, as occurs in the multipolar gauge and as is considered in Ref.~\cite{power_quantum_1982}, use of Eq.~(\ref{Del}) together with $-\nabla \cdot {\bf P}_g = \rho$ and integration by parts gives $H_\Delta = -V_{\rm image}$. This term therefore cancels exactly $V_{\rm image}$ appearing in the Coulomb gauge Hamiltonian. The multipolar Hamiltonian therefore involves no image-charge contributions and reads
\begin{align}\label{multcav}
H_1 =& {1\over 2m}\left[{\bf p}-q{\bf A}_1({\bf r}_-)\right]^2 + V_{\rm charges} + H_{{\bf P}_1,\rm self}\nonumber\\ &+H_{\rm pol,1}+H_c. 
\end{align}
where ${\bf A}_1$ and ${\bf P}_1$ are the multipolar vector potential and polarisation fields given by
\begin{align}
{\bf A}_1({\bf x}) &= -\int_0^1 d\lambda\, \lambda{\bf x}\times {\bf B}(\lambda {\bf x}),\\
{\bf P}_1({\bf x}) &= q\int_0^1d\lambda\, {\bf r}\delta({\bf x}-{\bf R}-\lambda{\bf r}),
\end{align}
in which ${\bf r}={\bf r}_--{\bf R}$. The cavity boundaries are accounted for entirely through the use of the appropriate mode-functions in Eq.~(\ref{multcav}). For the case of a rectangular parallelepiped the cavity mode functions are \cite{power_quantum_1982,milonni_quantum_1994,loudon_quantum_2000}
\begin{align}
f_{{\bf k}\lambda,1}({\bf x}) = \sqrt{8}{\bf e}_{\lambda,1} \cos k_1 x_1 \sin k_2 x_2 \sin k_3 x_3,\\
f_{{\bf k}\lambda,2}({\bf x}) = \sqrt{8}{\bf e}_{\lambda,2} \sin k_1 x_1 \cos k_2 x_2 \sin k_3 x_3,\\
f_{{\bf k}\lambda,3}({\bf x}) = \sqrt{8}{\bf e}_{\lambda,3} \sin k_1 x_1 \sin k_2 x_2 \cos k_3 x_3.
\end{align}
where ${\bf e}_\lambda,\,\lambda=1,2$ are unit polarisation vectors orthogonal to ${\bf k}$.

We conclude by remarking that in the EDA the multipolar Hamiltonian reads
\begin{align}
H_1={\tilde H}_m + {\bf d}\cdot {\bf \Pi}({\bf R}) + H_c
\end{align}
where ${\bf d}=q{\bf r}$ and
\begin{align}
{\tilde H}_m ={{\bf p}^2\over 2m}-{q^2\over 4\pi|{\bf r}|}+H_{{\bf P}_1,\rm self}
\end{align}
is the renormalised atomic Hamiltonian. Conventional models of cavity-matter coupling use the Coulomb and multipolar gauges within the EDA. However, image charges are often not taken into account, therefore, only within the multipolar gauge will such a model possess the same form as is given by the more fundamental derivation above.

\subsubsection{Leakage from an empty cavity}\label{leaky}

We now consider the case in which one of the cavity mirrors is semi-transparent, allowing light to leak out. We first consider the case of an empty cavity and for simplicity we will consider a one-dimensional model. There are a number of approaches to describing this situation, including scattering methods \cite{dutra_quantized_2000,dutra_cavity_2004}, psuedomodes \cite{barnett_quantum_1988,dalton_theory_2001}, and the theory of QED in material media \cite{viviescas_field_2003}. A well-known and simple phenomenological description is given by the so-called Gardiner-Collett Hamiltonian \cite{gardiner_input_1985}, in which the cavity modes are assumed to couple via a number-conserving interaction to external modes.
Leakage can subsequently be studied using input-output relations \cite{gardiner_input_1985,gardiner_quantum_2004}. Here we briefly review a description of leakage that uses an intuitive scattering method combined with the local conservation of energy \cite{dutra_cavity_2004}. We review the conditions under which the final result obtained reduces to the phenomenological Gardiner-Collett Hamiltonian \cite{dutra_cavity_2004,barnett_quantum_1988}.

We consider a one-dimensional cavity consisting of a perfectly reflecting wall at $x=-L$ and a semi-transparent wall at $x=0$. The imperfect mirror reflects a plane wave $e^{-i\omega x}$ incoming from $x=\infty$ as $re^{i\omega x}$, in which $r$ is called the reflection coefficient. It transmits the wave as $t e^{-i\omega x}$, where $t$ is the transmission coefficient. These coefficients satisfy $|r|^2+|t|^2=1$ and $rt^*+r^*t=0$, implying that the semi-transparent mirror does not incur any absorption. The perfect mirror at $x=-L$ reflects a plane wave $e^{-i\omega x}$ as $-e^{2i\omega L}e^{i\omega x}$ and does not transmit ($t=0$). From these elementary considerations it is possible to deduce the following mode expansions for the electric and magnetic fields inside and outside the cavity \cite{dutra_cavity_2004}
\begin{align}
&E_<(x) = -i\int_0^\infty d\omega \,\sqrt{2\omega}{\cal L}(\omega)a(\omega)\sin(\omega [x+L]) \nonumber \\ &\qquad \qquad +{\rm H.c.},\label{cave1}\\
&B_<(x) = -\int_0^\infty d\omega \,\sqrt{2\omega}{\cal L}(\omega)a(\omega)\cos(\omega [x+L]) \nonumber \\ &\qquad \qquad+{\rm H.c.},\\
&E_>(x) =\nonumber \\  &-\int_0^\infty d\omega \, \sqrt{2\omega} \, a(\omega)\left(e^{i\omega x} [r-te^{i\omega L}{\cal L}(\omega)] +e^{-i\omega x} \right) \nonumber \\ &\qquad \qquad +{\rm H.c.},\\
&B_>(x) =\nonumber \\  &-\int_0^\infty d\omega \, \sqrt{2\omega} \, a(\omega)\left(e^{i\omega x} [r-te^{i\omega L}{\cal L}(\omega)] -e^{-i\omega x} \right) \nonumber \\ &\qquad \qquad +{\rm H.c.},\label{cavb2}
\end{align}
where $[a(\omega),a^\dagger(\omega')]=\delta(\omega-\omega')$ and
\begin{align}\label{reso}
{\cal L}(\omega) &= te^{i\omega L}  \sum_{n=0}^\infty \left[-re^{2i\omega L} \right]^n = {t\over e^{i\omega L}+re^{-i\omega L}}\nonumber \\ &= te^{i\omega L}\left[{1\over 2}+\sum_{n=-\infty}^\infty {\cal L}_n(\omega)\right]
\end{align}
in which
\begin{align}
&{\cal L}_n(\omega) = {i\over 2(\omega L- n\pi) + \phi+\pi-i\ln |r|}, \\
&r=|r|e^{i\phi}.
\end{align}
The symbols $>$ and $<$ refer to the fields for $x>0$ (outside the cavity) and $x<0$ (inside the cavity) respectively. Eq.~(\ref{reso}) shows that $|{\cal L}(\omega)|^2$ is a sum of Lorentzians, the centres of which define the cavity resonances.

The modes defined by $a(\omega)$ are global, being sufficient to specify the fields both inside and outside the cavity. We may therefore define the Hamiltonian as
\begin{align}\label{Hcavst}
H = \int_0^\infty d\omega\, \omega\left[a^\dagger(\omega)a(\omega)+{1\over 2}\right].
\end{align}
The local energy and momentum densities inside and outside the cavity are given by
\begin{align}
&U_a(x)={1\over 2}\left[E_a(x)^2+B_a(x)^2\right],\\
&S_a(x) = {1\over 2}\left[E_a(x) B_a(x) - E_a(x) B_a(x)\right]\label{Sacav}
\end{align}
where $a$ labels the inside and outside fields. Poynting's theorem (local conservation of energy), ${\dot U}(t,x) = -\partial_x S(t,x)$, can be used to deduce the energy within the semi-transparent mirror. The rate of change of this energy is \cite{dutra_cavity_2004}
\begin{align}\label{poyntcav2}
{\dot H}_{\rm s-t}(t) = \lim_{\epsilon \to 0^+} \int_{-\epsilon}^\epsilon dx\, {\dot U}(t,x) = S_<(0)-S_>(0).
\end{align}
Using Eqs.~(\ref{cave1})-(\ref{cavb2}) within Eq.~(\ref{Sacav}), substituting the result into the right-hand-side of Eq.~(\ref{poyntcav2}), and making use of $a(t,\omega)=e^{-i\omega t}a(0,\omega)$ where $a(\omega):=a(0,\omega)$, yields upon integrating with respect to time, an expression for $H_{\rm s-t}(t)$ in terms of $a(\omega)$ and $a(\omega)^\dagger$. As a consistency check, one may then sum the energies inside and outside the cavity, and add the energy of the semi-transparent mirror, as $H=\int_{-L}^0 dx\, U_<(x) + \int_0^\infty dx\, U_>(x) + H_{\rm s-t}$, which yields Eq.~(\ref{Hcavst}), as required \cite{dutra_cavity_2004}.

The above derivation yields an expression for the Hamiltonian in terms of global modes, Eq.~(\ref{Hcavst}), as well as associated mode expansions for the electric and magnetic fields inside and outside the cavity. This allows {\em local} quantities of interest to be computed in terms of the global modes. The treatment is exact. We now look to ascertain under what conditions the exact result will coincide with a phenomenological Gardiner-Collett model of the form \cite{gardiner_input_1985}
\begin{align}\label{HGC}
H_{\rm G-C}=& \sum_n \omega_n a_n^\dagger a_n + \int_0^\infty d\omega\, \omega b^\dagger(\omega)b(\omega) \nonumber \\ &+ \left[\int_0^\infty d\omega \sum_{n=1}^\infty g_n(\omega)a_n b^\dagger(\omega)+{\rm H.c.}\right]
\end{align}
where $a_n$ and $b(\omega)$ are bosonic annihilation operators that define the cavity and external modes respectively. Sufficient conditions to obtain such a model are ascertained in Ref.~\cite{dutra_cavity_2004}, whose treatment we now briefly review.

We wish to provide a description in terms of complete sets of internal and external modes. Completeness requires that we must be able to expand the internal and external electric and magnetic fields in terms of these modes as
\begin{align}
&E_<(x)= 2\sum_{n=1}^\infty \sqrt{\omega_n}p_n\sin(\omega_n[x+L]),\\
&B_<(x)= -2\sum_{n=1}^\infty \sqrt{\omega_n}q_n\cos(\omega_n[x+L]),\\
&E_>(x)= 2\int_0^\infty d\omega \,\sqrt{\omega}p(\omega)\sin(\omega[x+L]),\\
&B_>(x)= -2\int_0^\infty d\omega \,\sqrt{\omega}q(\omega)\cos(\omega[x+L]),
\end{align}
where $q_n = (a_n^\dagger +a_n)\sqrt{2}$ and $p_n = i(a_n^\dagger - a_n)/\sqrt{2}$ are cavity mode quadratures satisfying $[q_n,p_m]=i\delta_{nm}$, and $q(\omega) = [b^\dagger(\omega)+b(\omega)]\sqrt{2}$ and $p_n = i[b^\dagger(\omega)-b(\omega)]\sqrt{2}$ are external mode quadratures satisfying $[q(\omega),p(\omega')]=i\delta(\omega-\omega')$. Comparison of these mode expansions with Eqs.~(\ref{cave1})-(\ref{cavb2}) then yields expressions for $a_n,\,a_n^\dagger,\,b(\omega)$ and $b^\dagger(\omega)$ in terms of $a(\omega)$ and $a^\dagger(\omega)$ \cite{dutra_cavity_2004}.

An expression for the Hamiltonian in terms of the internal and external mode operators can be deduced via the procedure of Fano diagonalisation \cite{barnett_methods_1997,barnett_quantum_1988} as follows. If the internal and external modes form a complete set then it must be possible to express the global mode operator $a(\omega)$ as a sum of discrete and continuous parts;
\begin{align}
&a(\omega) = a_{\rm d}(\omega) + a_{\rm c}(\omega),\label{atotcav}\\
&a_{\rm d}(\omega) = \sum_{n=1}^\infty \left[\alpha^1_n(\omega)a_n + \alpha^2_n(\omega) a_n^\dagger \right],\\
&a_{\rm c}(\omega) = \int_0^\infty d\omega' \, \left[\beta^1(\omega,\omega')b(\omega')+\beta^2(\omega,\omega')b^\dagger(\omega')\right].
\end{align}
Expressions for the coefficients $\alpha^j,\,\beta^j,\, j=1,2$, can be obtained by using the expressions for $a_n$ and $b(\omega)$ in terms of $a(\omega)$ and $a^\dagger(\omega)$, to compute the commutators of $a(\omega)$ with $a_n$ and $b(\omega)$. For example, 
\begin{align}
\alpha^1_n(\omega) &= [a(\omega),a_n^\dagger] \nonumber \\ &=\sqrt{\omega L \over\pi \omega_n}{\rm sinc}([\omega-\omega_n]L)e^{-i\omega L}{\cal L}(\omega)^*,
\end{align}
with similar expressions holding for the remaining coefficients \cite{dutra_cavity_2004}. If Eq.~(\ref{atotcav}) is valid, then by substituting the expressions obtained for the coefficients $\alpha^j,\,\beta^j,\,j=1,2$ into Eq.~(\ref{atotcav}) and using the result to compute the commutators $[a(\omega),a^\dagger (\omega')]$ and $[a(\omega),a(\omega')]$, one should obtain the results $\delta(\omega-\omega')$ and $0$ respectively. This however, is not the case, implying that Eq.~(\ref{atotcav}) does not hold in general. The reason for this is that, unlike the global mode operators, the internal and external mode operators cannot describe configurations of the system at the boundary. However, it can be shown that the internal and external mode operators are sufficient to describe high-$Q$ cavities. The latter are defined by the property that $|r|$ is close to unity and $|t|$ is close to zero. By letting \cite{dutra_cavity_2004}
\begin{align}
&r=\sqrt{1-\epsilon^2},\\
&t=i\epsilon,
\end{align}
where $\epsilon \ll 1$, then we can expand expressions in powers of $\epsilon$ and retain only the leading order terms, which should be sufficient to describe the case of a high-$Q$ cavity. In this way, one obtains the following approximate expressions correct to first order in $\epsilon$ \cite{dutra_cavity_2004}
\begin{align}
&{\cal L}(\omega) = -{\epsilon \over 2L}\sum_{n=0}^\infty {1 \over \omega-\omega_n+i\epsilon^2/(4L)},\\
&\alpha^1_n(\omega) ={\epsilon \over \sqrt{4\pi L}}{(-1)^{n+1}\over  \omega-\omega_n-i\epsilon^2/(4L)},\label{alpha1cav}\\
&\beta^1(\omega,\omega')=\delta(\omega-\omega')\nonumber \\ &\qquad \qquad +{\epsilon \over \sqrt{4\pi L}}\lim_{\eta \to 0^+} {1\over \omega'-\omega+i\eta}\sum_{n=0}^\infty (-1)^n\alpha^1_n(\omega),\\
&\alpha^2_n(\omega)=0,\\
&\beta^2(\omega,\omega')=0.\label{beta2}
\end{align}
If one substitutes these expressions into Eq.~(\ref{atotcav}) and computes the commutators $[a(\omega),a^\dagger (\omega')]$ and $[a(\omega),a(\omega')]$, one now obtains the correct results. Thus, Eq.~(\ref{atotcav}) together with the coefficients specified by Eqs.~(\ref{alpha1cav})-(\ref{beta2}), is valid up to first order in $\epsilon$, i.e., for a high-$Q$ cavity. By subsequently substituting Eq.~(\ref{atotcav}) into Eq.~(\ref{Hcavst}) one obtains, up to a constant vacuum energy, a Gardiner-Collett model of the form in Eq.~(\ref{HGC}) with coupling function \cite{dutra_cavity_2004,barnett_quantum_1988}
\begin{align}
g_n(\omega)=-{\epsilon\over \sqrt{4\pi L}} {\rm sinc}([\omega-\omega_n]L)e^{-i\omega L}.
\end{align}

In summary, we have reviewed an exact description of a leaky one-dimensional cavity based on applying Maxwell boundary conditions and using a scattering method \cite{dutra_cavity_2004}. The resulting energy and local fields are given in terms of global mode operators. For a high-$Q$ cavity the description can be reduced to a phenomenological Gardiner-Collett model, characterised by an interaction Hamiltonian that is bilinear in internal and external mode operators. It is also noteworthy that the interaction possesses a number conserving form without use of a rotating-wave approximation. This is because to first order in $\epsilon$, we have $\alpha^2_n(\omega)=0=\beta^2(\omega,\omega')$, that is, counter-rotating terms of the form $a_nb(\omega)$ and $a_n^\dagger b(\omega)^\dagger$ must possess coupling strengths of at least $O(\epsilon^2)$.

We conclude by remarking that linear-coupling models can also be arrived at through other approaches, for example, Ref.~\cite{viviescas_field_2003} uses the theory of QED in media (cf. Sec.~\ref{absm}) characterised by a non-constant permitivitty $\epsilon({\bf x})$ to provide a Hamiltonian written in terms of global mode operators as
\begin{align}
H =  \sum_{m=1}^M \int d\omega \,\omega \left[a_m^\dagger(\omega)a_m(\omega)+{1\over 2}\right],
\end{align}
Mode expansions for the canonical fields ${\bf A}_{\rm T}({\bf x})$ and ${\bf \Pi}({\bf x})=\epsilon({\bf x}){\dot {\bf A}}_{\rm T}({\bf x})$ are given in terms of the same global operators as
\begin{align}
&{\bf A}_{\rm T}({\bf x}) = \int d\omega\, {\bf f}(\omega,{\bf x})q(\omega),\\
&{\bf \Pi}({\bf x}) = \int d\omega\, {\bf f}^\dagger(\omega,{\bf x})p(\omega)
\end{align}
where $q(\omega)$ and $p(\omega)$ are respectively $M$-component row and column-vector quadratures defined in terms of $a_m(\omega)$ and $a^\dagger_m(\omega)$, such that $[q_m(\omega),p_{m'}(\omega')]=i\delta_{mm'}\delta(\omega-\omega')$. The mode-function ${\bf f}(\omega,{\bf x})$ is an $M$-component vector with components that are three-vector functions satisfying
\begin{align}
\nabla \times \nabla \times {\bf f}_m(\omega,{\bf x}) -\epsilon({\bf x})\omega^2 {\bf f}_m(\omega,{\bf x}) = {\bf 0},
\end{align}
[cf. Eq.~(\ref{grndyad})]. Using a Feshbach projection method the fields are divided into components that depend on cavity mode operators $a_n$ and external mode operators $b_m(\omega)$ respectively, with $[a_n,a_{n'}^\dagger] = \delta_{nn'}$ and $[b_m(\omega),b^\dagger_{m'}(\omega')] = \delta_{mm'}\delta(\omega-\omega')$. The internal and external modes define separate cavity and environment subsystems and the Hamiltonian expressed in terms of the corresponding mode operators is found to be of the linear-coupling form \cite{viviescas_field_2003}
\begin{align}
H =& \sum_n \omega_n a^\dagger_n a_n + \sum_m  \int d\omega\, \omega b^\dagger_m(\omega)b_m(\omega) \nonumber \\ &+\bigg[ \sum_m \sum_n \int d\omega\, g_{nm}(\omega)a_n b^\dagger_m(\omega) \nonumber \\ &\qquad ~~+ h_{nm}(\omega)a_n b_m(\omega) +{\rm H.c.}\bigg].
\end{align}
This Hamiltonian is of the Gardiner-Collett form but with the addition of counter-rotating terms [$\sim a_nb_m(\omega)$ and $\sim a^\dagger_n b^\dagger_m(\omega)$] between the cavity and environment. In Ref.~\cite{viviescas_field_2003} it is assumed that damping rates are much smaller than the frequencies of interest. The counter-rotating terms are then neglected in order to study input-output relations.

%\bibliography{review.bib}

\end{document}